\documentclass[
	reprint,
	superscriptaddress,
	showkeys,
	aps,
	pra,
	longbibliography,
	floatfix,
	nofootinbib
]{revtex4-2}

\usepackage[utf8]{inputenc}

\usepackage[hyperref]{xcolor}
	\definecolor{goethe-blau}{cmyk}{1.0,0.2,0.0,0.4}
	\definecolor{hellgrau}{cmyk}{0.04,0.04,0.05,0.02}
	\definecolor{sandgrau}{cmyk}{0.12,0.09,0.13,0.0}
	\definecolor{dunkelgrau}{cmyk}{0.25,0.25,0.30,0.75}
	\definecolor{emo-rot}{cmyk}{0.04,1.0,0.8,0.07}
	\definecolor{purple}{cmyk}{0.08,1.0,0.3,0.36}
	\definecolor{senfgelb}{cmyk}{0.01,0.25,1.0,0.05}
	\definecolor{gruen}{cmyk}{0.62,0.4,0.87,0.09}
	\definecolor{magenta}{cmyk}{0.08,0.86,0.12,0.12}
	\definecolor{orange}{cmyk}{0.0,0.7,1.0,0.04}
	\definecolor{sonnengelb}{cmyk}{0.0,0.12,0.95,0.0}
	\definecolor{helles-gruen}{cmyk}{0.4,0.17,0.81,0.07}
	\definecolor{lichtblau}{cmyk}{0.8,0.0,0.06,0.04}

\usepackage[%
colorlinks,%
pdfpagelabels,%
breaklinks,%
pdfstartview=FitH,%
bookmarksopen=true,%
bookmarksnumbered=true,%
bookmarksopenlevel=2,%
plainpages=false,%
hypertexnames=false,%
citecolor=emo-rot,%
linkcolor=goethe-blau,%
urlcolor=purple,%
pdftitle={{Numerical fluid dynamics for FRG flow equations: Zero-dimensional QFTs as numerical test cases. I. The O(N) model}},%
pdfauthor={{Adrian Koenigstein, Martin J. Steil, Nicolas Wink, Eduardo Grossi, Jens Braun, Michael Buballa and Dirk H. Rischke}},%
pdfkeywords={Functional Renormalization Group, conservation laws, numerical fluid dynamics, O(N) model, zero-dimensional QFT}%
]{hyperref}

\usepackage{amsmath} 
\usepackage{amssymb} 
\usepackage{bm} 

\usepackage{bookmark} 
\usepackage{comment} 
\usepackage{graphicx} 

\allowdisplaybreaks 

\begin{document}


\title{
	Numerical fluid dynamics for FRG flow equations:\texorpdfstring{\\}{ }Zero-dimensional QFTs as numerical test cases. \texorpdfstring{\\}{ }I. The \texorpdfstring{$O(N)$}{O(N)} model
}

\author{Adrian Koenigstein}
	\email{koenigstein@th.physik.uni-frankfurt.de}
	\affiliation{
		Institut f\"ur Theoretische Physik, Goethe University,\\
		Max-von-Laue-Stra{\ss}e 1, D-60438 Frankfurt am Main, Germany
	}

\author{Martin J.\ Steil}
	\email{msteil@theorie.ikp.physik.tu-darmstadt.de}
	\affiliation{
		Technische Universit\"at Darmstadt, Department of Physics, Institut f\"ur Kernphysik, Theoriezentrum,\\
		Schlossgartenstra{\ss}e 2, D-64289 Darmstadt, Germany
	}

\author{Nicolas Wink}
	\email{wink@thphys.uni-heidelberg.de}
	\affiliation{
		Institut f\"ur Theoretische Physik, University Heidelberg,\\
		Philosophenweg 16, D-69120 Heidelberg, Germany
	}

\author{Eduardo Grossi}
\email{eduardo.grossi@stonybrook.edu}
\affiliation{
	Center for Nuclear Theory, Department of Physics and Astronomy,\\
	Stony Brook University, Stony Brook, NY 11794, U.S.A.
}

\author{Jens Braun}
	\email{jens.braun@physik.tu-darmstadt.de}
	\affiliation{
		Technische Universit\"at Darmstadt, Department of Physics, Institut f\"ur Kernphysik, Theoriezentrum,\\
		Schlossgartenstra{\ss}e 2, D-64289 Darmstadt, Germany
	}
	\affiliation{
		Helmholtz Research Academy Hesse for FAIR, Campus Darmstadt,\\
		D-64289 Darmstadt, Germany
	}
	\affiliation{
		ExtreMe Matter Institute EMMI, GSI,\\ 
		Planckstraße 1, D-64291 Darmstadt, Germany
	}

\author{Michael Buballa}
	\email{michael.buballa@physik.tu-darmstadt.de}
	\affiliation{
		Technische Universit\"at Darmstadt, Department of Physics, Institut f\"ur Kernphysik, Theoriezentrum,\\
		Schlossgartenstra{\ss}e 2, D-64289 Darmstadt, Germany
	}
	\affiliation{
		Helmholtz Research Academy Hesse for FAIR, Campus Darmstadt,\\
		D-64289 Darmstadt, Germany
	}

\author{Dirk H.\ Rischke}
	\email{drischke@th.physik.uni-frankfurt.de}
	\affiliation{
		Institut f\"ur Theoretische Physik, Goethe University,\\
		Max-von-Laue-Stra{\ss}e 1, D-60438 Frankfurt am Main, Germany
	}
	\affiliation{
		Helmholtz Research Academy Hesse for FAIR, Campus Riedberg,\\
		Max-von-Laue-Stra{\ss}e 12, D-60438 Frankfurt am Main, Germany
	}

\date{\today}

\begin{abstract}
	The functional renormalization group (FRG) approach is a powerful tool for studies of a large variety of systems, ranging from statistical physics over the theory of the strong interaction to gravity.
	The practical application of this approach relies on the derivation of so-called flow equations, which describe the change of the quantum effective action under the variation of a coarse-graining parameter.
	In the present work, we discuss in detail a novel approach to solve such flow equations.
	This approach relies on the fact that RG equations can be rewritten such that they exhibit similarities with the conservation laws of fluid dynamics.
	This observation can be exploited in different ways.
	First of all, we show that this allows to employ powerful numerical techniques developed in the context of fluid dynamics to solve RG equations.
	In particular, it allows to reliably treat the emergence of non-analytic behavior in the RG flow of the effective action as it is expected to occur in studies of, \textit{e.g.}, spontaneous	symmetry breaking.
	Second, the analogy between RG equations and fluid dynamics offers the opportunity to gain novel insights into RG flows and their interpretation in general, including the irreversibility of RG flows.
	We work out this connection in practice by applying it to zero-dimensional quantum-field theoretical models.
	The generalization to higher-dimensional models is also discussed.
	Our findings are expected to help improving future FRG studies of quantum field theories in higher dimensions both on a qualitative and quantitative level.	
\end{abstract}

\keywords{Functional Renormalization Group, conservation laws, numerical fluid dynamics, $O(N)$ model, zero-dimensional QFT} 
\maketitle

\tableofcontents

\section{Introduction}
\label{sec:introduction}

	In statistical mechanics and quantum field theory (QFT) the central objective is to compute the expectation values of physical observables from a partition of probabilities among the various microscopic states of a given model or theory.
	On a technical level, the calculation of expectation values oftentimes corresponds to the evaluation of nested sums (for discrete systems) or complicated high-dimensional integrals (for continuous systems) in the framework of partition functions or functional integrals.
	In most cases such computations cannot be done analytically.
	Various methods were developed to overcome this difficulty.
	Focusing on high-energy physics, stochastic methods have been developed to study Quantum Chromodynamics from first principles (see Refs.~\cite{Philipsen:2012nu,Ding:2015ona,Guenther:2017grd,Attanasio:2020spv} for reviews), but also systematic approximation schemes such as (chiral) perturbation theory (see Refs.~\cite{Scherer:2002tk,Epelbaum:2008ga} for reviews) or the large-$N$ expansion \cite{tHooft:1973alw,Witten:1979kh,Moshe:2003xn} have been employed, where (at least) parts of the calculations can still be performed analytically.
	Within the last decades non-perturbative holographic and functional methods, such as the AdS/CFT correspondence \cite{Maldacena:1997re,Witten:1998qj}, Dyson-Schwinger equations (see Ref.~\cite{Fischer:2018sdj} for a review), and the (Functional) Renormalization Group ((F)RG) (see Ref.~\cite{Dupuis:2020fhh} for a recent review) have significantly gained importance and nowadays provide a viable complement to Monte-Carlo simulations and semi-analytic methods.
	However, despite great success within various areas of physics, holographic and functional methods are sometimes still criticized for the lack of providing reliable systematic and numerical error estimates.
	In this work, we will provide important steps to amend this shortcoming for the FRG approach.
	
	Although the mathematical formulation of the FRG approach is in principle exact, a first source of systematic errors is introduced by the fact that one has to make certain approximations (truncations) in order to actually perform calculations.
	However, since the method is non-perturbative, the identification of, \textit{e.g.}, a small expansion parameter is challenging, if at all possible.
	A lot of work has already been invested into this question, \textit{e.g.}, approximation errors can be evaluated by comparing different truncation schemes and truncation orders against each other~\cite{Cichutek:2020bli, Papp:1999he,Pawlowski:2014zaa}.
	Furthermore, the comparison with other non-perturbative methods \cite{Fischer:2006vf}, effective field theories \cite{Eser:2018jqo,Divotgey:2019xea,Eser:2019pvd,Cichutek:2020bli,Eser:2021ivo}, or with Monte-Carlo studies \cite{Bonanno:2004pq,Marko:2019rsl,Cyrol:2017ewj,Cyrol:2016tym} can provide estimates on the reliability of the results.
	
	A second source of systematic errors arises from the way the RG flow equations are solved in practice.
	In recent work by two of us and collaborators~\cite{Grossi:2019urj,Grossi:2021ksl}, it was pointed out that the possible appearance of non-analytic behavior in field space as well as the influence of the boundary conditions require great care in the numerical solution of RG flow equations.
	In particular, it was shown that these equations can be cast into a conservative form, such that analogies to fluid-dynamical flow equations become manifest and allow to access to the highly developed toolbox of numerical fluid dynamics, \textit{e.g.}, in the case of Refs.~\cite{Grossi:2019urj,Grossi:2021ksl} including the discontinuous Galerkin method.
	In consequence, this suggests that a systematic analysis of the quality of the different numerical methods to solve RG flow equations as well as an analysis of the structure of the RG flow equations themselves is in order.
	The question of numerical errors in FRG calculations was systematically addressed in Ref.~\cite{Grossi:2019urj} by a comparison of numerical results with analytically known solutions for the $O(N)$ model in the large-$N$ limit \cite{Tetradis:1995br,Litim:1995ex,DAttanasio:1997yph}.
	Furthermore, phenomena like shock waves in the derivative of the effective potential along the field space direction during the RG flow, which are directly related to phase transitions \cite{Grossi:2019urj,Grossi:2021ksl,Steil:2021cbu,Stoll:2021ori}, were resolved and interpreted in a fluid-dynamical framework.
	
	The goal of the present work is threefold.
	On the one hand, we will continue to elaborate on the analogies between RG flow equations and (numeric) fluid dynamics, including precision and stability tests for numerical schemes.
	On the other hand, we will contribute to the ongoing discussion on truncation schemes of the FRG framework.
	In addition, this article is supposed to provide a low-level introduction to the FRG method within the fluid-dynamic mindset also for non-experts and (under)graduate students.
	
	In order to provide reliable estimates of the precision of numerical methods and the quality of truncation schemes, the standard approach is to compare numerical results and/or results from truncations against analytically known results.
	However, analytically known results for non-trivial QFTs or statistical mechanics are scarce.
	Fortunately, there is a class of non-trivial QFTs, where either analytic results are known or numerical results can be easily obtained with arbitrary precision: zero-dimensional QFTs.
	In this work, we choose the zero-dimensional $O(N)$ model as a testing ground to systematically analyze the precision of the numerical methods which are used to solve the RG flow equations.
	Furthermore, we will use zero-dimensional QFT to demonstrate the similarities between RG flow equations and conservation laws from fluid dynamics (which also generalize to an arbitrary number of space-time dimensions and different field content).
	We will elucidate the different roles played by advective and diffusive contributions in the RG flow equations as partial differential equations (PDEs).
	Furthermore, we start a discussion of the relation between the RG time, entropy production in the RG flow, the dissipative character of the FRG equation, and the irreversibility of RG transformations during the RG flow. This discussion is deepened in part II and III of this series of publications \cite{Koenigstein:2021rxj,Steil:2021cbu}.
	
	In order to numerically solve the RG flow equations, in this work we apply the Kurganov-Tadmor scheme, a finite-volume method which is well-established in numerical fluid dynamics.
	We test the accuracy of the FRG results against direct evaluations of expectation values from the partition function, which can be calculated to in principle arbitrary precision in zero space-time dimensions.
	We note that the RG flow equations arising in the FRG framework for certain zero-dimensional models, and in particular the $O(N)$ model, are exact PDEs.
	Therefore, they do not involve any systematic error of the first kind mentioned above, namely truncation errors.
	Possible errors are therefore solely of the second kind, introduced by the numerical scheme used to solve the flow equations.
	
	As a next step, we will analyze the FRG Taylor expansion as a truncation to the FRG approach and contrast our findings with the general properties of the FRG equation as a non-linear PDE in zero space-time dimensions.
	In a follow-up publication, we will also introduce more elaborate zero-dimensional models including Grassmann numbers (mimicking fermionic degrees of freedom in $d=0$) \cite{Steil:partIV}.
	In this context, we will apply the methods developed in the present work to investigate several truncation schemes by comparing against exact results for a constructed fermion-boson-model.
	Generalizing our findings from zero dimensions to higher-dimensional QFTs is not necessarily trivial.
	Nevertheless, we will comment on this issue at various places throughout this work.
	We thus hope that this paper will contribute to ongoing debates on subtleties of the RG flow equations.
	Furthermore, we hope to establish reliable minimal requirements for numerical methods to solve RG flow equations, which can be used as benchmark tests for future numerical toolboxes.\\
	
	The length of this paper is explained by the fact that we have tried to make the presentation self-contained as much as possible.
	This should enable the reader not familiar with the FRG approach to understand all arguments and intermediate steps without resorting to the literature.
	The more experienced reader can certainly skip or skim over some parts, as indicated below.
	
	The remainder of this paper is organized as follows.
	In Sec.~\ref{sec:frg_introduction} we give an introduction to the FRG approach for zero-dimensional QFTs.
	In Sec.~\ref{sec:zero-dimensional_o(n)-model} we focus on the zero-dimensional $O(N)$ model and its respective RG flow equation.
	Readers familiar with the FRG approach and the $O(N)$ model can omit these two sections.
	The relationship between RG flow equations and fluid dynamics is discussed in Sec.~\ref{sec:frg-flow_equations_and_numerical_fluid_dynamics}.
	Readers familiar with fluid dynamics may be interested in the analogy between the FRG and fluid dynamics discussed in Sub.Sec.~\ref{subsec:conservative_form_of_frg_flow_equations}, but can skip over the remainder of this section that focuses on details of the numeric implementation.
	Section \ref{sec:results} presents our numerical results.
	Readers familiar with both the FRG approach and fluid dynamics should focus on this section and the Sub.Sec.~\ref{subsec:conservative_form_of_frg_flow_equations}.
	We conclude this work with a discussion and an outlook for future studies in Sec.~\ref{sec:conclusions}.
	In the Appendices, we list useful formulas for the calculation of numerical derivatives and present a discussion of the absence of spontaneous symmetry breaking in zero space-time dimensions.	

\section{The Functional Renormalization Group -- an introduction in zero dimensions}
\label{sec:frg_introduction}

	This section provides an introduction to the Functional Renormalization Group and a detailed derivation of the FRG equation~\cite{Wetterich:1992yh,Ellwanger:1993mw,Morris:1993qb} for a zero-dimensional QFT.
	Our discussion is geared towards non-experts.
	Readers who are familiar with the FRG method might still find this discussion instructive, because we will introduce the FRG without any direct reference to regularization and renormalization, only based on properties of (functional) integrals.
	This sheds light on the details and structure of the flow equations and the technical subtleties in their solution.
	In addition we use this introduction to establish some notation and special features of zero-dimensional field theory.\\
	
	As already mentioned in the introduction, the efficient and sufficiently precise calculation of correlation functions is key to understanding the properties of a particular model or theory.
	Usually this is done by introducing a partition function or functional integral that provides a probability distribution for the microstates of the model and serves as a generating functional for the $n$-point-correlation functions \cite{Weinberg:1996kr,Peskin:1995ev,ZinnJustin:2002ru,Kleinert:2004ev}.
	The partition function is based on an energy function that can be a discrete or continuous Hamilton function or an action, which determines the microscopic properties of the model.
	Another way of calculating the $n$-point correlation functions is to calculate the effective infrared action of the model, for example via the FRG equation.
	Both methods are discussed in this section.

\subsection{The partition function in zero dimensions}
\label{subsec:partition_function}

	Consider a zero-dimensional QFT with a single real bosonic scalar field or degree of freedom $\phi$.
	While all definitions generalize to arbitrary QFTs in zero or higher dimensions and arbitrary space-time backgrounds, in zero dimensions the field $\phi$ does not depend on the space-time position.
	The same applies to derivatives of the field or space-time integrals, which simply do not exist.
	This implies that the action $\mathcal{S}[\phi]$ of the model is identical to the Lagrangian $\mathcal{L}[\phi]$.
	The action, the Lagrangian, and also the Hamiltonian $\mathcal{H}[\phi]$ are simply functions of $\phi$ instead of functionals%
	\footnote{%
		Nevertheless, we will stick to the notation of functionals using square brackets, in order to facilitate the generalization to a nonzero number of space-time dimensions, as long as we do not focus on particular zero-dimensional examples.
	}%
	.
	Furthermore, because of the absence of a space-time derivative and thus of kinetic terms, $\mathcal{S}[\phi] = \mathcal{L}[\phi] = \mathcal{H}[\phi] = U(\phi)$, where $U(\phi)$ is the potential.
	Therefore, the only requirement for these functions is that they must be bounded from below, in order to exclude ``negative-energy states''%
	\footnote{
		We put ``negative-energy states'' in quotation marks, because all quantities in zero-dimensional field theory are dimensionless, hence bare numbers without physical dimensions.
		For convenience, we will still use the well-established notions from higher-dimensional QFT in our discussion.
	}
	and to obtain positive normalizable probability distributions.
	Apart from this requirement, for the moment we do not demand any additional properties, like symmetries (\textit{e.g.}, $\mathbb{Z}_2$, $\phi \rightarrow - \phi$) or analyticity.
	
	If we choose a specific model with action $\mathcal{S}[\phi]$ all expectation values of arbitrary functions $f(\phi)$ that do not grow exponentially in $\phi$ are defined and can be calculated via the following expression
		\begin{align}
			\langle f(\phi) \rangle \equiv \frac{\int_{-\infty}^{\infty} \mathrm{d} \phi \, f(\phi) \, \mathrm{e}^{ - \mathcal{S}[\phi] }}{\int_{-\infty}^{\infty} \mathrm{d} \phi \, \mathrm{e}^{ - \mathcal{S}[\phi] }} \, ,	\label{eq:expectation_value_1}
		\end{align}
	where $\mathrm{e}^{ - \mathcal{S}[\phi] }$ provides the partition of probabilities among the microstates.
	Note that due to the zero-dimensional nature all expectation values for such a model reduce to proper one-dimensional integrals over $\phi$.
	Such integrals can be computed to extremely high precision using standard techniques of numerical integration~\cite{Press:1992zz,PresTeukVettFlan92}.
	It is worth emphasizing that the current discussion holds also for non-analytic $\mathcal{S}[\phi]$ and/or $f(\phi)$.
	Some specific choices of $\mathcal{S}[\phi]$ and $f(\phi)$ even allow for an analytic evaluation of Eq.~\eqref{eq:expectation_value_1}, see, \textit{e.g.}, Ref.~\cite{Keitel:2011pn}.
	The possibility to compute expectation values to high precision makes zero-dimensional field theory of great interest as a testing ground for approximations and/or numerical methods.
	
	Some explicit examples of zero-dimensional field theories used as a testing ground for methods in statistical mechanics and QFT can be found in Refs.~\cite{Bessis:1980ss,Zinn-Justin:1998hwu,DiVecchia:1990ce,Hikami:1978ya,Nishigaki:1990sk,Schelstraete:1994sc,Keitel:2011pn,Pawlowski:talk,Moroz:2011thesis,Fl_rchinger_2010,SkinnerScript,Strocchi:2013awa,Kemler:2013yka,Rentrop_2015,Rosa:2016czs,Liang:2017whg,Millington:2019nkw,Alexander:2019cgw,Catalano:2019,Millington:2020Talk,Millington:2021ftp,Kades:2021hir,Fraboulet:2021amf}.
	In Ref.~\cite{Strocchi:2013awa}, for example, the asymptotic convergence and the vanishing convergence radius of perturbation theory of $\phi^4$-theory is discussed.
	Approximation schemes such as the large-$N$, the FRG vertex expansion, or the FRG Taylor expansion were analyzed in Ref.~\cite{Keitel:2011pn}.
	Zero-dimensional field theory was also used to study density-functional theory~\cite{Kemler:2013yka,Rentrop_2015,Liang:2017whg} and applied to fermionic fields~\cite{SkinnerScript}.
	Recently, it was used to study and visualize 2PI effective actions \cite{Millington:2019nkw} -- also in the FRG framework \cite{Alexander:2019cgw,Millington:2020Talk,Millington:2021ftp}.
	
	Usually the calculation of expectation values is facilitated by a suitably defined generating functional
		\begin{align}
			\mathcal{Z} [J] \equiv \mathcal{N} \int_{-\infty}^{\infty} \mathrm{d} \phi \, \mathrm{e}^{- \mathcal{S}[\phi] + J \, \phi} \, , \label{eq:partition_function}
		\end{align}
	from which one can derive all correlation functions by taking the corresponding number of derivatives with respect to the external source $J$,
		\begin{align}
			\langle f(\phi) \rangle = \frac{f(\tfrac{\delta}{\delta J}) \, \mathcal{Z}[J]}{\mathcal{Z}[J]} \bigg|_{J = 0} \, .	\label{eq:expectation_value_2}
		\end{align}
	One should note that if $f(\phi)$ is non-analytic, then Eq.~\eqref{eq:expectation_value_2} is to be understood symbolically.
	Otherwise, it is defined through a Taylor series in $\frac{\delta}{\delta J}$.
	Irrespective of that, Eqs.~\eqref{eq:expectation_value_1} and \eqref{eq:partition_function} are always well defined and Eq.~\eqref{eq:partition_function} can be always calculated for arbitrary $J$.
	One can even show in zero dimensions that $\mathcal{Z}[J] \in C^\infty$, hence, $\mathcal{Z}[J]$ is a smooth function, see Ref.~\cite{Moroz:2011thesis} and App.~\ref{app:mermin_wagner}.
	We shall come back to this crucial point later on in our discussion of the Coleman-Mermin-Wagner-Hohenberg theorem \cite{Mermin:1966,Hohenberg:1967,Coleman:1973ci}.
	
	The normalization $\mathcal{N}$ is not an observable quantity.
	For our purposes, it is convenient to choose
		\begin{align}
			&	\mathcal{Z} [ 0 ] \overset{!}{=} 1 \, ,	&&	\longleftrightarrow	&&	\mathcal{N}^{-1} = \int_{-\infty}^{\infty} \mathrm{d} \phi \, \mathrm{e}^{ - \mathcal{S} [\phi] } \, .	\label{eq:normalization}
		\end{align}
	
	As already mentioned above, calculating expectation values in a zero-dimensional QFT via Eq.~\eqref{eq:expectation_value_1} is (numerically) rather straightforward.
	In contrast, for higher-dimensional models or theories with non-trivial field-content \textit{etc.}\ calculating functional integrals similar to Eq.~\eqref{eq:expectation_value_1} with sufficient precision is usually extremely challenging or might even be impossible with limited computational resources.
	Therefore, alternative methods or approximation schemes apart from ``direct numerical integration'', like in lattice simulations, are of great interest.
	One of these alternatives, which is at the heart of this work, is the FRG.
	
	In the following, we will therefore focus on the FRG as a specific method for calculating $n$-point correlation functions in QFT and statistical mechanics.
	In contrast to the usual motivation of the FRG, arising in the discussion of renormalization and the integration of momentum shells from ultraviolet to infrared energy scales, we will take a different path to arrive at the FRG equation, which does not require any knowledge of renormalization.
	To this end, we will follow and extend the discussion in Refs.~\cite{Keitel:2011pn,Pawlowski:talk,Moroz:2011thesis,Fl_rchinger_2010,SkinnerScript} and discuss its technical properties as an alternative way of solving the integrals in Eqs.~\eqref{eq:expectation_value_1} and \eqref{eq:partition_function}.

\subsection{Solving integrals  with flow equations}
\label{subsec:integrals_with_equations}

	The starting point is the observation that there is one well-known non-trivial class of actions $\mathcal{S}[\phi]$ for which the calculation of integrals like Eq.~\eqref{eq:expectation_value_1} is straightforward, even in higher dimensions and even for more complicated field content.
	These actions are QFTs for ``(massive) free particles'' and correspond to Gaussian-type integrals.
	In the present case the Gaussian-type action takes the following simple form,
		\begin{align}
			\mathcal{S} [ \phi ] = \tfrac{m^2}{2} \, \phi^2 \, .	\label{eq:mass_term}
		\end{align}
	where $m$ is called a ``mass'' for convenience, although it is actually a dimensionless quantity in zero space-time dimensions.
	
	For non-trivial actions $\mathcal{S}[\phi]$, Eq.~\eqref{eq:expectation_value_1} can still be approximated by a Gaussian integral, as long as $\mathcal{S}[\phi]$ contains a mass term \eqref{eq:mass_term} with a coefficient $m^2$ that is much larger than all other scales contained in $\mathcal{S}[\phi]$.
	If this is the case, the Gaussian part of the integrand $\mathrm{e}^{-\mathcal{S}[\phi]}$ completely dominates the integrals in Eqs.~\eqref{eq:expectation_value_1} and \eqref{eq:partition_function}.
	The reason is that the mass term $\sim\phi^2$ is dominant for small and moderate $\phi$, and most of the area under the curve $\mathrm{e}^{-\mathcal{S}[\phi]}$ lies in the region of small $\phi$, similar to a pure Gaussian integral.
	For very large values of $\phi$ other terms in the action $\mathcal{S}[\phi]$ may become more important.
	Nevertheless, if $m^2$ is large enough, the corresponding area under the curve $\mathrm{e}^{-\mathcal{S}[\phi]}$ is completely negligible in regions where $\phi$ is large, because $\mathcal{S}[\phi]$ is bounded from below such that $\mathrm{e}^{-\mathcal{S}[\phi]}$ tends to zero exponentially fast for $\phi\to\infty$.
	In summary, the Gaussian part with the huge mass term dominates the integral and even non-trivial $\mathcal{S}[\phi]$ can be approximated by Gaussian integrals.
	
	This observation generalizes to higher dimensions and arbitrary field content, but is more apparent in a zero-dimensional field theory with one degree of freedom.
	This is illustrated in Figs.\ \ref{fig:rg_flow_integrand_z_1} and \ref{fig:rg_flow_integrand_z_2}, which are discussed in the following subsubsection.

\subsubsection{The scale-dependent partition function}

	Based on the above observation, let us now introduce the following quantity:
		\begin{align}
			\mathcal{Z}_t [ J ] \equiv \mathcal{N} \, \int_{- \infty}^{\infty} \mathrm{d} \phi \, \mathrm{e}^{- \mathcal{S} [ \phi ] - \Delta \mathcal{S}_t [ \phi ] + J \, \phi } \, ,	\label{eq:scale_dependent_z}
		\end{align}
	which is called the \textit{scale-dependent generating functional} or \textit{scale-dependent partition function}.
	It differs from the usual partition function \eqref{eq:partition_function} only by a scale-dependent mass term
		\begin{align}
			\Delta \mathcal{S}_t [ \phi ] \equiv \tfrac{1}{2} \, r ( t ) \, \phi^2 \, .	\label{eq:regulator_insertion}
		\end{align}
	We directly adopt the common notation from the FRG community and call $r(t)$ the \textit{regulator (shape) function}, which depends on the \textit{RG scale} (``\textit{time}'') $t \in [ \, 0 , \infty )$, see, \textit{e.g.}, Refs.~\cite{Litim:2000ci,Pawlowski:2015mlf}.
	We will discuss this interpretation of $r(t)$ and $t$ in Sub.Sec.~\ref{subsec:interpretation_frg}.
	For now, we only demand that the function $r(t)$ has such properties that $\mathcal{Z}_t [J]$ interpolates between an almost Gaussian-type partition function%
	\footnote{%
		This is also why the UV fixed point of RG flows is denoted as the \textit{trivial} or \textit{Gaussian fixed point}.
	}
	with extremely massive free fields at $t = 0$ and the actual partition function $\mathcal{Z}[J]$ that we are interested in at $t \rightarrow \infty$.
	In order to achieve this behavior, $r ( t )$ has to have the following properties:
		\begin{enumerate}
			\item	In the limit of $t \rightarrow 0$, $r(t)$ ($\mathcal{S} [ \phi ]$) should behave like a mass (term), similar to what we discussed at the beginning of this section, and be much larger than all other scales in $\mathcal{S}[\phi]$. Oftentimes in the literature $r(t)$ is set to infinity at $t = 0$.
			We will see, \textit{cf.}\ Sub.Sec.~\ref{subsec:exact_rg_equation}, that this is not suitable.
			
			\item	For $t \rightarrow \infty$, $r(t)$ is supposed to vanish, such that $\lim_{t \rightarrow \infty} \mathcal{Z}_t [J]=\mathcal{Z}[J]$.
			The same applies to expectation values calculated from $\mathcal{Z}_t[J]$, which become expectation values of $\mathcal{Z}[J]$.
			For practical calculations it is sufficient to assume that, for $t \rightarrow \infty$, $r(t)$ becomes much smaller than all scales in $\mathcal{S}[\phi]$, because then the contribution $\Delta \mathcal{S}_t [\phi]$ to the whole integrand $\mathrm{e}^{- \mathcal{S}[\phi] - \Delta \mathcal{S}_t [\phi]}$ is negligible and the integrand is almost identical to $\mathrm{e}^{- \mathcal{S}[\phi]}$.
			The value $\lim_{t \rightarrow \infty} r ( t ) = r_\mathrm{IR} \gtrsim 0$ is usually referred to as \textit{(numerical) infrared (IR) cutoff}.
			
			\item	The interpretation of $r(t)$ ($\mathcal{S} [ \phi ]$) as a mass (term) is guaranteed by further demanding monotonicity, ${\partial_t r(t) \leq 0 \quad \forall t}$.
			We will provide additional arguments for monotonicity in Sub.Sub.Sec.~\ref{subsec:flow_equation_of_the_partition_function}.
					
			\item	In order to be able to smoothly deform the integral in Eq.~\eqref{eq:partition_function} and for the following derivation of evolution equations, we further require $r ( t ) \in C^1$.
		\end{enumerate}
	Apart from these four properties there are no further requirements on $r(t)$ in zero dimensions.%
	\footnote{%
		For the subtleties associated with the choice of regulators in higher-dimensional theories, we refer the interested reader to Refs.~\cite{Litim:2000ci,Pawlowski:2015mlf,Braun:2020bhy,Pawlowski:2005xe,Rosten:2010vm,Osborn:2011kw}.
		Note that for higher-dimensional field theories the fourth requirement turns into $\Delta \mathcal{S}_t [ \phi ] \in C^1$.
	}%
	A specific choice which is used in large parts of our work is the so-called \textit{exponential regulator (shape) function}
		\begin{align}
			r ( t ) = \Lambda \, \mathrm{e}^{- t} \, ,	\label{eq:exponential_regulator}
		\end{align}
	with an \textit{ultraviolet (UV) cutoff} $\Lambda$, which must be chosen much larger than all scales in $\mathcal{S} [ \phi ]$.
		\begin{figure}
			\centering
			\includegraphics{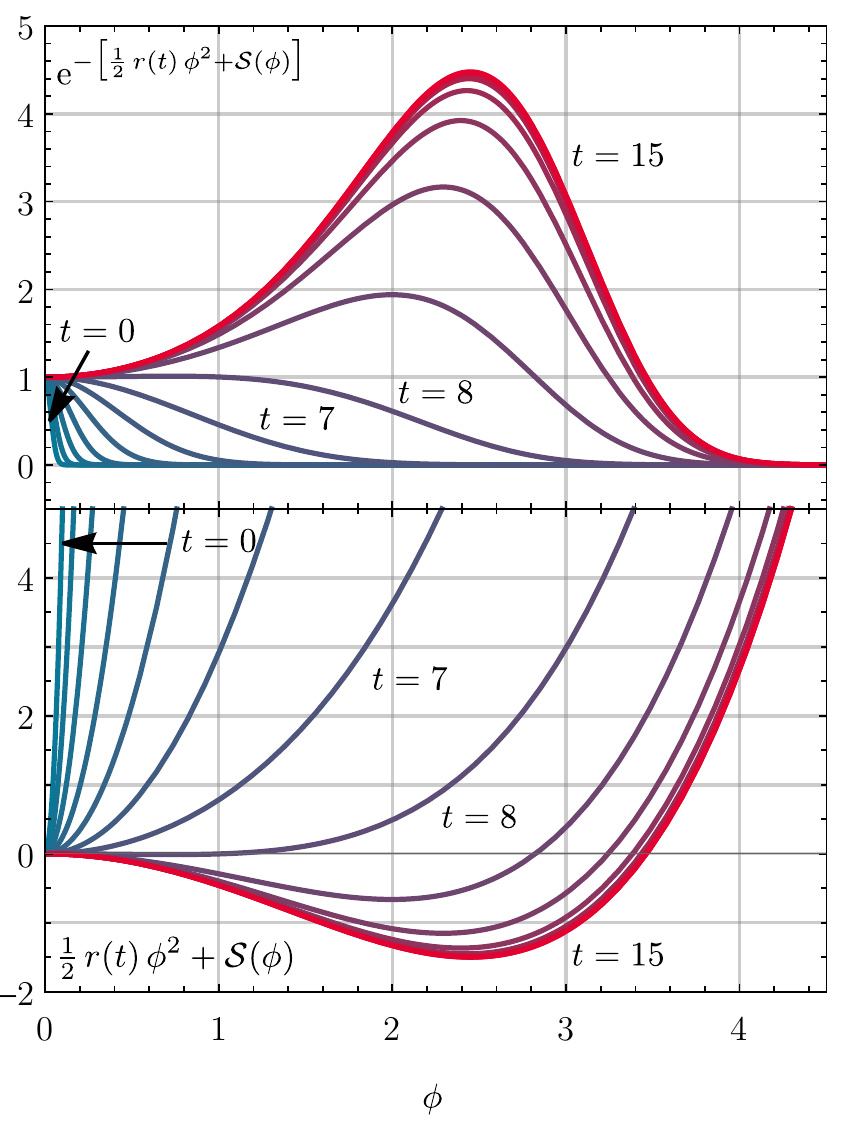}
			\caption{\label{fig:rg_flow_integrand_z_1}%
				The integrand (upper panel) and exponent (lower panel) from Eq.~\eqref{eq:scale_dependent_z} (at $J = 0$) as a function of the field variable $\phi$ for various RG times $t = 0, 1, 2, \ldots, 15$ and for the action \eqref{eq:example_analytic_action}.
				We choose the exponential regulator \eqref{eq:exponential_regulator} with UV cutoff $\Lambda = 10^3$, which is notably larger than the absolute value of the mass term and the quartic coupling.
				The IR cutoff scale $r_\mathrm{IR}$ was chosen at $t = 15$, which corresponds to $r_\mathrm{IR} \simeq 3.06 \cdot 10^{-4}$.
				This value is significantly smaller than all scales in $\mathcal{S}[\phi]$.
			}
		\end{figure}
	
\begin{figure}
			\centering
			\includegraphics{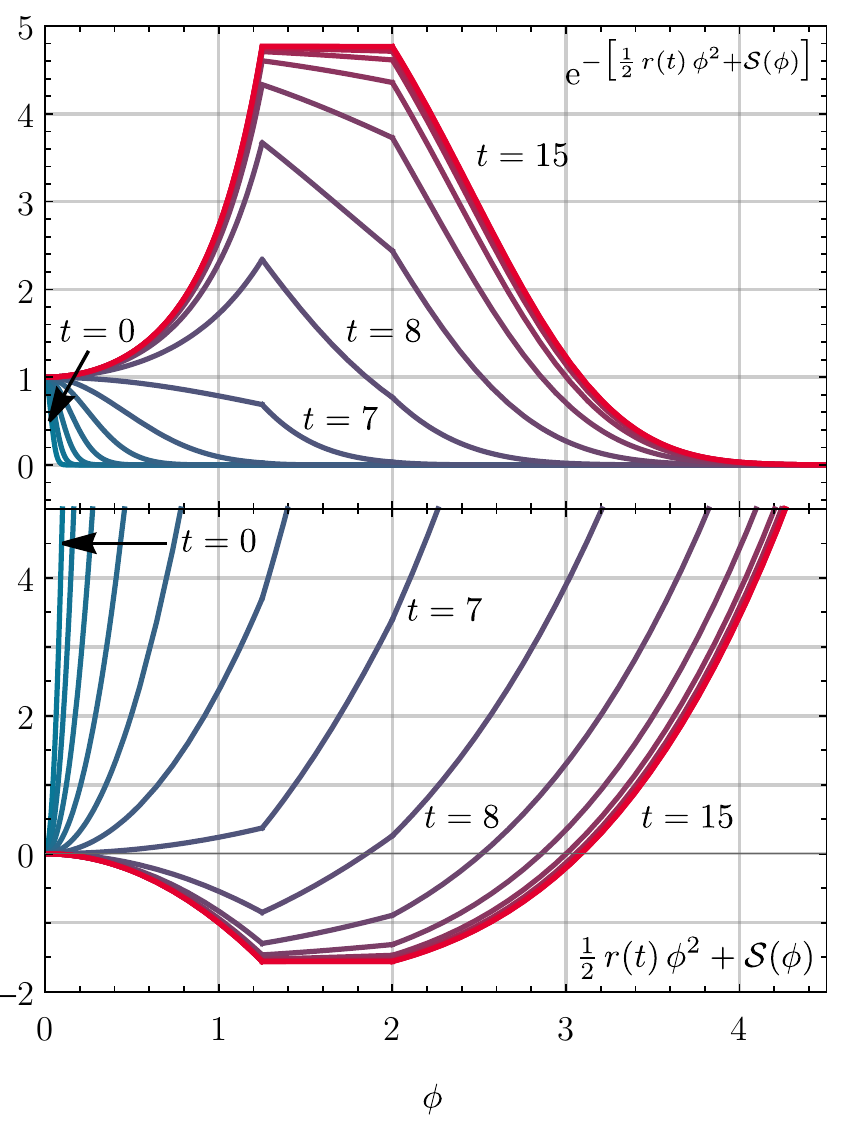}
			\caption{\label{fig:rg_flow_integrand_z_2}%
				The same as in Fig.\ \ref{fig:rg_flow_integrand_z_1}, but for the action \eqref{eq:example_non-analytic_action}.
			}
		\end{figure}

	In order to get a better intuition of the effect of $r(t)$ on the integral \eqref{eq:scale_dependent_z}, in Fig.~\ref{fig:rg_flow_integrand_z_1} we show the integrand at $J=0$, $\mathrm{e}^{- \mathcal{S}[\phi] - \Delta \mathcal{S}_t [\phi]}$, and the respective exponent for different values of $t$ for the analytic action
		\begin{align}
			\mathcal{S} ( \phi ) = - \tfrac{1}{2} \, \phi^2 + \tfrac{1}{4!} \, \phi^4 \label{eq:example_analytic_action}
		\end{align}
	and in Fig.~\ref{fig:rg_flow_integrand_z_2} the same quantities for the non-analytic action
		\begin{align}
			\mathcal{S} ( \phi ) =
			\begin{cases}
				- \phi^2 \, ,								&	\text{if} \quad | \phi | \leq \tfrac{5}{4} \, ,	\vphantom{\bigg(\bigg)}
				\\
				- \big( \tfrac{5}{4} \big)^2 \, ,			&	\text{if} \quad \tfrac{5}{4} < | \phi | \leq 2 \, ,	\vphantom{\bigg(\bigg)}
				\\
				\tfrac{1}{48} \, \big( \phi^4 - 91 \big)	&	\text{if} \quad 2 < | \phi | \, .	\vphantom{\bigg(\bigg)}
				\\
			\end{cases}	\label{eq:example_non-analytic_action}
		\end{align}
	The figures show how the integrands are deformed from Gaussian-shaped integrands to the integrands $\mathrm{e}^{- \mathcal{S}[\phi]}$.
	One observes that, as long as $r(t)$ is much larger than all other parameters in $\mathcal{S}[\phi]$, the Gaussian-like mass term dominates, while for increasing $t$ the regulator $r(t)$ becomes negligible.
	The most interesting part, where the integrands change their shapes significantly, is where $r(t)$ is of the same order as the scales in $\mathcal{S} ( \phi )$.

\subsubsection{A flow equation for the scale-dependent partition function}
\label{subsec:flow_equation_of_the_partition_function}

	The change of the integrals with $t$ between the two limiting cases at $t=0$ and $t\rightarrow\infty$ is called \textit{RG flow}.
	If this RG flow is known, we can obtain the function ${\mathcal{Z}(J) \equiv \lim_{t \rightarrow \infty} \mathcal{Z}_t [J] = \mathcal{Z}[J]}$ right from the Gaussian-like partition function $\mathcal{Z}_{t = 0}[J]$ without the need to calculate the $\phi$-integral in the partition function~\eqref{eq:partition_function} directly.
	For zero dimensions this does not seem to be an advantage, because the integrals in field space are (at least numerically) simple to compute.
	For higher dimensions, however, circumventing the challenging functional integration is a tremendous benefit.
	
	The RG flow of $\mathcal{Z}_t [J]$ is characterized by taking the derivative with respect to the RG time $t$,
		\begin{align}
			& \partial_t \mathcal{Z}_t[J] =	\vphantom{\bigg(\bigg)}	\label{eq:polchinski_equation}
			\\
			= \, & - \big[ \tfrac{1}{2} \, \partial_t r ( t ) \big] \, \mathcal{N} \int_{-\infty}^{\infty} \mathrm{d} \phi \, \phi^2 \, \mathrm{e}^{ - \mathcal{S}[\phi] - \Delta \mathcal{S}[\phi] + J \, \phi} =	\vphantom{\bigg(\bigg)}	\nonumber
			\\
			= \, & - \big[ \tfrac{1}{2} \, \partial_t r ( t ) \big] \, \frac{\delta^2 \mathcal{Z}_t [J]}{\delta J \, \delta J} \equiv	\vphantom{\bigg(\bigg)}	\nonumber
			\\
			\equiv \, & - \big[ \tfrac{1}{2} \, \partial_t r ( t ) \big] \, \mathcal{Z}^{(2)}_{t, J J} [J] \, ,	\vphantom{\bigg(\bigg)}	\nonumber
		\end{align}
	which is a PDE for a function $\mathcal{Z} ( t, J )$ in the $( t, J)$-plane,
		\begin{align}
			\partial_t \mathcal{Z}(t,J) = - \big[ \tfrac{1}{2} \, \partial_t r ( t ) \big] \, \partial_J^2 \mathcal{Z}(t,J) \, .	\label{eq:pde_z}
		\end{align}
	With slight modifications, this also applies to higher-dimensional QFTs.
	Solving this equation with appropriate initial and boundary conditions results in a function $\mathcal{Z} ( J )$ from which one can calculate expectation values by taking ordinary (numerical) derivatives with respect to $J$ at $J = 0$, \textit{cf.}\ Eq.~\eqref{eq:expectation_value_2}.
	
	The structure of this equation is that of a linear one-dimensional diffusion equation (\textit{heat equation}) \cite{Rosten:2010vm,SkinnerScript,Salmhofer:2020Talk,Cannon:1984}, where $t$ corresponds to the temporal direction, while $J$ corresponds to the spatial direction.
	The term $- \tfrac{1}{2} \, \partial_t r ( t )$ corresponds to a time-dependent (positive definite) diffusion coefficient%
	\footnote{%
		Note that in zero dimensions one can get rid of $\partial_t r ( t )$ by an appropriate reparametrization of the time coordinate $t$, which nevertheless keeps the structure of the equation unchanged.
		In higher dimensions this elimination of $r ( t )$ is in general not possible.
		The positivity of the diffusion coefficient is directly related to the stability of solutions of the heat equation \cite{LeVeque:1992,LeVeque:2002} and positivity -- here guaranteed by the regulator properties -- is necessary for a stable solution \cite{Rosten:2010vm,Osborn:2011kw}.
	}%
	.
	This also motivates the name RG ``time'' for the parameter $t$.
	We will come back to the concept of RG ``time'' in the true sense of the word and the diffusive, irreversible character of RG flows in Sub.Sec.~\ref{subsec:conservative_form_of_frg_flow_equations}.
	
	In zero dimensions, the Eq.~\eqref{eq:pde_z} is a PDE in two variables.
	For the remainder of this subsection we will discuss properties and practical issues considering this exact PDE.
	We will neither discuss any kind of expansions in $J$ nor its application in higher dimensions.
	However, some of the issues and questions raised in the following are also relevant for higher-dimensional theories.
	
	Finding the correct initial and boundary conditions for numerical solutions of Eq.~\eqref{eq:pde_z} as an exact PDE is challenging.
	By construction $\mathcal{Z}_{t = 0} [ J ]$ approaches a Gaussian integral,
		\begin{align}
			& \mathcal{Z}_{t = 0} [ J ] =	\vphantom{\bigg(\bigg)}	\label{eq:initial_condition_z}
			\\
				= \, & \mathcal{N} \, \int_{- \infty}^{\infty} \mathrm{d} \phi \, \mathrm{e}^{- \frac{1}{2} r ( 0 ) \, \phi^2 + J \, \phi} \, \mathrm{e}^{- \mathcal{S} ( \phi )}=	\vphantom{\bigg(\bigg)}	\nonumber
			\\
			= \, & \mathcal{N} \, \int_{- \infty}^{\infty} \tfrac{\mathrm{d}\tilde{\phi}}{\sqrt{r(0)}} \, \mathrm{e}^{- \frac{1}{2} \, \tilde{\phi}^2 + J \, \frac{\tilde{\phi}}{\sqrt{r(0)}}} \, \Big[ 1- \mathcal{O} \big( \mathcal{S} \big( r ( 0 )^{-\frac{1}{2}} \big) \big) \Big] \nonumber
			\\
			= \, & \mathcal{N} \, \sqrt{\tfrac{2\pi}{r(0)}} \, \mathrm{e}^{\frac{J^2}{2 r(0)}} \, \Big[ 1- \mathcal{O} \big( \mathcal{S} \big( r(0)^{-\frac{1}{2}} \big) \big) \Big] \, ,	\vphantom{\bigg(\bigg)}	\nonumber
		\end{align}
	with $\tilde{\phi}\equiv \sqrt{r(0)}\,\phi$ and independent of the explicit shape of $\mathcal{S}[\phi]$.
	Considering different actions $\mathcal{S}[\phi]$ with couplings of the same order of magnitude we can choose the same regulator with an $r(0)$ larger than all internal scales involved in the different actions.
	The initial condition $\mathcal{Z}(0,J)$ is then independent of the explicit action under consideration.
	
	According to the integral formulation~\eqref{eq:scale_dependent_z}, $\mathcal{Z}(t,J)$ changes for different actions when $t > 0$.
	In the differential formulation of the Eq.~\eqref{eq:pde_z} those changes are generated by the diffusion term on the right-hand side.
	However, we argued that it is permissible to use identical initial conditions $\mathcal{Z} ( 0, J)$ for different actions involving similar scales (as long as these are much smaller than $r ( 0 )$).
	This then results in an identical diffusion on the right-hand side of Eq.~\eqref{eq:pde_z} when the latter is computed by means of a second derivative of $\mathcal{Z}(0,J)$.
	If one uses identical large-$J$ boundary conditions for the solution of the PDE \eqref{eq:pde_z} for different actions, this would imply that, despite different $\mathcal{S}[\phi]$, the RG time evolution leads to identical $\mathcal{Z}(J)$ for $t\rightarrow\infty$, which in general cannot be correct.
	
	In order to resolve this problem, particular action-dependent spatial boundary conditions seem to be necessary for a direct numerical solution starting at $t = 0$ with a Gaussian for $\mathcal{Z} ( 0, J )$.
	It is not obvious how to derive or formulate such boundary conditions from the asymptotics of Eq.~\eqref{eq:pde_z} alone.
	In light of this, a numerical solution of Eq.~\eqref{eq:pde_z} in the $( t, J )$ plane by means of a spatial discretization in $J$ direction and an integration in $t$ direction appears to be conceptually questionable.
	
	However, this invalidates by no means the flow equation for $\mathcal{Z} ( t, J )$ in general.
	Augmenting it (at $t = 0$) with information from the integral formulation~\eqref{eq:scale_dependent_z} or, equivalently, other additional information, could enable practical computations using the PDE \eqref{eq:pde_z}.
	But it is at this point (at least to us) not obvious how one would implement a numerical solution strategy for the PDE \eqref{eq:pde_z} avoiding integrals of the action.\\
	
	There is another well-known drawback in using the partition function $\mathcal{Z} [ J ]$ for calculating $n$-point correlation functions (or expectation values) $\langle \phi^n \rangle$.
	The latter are rather inefficient in storing information, because they contain redundant information in the form of disconnected and reducible terms, see Refs.~\cite{Iliopoulos:1974ur,ZinnJustin:2002ru,Weinberg:1996kr,Peskin:1995ev} or the mathematical theory of moment- and cumulant-generating functionals in statistics for details \cite{McCullagh:2009}.
	This is further discussed in Sub.Sec.~\ref{subsec:n_pt_functions}.
	However, the redundant information in $\langle \phi^n \rangle$ is not necessarily a strong argument against the use of the flow equation \eqref{eq:pde_z} in practical computations, since the irreducible information can be extracted from the correlation functions $\langle \phi^n \rangle$.\\
	
	In order to resolve both the problem of initial and boundary conditions for $\mathcal{Z}(t,J)$ as well as the issue of redundant information in $\langle \phi^n \rangle$, we now consider two different generating functionals, which are better suited for practical calculations of $n$-point correlation functions or expectation values, respectively.
	To this end, we employ the \textit{Schwinger functional},
		\begin{align}
			&	\mathcal{W}[J] \equiv \ln \mathcal{Z}[J] \, ,	&&	\mathcal{W} [ 0 ] = 0 \, ,	\label{eq:schwinger_functional}
		\end{align}
	and its Legendre transform, the \textit{effective action}, 
		\begin{align}
			\Gamma [\varphi] \equiv \mathop{\mathrm{sup}}_J \big\{ J \, \varphi - \mathcal{W} [J] \big\} \, .	\label{eq:gamma_legendre}
		\end{align}
	Here, ``$\mathrm{sup}$'' denotes the supremum with respect to the source $J$.
	The Schwinger functional generates all connected $n$-point correlation functions while the effective action generates all one-particle irreducible (1PI) $n$-point vertex (correlation) functions, see Sub.Sec.~\ref{subsec:n_pt_functions} or Refs.~\cite{Iliopoulos:1974ur,ZinnJustin:2002ru,Weinberg:1996kr,Peskin:1995ev} for details.
	
	In general $\mathcal{W}[J]$ is convex with a positive definite Hessian $\mathcal{W}^{(2)}_{J J} [ J ]$, which implies convexity for $\Gamma [ \varphi ]$, since the Legendre transform of a convex function is convex by definition, see \textit{e.g.}, Refs.~\cite{Fujimoto:1982tc,Wipf:2013vp} for details.
	In the present case the convexity of $\mathcal{W} [ J ] = \mathcal{W} ( J )$ becomes apparent considering its second derivative,
		\begin{align}
			\partial_J^2 \mathcal{W}(J) = \langle \phi^2 \rangle_J -\langle \phi \rangle_J \langle \phi \rangle_J= \langle (\phi-\langle \phi \rangle_J)^2\rangle_J \, ,	\label{eq:jacobian_w}
		\end{align}
	which, as the expectation value of a positive quantity, is always positive.%
	\footnote{%
		Note that also $\mathcal{Z}[J]$ is convex, which can be seen by investigating its second derivative.
	}
	In zero dimensions, also smoothness, $\mathcal{Z} [ J ] \in C^\infty$, directly translates to $\mathcal{W} [ J ] \in C^\infty$ and $\Gamma [\varphi] \in C^\infty$, because all derivatives $\mathcal{W} [ J ]$ and $\Gamma [ \varphi ]$ can be entirely expressed in terms of derivatives of $\mathcal{Z} [J]$, see Sub.Sec.~\ref{subsec:n_pt_functions}.
	We will need both properties several times during our discussion, see also the discussion in App.~\ref{app:mermin_wagner}.
	
	Having these definitions at hand, we shall start the next section by defining scale-dependent generating functionals $\mathcal{W}_t[J]$ and $\Gamma_t [\varphi]$.
	From these, we will also derive and discuss two flow equations which are similar to Eq.~\eqref{eq:polchinski_equation}.
	The final result of the next subsection is the FRG equation (known as Wetterich equation), which is the exact analogue to Eq.~\eqref{eq:polchinski_equation} on the level of $\Gamma_t [\varphi]$.
	It provides the opportunity to circumvent the direct calculation of integrals of type~\eqref{eq:expectation_value_1}.

\subsection{The Functional Renormalization Group equation}
\label{subsec:exact_rg_equation}

	In this subsection we derive and discuss the Functional Renormalization Group (FRG) equation~\cite{Wetterich:1992yh, Ellwanger:1993mw, Morris:1993qb} (also known as Exact Renormalization Group equation) for our zero-dimensional toy-model QFT.
	All formulas presented in this section can be generalized to higher dimensions and arbitrary field content, see \textit{e.g.}, Refs.~\cite{Reuter:1993kw,Reuter:1996cp,Reuter:2001ag,Berges:2000ew, Pawlowski:2005xe, Kopietz:2010zz}.
	
\subsubsection{The scale-dependent Schwinger functional}
\label{subsec:scale_dependent_schwinder_functional}

	We begin the derivation by introducing the \textit{scale-dependent Schwinger functional} starting from definition~\eqref{eq:scale_dependent_z},
		\begin{align}
			\mathcal{W}_t [J] \equiv \ln \mathcal{Z}_t [J] \, .	\label{eq:scale_dependent_w}
		\end{align}
	It follows from our previous discussion that for $t \rightarrow \infty$ the Schwinger functional \eqref{eq:schwinger_functional} is recovered,
		\begin{align}
			\lim\limits_{t \rightarrow \infty} \mathcal{W}_t [J] = \mathcal{W} [J] \, ,
		\end{align}
	while $\mathcal{W}_{t = 0} [J]$ is given by the logarithm of Eq.~\eqref{eq:initial_condition_z}.
	
	The insertion of the regulator \eqref{eq:regulator_insertion} into $\mathcal{Z}_t [J]$ does not spoil the convexity and smoothness (in zero dimensions) of the Schwinger functional: $\mathcal{W}_t [J]$ and $\mathcal{Z}_t [J]$ are convex and smooth for all $t$.
	
	Completely analogous to Eq.~\eqref{eq:polchinski_equation} one can derive a PDE for $\mathcal{W}_t[J] = \mathcal{W}(t,J)$ in the $(t,J)$ plane,
		\begin{align}
			& \partial_t \mathcal{W}(t,J) =	\vphantom{\bigg(\bigg)}	\label{eq:pde_w}
			\\
			= \, & - \big[ \tfrac{1}{2} \, \partial_t r ( t ) \big] \, \Big( \partial_J^2 \mathcal{W}(t,J) + \big[ \partial_J \mathcal{W}(t,J) \big]^2 \Big) \, .	\vphantom{\bigg(\bigg)}	\nonumber
		\end{align}
	which describes the flow of $\mathcal{W}(t,J)$ from $t = 0$ to $t \rightarrow \infty$\footnote{In terms of its structure Eq.~\eqref{eq:pde_w} is also known as the \textit{Polchinski equation} in the context of the RG for higher-dimensional QFTs. However, in the original work \cite{Polchinski:1983gv} an effective action $L ( \Lambda, \phi )$ takes the role of $\mathcal{W}$ and it is formulated in terms of the fields $\phi$ instead of the sources $J$. For relations between the original Polchinski equation and the flow equations studied in this work and selected applications of the Polchinski equation, see, \textit{e.g.}, Refs.~\cite{Litim:2005us,Yabunaka:2018mju,Litim:2018pxe,Cotler:2022fze}.}.
	
	We could now repeat the discussion about the issues of initial and boundary conditions for the solution of this PDE.
	However, the problems are almost identical to those of Eq.~\eqref{eq:pde_z}, because on the level of the PDE, we only substituted the function $\mathcal{Z} ( t, J )$ by $\mathcal{W} ( t, J )$ via the logarithm, which does not change the structure of the problem fundamentally.
	Formulating appropriate initial and boundary conditions in the spatial $J$ direction therefore remains as complicated as before.
	Note that the PDE \eqref{eq:pde_w} became more complicated when compared to Eq.~\eqref{eq:pde_z} due to the non-linear term on the right-hand side.
	In summary, the scale-dependent Schwinger functional is, from a practical point of view, as badly suited as $\mathcal{Z}(t,J)$ to perform the (numeric) calculation of the functional integral via a flow equation starting from a Gaussian-type integral.\\
	
	In the following we will focus on the scale-dependent effective (average) action and its respective flow equation, which does not suffer from the issues of particular initial and boundary conditions.
	As an added benefit, the effective action is also the most efficient functional in terms of storing information of a theory at hand.
	Formulating proper initial and boundary conditions for the flow equations for $\mathcal{Z} ( t, J )$ and $\mathcal{W} ( t, J )$ and if possible implementing adequate numerical schemes in the context of zero-dimensional field theories would certainly be interesting from an academic point of view.
	Translating the initial and boundary conditions for the scale-dependent effective (average) action to $\mathcal{Z} ( t, J )$ and $\mathcal{W} ( t, J )$ could be a possible and potentially feasible strategy.
	A comparison of the flows of $\mathcal{Z} ( t, J )$, $\mathcal{W} ( t, J )$, and $\Gamma ( t, \varphi )$, both conceptually and for explicitly specified actions, is a worthwhile subject of future work.

\subsubsection{The scale-dependent effective action}
\label{subsec:scale_dependent_effective_action}

	We now define the \textit{scale-dependent effective action} $\Gamma_t[\varphi]$ via the Legendre transform of Eq.~\eqref{eq:scale_dependent_w} with respect to the sources $J$ at a RG timescale $t$,
		\begin{align}
			\Gamma_t [\varphi] \equiv \, & \mathop{\mathrm{sup}}_J \big\{ J \, \varphi - \mathcal{W}_t[J] \big\} \equiv 	\vphantom{\bigg(\bigg)} 	\label{eq:scale_dependent_effective_action}
			\\
			\equiv  \, & J_t(\varphi) \, \varphi - \mathcal{W}_t[J_t(\varphi)] \, ,	\vphantom{\bigg(\bigg)}	\label{eq:scale_dependent_effective_action_Jt}
		\end{align}
	where we introduced the source $J_t(\varphi)$ which realizes the supremum.
	
	Note that, analogous to $\mathcal{Z}_t [J]$ and $\mathcal{W}_t [J]$, the convexity and smoothness (in zero dimensions) of $\Gamma_t [\varphi]$ is not spoiled by the $t$ dependence, because the properties of the Legendre transformation still ensure both.
	
	To obtain an explicit relation for the scale-dependent source $J_t ( \varphi )$, which realizes the supremum in Eq.~\eqref{eq:scale_dependent_effective_action}, we consider the functional derivative of Eq.~\eqref{eq:scale_dependent_effective_action} at the supremum to find the important relation
		\begin{align}
			\mathcal{W}_{t,J}^{(1)}[J_t(\varphi)] \equiv \frac{\delta \mathcal{W}_t [J]}{\delta J}\bigg|_{J=J_t(\varphi)} = \varphi \, ,	\label{eq:definition_phi_t}
		\end{align}
	which will be used frequently in the following.
	Taking the functional derivative of Eq.~\eqref{eq:scale_dependent_effective_action_Jt} with respect to $\varphi$ and using Eq.~\eqref{eq:definition_phi_t} we ultimately find
		\begin{align}
			\Gamma^{(1)}_{t, \varphi}[\varphi] \equiv \frac{\delta \Gamma_t [\varphi]}{\delta \varphi} = J_t(\varphi) \, ,	\label{eq:definition_j_t}
		\end{align}
	which is referred to as \textit{quantum equation of motion}.
	Due to the strict convexity of $\Gamma_t [ \varphi ]$ the function $J_t ( \varphi )$ is bijective and as such can be inverted, which can be achieved by considering Eq.~\eqref{eq:definition_phi_t} at fixed value $J$ for $J_t$:
		\begin{align}
			\varphi_t ( J ) \equiv \frac{\delta \mathcal{W}_t [ J ]}{\delta J} \, ,	\label{eq:definition_phi_t_explicit}
		\end{align}
	where $\varphi_t ( J )$ is the so-called \textit{scale-dependent classical field} (sometimes also referred to as \textit{scale-dependent mean field}).
	
	The subtle relations between, and scale dependences of, $\varphi_t ( J )$ and $J_t ( \varphi )$ are rarely discussed in literature and usually suppressed in the notation.
	The relation between $\varphi_t$ and $J_t$ will be of particular importance in the discussion of $n$-point correlation functions in Sub.Sec.~\ref{subsec:n_pt_functions}.
	The scale dependence of $\varphi_t(J)$ from Eq.~\eqref{eq:definition_phi_t_explicit} is not related to a rescaling (RG transformation) using, \textit{e.g.}, a wave-function renormalization for $\varphi$.
	
	Before we derive the FRG equation, which is the flow equation for $\Gamma_t [ \varphi ]$ and a PDE for the function $\Gamma ( t, \varphi )$ in the $( t, \varphi )$ plane, we check whether we will run into the same issues (related to initial and boundary conditions) as before.
	Hence, first of all, we must derive the initial condition for the PDE for $\Gamma ( t, \varphi )$.
	To this end, we study the limit $t \rightarrow 0$ of $\Gamma_t [ \varphi ]$.
	We use the definitions~\eqref{eq:scale_dependent_z}, \eqref{eq:scale_dependent_w}, \eqref{eq:scale_dependent_effective_action}, and \eqref{eq:scale_dependent_effective_action_Jt} to obtain
		\begin{align}
			\mathrm{e}^{ - \Gamma_t [ \varphi ] } = \, & \mathrm{e}^{ - \mathop{\mathrm{sup}}_J \{ J \, \varphi - \mathcal{W}_t [ J ] \} } =	\vphantom{\bigg(\bigg)}	\label{eq:derivation_uv_condition}
			\\
			= \, & \mathrm{e}^{ \ln \mathcal{Z}_t [ J_t ( \varphi )] - J_t ( \varphi ) \, \varphi } =	\vphantom{\bigg(\bigg)}	\nonumber
			\\
			= \, & \mathcal{N} \int_{- \infty}^{\infty} \mathrm{d} \phi \, \mathrm{e}^{ - \mathcal{S} [ \phi ] - \Delta \mathcal{S}_t [ \phi ] + J_t ( \varphi ) \, ( \phi - \varphi ) } \, .	\vphantom{\bigg(\bigg)}	\nonumber
		\end{align}
	We now shift the integration variable%
	\footnote{%
		It is the same shift that is used in the background field formalism \cite{DeWitt:1965jb,Abbott:1981ke}, where the full fluctuating quantum field $\phi$ is split into a background field configuration $\varphi$ and additional fluctuations $\phi^\prime$ about the background field.
		This is why $\varphi$ is called the classical or mean field.
	}
	$\phi \mapsto \phi^\prime = \phi - \varphi$. Using Eq.~\eqref{eq:regulator_insertion}, we find
		\begin{align}
			& \mathrm{e}^{- \Gamma_t [ \varphi ] + \Delta \mathcal{S}_t [ \varphi ]} =	\vphantom{\bigg(\bigg)}	\label{eq:derivation_uv_condition_3}
			\\
			= \, & \mathcal{N} \int_{- \infty}^{\infty} \mathrm{d} \phi^\prime \, \mathrm{e}^{ - \mathcal{S} [ \phi^\prime + \varphi ] - \Delta \mathcal{S}_t [ \phi^\prime ] - r ( t ) \, \phi^\prime \, \varphi + \Gamma^{(1)}_{t, \varphi} [ \varphi ] \, \phi^\prime } \, .	\vphantom{\bigg(\bigg)}	\nonumber
		\end{align}
	In the next step, we introduce the \textit{scale-dependent effective average action},
		\begin{align}
			\bar{\Gamma}_t [ \varphi ] \equiv \Gamma_t [ \varphi ] - \Delta \mathcal{S}_t [ \varphi ] \, ,	\label{eq:scale_dependent_effective_average_action}
		\end{align}
	which also tends to the effective action $\Gamma [ \varphi ]$ for $t \rightarrow \infty$, because the second term vanishes in this limit, \textit{cf.}~Eq.~\eqref{eq:exponential_regulator}.
	
	At any finite value of $t$ (including $t=0$), $\bar{\Gamma}_t [ \varphi ]$ differs from $\Gamma_t [ \varphi ]$ and is no longer guaranteed to be convex, which can be seen directly from the second term in Eq.~\eqref{eq:scale_dependent_effective_average_action}.
	Convexity is only recovered for $t \rightarrow \infty$.
	However, the second term in Eq.~\eqref{eq:scale_dependent_effective_average_action} does not violate the smoothness of $\bar{\Gamma}_t[\varphi]$ in zero dimensions for all $t$, because $\Delta \mathcal{S}_t [\varphi] \equiv \mathcal{S}_t (\varphi) \in C^\infty$ in $\varphi$.
	
	We express Eq.~\eqref{eq:derivation_uv_condition_3} in terms of the scale-dependent effective average action \eqref{eq:scale_dependent_effective_average_action} and, for the sake of convenience, revert the notation $\phi^\prime \rightarrow \phi$,
		\begin{align}
			& \mathrm{e}^{- \bar{\Gamma}_t [ \varphi ]} =	\mathcal{N} \, \int_{- \infty}^{\infty} \mathrm{d} \phi \, \mathrm{e}^{ - \mathcal{S} [ \phi + \varphi ] - \Delta \mathcal{S}_t [ \phi ] + \bar{\Gamma}^{(1)}_{t, \varphi} [ \varphi ] \, \phi } \, .	\label{eq:exp_gamma_bar_int}
		\end{align}
	In the next step one formally introduces the normalization of a Gaussian integral with mass $r ( t )$ and takes the logarithm, which results in
		\begin{align}
			& \bar{\Gamma}_t [ \varphi ] =	\vphantom{\bigg(\bigg)}	\label{eq:derivation_uv_condition_2}
			\\
			= \, & - \ln \int_{- \infty}^{\infty} \mathrm{d} \phi \, \sqrt{\tfrac{r ( t )}{2 \pi}} \, \mathrm{e}^{ - \mathcal{S} [ \phi + \varphi ] - \frac{1}{2} r ( t ) \, \phi^2 + \bar{\Gamma}^{(1)}_{t, \varphi} [ \varphi ] \, \phi} -	\vphantom{\bigg(\bigg)}	\nonumber
			\\
			& - \ln \Big[ \mathcal{N} \sqrt{\tfrac{2\pi}{r ( t )}} \Big] \, .	\vphantom{\bigg(\bigg)}	\nonumber
		\end{align}
	We are now ready to study the limit $t \rightarrow 0$, which corresponds to the initial condition for a possible flow equation for $\Gamma_t [ \varphi ]$ or $\bar{\Gamma}_t [ \varphi ]$, respectively.
	Focusing on the $\phi$ integral in the first term on the right-hand side of Eq.~\eqref{eq:derivation_uv_condition_2}, we employ the fact that the regulator terms act like a Gaussian representation of the Dirac delta distribution,
		\begin{align}
			\lim\limits_{t \rightarrow 0} \sqrt{\tfrac{r ( t )}{2\pi}} \, \mathrm{e}^{- \frac{1}{2} r ( t ) \, \phi^2} \approx \delta (\phi) \, ,	\label{eq:delta_distribution}
		\end{align}
	as long as $r ( t )$ is much larger than all scales in $\mathcal{S} [ \phi ]$.
	Thus, denoting
		\begin{align}
			c ( t ) \equiv - \ln \Big[ \mathcal{N} \, \sqrt{\tfrac{2\pi}{r ( t )}} \, \Big] \, ,
		\end{align}
	we find as $t \rightarrow 0$
		\begin{align}
			\bar{\Gamma}_t [\varphi] \rightarrow \, & - \ln \int_{-\infty}^{\infty} \mathrm{d} \phi \, \delta(\phi) \, \mathrm{e}^{ - \mathcal{S}[\phi + \varphi] + \bar{\Gamma}^{(1)}_{t,\varphi}[\varphi] \, \phi } + c ( t ) =	\vphantom{\bigg(\bigg)}	\nonumber
			\\
			= \, & \mathcal{S}[\varphi] + c ( t ) \, ,	\qquad	( t \rightarrow \infty )	\vphantom{\bigg(\bigg)}	\label{eq:initial_condition_gamma}
		\end{align}
	This means that the initial condition for a flow of $\bar{\Gamma}_t [ \varphi ]$ is given by the classical action $\mathcal{S}$ evaluated for the classical field $\varphi$ and some additional $t$ dependent, but $\varphi$ independent term $c ( t )$.
	This choice for an initial condition of a PDE for $\bar{\Gamma}_t [ \varphi ]$ has subtle consequences:
	
	Although $c ( t )$ does not depend on $\varphi$, it is large, $c ( t ) \sim  \frac{1}{2} \ln r ( t )$.
	Consequently, as far as the initial condition for the PDE for $\Gamma_t[\varphi]$ or $\bar{\Gamma}_t[\varphi]$ is concerned, it seems as if we run into the same problem as before:
	The initial condition is dominated by the artificial mass of the regulator $r ( t )$, independent of the specific action $\mathcal{S} [ \phi ]$, and differences in the specific choice for $\mathcal{S} [ \phi ]$ enter the initial condition only as small deviations from the large term $c ( t )$.
	Furthermore, $c ( t )$ contains the normalization constant $\mathcal{N}$, which was fixed according to Eq.~\eqref{eq:normalization}.
	
	However, precisely because $c ( t )$ appears like the normalization $\mathcal{N}$, it should be irrelevant for all physical observables.
	Indeed this is the case, because all $\varphi$ independent terms in $\Gamma_t [ \varphi ]$ do not enter the $n$-point correlation functions, since the latter are calculated as derivatives of $\Gamma [ \varphi ]$ with respect to $\varphi$ at $t \rightarrow \infty$, see Sub.Sec.~\ref{subsec:n_pt_functions}.
	This implies that an additive, $\varphi$ independent term in the three effective actions $\Gamma [ \varphi ]$, $\Gamma_t [ \varphi ]$, and $\bar{\Gamma}_t [ \varphi ]$ is irrelevant and only relative differences in the effective actions are observable.
	Therefore, we can simply omit $c ( t )$ and take as initial condition for the PDE for $\bar{\Gamma}_t [ \varphi ]$ the value $\mathcal{S} [ \varphi ]$, which perfectly incorporates the difference between different models with distinct actions $\mathcal{S} [ \phi ]$.
	
	One problem in disregarding $c ( t )$ remains: one has to ensure that a PDE for $\bar{\Gamma}_t [ \varphi ]$ must not contain any terms without field derivatives of $\bar{\Gamma}_t [ \varphi ]$.
	Otherwise $c ( t )$ would influence the flow in a time-dependent manner.
	Fortunately, this does not happen, as we will see later, and the FRG equation \eqref{eq:wetterich_equation} does not contain terms without field derivatives of $\bar{\Gamma}_t [ \varphi ]$ on the right-hand side.
	
	This, however, brings up another question: After Eq.~\eqref{eq:scale_dependent_effective_average_action} we argued that $\bar{\Gamma}_t [ \varphi ]$ does not need to be convex, but must still be smooth for all $t$.
	Let us for example consider the non-analytic action \eqref{eq:example_non-analytic_action} as an initial condition, $\bar{\Gamma}_{t = 0} [ \varphi ] = \mathcal{S} [ \varphi ]$.
	This action does not cause any problems for the convexity and the smoothness of $\mathcal{Z}_t [ J ]$ and $\mathcal{W}_t [ J ]$ at arbitrary $t$, see for example App.~\ref{app:mermin_wagner} and Fig.~\ref{fig:w_z_j_c_infinity}.
	The non-convexity of $\mathcal{S} [ \varphi ]$ is also not a problem for $\bar{\Gamma}_t [ \varphi ]$, which does not necessarily need to be convex at finite $t$.
	Nevertheless, the smoothness of $\bar{\Gamma}_t [ \varphi ]$ is violated by this choice of $\mathcal{S} [ \varphi ]$ at $t = 0$.
	This issue originates from relation \eqref{eq:delta_distribution}, which is exactly fulfilled only in the limit $\Lambda \rightarrow \infty$ for the UV cutoff.
	This, however, leads to a trivial theory of infinitely massive particles at $t = 0$, \textit{cf.}\ Eq.~\eqref{eq:scale_dependent_z}.
	If one chooses a reasonably large but finite $\Lambda$ and does not use Eq.~\eqref{eq:delta_distribution}, one would ensure that $\bar{\Gamma}_t [ \varphi ]$ is also smooth at $t = 0$.
	However, then the initial condition is not exactly $\mathcal{S} [ \varphi ]$, but rather an extremely complicated expression.
	In consequence, if we use the approximation \eqref{eq:delta_distribution} even for finite $\Lambda$, one has to pay the price of introducing errors into the initial condition as well as violating the smoothness of $\bar{\Gamma}_t [ \varphi ]$ at $t = 0$.
	In return one has a well-defined initial condition $\mathcal{S} [ \varphi ]$ for the PDE for $\bar{\Gamma}_t [ \varphi ]$.
	However, if $\Lambda$ is chosen to be much larger than all scales in $\mathcal{S} [ \phi ]$, the errors from the initial condition are minor and expected to be of magnitude
		\begin{align}
			\mathrm{error} \approx \frac{\text{largest scale in\,\,} \mathcal{S}}{\Lambda} \, ,	\label{eq:error_scaling_uv_cutoff}
		\end{align}
	We will come back to this issue in Sec.~\ref{sec:results} in the context of RG consistency \cite{Braun:2018svj,Braun:2003ii,Herbst:2013ufa,Springer2017,PhysRevD.87.076004}.
	
	Additionally, we will find that also the smoothness of $\bar{\Gamma}_t [ \varphi ]$ is recovered automatically for all $t > 0$ by the structure of the PDE for $\bar{\Gamma}_t [ \varphi ]$, because it always contains diffusive contributions which immediately smear out kinks in the initial condition right in the first time step.
	We will also come back to this issue later on, after we have derived the FRG equation \eqref{eq:wetterich_equation} and discussed its diffusive, irreversible character.

\subsubsection{The Exact Renormalization Group equation}
\label{subsubsec:exact_rg_equation}

	In analogy to the previous flow equations, the FRG equation, which is the flow equation for $\bar{\Gamma}_t [ \varphi ]$, is obtained by taking the derivative of $\bar{\Gamma}_t [ \varphi ]$ with respect to $t$ and using the definitions~\eqref{eq:scale_dependent_effective_action} and \eqref{eq:scale_dependent_effective_average_action} to express the derivative of $\bar{\Gamma}_t [ \varphi ]$ by the scale-dependent Schwinger functional,
		\begin{align}
			\partial_t \bar{\Gamma}_t [\varphi] = \, & \partial_t \, \big( \Gamma_t [\varphi] - \Delta \mathcal{S}_t [\varphi] \big) =	\vphantom{\bigg(\bigg)}	\label{eq:wetterich_mix_w_gamma}
			\\
			= \, & \partial_t \, \big( J_t(\varphi) \, \varphi - \mathcal{W}_t[J_t(\varphi)] - \Delta \mathcal{S}_t [\varphi] \big) =	\vphantom{\bigg(\bigg)}	\nonumber
			\\
			= \, & [ \partial_t J_t (\varphi) ] \, \varphi - \partial_t \mathcal{W}_t [J_t (\varphi)] - \vphantom{\bigg(\bigg)}	\nonumber
			\\
			& - [ \partial_t J_t (\varphi) ] \, \mathcal{W}^{(1)}_{t,J_t} [J_t] - \big[\tfrac{1}{2} \, \partial_t r ( t ) \big]\, \varphi^2 = \vphantom{\bigg(\bigg)}	\nonumber
			\\
			= \, & - \partial_t \mathcal{W}_t [J_t (\varphi)] - \big[\tfrac{1}{2} \, \partial_t r ( t ) \big] \, \varphi^2 \, ,	\vphantom{\bigg(\bigg)}	\nonumber
		\end{align}
	where we used the chainrule and Eq.~\eqref{eq:definition_phi_t}.
	
	We now use the flow equation for the Schwinger functional \eqref{eq:pde_w} to substitute the first term on the right-hand side.
	Again employing the identity \eqref{eq:definition_phi_t}, the last term in the last line of Eq.~\eqref{eq:wetterich_mix_w_gamma} cancels with the non-linear term in Eq.~\eqref{eq:pde_w}, such that
		\begin{align}
			\partial_t \bar{\Gamma}_t [ \varphi ] = \, & \big[\tfrac{1}{2} \, \partial_t r ( t ) \big] \, \mathcal{W}_{t,J J}^{(2)} [ J_t ( \varphi ) ] \, .	\label{eq:wetterich_mix_w_gamma_2}
		\end{align}
	It remains to replace the second derivative of the scale-dependent Schwinger functional by a corresponding derivative of $\bar{\Gamma}_t [\varphi]$.
	This is done via the identity
		\begin{align}
			1 = \frac{\delta J_t(\varphi)}{\delta \varphi} \, \frac{\delta \varphi}{\delta J_t ( \varphi )} = \Gamma^{(2)}_{t ,\varphi \varphi} [ \varphi ] \, \mathcal{W}^{(2)}_{t , J J} [ J_t ( \varphi ) ] \, ,	\label{eq:g2w2_identity}
		\end{align}
	which follows from Eqs.~\eqref{eq:definition_phi_t} and \eqref{eq:definition_j_t}.
	Plugging this into Eq.~\eqref{eq:wetterich_mix_w_gamma_2} and using Eq.~\eqref{eq:scale_dependent_effective_average_action} with Eq.~\eqref{eq:regulator_insertion} we obtain the \textit{FRG equation}, \textit{Exact Renormalization Group equation} or \textit{Wetterich equation} \cite{Wetterich:1992yh,Ellwanger:1993mw,Morris:1993qb}
		\begin{align}
			\partial_t \bar{\Gamma}_t [\varphi] = \big[\tfrac{1}{2} \, \partial_t r ( t ) \big] \, \big[ \bar{\Gamma}^{(2)}_{t , \varphi \varphi} [\varphi] + r ( t ) \big]^{-1} \, ,	\label{eq:wetterich_equation}
		\end{align}
	which is a flow equation -- a PDE -- for the scale-dependent effective average action $\bar{\Gamma} ( t, \varphi )$ in the $( t, \varphi )$ plane,
		\begin{align}
			\partial_t \bar{\Gamma}(t,\varphi) = \frac{\frac{1}{2} \, \partial_t r ( t )}{\partial_\varphi^2 \bar{\Gamma}(t,\varphi) + r ( t )} \, ,	\label{eq:pde_gamma}
		\end{align}
	with the initial condition $\bar{\Gamma} ( t = 0, \varphi ) = \mathcal{S} [ \varphi ]$.
	Some remarks are in order:
		\begin{enumerate}
			\item	In contrast to the PDEs for $\mathcal{Z} ( t, J )$ and $\mathcal{W} ( t, J )$ the FRG equation can be initialized with a suitable initial condition at $t = 0$ that produces distinct flows for different actions $\mathcal{S}[\phi]$, as was discussed in the previous subsubsection.
			
			\item	The spatial boundary conditions, \textit{i.e.}, for $\varphi \rightarrow \pm \infty$ are provided by the asymptotics of the FRG equation \eqref{eq:pde_gamma} itself and by the requirement that $\mathcal{S} [ \varphi ]$ must be bounded from below: The action $\mathcal{S} [ \varphi ]$ of an (interacting) field theory must at least grow like $\varphi^2$ for large $| \varphi |$ and the dominant contribution for large $| \varphi |$ must be even in $\varphi$.
			For actions $\mathcal{S} [ \varphi ]$ that grow asymptotically faster than $\varphi^2$ the denominator on the right-hand side of the PDE \eqref{eq:pde_gamma} already diverges at $t \approx 0$, such that
				\begin{align}
					\lim\limits_{| \varphi | \rightarrow \infty} \partial_t \bar{\Gamma} ( t, \varphi ) \approx 0 \, .
				\end{align}
			It follows that for $| \varphi | \rightarrow \infty$ the function $\bar{\Gamma} ( t, \varphi )$ does not change at all, but keeps its initial value $\mathcal{S} [ \varphi ]$.
			These are perfectly valid boundary conditions for a PDE.
			The scenario for initial conditions with $\lim\limits_{| \varphi | \rightarrow \infty} \mathcal{S} [ \varphi ] \sim \varphi^2$ is more delicate.
			We will return to this issue and a detailed discussion of boundary conditions, when we discuss the numerical implementation and solution of Eq.~\eqref{eq:pde_gamma} in Sub.Sec.~\ref{subsec:boundary_conditions_finite_volume} in the context of numerical fluid dynamics.
			
			\item	The structure of the PDE \eqref{eq:pde_gamma} is again a diffusion equation.
			In contrast to the PDEs \eqref{eq:pde_z} and \eqref{eq:pde_w} it is non-linear in the second-order spatial derivatives of $\Gamma ( t, \varphi )$ that appear in the denominator.
			By applying the same formalism to models with different field content, the FRG equation can also acquire convective/advective terms and source terms.
			We will thus find that the FRG equation shares many properties with other notable advection-diffusion equations, \textit{e.g.}, the Navier-Stokes equation \cite{navier-stokes}.
			This is discussed in Sec.~\ref{sec:frg-flow_equations_and_numerical_fluid_dynamics}, where our numerical approach to the FRG equation is presented in more detail.
			However, it should be already mentioned at this point that analyzing and solving non-linear advection-diffusion-source/sink equations like Eq.~\eqref{eq:pde_gamma} is a state-of-the-art problem in numerical mathematics.
			Thus, some care is required in the search for well-established numerical solution schemes for PDEs of this type.
			
			\item	In zero dimensions, similar to the flow equations for $\mathcal{Z} ( t, J )$ and $\mathcal{W} ( t, J )$, one can reparameterize the flow time $t$ in terms of $r$ in Eq.~\eqref{eq:pde_gamma} and get rid of the prefactor $\partial_t r ( t )$.
			Additionally, one could eliminate $r ( t )$ in the denominator in Eq.~\eqref{eq:pde_gamma} by shifting $\bar{\Gamma} ( t, \varphi ) \rightarrow \bar{\Gamma} ( r, \varphi ) - \tfrac{1}{2} \, r \, \varphi^2$ and switching from $t$ to $r$ as flow parameter, which corresponds to the zero-dimensional analogue of the rescaled  ``dimensionless'' flow equation in fixed-point form, but is not suited for the practical calculations in this work.
			
			This reparameterization effectively corresponds to different choices of regulator (shape) functions in zero dimensions.
			However, for higher-dimensional problems, different choices of regulators do not need to be related to each other via simple reparametrization of the RG time.
			In any case, the effective dynamics in the PDE during the RG flow strongly depends on the parametrization of the RG scale as well as the explicit choice of regulator, which has two direct consequences: First, although the dynamics and $t$ evolution of observables (the $n$-point correlation functions) during the RG flow might be highly interesting and must also be studied to ensure that the UV and IR cutoff scales are chosen appropriately, one must clearly state that only the IR value of $\Gamma [ \varphi ]$ is mathematically and physically meaningful and suitable for extracting information on the $n$-point correlation functions.
			This is demonstrated and discussed again in the context of numerical precision tests of the $O(N)$ model in Sec.~\ref{sec:results}.
			Second, from a numerical point of view, some parametrizations or choices of regulators might be more challenging for the numerical integrators than others and must be adopted to the specific problems at hand.
			On the level of the PDE this corresponds to the time-dependent strength of the diffusion, see below.
			
			\item Unrelated to the present discussion, a formulation of the FRG equation using mean fields carrying an explicit scale dependence (in higher dimensions often related to a running wave-function renormalization) is also possible with a careful consideration and distinction between total and partial derivatives with respect to $t$.
			Generalizations including composite mean fields are also possible, see, \textit{e.g.}, Ref.~\cite{Pawlowski:2005xe}.
		\end{enumerate}
	Using a zero-dimensional field theory with one degree of freedom, we have therefore demonstrated that it is possible to transform the problem of solving functional integrals like Eqs.~\eqref{eq:expectation_value_1} and \eqref{eq:partition_function} for a model with action $\mathcal{S}[\phi]$ into solving the PDE \eqref{eq:pde_gamma} in $t$ and $\varphi$ with initial condition $\mathcal{S} [ \varphi ]$.
	The FRG equation thus directly implements the idea of transforming Gaussian-type functional integrals into arbitrary functional integrals, but on the level of the effective action $\Gamma[\varphi]$ rather than the partition function $\mathcal{Z}[J]$.
	Both formulations of the problem of calculating $n$-point correlation functions -- the functional-integral formulation and the FRG formulation -- are mathematically equivalent.
	This, however, is, as we have seen, a highly non-trivial statement and demands numerical precision tests, which are part of this work.
	
	In Ref.~\cite{Pawlowski:2005xe} it is shown that the FRG framework can be generalized to models or theories with arbitrary field content in arbitrary dimensions and space-time background (even a formulation for space-time itself, \textit{i.e.}, quantum gravity is possible \cite{Reuter:1996cp,Reuter:2001ag}, see Ref.~\cite{Dupuis:2020fhh} for a recent review).
	
	Before we introduce the zero-dimensional $O(N)$ model and explain the relation of the FRG to fluid dynamics, followed by our main discussion of zero-dimensional QFTs as a testing ground for numerical methods and truncation schemes, we discuss two further issues.
	The first contextualizes our previous discussion with an interpretation of the FRG from the RG perspective (also for higher-dimensional field theories).
	Furthermore, it briefly discusses the generalization of the FRG equation to different field content.
	This can also be found in Refs.~\cite{Pawlowski:2005xe,Rosten:2010vm,Dupuis:2020fhh,Berges:2000ew,Kopietz:2010zz,Gies:2006wv,PawlowskiScript,Delamotte:2007pf}.
	The second issue discusses the relation between the $n$-point correlation functions of the different generating functionals $\mathcal{Z}_{( t )} [ J ]$, $\mathcal{W}_{( t )} [ J ]$, and $\Gamma_{( t )} [ \varphi ]$.
	This is needed for a comparison of the exact results from the partition function $\mathcal{Z} [ J ]$ with our results from the FRG based on $\Gamma [ \varphi ]$.
	Readers familiar with these issues may skip the following two sections.

\subsection{Contextualization with FRG in higher-dimensional space-time}
\label{subsec:interpretation_frg}

	The structure of the FRG equation \eqref{eq:pde_gamma} is already very general and extends with only minor modifications to arbitrary fields and dimensions.
	Derivations can be found in, \textit{e.g.}, Refs.~\cite{Pawlowski:2005xe,Gies:2006wv,Kopietz:2010zz}.
	The FRG equation reads
		\begin{align}
			\partial_t \bar{\Gamma}_t [ \Phi ] = \mathrm{STr} \Big[ \big( \tfrac{1}{2} \, \partial_t R_t \big) \, \big( \bar{\Gamma}^{(2)}_t [ \Phi ] + R_t \big)^{-1} \Big] \, .	\label{eq:wetterich_general}
		\end{align}
	The supertrace in Eq.~\eqref{eq:wetterich_general} entails sums over internal indices and different fields and integrals over momenta, taking minus signs for fermionic fields properly into account. The fundamental difference between ERG Eq.~\eqref{eq:wetterich_general} and its counterpart in zero dimensions \eqref{eq:wetterich_equation}, is that the ERG equation in $d>0$ is a functional differential equation for the classical fields $\Phi$. It does not naturally present as a PDE which necessitates truncations in practical computation to project the ERG equation onto a finite set of coupled ODEs and/or PDEs. The regulator $R_t$ for computations in $d>0$ is no longer a simple scalar function but an operator with a particular, non-trivial structure in position/momentum space. While different regulator choices are still possible in higher dimensions, corresponding RG flows are no-longer related by simple rescaling and a suitable regulator choice for the problem at hand becomes particularity important when considering explicit truncated FRG flow equations see, \textit{e.g.}, Refs.~\cite{Litim:2000ci,Pawlowski:2015mlf}. More details can be found in, \textit{e.g.}, Refs.~\cite{Pawlowski:2005xe,Rosten:2010vm,Berges:2000ew,Gies:2006wv,PawlowskiScript,Delamotte:2007pf,Kopietz:2010zz,Dupuis:2020fhh,Blaizot:2021ikl}.
	The equation is based upon momentum locality, \textit{i.e.}, the integrand of the momentum integral on the right-hand side is peaked around the RG scale $k \approx q$ (for conventional regulators), see, \textit{e.g.}, Fig.~1 in Ref.~\cite{Gies:2006wv} or Fig.~3.1 of Ref.~\cite{Rennecke:2015lur}, where $q$ is the loop momentum and
		\begin{align}
			t = - \ln \big( \tfrac{k}{\Lambda} \big) \, .
		\end{align}
	The FRG equation can be interpreted as a direct implementation of Wilson's approach to the RG \cite{Wilson:1971bg,Wilson:1971dh,Wilson:1979qg}.
	
	In general, the space-time dimensionality has to be taken into account when considering the convergence properties of different expansion schemes. For example, the vertex expansion is believed to work very well for QCD in $d=4$ dimensions (see, \textit{e.g.}, Ref.~\cite{Fu:2019hdw} for a recent overview), however, as we will discuss below, the convergence of the expansion is in general not guaranteed.
	The vertex expansion is an expansion in terms of moments of the quantum effective action, explained in detail in Sub.Sub.Sec.~\ref{subsubsec:vertex_expansion}.
	Here, the moments are the irreducible parts of scattering kernels.
	
	The convergence of this expansion is given by two main ingredients,
		\begin{enumerate}
			\item phase-space suppression,
				
			\item finite couplings.
		\end{enumerate}
	The first point means that higher-order vertices, which originate from quantum effects and are typically not present in the classical action, come with increasing suppression factors, \textit{e.g.}, due to the angular integrations.
	The second point simply relates to the fact that all couplings have to stay finite.
	Otherwise the argument related to phase-space suppression simply does not work.
	There are several scenarios where this can be the case.
	The main one being the presence of resonances, where couplings can be divergent.
	Also large densities might circumvent the effect of phase-space suppression, but are not our main concern in this work.
	The last, and for this work most important effect, is that of the dimension.
	
	In particular, for zero-dimensional space-time the angular integrations are not present, and hence the entire argument of phase-space suppression does not work.
	Zero-dimensional QFT is ultra-local -- defined only in a single point -- and thus extremely coupled in field space.
	This, of course, has to be kept  in mind when considering convergence properties of vertex expansions.
	
	Still, also a parallel work in $1 + 1$ spacetime dimensions by some of us and collaborators \cite{Stoll:2021ori} generically supports these statements and the increasing importance of local interactions in low spacetime dimensions.

\subsection{\texorpdfstring{$n$}{n}-point correlation functions}
\label{subsec:n_pt_functions}

	In this section we discuss the scale-dependent correlation functions, which can be extracted from the (scale-)dependent generating functionals $\mathcal{Z}_{( t )}$, $\mathcal{W}_{( t )}$, and $\Gamma_{( t )}$.
	We restrict the discussion to a zero-dimensional quantum theory with a single real scalar.
	The concepts and expressions can be generalized to theories including arbitrary fields and generalize to higher dimensions.
	For a broader discussion in the context of QTFs we refer the interested reader to the textbooks \cite{Peskin:1995ev,Weinberg:1996kr,DeWitt:1965jb,Greiner:1996zu,ZinnJustin:2002ru}.
	For a comprehensive discussion of correlation functions and their relations in the FRG see, \textit{e.g.}, Refs.~\cite{Wipf:2013vp,PawlowskiScript}.\\
	
	Correlation functions can be extracted by taking successive functional derivatives of the generating functional, \textit{cf.}\ Eq.~\eqref{eq:expectation_value_1}:
		\begin{align}
			\langle \phi^n \rangle_{t,J} \equiv \frac{\mathcal{Z}^{(n)}_{t,J \cdots J} [J] }{\mathcal{Z}_t [J] }\, .
		\end{align}
	(The non-observable normalization, which we fixed by means of Eq.~\eqref{eq:normalization} cancels.)
	
	According to Eq.~\eqref{eq:definition_phi_t_explicit}, the one-point correlation function
		\begin{align}
			\langle \phi \rangle_{t,J} = \frac{\mathcal{Z}^{(1)}_{t,J} [J]}{\mathcal{Z}_t [J]} = \mathcal{W}^{(1)}_{t,J} [J]
		\end{align}
	equals the scale-dependent classical field $\varphi_t(J)$.
	
	The two-point correlation function
		\begin{align}
			\langle \phi^2 \rangle_{t,J} = \frac{\mathcal{Z}^{(2)}_{t,JJ} [J]}{\mathcal{Z}_t [J]} = \mathcal{W}^{(2)}_{t,J J} [J] + \langle \phi \rangle_{t,J}^2
		\end{align}
	is of particular interest in QFT since it is related to the transition amplitude between two states.
	In $d > 0$ such an amplitude between $\phi ( x_1 )$ and $\phi ( x_2 )$ encodes the particle motion between the space-time points $x_1$ and $x_2$.
	$\langle \phi^2 \rangle_{t, J}$ includes the disconnected%
	\footnote{%
		``Connected'' and ``disconnected'' in this context refers to the connectivity of the Feynman-diagram representation of the correlation functions.
		In a connected Feynman diagram all external lines are connected in the diagram through at least one internal line.
	}
	contribution $\langle \phi \rangle_{t, J}^2$.
	This information is already stored in the $1$-point correlation function.
	Higher-order $n$-point correlation functions include disconnected parts consisting of products of lower $m$-point functions with $m < n$ \cite{McCullagh:2009}.
	The disconnected contributions correspond to scattering processes where only a subset of the fields interact with each other and are as such irrelevant for observables.
	Loosely speaking, $\mathcal{Z}_t [ J ]$ contains redundant information in the form of these disconnected diagrams.
	
	The Schwinger functional $\mathcal{W}_t [ J ]$ does not contain this redundant information.
	Functional derivatives of $\mathcal{W}_t [ J ]$ generate connected $n$-point functions:
		\begin{align}
			\langle \phi^n \rangle^{c}_{t,J} \equiv \mathcal{W}^{(n)}_{t,J\ldots J} [J] \, .	\label{eq:def_conn_corr_func}
		\end{align}
	The first two connected $n$-point functions are
		\begin{align}
			\langle \phi \rangle^{c}_{t, J} = \, & \langle \phi \rangle_{t, J} = \mathcal{W}^{(1)}_{t, J} [ J ] \, ,	\vphantom{\bigg(\bigg)}
			\\
			\langle \phi^2 \rangle^{c}_{t, J} = \, & \langle \phi^2 \rangle_{t, J} - \langle \phi \rangle_{t, J}^2 = \mathcal{W}^{(2)}_{t, J J} [ J ] \, .	\vphantom{\bigg(\bigg)}
		\end{align}
	Higher-order $n$-point functions are interpreted as interaction vertices.
	For example, the connected three-point correlation function is given by
		\begin{align}
			\langle \phi^3 \rangle^{c}_{t, J} = \, & \langle \phi^3 \rangle_{t, J} - 3 \, \langle \phi^2 \rangle_{t, J} \, \langle \phi \rangle_{t, J} + 2 \, \langle \phi \rangle_{t,J}^3 \, .	\label{eq:conn_3ptfct}
		\end{align}
	The Schwinger functional, as the generating functional of connected correlation functions, still contains redundant information since connected correlation functions can be decomposed into 1PI%
	\footnote{%
		One-particle irreducible (1PI) in this context refers to Feynman diagrams, which cannot be split into two disconnected diagrams by cutting a single internal line.
	}
	vertex functions.
	1PI vertex functions encode all information about a~QFT.
	
	The effective action $\Gamma_t[\varphi]$ is the generating functional of 1PI vertex functions \cite{Goldstone:1962es,DeWitt:1965jb,Iliopoulos:1974ur,Greiner:1996zu,Weinberg:1996kr,Peskin:1995ev,ZinnJustin:2002ru,Wetterich:2001kra}.
	We now introduce a central object in functional approaches to QFT: the full scale-dependent propagator

		\begin{align}
			G_t^{\varphi \varphi} [ \varphi_t ] \equiv \mathcal{W}^{(2)}_{t , J_t J_t} [ J_t ] = \big( \Gamma^{(2)}_{t , \varphi_t \varphi_t} [ \varphi_t ] \big)^{-1} \, ,	\label{eq:fullprop}
		\end{align}
	where the last equality follows from Eq.~\eqref{eq:g2w2_identity}.
	Recalling Eqs.~\eqref{eq:definition_j_t} and\ \eqref{eq:definition_phi_t_explicit} we then obtain
		\begin{align}
			\frac{\delta}{\delta J_t} = \frac{\delta \varphi_t}{\delta J_t} \, \frac{\delta}{\delta\varphi_t} = \, & \mathcal{W}^{(2)}_{t , J_t J_t} [ J_t ] \, \frac{\delta}{\delta \varphi_t} \equiv G_t^{\varphi \varphi} [ \varphi_t ] \, \frac{\delta}{\delta \varphi_t} \, ,	\label{eq:j_varphi_chainrule}
		\end{align}
	Here we dropped the explicit $\varphi$ ($J$) dependence of the source realizing the supremum $J_t$ (the scale-dependent mean field $\varphi_t$) for readability only and will do so for the remainder of this section.
	Equation~\eqref{eq:j_varphi_chainrule} is basically a chain rule, which allows to convert functional $J_t$ derivatives into $\varphi_t$ derivatives.
	The correlation function $\langle \phi^n \rangle_{t,J_t}$ for $n\geq 1$ can be rewritten by successively pulling out functional $J_t$ derivatives,
		\begin{align}
			\langle \phi^n \rangle_{t, J_t} = \, & \frac{\mathcal{Z}^{( n )}_{t, J_t \cdots J_t} [ J_t ]}{\mathcal{Z}_t [ J_t ]} = \bigg( \frac{\delta}{\delta J_t} + \varphi_t \bigg) \, \frac{\mathcal{Z}^{( n - 1 )}_{t, J_t \ldots J_t} [ J_t ]}{\mathcal{Z}_t [ J_t ]} =	\vphantom{\bigg(\Bigg)}	\nonumber
			\\
			= \, & \Bigg(\prod_{i = 1}^{n - 1} \bigg( \frac{\delta}{\delta J_t} + \varphi_t \bigg)\Bigg) \, \varphi_t \, , \label{eq:exp_phi_n_t_Jt_prod1}	\vphantom{\bigg(\bigg)}
		\end{align}
	where the $\varphi_t$ terms account for the derivatives of the normalization $1/\mathcal{Z}_t [ J_t ]$.
	Using the chain rule \eqref{eq:j_varphi_chainrule} in Eq.~\eqref{eq:exp_phi_n_t_Jt_prod1} we arrive at
		\begin{align}
			\langle \phi^n \rangle_{t, J_t} = \Bigg(\prod_{i = 1}^{n - 1} \bigg( G_t^{\varphi\varphi} [ \varphi_t ] \, \frac{\delta}{\delta \varphi_t} + \varphi_t \bigg)\Bigg) \, \varphi_t \, ,	\label{eq:exp_phi_n_t_Jt_prod2}
		\end{align}
	which expresses the correlation function $\langle \phi^n \rangle_{t, J_t}$ completely in terms of $\varphi_t$, $G_t^{\varphi \varphi} [ \varphi_t ]$, and 1PI vertices for $n \geq 3$.
	The higher ($n \geq 3$) 1PI vertices $\Gamma^{( n )}_{t , \varphi_t \cdots \varphi_t} [ \varphi_t ]$ emerge in Eq.~\eqref{eq:exp_phi_n_t_Jt_prod2} from the functional derivatives of the propagator.
	Taking the $\varphi_t$ derivative of Eq.~\eqref{eq:g2w2_identity} (for $\varphi \equiv \varphi_t$, $J \equiv J_t$), we derive
		\begin{align}
			\frac{\delta}{\delta \varphi_t} \, G_t^{\varphi \varphi} [ \varphi_t ] = \, & \frac{\delta}{\delta \varphi_t} \, \big( \Gamma^{(2)}_{t , \varphi_t \varphi_t} [ \varphi_t ] \big)^{- 1}	\vphantom{\bigg(\bigg)}
			\\
			= \, & - G_t^{\varphi \varphi} [ \varphi_t ] \, \Gamma^{(3)}_{t , \varphi_t \varphi_t \varphi_t} [ \varphi_t ] \, G_t^{\varphi \varphi} [ \varphi_t ] \, ,	\vphantom{\bigg(\bigg)}	\nonumber
		\end{align}
	where we have used Eq.~\eqref{eq:fullprop} and where
		\begin{align}
			\frac{\delta}{\delta \varphi_t} \, \Gamma^{(n)}_{t , \varphi_t \cdots \varphi_t} [\varphi_t] = \Gamma^{(n+1)}_{t , \varphi_t \cdots \varphi_t\varphi_t} [\varphi_t]\, .
		\end{align}
	From the definition \eqref{eq:def_conn_corr_func} and Eq.~\eqref{eq:j_varphi_chainrule} it is even simpler to derive
		\begin{align}
			\langle \phi^n \rangle_{t,J_t}^{c} = \, & \Bigg(\prod_{i=1}^{n-1} \bigg( G_t^{\varphi\varphi} [\varphi_t] \, \frac{\delta}{\delta \varphi_t} \bigg)\Bigg) \, \varphi_t \, ,	\label{eq:exp_phi_n_c_t_Jt_prod}
		\end{align}
	which establishes a decomposition of connected correlation functions in terms of $\varphi_t$, $G_t^{\varphi \varphi} [ \varphi_t ]$, and 1PI vertices for $n \geq 3$.
	Equation~\eqref{eq:exp_phi_n_c_t_Jt_prod} is simpler than Eq.~\eqref{eq:exp_phi_n_t_Jt_prod2} because disconnected contributions arising from the term $\sim \varphi_t$ in the parenthesis in Eq.~\eqref{eq:exp_phi_n_t_Jt_prod2} are absent.
	
	In terms of $\Gamma_t [ \varphi_t ]$ the first three (connected) correlation functions are given by
		\begin{align}
			\langle \phi^1 \rangle_{t, J_t} = \, & \langle \phi^1 \rangle_{t, J_t}^{c} = \varphi_t \, ,	\vphantom{\bigg(\bigg)}
			\\
			\langle \phi^2 \rangle_{t, J_t}^{c} = \, & G_t^{\varphi \varphi} [ \varphi_t ] \, ,	\vphantom{\bigg(\bigg)}
			\\
			\langle \phi^2 \rangle_{t, J_t} = \, & G_t^{\varphi \varphi} [ \varphi_t ] + \varphi_t^2 \, ,	\vphantom{\bigg(\bigg)}
			\\
			\langle \phi^3 \rangle_{t, J_t}^{c} = \, & - \big( G_t^{\varphi \varphi} [ \varphi_t ] \big)^3 \, \Gamma^{(3)}_{t , \varphi_t \varphi_t \varphi_t} [ \varphi_t ] \, ,	\vphantom{\bigg(\bigg)}
			\\
			\langle \phi^3 \rangle_{t, J_t} = \, & - \big( G_t^{\varphi \varphi} [ \varphi_t ] \big)^3 \, \Gamma^{(3)}_{t , \varphi_t \varphi_t \varphi_t} [ \varphi_t ] +	\vphantom{\bigg(\bigg)}
			\\
			& + 3 \, G_t^{\varphi \varphi} [\varphi_t] \varphi_t + \varphi_t^3 \, .	\vphantom{\bigg(\bigg)}	\nonumber
		\end{align}
	We will need these relations among the different $n$-point correlation functions to compare our numerical results from solving the RG flow equation with fluid-dynamical methods to the direct computation of the correlation functions from the partition function $\mathcal{Z} [ J ]$.

\section{The \texorpdfstring{$O(N)$}{O(N)} model in zero dimensions and its treatment within the FRG}
\label{sec:zero-dimensional_o(n)-model}

	Zero-dimensional $O(N)$ models are predominantly studied for pedagogical and conceptual purposes \cite{Bessis:1980ss,Zinn-Justin:1998hwu,DiVecchia:1990ce,Hikami:1978ya,Nishigaki:1990sk,Schelstraete:1994sc,Catalano:2019,Fl_rchinger_2010,Keitel:2011pn,SkinnerScript,Moroz:2011thesis,Pawlowski:talk,Strocchi:2013awa,Kemler:2013yka,Rosa:2016czs,Millington:2019nkw,Millington:2020Talk,Millington:2021ftp}.
	In Ref.~\cite{Keitel:2011pn} the model was used to compare the quality of perturbation theory, the large-$N$ expansion, and the FRG vertex/Taylor expansion with the exact result.
	The primary focus of the present work is to push this analysis even further and to study the limits of untruncated RG flow equations as well as the FRG Taylor expansion.
	
	$O(N)$ models in higher dimensions play an important role in understanding spin systems, like the Ising model \cite{Ising:1925em,Canet:2003qd,Delamotte:2007pf}, and magnetization phenomena.
	Furthermore, they are often used as toy models and are of utmost importance for understanding the Anderson-Brout-Englert-Guralnik-Hagen-Higgs-Kibble mechanism and the formation of a chiral condensate in strong-interaction matter.
	In the context of numerical methods for the FRG, two of us used the $O(N)$ model in $d = 3$ to study numerical solutions of RG flow equations in the large-$N$ limit \cite{Grossi:2019urj}.
	
	This section is structured as follows: In Sub.Sec.~\ref{subsec:ON_model} we introduce the $O(N)$ model on the level of the classical action and the functional integral.
	We further comment on the calculation of expectation values and 1PI vertex functions from the functional integral, which are our observables of interest.
	Thereafter, in Sub.Sec.~\ref{subsec:symmetry_restoration}, we comment on symmetry restoration during the RG flow, for scenarios in which the classical action $\mathcal{S} [ \vec{\varphi} \, ] = U ( t_\mathrm{UV}, \vec{\varphi} \, )$ possesses a non-trivial minimum.
	In Sub.Sec.~\ref{subsec:FRG-formulationONmodel}, we introduce the exact FRG formulation of the model, which includes the derivation of the RG flow equation as an exact PDE and generalization of Eq.~\eqref{eq:pde_gamma}.
	We close this section by deriving the FRG Taylor expansion for the $O(N)$ model, which is a commonly used expansion scheme in FRG studies.

\subsection{The zero-dimensional \texorpdfstring{$O(N)$}{O(N)} model}
\label{subsec:ON_model}

	Consider a zero-dimensional theory of $N$ bosonic scalars $\phi_a$, which transform according to
		\begin{align}
			\phi_a \mapsto \phi^\prime_a = O_{a b} \, \phi_b \, ,
		\end{align}
	where $O \in O(N)$ and $a, b \in \{ 1, \, \ldots , \, N \}$.
	In vector notation, this reads
		\begin{align}
			\vec{\phi} \mapsto \vec{\phi}^{\, \prime} = O \, \vec{\phi} \, ,
		\end{align}
	where $\vec{\phi} = ( \phi_1 , \, \phi_2 , \, \ldots, \, \phi_N )$.
	If the action $\mathcal{S} [ \vec{\phi} \, ]$ of the model possesses an $O(N)$ symmetry, it can contain all possible terms that are functions of the $O(N)$ invariant
		\begin{align}
			\rho \equiv \tfrac{1}{2} \, \phi_a \, \phi_a \equiv \tfrac{1}{2} \, \vec{\phi}^{\, 2}	\label{eq:definition_so_3_invariant}
		\end{align}
	This implies that the most general action obeying this symmetry is given by
		\begin{align}
			\mathcal{S}[ \vec{\phi} \, ] = U ( \vec{\phi} \, ) = U ( \rho ) \, ,
		\end{align}
	where $U ( \rho )$ is the effective potential, in analogy to models from higher-dimensional space-times.
	This effective potential might for example include a bosonic ``mass term'' $m^2 \rho$ as well as other interaction terms containing arbitrary powers of $\rho$.
	Although one may now be tempted to assume that the effective potential $U ( \rho )$ must be a power series or an analytic function of $\rho$, as long as it fulfills all symmetries it can be any continuous function of $\rho$ which is bounded from below, \textit{cf.}\ the discussion in Sec.~\ref{sec:frg_introduction} for the special case of the $O ( 1 )$ model.\\
	
	In the remainder of this section we will summarize relevant relations for the $O(N)$ model.
	For a more detailed discussion, we refer the interested reader to Ref.~\cite{Keitel:2011pn} and references therein.\\
	
	All generating functionals of the theory retain the $O(N)$ symmetry of the action, which makes them functionals of the invariants $\tfrac{1}{2} \, \vec{J}^{\, 2}$ for $\mathcal{Z}$ and $\mathcal{W}$ and $\varrho \equiv \frac{1}{2} \, \vec{\varphi}^{\, 2}$ for $\Gamma$.
	This entails that all $n$-point correlation functions for odd $n$ vanish by symmetry and all $n$-point correlation functions of a given order of even $n$ are proportional to each other, \textit{e.g.}, for the four-point function we find
		\begin{align}
			\langle \phi_i \, \phi_i \, \phi_j \, \phi_j \rangle = \tfrac{1}{3} \, \langle \phi_i \, \phi_i \, \phi_i \, \phi_i \rangle\, ,
		\end{align}
	for $i\neq j$ and $i, j \in \{ 1, \ldots, N \}$ (no summation over repeated indices implied here).
	For the proof, use that $\tfrac{\delta}{\delta J_i} \mathcal{Z} \big( \tfrac{1}{2} \vec{J}^{\, 2} \big) = J_i \, \mathcal{Z}^\prime \big( \tfrac{1}{2} \, \vec{J}^{\, 2} \big)$ and set the source $\vec{J} = 0$ at the end of the calculation.
	Using the $O(N)$ symmetry on the right-hand side of
		\begin{align}
			& \langle \phi_{i_1} \, \cdots \, \phi_{i_n} \rangle =	\vphantom{\bigg(\bigg)}	\label{eq:def_corr_func_ON}
			\\
			= \, & \frac{1}{\mathcal{Z} [0]} \, \int_{-\infty}^{\infty} \mathrm{d}^N \phi \, \phi_{i_1} \, \cdots \, \phi_{i_n} \, \mathrm{e}^{- U ( \vec{\phi}^{\, 2}/2 ) } \, ,	\vphantom{\bigg(\bigg)}	\nonumber
		\end{align}
	one can relate correlation functions of even order $2n$ to the expectation value $\langle ( \vec{\phi}^{\, 2} )^n \rangle$.
	For the two-, four-, and six-point functions, which are studied in this work, we find
	\begin{widetext}
		\begin{align}
			\langle \phi_i \, \phi_j \, \rangle = \, & \frac{1}{N} \, \delta_{ij} \, \langle \vec{\phi}^{\, 2} \rangle \, ,	\vphantom{\bigg(\bigg)}	\label{eq:on-model_relation_2pfi}
			\\
			\langle \phi_i \, \phi_j \, \phi_k \, \phi_l \rangle = \, & \frac{1}{N ( N + 2 )} \, ( \delta_{ij} \, \delta_{kl} + \delta_{ik} \, \delta_{jl} + \delta_{il} \, \delta_{jk} ) \, \langle ( \vec{\phi}^{\, 2} )^2 \rangle \, ,	\vphantom{\bigg(\bigg)}	\label{eq:on-model_relation_4pfi}
			\\
			\langle \phi_i \, \phi_j \, \phi_k \, \phi_l \, \phi_m \, \phi_n \rangle = \, & \frac{1}{N ( N + 2 ) ( N + 4)} \, ( \delta_{ij} \, \delta_{kl} \, \delta_{mn} + \mathrm{all\ permutations} ) \, \langle ( \vec{\phi}^{\, 2} )^3 \rangle \, .	\vphantom{\bigg(\bigg)}	\label{eq:on-model_relation_6pfi}
		\end{align}
	Connected correlation functions and 1PI vertex functions are related to correlation functions as outlined in Sub.Sec.~\ref{subsec:n_pt_functions}.
	Using the fact that, for odd $n$, all $n$-point correlation functions and all $n$-point 1PI vertex functions vanish by symmetry, the following relations hold for the two-, four-, and six-point functions (no summation over repeated indices):
		\begin{align}
			\langle \phi_i \, \phi_i \rangle^{c} = \, & \langle \phi_i \, \phi_i \rangle = \big( \Gamma^{(2)}_{\varphi_i\varphi_i} \big)^{-1} \, ,	\vphantom{\bigg(\bigg)}	\label{eq:on-model_relation_2pf}
			\\
			\langle \phi_i \, \phi_i \, \phi_i \, \phi_i \rangle^{c} = \, & \langle \phi_i \, \phi_i \, \phi_i \, \phi_i \rangle - 3 \, \langle \phi_i \, \phi_i \rangle^2 = - \langle \phi_i \phi_i \rangle^4 \, \Gamma^{(4)}_{\varphi_i \varphi_i \varphi_i \varphi_i} \, ,	\vphantom{\bigg(\bigg)}	\label{eq:on-model_relation_4pf}
			\\
			\langle \phi_i \, \phi_i \, \phi_i \, \phi_i \, \phi_i \, \phi_i \rangle^{c} = \, & \langle \phi_i \, \phi_i \, \phi_i \, \phi_i \, \phi_i \, \phi_i \rangle -	15 \, \langle \phi_i \, \phi_i \, \phi_i \, \phi_i \rangle \, \langle \phi_i \, \phi_i \rangle + 30 \, \langle \phi_i \, \phi_i \rangle^3 = \vphantom{\bigg(\bigg)}	\label{eq:on-model_relation_6pf}
			\\
			= \, & - \langle \phi_i \, \phi_i \rangle^6 \, \Gamma^{(6)}_{\varphi_i \varphi_i \varphi_i \varphi_i \varphi_i \varphi_i} + 10 \, \langle \phi_i \, \phi_i \rangle^{-1} \, ( \langle \phi_i \, \phi_i \, \phi_i \, \phi_i \rangle^{c} )^2 \, . \vphantom{\bigg(\bigg)}	\nonumber
		\end{align}
	Inserting Eqs.~\eqref{eq:on-model_relation_2pfi} -- \eqref{eq:on-model_relation_6pfi} into Eqs.~\eqref{eq:on-model_relation_2pf} -- \eqref{eq:on-model_relation_6pf} and solving for the 1PI vertex functions yields
		\begin{align}
			\Gamma^{(2)} \equiv \, & \Gamma^{(2)}_{\varphi_i \varphi_i} = N \, \frac{1}{ \langle \vec{\phi}^{\, 2} \rangle} \, ,	\label{eq:on-model_relation_2pf_phi2}	\vphantom{\bigg(\bigg)}
			\\
			\Gamma^{(4)} \equiv \, & \Gamma^{(4)}_{\varphi_i \varphi_i \varphi_i \varphi_i} = 3 \, N^2 \, \frac{1}{\langle \vec{\phi}^{\, 2} \rangle^2} \, \bigg[ 1 - \frac{N}{N + 2} \, \frac{\langle ( \vec{\phi}^{\, 2} )^2 \rangle}{\langle \vec{\phi}^{\, 2} \rangle^2} \bigg] \, ,	\label{eq:on-model_relation_4pf_phi2}	\vphantom{\bigg(\bigg)}
			\\
			\Gamma^{(6)} \equiv \, & \Gamma^{(6)}_{\varphi_i \ldots \varphi_i} = 60 \, N^3 \, \frac{1}{ \langle \vec{\phi}^{\, 2} \rangle^3} \bigg[ 1 - \frac{9 \, N}{ 4 \, ( N + 2 )} \, \frac{\langle ( \vec{\phi}^{\, 2} )^2 \rangle}{\langle \vec{\phi}^{\, 2} \rangle^2} + \frac{3 \, N^2}{2 \, ( N + 2 )^2} \, \frac{\langle ( \vec{\phi}^{\, 2} )^2 \rangle^2}{\langle \vec{\phi}^{\, 2} \rangle^4} - \frac{N^2}{4 \, ( N + 2 ) \, ( N + 4 )} \, \frac{\langle ( \vec{\phi}^{\, 2} )^3 \rangle}{\langle \vec{\phi}^{\, 2} \rangle^3} \bigg] \, .	\label{eq:on-model_relation_6pf_phi2}	\vphantom{\bigg(\bigg)}
		\end{align}
\end{widetext}
	In summary, computing arbitrary correlation functions (or 1PI vertex functions) of the zero-dimensional $O(N)$ model boils down to computing expectation values $\langle ( \vec{\phi}^{\, 2} )^n \rangle$.
	The latter can be computed using Eq.~\eqref{eq:def_corr_func_ON}.
	Because of the $O(N)$ symmetry of the integrand, this is most easily done in spherical coordinates.
	Performing the integration over spherical coordinates, we have
		\begin{align}
			\int_{-\infty}^{\infty} \mathrm{d} \phi_1 \cdots \int_{-\infty}^{\infty} \mathrm{d} \phi_N 	
		= \frac{2 \, \pi^{\frac{N}{2}}}{\Gamma \big( \frac{N}{2} \big)} \int_{0}^{\infty} \mathrm{d} \rho \, ( 2 \rho )^{\frac{N}{2} - 1} \, ,
		\end{align}
	Then the expectation value is a simple one-dimensional integral,
		\begin{align}
			\langle ( \vec{\phi}^{\, 2} )^n \rangle = \frac{2^n \int_0^\infty \mathrm{d}\rho \, \rho^{\frac{N}{2} - 1} \,  \rho^n \, \mathrm{e}^{-U ( \rho )}}{\int_0^\infty \mathrm{d} \rho \, \rho^{\frac{N}{2} - 1} \, \mathrm{e}^{- U ( \rho )}} \, .	\label{eq:ON_expectation_value}
		\end{align}
	For certain potentials $U(\rho)$, the integral \eqref{eq:ON_expectation_value} can even be computed symbolically in terms of known functions \cite{Keitel:2011pn,Kemler:2013yka,Steil:2021cbu}, whereas for general $U ( \rho )$ a numerical evaluation to high precision is straightforward using standard methods \cite{Press:1992zz,PresTeukVettFlan92}.
	Thus, the zero-dimensional $O(N)$ model is an ideal testing ground for alternative methods to calculate correlation functions, such as, \textit{e.g.}, the FRG.

\subsection{Symmetry restoration during the RG flow}
\label{subsec:symmetry_restoration}

	Besides being invariant under $O(N)$ transformations the classical action (potential) $\mathcal{S} [ \vec{\phi} \, ] = U ( \vec{\phi} \, )$ is also invariant under the discrete $\mathbb{Z}_2$ transformation
		\begin{align}
			\phi_a \rightarrow - \phi_a \, ,
		\end{align}
	which, as already mentioned above, implies that all $n$-point functions with odd $n$ vanish, \textit{e.g.}, the one-point function $\varphi_a=\langle\phi_a\rangle=0$.
	
	However, it is possible to consider actions (potentials) $\mathcal{S} [ \rho ] = U ( \rho )$ which possess non-trivial minima $\rho_0 \neq 0$.
	This means that the RG flow of $\bar{\Gamma}_t [ \vec{\varphi} \, ]$ of such models is initialized in a symmetry-broken regime in the UV, where the $O(N)$ symmetry is broken to its $O ( N - 1 )$ subgroup.
	(For the $O ( 1 )$ model, this reduces to a breaking of the $\mathbb{Z}_2$ symmetry.)
	Following the discussion in App.~\ref{app:mermin_wagner}, this property of the classical action neither translates to the full quantum effective action $\Gamma [ \vec{\varphi} \, ]$ in the IR nor to the $n$-point functions, due to a limiting case of the Coleman-Mermin-Wagner-Hohenberg theorem \cite{Mermin:1966,Hohenberg:1967,Coleman:1973ci}.
	The theorem states that there is no long-range order in $d \leq 2$ dimensions if the interactions between the constituents are sufficiently short of range.
	Therefore, there is no breaking of a (continuous) symmetry in such systems in the IR, \textit{i.e.}, after integrating out all quantum fluctuations, even when starting with a classical action in the UV that has non-trivial minima.
	This is the equivalent of the statement that $\varphi_a = \langle \phi_a \rangle = 0$.
	The ``Nambu-Goldstone modes'' \cite{Nambu:1960tm,Goldstone:1961eq,Goldstone:1962es}%
	\footnote{%
		We put the term ``Nambu-Goldstone modes'' in quotation marks, because in zero dimensions the concept of ``massless modes'' can only refer to the curvature masses in the corresponding bosonic field direction, which are obtained from the effective potential $U(\rho)$.
		But the actual particle masses in a higher-dimensional QFT are derived from the poles of the real-time propagators, which simply do not exist in zero dimensions.
	}%
	, which we will also call pions%
	\footnote{%
		We adopt the high-energy terminology. Condensed-matter physicists  associate the pions with quasiparticles -- the Anderson-Bogoliubov modes.%
	}	$\vec{\pi}$ in the zero-dimensional $O(N)$ model, and the radial $\sigma$ mode ``vaporize" any condensate and smear out all cusps in $\bar{\Gamma}_t [ \vec{\varphi} \, ]$ during the RG flow.
	In the IR all modes are then ``massive'' again.
	
	There are two reasons, why this feature of symmetry restoration on the level of $\bar{\Gamma}_t [ \vec{\varphi} \, ]$ is desirable for our numerical tests:
		\begin{enumerate}
			\item	Symmetry breaking/restoration associated with condensation/``vaporization" is an essential property of all kinds of QFTs \cite{Weinberg:1996kr,Peskin:1995ev,ZinnJustin:2002ru} and we have to show that it is correctly captured by our numerical tools.
			This is especially important, because it was shown by two of us and collaborators \cite{Grossi:2019urj,Grossi:2021ksl} that non-analytic behavior in the effective potential $U ( t, \vec{\varphi} \, )$, \textit{cf.}\ Refs.~\cite{Bonanno:2004pq,Aoki:2017rjl}, which is directly associated with dynamical symmetry breaking/restoration, is realized as shock and rarefaction waves in field space during the RG flow.
			
			\item	The possibility of dynamical symmetry restoration on the level of $\bar{\Gamma}_t [ \vec{\varphi} \, ]$ is also a desired feature in order to demonstrate that it is of utmost importance to choose the UV cutoff $\Lambda$ and the IR cutoff $r_\mathrm{IR}$ as well as initial and boundary conditions in numerical FRG-flow calculations carefully.
			For our example it is expected that if the IR cutoff time $t_\mathrm{IR}$ is chosen too small, such that the regulator $r ( t )$ is still too large, the system might still be in the symmetry-broken phase (indicated by a non-trivial minimum).
			This means that the scale-dependent effective average action $\bar{\Gamma}_{t_\mathrm{IR}} [ \vec{\varphi} \, ]$ at this RG scale cannot be interpreted as the full quantum effective action $\Gamma [ \vec{\varphi} \, ]$, because the Coleman-Mermin-Wagner-Hohenberg theorem is still violated.
			The same applies to a problematic implementation boundary conditions, especially at $\varrho =  0$, which can lead to a violation of the Coleman-Mermin-Wagner-Hohenberg theorem, such that the system is not in the restored phase in the IR.
			
			For a direct physical consequences of these subtleties, we refer to the parallel works \cite{Stoll:2021ori,Steil:2021cbu} by two of us and collaborators.
		\end{enumerate}
	In a follow-up publication \cite{Steil:partIV}, we will generalize the zero-dimensional $O(N)$ model to a model involving fermions (Grassmann numbers) and bosons.
	The more complicated interactions may also allow for dynamical symmetry breaking via attractive fermion interactions during the RG flow.
	Of course, the system must return to the restored phase in the limit~$t\to\infty$.

\subsection{FRG formulation and flow equations}
\label{subsec:FRG-formulationONmodel}

	This subsection is dedicated to the FRG formulation of the $O(N)$ model of the previous Sub.Sec.~\ref{subsec:ON_model}.
	To this end, we demonstrate how to arrive at the exact untruncated RG flow equation of the $O(N)$ model.
	Furthermore, we introduce a commonly used truncation scheme for RG flow equations -- the FRG Taylor expansion, see, \textit{e.g.}, Refs.~\cite{Berges:2000ew,Gies:2006wv,Delamotte:2007pf,Kopietz:2010zz,Keitel:2011pn,Rosten:2010vm,Dupuis:2020fhh}.
	We start our discussion with general remarks on the derivation of RG flow equations and truncation schemes.\\
	
	From Sec.~\ref{sec:frg_introduction} and especially Sub.Secs.~\ref{subsec:exact_rg_equation} and \ref{subsec:interpretation_frg} we have learned that the FRG equation \eqref{eq:wetterich_general} constitutes an exact PDE for the RG time evolution of the full field-dependent effective average action $\bar{\Gamma}_t [ \Phi ]$ with initial condition $\bar{\Gamma}_{t = 0} [ \Phi ] = \mathcal{S} [ \Phi ]$.
	Here, $\Phi$ stands for the field space vector of all fields of the specific model under consideration.
	However, if there is more than one field space degree of freedom, the direct (numerical) solution of the FRG equation \eqref{eq:wetterich_general} as a PDE is exceedingly difficult, because of the high dimensionality of the field space.
	In higher space-time dimensions, space-time or momentum dependences of the fields complicate this issue and promote Eq.~\eqref{eq:wetterich_general} to a functional integro-partial-differential equation with a functional $\mathcal{S} [ \Phi (x) ]$ or $\mathcal{S} [ \tilde{\Phi} (p) ]$ as initial condition.
	
	Instead of solving Eq.~\eqref{eq:wetterich_general} directly (independent of the dimensionality and the field content), one usually specifies some ansatz function for the effective average action $\bar{\Gamma}_t [ \Phi ]$, which involves only a finite number of $t$ dependent couplings (vertices).
	The ansatz function for $\bar{\Gamma}_t [ \Phi ]$ must respect all symmetries of the model and the functional integral.
	Afterwards, one works out a projection prescription, which extracts these couplings from $\bar{\Gamma}_t [ \Phi ]$.
	Usually this is done by	
		\begin{enumerate}
			\item	Taking a suitable number of (functional) derivatives in field (and/or momentum) space,
			
			\item	Evaluating the resulting expression on a specific (usually constant) field configuration (and/or at specific external momenta, energies \textit{etc.}),
			
			\item	Applying contractions of open field space and space-time indices with suitable tensors.
		\end{enumerate}
	Thus, inserting the ansatz for $\bar{\Gamma}_t [ \Phi ]$ into the FRG equation \eqref{eq:wetterich_general} and applying these projection rules to both sides of the equation yields a coupled set of PDEs and/or ODEs for the couplings.
	This system of differential equations must be initialized at $t = 0$ with the values of the couplings taken from the specific choice of the classical action $\mathcal{S} [ \Phi ]$.
	The system for the $t$ dependent couplings is then evolved to $t \rightarrow \infty$.
	If needed, the values of the couplings at $t \rightarrow \infty$ can afterwards be reinserted in $\bar{\Gamma}_t [ \Phi ]$ to obtain the effective action $\Gamma [ \Phi ]$ in the IR.
	We will present this procedure explicitly for the zero-dimensional $O(N)$ model in the next paragraphs.
	
	However, by considering an ansatz function for $\bar{\Gamma}_t [ \Phi ]$, which consists of a finite number (of usually an infinite set) of all the possible interaction terms that respect the symmetries of the system, one effectively introduces an approximation.
	In the context of the FRG this is called a \textit{truncation}.
	The concept of a truncation of the system can directly be seen from Eq.~\eqref{eq:wetterich_general}:
	Taking an appropriate number of field space derivatives of this equation to project on a specific coupling, the right-hand side of this equation depends on higher-order interaction vertices.
	These are up to two orders higher than the ones on the left-hand side, because of $\bar{\Gamma}_t^{(2)} [ \Phi ]$ already involves two field space derivatives.
	The highest-order couplings in the system of PDEs for the couplings are, however, set to zero by definition via the ansatz for $\bar{\Gamma}_t [ \Phi ]$, because only a finite number of couplings is evolved with $t$.
	As a result Eq.~\eqref{eq:wetterich_general}, which originally corresponds to a coupled system of infinitely many ODEs and PDEs for couplings of all orders in field and momentum or position space, is reduced to a finite set of PDEs for the couplings involved in the ansatz for $\bar{\Gamma}_{t} [ \Phi ]$, see Refs.\ \cite{Berges:2000ew,Gies:2006wv,Delamotte:2007pf,Kopietz:2010zz,Rosten:2010vm,Dupuis:2020fhh} for general discussions or, \textit{e.g.}, Refs.~\cite{Braun:2017srn,Braun:2018bik,Braun:2019aow,Eser:2018jqo,Cichutek:2020bli} for specific applications.
	After all, the quality of the ansatz completely determines the quality of the approximation to the actual IR effective action $\Gamma [ \Phi ]$ after the RG flow of the truncated system is solved.
	
	In general, finding reliable truncations for a given problem is a challenging problem.
	In particular, the identification of a small parameter to justify the truncations is a difficult task.
	In fact, such a parameter may not even exist.
	It may also turn out that a given truncation yields reliable results for one observable but not for another.
	The latter observation may even be considered a feature as it allows to identify mechanisms underlying specific phenomena.
	In any case, there are construction schemes for systematic ans\"atze for the effective action.
	Commonly used truncation schemes are for example the derivative expansion \cite{Berges:2000ew,Canet:2002gs,Canet:2003qd,Balog:2019rrg}, which relies on the expansion of $\bar{\Gamma}_t [ \Phi ]$ in powers of derivatives (momenta) but includes all orders of field-dependent vertices at the same momentum order.
	Another expansion scheme is the vertex expansion, which expands $\bar{\Gamma}_t [ \Phi ]$ in terms of (momentum-dependent) $n$-point functions.
	Oftentimes different expansion schemes are combined, in order to keep the system of PDEs tractable \cite{Eser:2018jqo,Divotgey:2019xea,Eser:2019pvd,Cichutek:2020bli}.
	Moreover, truncations can always be benchmarked against perturbative studies, see, \textit{e.g.}, Refs.~\cite{Papenbrock:1994kf,Gies:2006wv} for instructive examples.
	
	One measure for the quality of these expansion schemes is comparing terms of different order.
	It is expected and can also be observed for certain systems and situations, see \textit{e.g.}, Refs.~\cite{Tetradis:1993ts,Papp:1999he,Bonanno:2000yp,Bervillier:2007rc,Litim:2010tt,Pawlowski:2014zaa,Cichutek:2020bli}, that the expansions seem to converge and deviations in the observables are decreasing by increasing the expansion order.
	In the FRG community, this is often referred to as \textit{apparent convergence}.
	Another indication for the quality of the truncation is the comparison of FRG results with results from other methods \cite{Marko:2019rsl,Caillol:2012zz,Bonanno:2004pq,Pangon:2009pj,Pangon:2010uf,Fischer:2006vf}, \textit{e.g.}, Monte-Carlo simulations, or the comparison of critical exponents derived from the FRG and other methods.
	
	In this context, zero-dimensional QFTs play a very special role: Due to the absence of space-time and momentum dependences of the fields, the effective average action $\bar{\Gamma}_t [ \Phi ] = \bar{\Gamma}_t ( \Phi )$ is merely a function (not a functional)  of the fields $\Phi$ and of the $t$ dependent couplings accompanying all possible terms which respect the symmetry of the model.
	This structure can, however, be summarized in terms of effective $\Phi$ and $t$ dependent terms.
	It is therefore possible to express the effective average action in terms of a finite amount of terms, which nevertheless incorporate all possible interactions to all orders in the fields and do not even need to be analytic functions of the fields.
	In consequence, truncating the system is superfluous and the PDEs, which are derived via projections from the FRG equation, constitute an exact and complete system.
	Solving this system must therefore lead to the exact effective action $\Gamma [ \Phi ]$ in the IR and is therefore completely equivalent to solving the functional integral.
	In other words, calculating $n$-point correlation functions via the (functional) integral or via the FRG equation (if done properly) must yield identical results without truncation errors.
	
	This feature makes zero-dimensional QFT particularly interesting for several reasons:
		\begin{enumerate}
			\item	It can be used to test the quality of numerical schemes which are used to solve the flow equations.
			
			\item	It can be used to estimate the errors resulting from the choices of various parameters entering the RG flow equations like UV and IR cutoff scales, \textit{etc.}.
			
			\item	It can be used to test commonly used truncation schemes by artificially truncating the system to a non-complete set of ordinary first-order differential equations.
		\end{enumerate}
	All these tests can be performed on a quantitative level, by studying the relative errors of the FRG results for $n$-point correlation functions compared to the exact results from the functional integral.
	We provide results for various precision tests in Sec.~\ref{sec:results}.
	
	For the remainder of this section, we will proceed as follows: First, we will derive the untruncated exact RG flow equation for the zero-dimensional $O(N)$ model.
	Afterwards, we introduce a commonly used truncation scheme -- the FRG Taylor (vertex) expansion.

\subsubsection{The exact RG flow equation of the zero-dimensional \texorpdfstring{$O(N)$}{O(N)} model}
\label{subsubsec:exact_flow_equation_potential}

	For the special case of the zero-dimensional $O(N)$ model, the most general ansatz for the effective average action is given by a scale-dependent effective potential
		\begin{align}
			\bar{\Gamma}_t [ \vec{\varphi} \, ] = U ( t, \vec{\varphi} \, ) = U ( t, \varrho )  \, .	\label{eq:ansatz_effective_average_action}
		\end{align}
	This ansatz can describe arbitrary $O(N)$ invariant effective actions and can include terms at all orders of $\varrho = \tfrac{1}{2} \, \vec{\varphi}^{\, 2}$.
	However, it is in principle not restricted to analytic (Taylor-expandable) functions.
	Truncations of $\bar{\Gamma}_t [ \vec{\varphi} \, ]$ are not required.
	
	In order to arrive at the exact flow equation for $U ( t, \vec{\varphi} \, )$ one has to perform the following steps:
		\begin{enumerate}
			\item	Insert the function \eqref{eq:ansatz_effective_average_action} into the FRG equation \eqref{eq:wetterich_general}.
			
			\item	Invert the full field-dependent two-point function
				\begin{align}
					\big( \bar{\Gamma}^{(2)}_{t , \varphi \varphi} [ \vec{\varphi} \, ] + R_t \big)_{i j} \, .	\label{eq:full_two-point_function}
				\end{align}
			
			\item	Take the trace in field space.
			
			\item	Remove the redundant $N - 1$ field space directions in $\vec{\varphi}$.
		\end{enumerate}
	For the last step, the RG flow equation can be evaluated on a constant background field configuration%
	\footnote{%
		Here we adopt terminology from higher-dimensional FRG: The word ``constant'' is therefore somewhat misleading in a QFT which cannot vary in space-time, but it is used anyhow.
	}
	$\varphi_1 = \ldots = \varphi_{N - 1} = 0$ and $\varphi_N = \sigma$.
	Without loss of generality, the $\varphi_N$ direction was singled out as the direction of the radial $\sigma$ mode and the constant background field.
	
	The inversion of the full field-dependent two-point function \eqref{eq:full_two-point_function} can be performed analytically \cite{Tetradis:1995br,Wetterich:1991be,PawlowskiScript,Delamotte:2007pf} by introducing the complete, orthogonal, and idempotent field space projection operators
		\begin{align}
			&	\mathcal{P}^\perp_{ij} ( \vec{\varphi} \, ) \equiv \delta_{ij} - \frac{\varphi_i\, \varphi_j}{\vec{\varphi}^{\, 2}} \, ,	&&	\mathcal{P}^\parallel_{ij} ( \vec{\varphi} \, ) \equiv \frac{\varphi_i\, \varphi_j}{\vec{\varphi}^{\, 2}} \, .	\label{eq:field_space_projection_operators}
		\end{align}
	The projection operators are used to decompose the full field-dependent two-point function \eqref{eq:full_two-point_function} into components perpendicular ($\perp$) and parallel ($\parallel$) to $\vec{\varphi}$, which can be inverted separately.
	The regulator $R_t$ is matrix-valued and diagonal in field space,
		\begin{align}
			( R_t )_{ij} = \delta_{ij} \, r ( t ) \, ,
		\end{align}
	where $r ( t )$ again is denoted as \textit{regulator shape function}, \textit{cf.}\ Eqs.~\eqref{eq:regulator_insertion} and \eqref{eq:exponential_regulator}.
	One finds that
		\begin{align}
			& \big( \bar{\Gamma}^{(2)}_{t,\varphi \varphi} [ \vec{\varphi} \, ] + R_{t} \big)^{-1}_{i j} =	\vphantom{\bigg(\bigg)}
			\\
			= \, &  \mathcal{P}^\parallel_{ij} ( \vec{\varphi} \, ) \, \frac{1}{ r ( t ) + \partial_\varrho U ( t, \varrho ) + 2 \varrho \, \partial_\varrho^2 U ( t, \varrho ) } +	\vphantom{\bigg(\bigg)}	\nonumber
			\\
			& + \mathcal{P}^\perp_{ij} ( \vec{\varphi} \, ) \, \frac{1}{ r(t) + \partial_\varrho U ( t, \varrho ) } \, ,	\vphantom{\bigg(\bigg)}	\nonumber
		\end{align}
	which can be inserted directly into the FRG equation \eqref{eq:wetterich_general}.
	
	After taking the field space trace and evaluating the resulting equation on the constant background field configuration, we arrive at the RG flow equation for the effective potential
		\begin{align}
			\partial_t U ( t, \sigma ) = \, & \big[ \tfrac{1}{2} \, \partial_t r ( t ) \big] \, \frac{N - 1}{r ( t ) + \frac{1}{\sigma} \, \partial_\sigma U ( t, \sigma )} +	\vphantom{\Bigg(\Bigg)}	\label{eq:flow_equation_effective_potential}
			\\
			& + \big[ \tfrac{1}{2} \, \partial_t r ( t ) \big] \,  \frac{1}{r ( t ) + \partial_\sigma^2 U ( t, \sigma )} = \vphantom{\Bigg(\Bigg)}	\nonumber
			\\
			= \, &
			\begin{gathered}
				\includegraphics{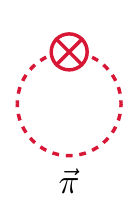}
			\end{gathered}
			+
			\begin{gathered}
				\includegraphics{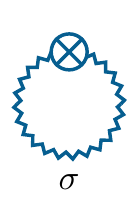}
			\end{gathered} \, .	\nonumber
		\end{align}
	This RG flow equation is an exact non-linear PDE for the effective potential $U ( t, \sigma )$, which is of first order in RG time $t$ and of first and second order in the field space direction $\sigma$.
	It also includes an explicit $\sigma$ dependence.
	A detailed analysis of the structure of this PDE, including its relation to conservation equations and fluid dynamics is provided in Sub.Sec.~\ref{subsec:conservative_form_of_frg_flow_equations}.
	
	For now, we conclude this section with a few comments on the widely used diagrammatic notation of the PDE and its relation to the RG flow equation \eqref{eq:pde_gamma} from Sec.~\ref{sec:frg_introduction}:
	Similar to Feynman diagrams which are commonly used in perturbation theory, the propagators%
	\footnote{%
		The term ``propagator'' is of course misleading for a QFT in a single point, where ``propagation'' in the true sense of the word is not possible.
		Nevertheless, we again adopt the notation from higher-dimensional QFT and statistical mechanics.
	}
	are depicted as lines; blue-jagged lines for the $\sigma$ propagator,
		\begin{align}
			\frac{1}{r ( t ) + \partial_\sigma^2 U ( t, \sigma )} \, ,
		\end{align}
	and red-dashed lines for the $\pi$ propagators
		\begin{align}
			\frac{1}{r ( t ) + \frac{1}{\sigma} \, \partial_\sigma U ( t, \sigma )} \, .
		\end{align}
	The crossed circle ($\otimes$) stands for the \textit{regulator insertion} $\frac{1}{2} \, \partial_t r ( t )$.
	(The factor $\frac{1}{2}$ is often not included in the regulator insertion, but written in front of the diagrams.
	See, \textit{e.g.}, Refs.\ \cite{Cyrol:2017ewj,Pawlowski:2014zaa,Pawlowski:2017gxj,Tripolt:2013jra,Gies:2006wv,Delamotte:2007pf,Keitel:2011pn} for different notations.)
	The factor $N - 1$ is the multiplicity of the pion-loop contribution (indicated by the vector over the pion field in the diagram, \textit{cf.} Eq.~\eqref{eq:flow_equation_effective_potential}) and corresponds to the number of pions in the system.
	
	For the special case $N = 1$, the $O(N)$ model reduces to the $O(1)$ model.
	Such a theory of a single scalar field in zero dimensions, was used in the introductory section \ref{sec:frg_introduction} on FRG.
	In this limit, the pion contributions to the flow equation vanish.
	As already stated in Sec.~\ref{sec:frg_introduction}, we find that for non-zero pion contributions ($N > 1$) the flow equation for $U ( t, \sigma )$ acquires a term that is of first order in the spatial derivative, $\partial_\sigma U ( t, \sigma )$, which no longer has diffusive character, but corresponds to advection in field space.
	This is further discussed in Sec.~\ref{sec:frg-flow_equations_and_numerical_fluid_dynamics}.

\subsubsection{FRG Taylor (vertex) expansion of the \texorpdfstring{$O(N)$}{O(N)} model}
\label{subsubsec:vertex_expansion}

	The FRG Taylor (vertex) expansion is based on the assumption that the effective (average) action $\bar{\Gamma}_t [ \vec{\varphi} \, ]$ can be expanded in a series in field space with RG-time dependent expansion coefficients \cite{Berges:2000ew}.
	In zero dimensions, this effectively reduces to an expansion of the effective potential $U ( t, \varrho )$, \textit{cf.}\ Eq.~\eqref{eq:ansatz_effective_average_action}.
	Consequently, it is also equivalent to a Taylor expansion of the effective potential, which is well-known from higher-dimensional truncation schemes \cite{Adams:1995cv,Berges:2000ew,Pawlowski:2017gxj,Pawlowski:2014zaa,Papp:1999he,Schaefer:1999em,Cichutek:2020bli,Eser:2018jqo,Eser:2019pvd,Divotgey:2019xea,Canet:2002gs,Delamotte:2007pf}.
	Throughout this work, we will therefore use the term ``FRG Taylor expansion'' to refer to this approach.
	The RG-scale dependent expansion coefficients $\bar{\Gamma}^{(2n)} ( t )$ correspond directly to the scale-dependent vertex functions $\bar{\Gamma}^{(2n)}_{t,\varphi_i \ldots \varphi_j}$ of the QFT.
	For $d > 0$, these expansion coefficients are usually position or momentum dependent whereas in $d = 0$ the coefficients depend only on the RG time $t$.
	
	The assumption of expandability and thus differentiability significantly restricts the form of the effective action $\bar{\Gamma}_t [ \vec{\varphi} \, ] = U ( t, \vec{\varphi} \, )$, \textit{cf.}\ Refs.~\cite{Pangon:2009pj,Pangon:2010uf}.
	In fact, it neither allows for the formation of any non-analytic behavior throughout the RG flow nor for any non-analytic initial conditions.
	However, non-analytic initial conditions are not forbidden, as we will see in Sec.~\ref{sec:results}.
	Furthermore, it is well known that non-analyticities can (and in some models have to) form in the effective potential during the RG flow \cite{Aoki:2017rjl,Grossi:2019urj,Grossi:2021ksl,Borchardt:2016pif}.
	Considering these caveats, an expansion in vertices of a given theory has always to be considered with care.
	Still, this expansion scheme is used in certain applications.
	
	In our work, we restrict our analysis of the precision of this truncation scheme to RG flows with rather specific properties: We study initial conditions that are analytic.
	Furthermore, we know, \textit{cf.}\ App.~\ref{app:mermin_wagner}, that the IR effective action is smooth for the special case of zero dimensions, which is a necessary condition for the convergence of a (Taylor) series.
	It should, however, be noted that smoothness is only a necessary but not a sufficient condition for the convergence of a Taylor series%
	\footnote{%
		A textbook example for a smooth function which has a non-converging Taylor series around $x = 0$ is $$f ( x ) = \begin{cases} \mathrm{e}^{-1/x} & \text{if } x>0, \\ 0 &\text{else.}\end{cases}$$
	}%
	.
	Only analyticity would formally imply the convergence of a Taylor series at all $\vec{\varphi}$.
	Additionally, we argue that for sufficiently small $N$, the diffusive contributions to the RG flow are important, which smear out any possible cusps.
	In summary, we expect that for these extremely special scenarios it is unlikely that non-analyticities will form and disappear again during the RG flow.
	Nevertheless, we do not know if a small finite number of expansion coefficients is always enough to reach a reliable approximation of $\bar{\Gamma}_t [ \vec{\varphi} \, ]$ during the RG flow or if it is always necessary to flow the effective potential as a PDE without additional assumptions.
	This (rather limited) applicability of the FRG Taylor expansion to analytic initial conditions will be tested by calculating the relative errors of 1PI $n$-point vertex functions in the FRG Taylor expansion in comparison with the exact results and the results from the flows of a full field-dependent $U ( t, \sigma )$ in Sec.~\ref{sec:results}.\\
	
	The FRG Taylor expansion of the zero-dimensional $O(N)$ model is given by the following ansatz \cite{Keitel:2011pn,Moroz:2011thesis,Pawlowski:talk,Kemler:2013yka},
		\begin{align}
			\bar{\Gamma}_t [ \vec{\varphi} \, ] =\, & \sum_{n = 0}^{m} \frac{\bar{\Gamma}^{(2n)}(t)}{(2n - 1)!!}\, \frac{1}{n!}\, \bigg( \frac{\vec{\varphi}^{\, 2}}{2} \bigg)^n = \vphantom{\bigg(\bigg)^2}	\label{eq:vertex_expansion_varphi}
			\\
			=\, & \bar{\Gamma}^{(0)}(t) + \bar{\Gamma}^{(2)}(t)\,  \frac{\vec{\varphi}^{\, 2}}{2} + \frac{\bar{\Gamma}^{(4)}(t)}{3}\, \frac{1}{2}\, \bigg( \frac{\vec{\varphi}^{\, 2}}{2} \bigg)^2 + \ldots \, ,	\vphantom{\bigg(\bigg)^2} \nonumber
		\end{align}
	where $\bar{\Gamma}^{(2n)}(t)$ are $t$ dependent expansion coefficients and $m$ is the truncation order.
	The factors of $(2n - 1)!!$ and $n!$ were introduced in order to have $\bar{\Gamma}^{(2n)}(t_\mathrm{IR}) = \Gamma^{(2n)}_{\varphi_i \ldots \varphi_i}$ in the IR, where $\Gamma^{(2n)}_{\varphi_i \ldots \varphi_i}$ are the 1PI $2n$-point vertex functions in the IR, with all indices being identical (no summation over $i$ here), see also Eqs.~\eqref{eq:on-model_relation_2pf_phi2} -- \eqref{eq:on-model_relation_6pf_phi2}.
	In order to arrive at the corresponding flow equations, we proceed in a similar manner as before in Sub.Sub.Sec.~\ref{subsubsec:exact_flow_equation_potential}:
	We insert our ansatz \eqref{eq:vertex_expansion_varphi} into the full field-dependent two-point function \eqref{eq:full_two-point_function} and use the field space projection operators \eqref{eq:field_space_projection_operators} to invert the latter.
	We obtain
		\begin{align}
			& \big( \bar{\Gamma}^{(2)}_{t,\varphi \varphi} [ \vec{\varphi} \, ] + R_t \big)^{-1}_{i j} = \vphantom{\bigg(\bigg)}
			\\
			=\, & \mathcal{P}^\perp_{ij} ( \vec{\varphi} \, ) \, G^{\pi\pi}_t ( \vec{\varphi} \, ) + \mathcal{P}^\parallel_{ij} ( \vec{\varphi} \, ) \, G^{\sigma\sigma}_t ( \vec{\varphi} \, ) \, , \vphantom{\bigg(\bigg)} \nonumber
		\end{align}
	where
		\begin{align*}
			G^{\pi \pi}_t ( \vec{\varphi} \, ) \equiv\, & \Bigg[ r ( t ) + \sum_{n = 1}^{m+1} \frac{\bar{\Gamma}^{(2n)}(t)}{(2n - 1)!!}\, \frac{1}{(n - 1)!}\, \bigg( \frac{\vec{\varphi}^{\, 2}}{2} \bigg)^{n-1} \Bigg]^{-1} \, ,
			\\
			G^{\sigma \sigma}_t ( \vec{\varphi} \, ) \equiv\, & \Bigg[ r ( t ) + \sum_{n = 1}^{m+1} \frac{\bar{\Gamma}^{(2n)}(t)}{(2n - 3)!!}\, \frac{1}{(n - 1)!}\, \bigg( \frac{\vec{\varphi}^{\, 2}}{2} \bigg)^{n-1} \Bigg]^{-1} \, ,
		\end{align*}
	are the field-dependent propagators of the pion and sigma field in the Taylor expansion.
	
	This result can be inserted into the FRG equation \eqref{eq:wetterich_general}, where the trace in field space is evaluated to
		\begin{align}
			&	\partial_t\, \bar{\Gamma}_t [ \vec{\varphi} \, ] =	\vphantom{\bigg(\bigg)}	\label{eq:wetterich_equation_vertex_expansion}
			\\
			= \, & \big[ \tfrac{1}{2} \, \partial_t r ( t ) \big] \, \big[  ( N - 1 ) \, G^{\pi \pi}_t ( \vec{\varphi} \, ) + G^{\sigma \sigma}_t ( \vec{\varphi} \, ) \big] \, .	\vphantom{\bigg(\bigg)}	\nonumber
		\end{align}
	Finally, we insert the ansatz \eqref{eq:vertex_expansion_varphi} for the effective average action into the left-hand side of this equation and expand the propagators $G^{\circ \circ}_t ( \vec{\varphi} \, )$ up to order $n=m$ in the expansion coefficients $\bar{\Gamma}^{(2n)} ( t )$.
	This can also be achieved by successively taking derivatives with respect to the fields and setting $\vec{\varphi} = 0$ afterwards.
	By comparing the expansion coefficients on the left- and right-hand sides of the equation, one arrives at a coupled set of ordinary differential equations for the $\bar{\Gamma}^{(2n)} ( t )$ with $0 \leq n \leq m$.
	The flow equation for $\bar{\Gamma}^{(2m)} ( t )$ contains $\bar{\Gamma}^{(2m+2)} ( t )$ on the right-hand side.
	We truncate the system by neglecting the flow of $\bar{\Gamma}^{(2m+2)} ( t )$, taking $\partial_t \bar{\Gamma}^{( 2 m + 2 )} ( t ) = 0$.
	
	For an automatization of the derivation of the flow equations (the system of ODEs) via computer algebra routines such as \texttt{Mathematica} \cite{Mathematica:12.1}, it is advisable to formulate the FRG Taylor expansion in the invariant ${\varrho = \tfrac{1}{2} \, \vec{\varphi}^{\, 2}}$,
		\begin{align}
			\bar{\Gamma}_t [ \varrho ] =\, & \sum_{n = 0}^{m} \frac{\bar{\Gamma}^{(2n)} (t)}{(2n - 1)!!}\, \frac{\varrho^n}{n!} \, . \vphantom{\bigg(\bigg)}
		\end{align}
	Equation \eqref{eq:wetterich_equation_vertex_expansion} becomes
		\begin{align}
			\partial_t\, \bar{\Gamma}_t [ \varrho ] =\, & \big[ \tfrac{1}{2} \, \partial_t r ( t ) \big] \, \big[ ( N - 1 ) \, G^{\pi\pi}_t ( \varrho ) + G^{\sigma\sigma}_t ( \varrho ) \big] \, ,
		\end{align}
	while
		\begin{align}
			G^{\pi \pi}_t ( \varrho ) \equiv\, & \bigg[ r ( t ) + \sum_{n = 1}^{m+1} \frac{\bar{\Gamma}^{(2 n)} (t)}{( 2 n - 1)!!}\, \frac{\varrho^{n-1}}{(n - 1)!} \bigg]^{-1} \, ,
			\\
			G^{\sigma \sigma}_t ( \varrho ) \equiv\, & \bigg[ r ( t ) + \sum_{n = 1}^{m+1} \frac{\bar{\Gamma}^{(2 n)} (t)}{( 2 n - 3)!!}\, \frac{\varrho^{n - 1}}{(n - 1)!} \bigg]^{-1} \, .
		\end{align}
	The coupled set of ODEs for the expansion coefficients $\bar{\Gamma}^{(2n)} ( t )$ is given by \cite{Keitel:2011pn,Moroz:2011thesis}%
	\footnote{%
		We do not indicate $t$ dependences of the $\bar{\Gamma}^{(2n)} ( t )$ for reasons of readability.
	}%
	,
		\begin{align}
			\partial_t \bar{\Gamma}^{(0)} = \, & \frac{N}{2} \,\, \frac{\partial_t r ( t )}{r ( t ) + \bar{\Gamma}^{(2)} } \, ,	\vphantom{\Bigg(\Bigg)}	\label{eq:example_vertex_expansion}
			\\
			\partial_t \bar{\Gamma}^{(2)} = \, & - \frac{N + 2}{6} \, \frac{\partial_t r ( t )}{\big[ r ( t ) + \bar{\Gamma}^{(2)} \big]^2}  \, \bar{\Gamma}^{(4)} \, ,	\vphantom{\Bigg(\Bigg)}	\nonumber
			\\
			\partial_t \bar{\Gamma}^{(4)} = \, & \frac{N + 8}{3} \, \frac{\partial_t r ( t )}{\big[ r ( t ) + \bar{\Gamma}^{(2)} \big]^3} \, \big[ \bar{\Gamma}^{(4)} \big]^2 -	\vphantom{\Bigg(\Bigg)}	\nonumber
			\\
			& - \frac{N + 4}{10} \, \frac{\partial_t r ( t )}{\big[ r (t) + \bar{\Gamma}^{(2)} \big]^2} \, \bar{\Gamma}^{(6)} \, ,	\vphantom{\Bigg(\Bigg)}	\nonumber
			\\
			\vdots \, \, &	\vphantom{\Bigg(\Bigg)}	\nonumber
		\end{align}
	Recall that
			\begin{align}
				\partial_t \bar{\Gamma}^{(n)} = 0 \,  
			\end{align}
	for $n \geq 2m+2$ in this approximation.
	
\section{FRG flow equations and (numerical) fluid dynamics}
\label{sec:frg-flow_equations_and_numerical_fluid_dynamics}

	In this section, we discuss the formulation of the RG flow equation as an advection-diffusion equation, as well as its interpretation in the context of fluid dynamics, including its numerical implementation.\\

	The fluid-dynamical formulation of the exact RG flow equation for the effective potential $U ( t, \varrho )$ of models of $O(N)$ type (in the large-$N$ limit \cite{Tetradis:1995br}) is also presented in a recent and a parallel publication by some of us and collaborators \cite{Grossi:2019urj,Grossi:2021ksl}.
	It was shown that the RG flow equation can be recast in the form of a pure advection equation (a hyperbolic conservation law) for the derivative of the effective potential $u ( t, \varrho ) = \partial_\varrho U ( t, \varrho )$, where $u ( t, \varrho )$ serves as the conserved quantity (the fluid), the RG time $t$ as a temporal coordinate and $\varrho$ as a spatial coordinate.
	In this section, we generalize this result and discuss various consequences for the numerical implementation and interpretation of FRG flow equations.%
		\footnote{%
			Generalizations of the fluid-dynamical picture of FRG flow equations from the large-$N$ results of Ref.~\cite{Grossi:2019urj} to systems with finite $N$ as well as the inclusion of fermions were already presented by us in various talks (see, \textit{e.g.}, Refs.~\cite{WinkHirschegg,Koenigstein:2020Talk}) and discussed in a master thesis \cite{Ihssen2020} co-supervised by some of us, as well as a PhD thesis by one of us~\cite{Wink:2020tnu}, see also Ref.~\cite{Grossi:2021ksl}.
			Furthermore, also in Ref.\ \cite{Aoki:2017rjl} a formulation of the flow equation as a conservation law and a discussion of shock waves based on the characteristics is presented, however, without really elaborating on a fluid-dynamical interpretation and its consequences.
		}%

\subsection{Conservative form of FRG flow equations -- advection-diffusion equations}
\label{subsec:conservative_form_of_frg_flow_equations}
	
	The formulation of FRG flow equations in terms of a fluid-dynamical language has two  major advantages:
	\begin{enumerate}
		\item It provides an intuitive explanation for all kinds of phenomena observed in FRG flow equations, \textit{e.g.}, the flattening of the effective potential for small $\sigma$ in the IR, which occurs in conjunction with a non-differentiable point of the effective potential at the ground state.
		Such non-analytic behavior cannot be handled and systematically analyzed by commonly used numerical schemes such as the Taylor expansion or related discretization schemes for the effective potential, since the latter strongly rely on differentiability.
		However, these phenomena have a direct impact on the physics, for instance on the occurrence of phase transitions \cite{Aoki:2017rjl,Bonanno:2004pq,Pangon:2009pj,Pangon:2010uf,Grossi:2019urj,Wipf:2013vp,Ehrenfest1933,Grossi:2021ksl}, and therefore must be resolved and analyzed accurately also on a numerical level.
		
		\item The formulation of the FRG flow equations in terms of fluid-dynamical concepts provides access to the highly developed and extremely powerful toolbox of numerical fluid dynamics, which finds applications in a wide area of fields, ranging from the natural sciences and engineering all the way to economics.
		How to adopt these methods to flow equations arising in the FRG framework is discussed in detail in Sub.Secs.~\ref{subsec:finite_volume_method} and \ref{subsec:kt_scheme}.
	\end{enumerate}	
	Interestingly, the idea of interpreting RG flow equations as ``flow'' equations in the true sense of the word is not new and explains the term ``RG flow equations'': A discussion of analogies between ``RG flow'' and hydro-dynamical flow can be found in widely used textbooks \cite{Peskin:1995ev,Coleman:1985rnk} and is discussed via the example of field-independent coupling constants in the context of perturbative renormalization.
	Furthermore, the RG flow was already associated with gradient flow and dissipative processes in Refs.~\cite{Wallace:1974dx,Wallace:1974dy,Zamolodchikov:1986gt,Zumbach:1994vg,Zumbach:1994kc,Zumbach:1994vg,Rosten:2010vm}, even though a stringent fluid-dynamical interpretation and formulation was not presented. 
	
	It is therefore also not accidental that the (F)RG community has chosen the term ``RG time'' for the logarithm of the RG scale $k$ over the UV cutoff $\Lambda$, $\tilde{t} = \ln \big( \tfrac{k}{\Lambda} \big)$. 
	In contrast, we find that $t = - \tilde{t} \in [ 0, \, \infty )$ can be naturally identified as a temporal coordinate in the fluid-dynamical picture of (F)RG flow equations, see below.
	
	It was also discussed, see, \textit{e.g.}, Refs.~\cite{Rosten:2010vm,Salmhofer:2020Talk,SkinnerScript}, that -- on the level of the scale-dependent generation functionals $\mathcal{Z}_t [ J ]$ or $\mathcal{W}_t [ J ]$ -- the corresponding PDEs can be considered as a (non-linear) functional diffusion equations for the source fields $J$ (\textit{cf.}\ Eqs.~\eqref{eq:polchinski_equation} and \eqref{eq:pde_w} for the respective zero-dimensional versions). 
	Sometimes Eq.~\eqref{eq:polchinski_equation} is even explicitly denoted as a (non-linear) heat equation, which is also a specific fluid-dynamical problem \cite{Cannon:1984,LeVeque:1992,LeVeque:2002,RezzollaZanotti:2013}.
	
	Considering the obvious analogies between RG flow equations arising in the FRG framework and fluid-dynamical equations, it is remarkable that the FRG equation \eqref{eq:wetterich_general} was so far not more systematically investigated and compared to equations well-known from fluid dynamics.
	For the related RG flow equations the situation is slightly different and the mathematical analysis on the level of PDEs was more systematic, see, \textit{e.g.}, Refs.~\cite{Felder:1987,Hasenfratz:1985dm,Zumbach:1994kc,Zumbach:1994vg,Rosten:2010vm}.
	Furthermore, certain phenomena well-known in fluid dynamics, such as discontinuities (shock waves), rarefaction waves, or cusps, occur in the solution of such PDEs.
	These require a careful numerical treatment to resolve them, but their occurrence was very often ignored by numerical approaches to solve the FRG equations by erroneously assuming that the solution $U ( t, \sigma )$ is continuous and differentiable.
	Still, there are some publications which use numerical schemes to systematically capture non-analytic behavior or discuss the limitations of numerical methods in the presence of these effects, see, \textit{e.g.}, Refs.~\cite{Borchardt:2016pif,Aoki:2017rjl}.
	
	In order to make the fluid-dynamical analogy more apparent, we present a formulation of the RG flow equation \eqref{eq:flow_equation_effective_potential} for the effective potential $U ( t, \sigma )$ in terms of a conservation law.
	Furthermore, we discuss its fluid-dynamical interpretation on a qualitative level and classify the various contributions to the PDE (the RG flow) in the fluid-dynamical picture.
	This sets the stage for an adequate qualitative interpretation of the RG flow equation and possible numerical approaches, which are presented in the next two Sub.Secs.\ \ref{subsec:finite_volume_method} and \ref{subsec:kt_scheme}.\\
	
\subsubsection{The conservative form}
\label{subsec:conservation_equation}

	Starting from the RG flow equation \eqref{eq:flow_equation_effective_potential} of the effective potential $U ( t, \sigma )$, we have several options to recast the flow equation in a conservative form, two of which are:
	\begin{widetext}
		\begin{enumerate}
			\item	Following Refs.~\cite{Aoki:2017rjl,Grossi:2019urj,Grossi:2021ksl,WinkHirschegg,Wink:2020tnu,Ihssen2020}, we can take an overall derivative of Eq.~\eqref{eq:flow_equation_effective_potential} with respect to the $O(N)$ invariant $\varrho = \tfrac{1}{2}\,\sigma^2$ and express the resulting equation in terms of $\varrho$ and $u ( t, \varrho ) \equiv \partial_\varrho U ( t, \varrho )$,
				\begin{align}
					\partial_t u ( t, \varrho ) = \frac{\mathrm{d}}{\mathrm{d} \varrho} \, \bigg( \big[ \tfrac{1}{2} \, \partial_t r ( t ) \big] \, \frac{N - 1}{r ( t ) + u ( t, \varrho )} + \big[ \tfrac{1}{2} \, \partial_t r ( t ) \big] \, \frac{1}{r ( t ) + u ( t, \varrho ) + 2 \varrho \, \partial_\varrho u ( t, \varrho )} \bigg) \, .	\label{eq:conservation_law_u_rho}
				\end{align}

			\item	Another option is to formulate the problem on the level of the background field $\sigma$ itself \cite{Koenigstein:2020Talk} and by alternatively defining $u ( t, \sigma ) \equiv \partial_\sigma U ( t, \sigma )$. 
			Taking an overall derivative of Eq.~\eqref{eq:flow_equation_effective_potential} with respect to $\sigma$ yields,
				\begin{align}
					\partial_t u ( t, \sigma ) = \frac{\mathrm{d}}{\mathrm{d} \sigma} \, \bigg( \big[ \tfrac{1}{2} \, \partial_t r ( t ) \big] \, \frac{N - 1}{r ( t ) + \frac{1}{\sigma} \, u ( t, \sigma )} + \big[ \tfrac{1}{2} \, \partial_t r ( t ) \big] \, \frac{1}{r ( t ) + \partial_\sigma u ( t, \sigma )} \bigg) \, .	\label{eq:conservation_law_u_phi}
				\end{align}
		\end{enumerate}
	\end{widetext}
	In both cases one ends up with a one-dimensional conservation law, where $u$ plays the role of the conserved quantity (the fluid), $t$ can be identified with the time variable and $\varrho$ or $\sigma$ are identified as the spatial variable.
	
	The conservative form of the RG flow equation \eqref{eq:flow_equation_effective_potential} for the effective potential $U$ on the level of its derivative $u$ is not restricted to zero space-time dimensions or models with purely bosonic field content, see also Refs.~\cite{Aoki:2017rjl,Grossi:2019urj,Grossi:2021ksl,WinkHirschegg,Koenigstein:2020Talk,Wink:2020tnu,Ihssen2020}.
	As a matter of fact, this formulation generalizes to arbitrary dimensions and also to models which include fermionic degrees of freedom on the level of the local potential approximation (LPA).
	In particular, the flow equation for the effective potential for models of strong-interaction matter, such as the quark-meson, the Nambu-Jona-Lasinio, and the Gross-Neveu(-Yukawa) model can be formulated in this fashion\footnote{%
		Meanwhile, we and our collaborators \cite{Ihssen2020,Grossi:2021ksl,Koenigstein:2021,Stoll:2021ori} were also working on the conservative formulation of (F)RG flow equations in higher dimensions in more advanced truncations as well as on conservative formulations of (F)RG flow equations for zero-dimensional systems involving fermions (Grassmann numbers) \cite{Steil:partIV}.
	}.

	In this context, it is also worthwhile to note that Eq.~\eqref{eq:conservation_law_u_phi} can be derived not only by taking a derivative of the FRG flow equation for the effective potential $U ( t, \sigma )$ with respect to the background field $\sigma$.
	It is also possible to already start by directly deriving the flow equation for $u ( t, \sigma )$ via a projection on the one-point function $\bar{\Gamma}_t^{(1)} ( \sigma \, )$,
		\begin{align}
			& \partial_t u ( t, \sigma ) =	\vphantom{\Bigg(\Bigg)}	\label{eq:flow_equation_derivative_effective_potential}
			\\
			= \, & \big( \partial_t \bar{\Gamma}_t^{(1)} [ \vec{\varphi} \, ] \big)_{\varphi_1 = \ldots = \varphi_{N - 1} = 0 , \, \varphi_{N} = \sigma}	=	\vphantom{\Bigg(\Bigg)}	\nonumber
			\\
			= \, & \big( \delta_{\varphi_N} \, \partial_t \bar{\Gamma}_t [ \vec{\varphi} \, ] \big)_{\ldots} =	\vphantom{\Bigg(\Bigg)}	\nonumber
			\\
			=\, & \Big( \delta_{\varphi_N} \, \mathrm{Tr} \Big[ \big( \tfrac{1}{2} \, \partial_t R_t \big) \big( \bar{\Gamma}_t^{(2)}[ \vec{\varphi} \, ] + R_t \big)^{-1} \Big] \Big)_{\ldots} =	\vphantom{\Bigg(\Bigg)}	\nonumber
			\\
			=\, &
			-
			\begin{gathered}
				\includegraphics{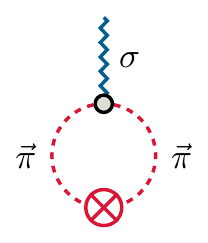}
			\end{gathered}
			-
			\begin{gathered}
				\includegraphics{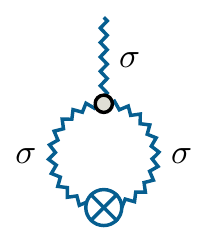}
			\end{gathered}
			= \nonumber
			\\
			=\, & - \big[ \tfrac{1}{2} \, \partial_t r ( t ) \big] \, \frac{N - 1}{\big[ r ( t ) + \frac{1}{\sigma} \, u ( t, \sigma ) \big]^2} \, \partial_\sigma \big[ \tfrac{1}{\sigma} \, u ( t, \sigma ) \big] -	\vphantom{\Bigg(\Bigg)}	\nonumber
			\\
			& - \big[ \tfrac{1}{2} \, \partial_t r ( t ) \big] \,  \frac{1}{\big[ r ( t ) +  \partial_\sigma u ( t, \sigma ) \big]^2} \, \partial_\sigma^2 u ( t, \sigma ) =	\vphantom{\Bigg(\Bigg)}	\nonumber
			\\
			=\, & \tfrac{\mathrm{d}}{\mathrm{d} \sigma} \, \left(
			+
			\begin{gathered}
				\includegraphics{diagrams/potential_pi.pdf}
			\end{gathered}
			+
			\begin{gathered}
				\includegraphics{diagrams/potential_sigma.pdf}
			\end{gathered}
			\right) \, . \nonumber
		\end{align}
	This corresponds to an interchange in the order of operations (evaluating the FRG equation on the background field configuration and taking derivatives with respect to the background field versus taking functional derivatives of the FRG equation and afterwards evaluating on the background field) and it is non-trivial (especially for flow equations for more complex models in higher dimensions and with truncation beyond LPA) that the resulting equations are identical.
	
	Before we turn to the fluid-dynamical interpretation of the conservation laws \eqref{eq:conservation_law_u_rho} and \eqref{eq:conservation_law_u_phi}, we comment on the question whether one of the two formulations \eqref{eq:conservation_law_u_rho} and \eqref{eq:conservation_law_u_phi} is preferable or even others should be considered.
	The answer to this question is not yet settled. 
	From our present understanding, a formulation of the conservation equation in terms of $\sigma$ is preferable, for reasons of numerical implementability.
	This is discussed at length in the context of the PDE boundary conditions for the RG flow equation in Sub.Sec.~\ref{subsec:boundary_conditions_finite_volume} and also in part III of this series of publications \cite{Steil:2021cbu}.
	Therefore, our discussion in the next sections is based on Eq.~\eqref{eq:conservation_law_u_phi}, and hence we identify $\sigma$ with the spatial coordinate $x$ and $u ( t, \sigma ) \equiv \partial_\sigma U ( t, \sigma )$ as the conserved quantity.

\subsubsection{Advection-diffusion equation, irreversibility of RG flows, and 
entropy production}
\label{subsec:advection_diffusion_equation}

	This section is dedicated to the fluid-dynamical interpretation of the RG flow equation \eqref{eq:conservation_law_u_phi}.
	To this end, we split the flux (current) on the right-hand side of the conservation law \eqref{eq:conservation_law_u_phi} and rewrite the whole equation in terms of an advection-diffusion equation in one spatial dimension $x = \sigma$ and one temporal dimension $t$,
		\begin{align}
			\partial_t u ( t, x ) + \tfrac{\mathrm{d}}{\mathrm{d} x} \, F [ t, x, u ( t, x ) ] = \tfrac{\mathrm{d}}{\mathrm{d} x} \, Q [ t, \partial_x u ( t, x ) ] \, .	\label{eq:advection_diffusion_equation_zero-dimensional}
		\end{align}
	The pionic contributions to the RG flow,
		\begin{align}
			F [ t, x, u ( t, x ) ] = \, & - \big[ \tfrac{1}{2} \, \partial_t r ( t ) \big] \, \frac{N - 1}{r ( t ) + \frac{1}{x} \, u ( t, x )} \, ,	\vphantom{\Bigg(\Bigg)}	\label{eq:advection_flux_pion_propagator}
		\end{align}
	are identified with a non-linear, position-dependent advection flux, while the contribution of the radial $\sigma$ mode,
		\begin{align}
			Q [ t, \partial_x u ( t, x ) ] = \, & + \big[ \tfrac{1}{2} \, \partial_t r ( t ) \big] \,  \frac{1}{r ( t ) + \partial_x u ( t, x )}	\, ,	\vphantom{\Bigg(\Bigg)}	\label{eq:diffusion_flux_sigma_propagator}
		\end{align}
	corresponds to a non-linear diffusion flux.
	This decomposition in~$F$ and~$Q$ can be understood as follows:\\
	
\paragraph{Advection:}
	If we ignore the contribution of the $\sigma$ mode for a moment (which -- after rescaling -- corresponds to the large-$N$ limit of the $O(N)$ model \cite{Grossi:2019urj,Tetradis:1995br,Grossi:2021ksl,Steil:2021cbu}), we can rewrite the left-hand side of Eq.~\eqref{eq:advection_diffusion_equation_zero-dimensional} as follows,
		\begin{align}
			& \partial_t u ( t, x ) + \tfrac{\mathrm{d}}{\mathrm{d} x} \, F [ t, x, u ( t, x ) ] =	\vphantom{\bigg(\bigg)}	\label{eq:advection_zero-dimensional}
			\\
			= \, & \partial_t u ( t, x ) + \partial_u F [ t, x, u ( t, x ) ] \, \partial_x u ( t, x ) +	\vphantom{\bigg(\bigg)}	\nonumber
			\\
			& + \partial_x F [ t, x, u ( t, x ) ] = 0 \, .	\vphantom{\bigg(\bigg)}	\nonumber
		\end{align}
	This is a non-linear advection equation for $u ( t, x )$ in its \textit{primitive form} including an internal source term, where $\partial_u F [ t, x, u ( t, x ) ]$ is identified with the velocity of the characteristics (the local $u$-dependent flow velocity of the quantity $u$) and $\partial_x F [ t, x, u ( t, x ) ]$ acts like an $x$ and $u$ dependent internal source term.  
	Hence $F [ t, x, u ( t, x ) ]$ is not purely advective nevertheless we will continue to refer to it as advection term.
	
	The identification of Eq.~\eqref{eq:advection_zero-dimensional} as an advection equation is easily understood by comparison with the one-dimensional linear advection equation,
		\begin{align}
			\partial_t u ( t, x ) + v \, \partial_x u ( t, x ) = 0 \, .
		\end{align}
	Here $v$ corresponds to the a (constant) fluid velocity.
	Another prominent example is the inviscid Bateman-Burgers equation \cite{Bateman1915,Burgers1948,LeVeque:1992,LeVeque:2002,RezzollaZanotti:2013,Ames:1992}, where the velocity is itself proportional to $u$.
	
	However, for our RG flow, the local flow velocity is highly non-linear in $t$, $x$ and $u$ and explicitly reads
		\begin{align}
			\partial_u F [ t, x, u ( t, x ) ] = \, & \frac{( N - 1 ) \, \tfrac{1}{2} \, \partial_t r ( t )}{x \, \big[ r ( t ) + \frac{1}{x} \, u ( t, x ) \big]^2} \, .	\vphantom{\Bigg(\Bigg)}	\label{eq:advection_velocity}
		\end{align}
	Considering for example the exponential regulator shape function \eqref{eq:exponential_regulator}, one finds that the advection velocity $\partial_u F [ t, x, u ( t, x ) ]$ is always negative (positive) for $x > 0$ ($x < 0$).
	In a fluid-dynamical picture, this means that the conserved quantity $u ( t, x )$ is always propagated from larger values of $|x|$ towards the point $x = 0$ by advection.
	Furthermore, the closer the fluid $u ( t, x )$ is to $x = 0$, the faster the fluid moves, due to the factor $\frac{1}{x}$. Since $u ( t, x )$ is anti-symmetric in $x$ [because of the $O(N)$ symmetry of $U ( t, \vec{\varphi} \, )$], this implies that ``waves'' of positive and negative $u ( t, x )$ collide with huge velocity at $x = 0$ and annihilate.
	At large $|x|$, the fluid velocity tends to zero.
	
	We also observe that the advection velocity \eqref{eq:advection_velocity} is proportional to the number of pions, $ N - 1 $.
	Hence, in the large-$N$ limit the system is completely advection driven, while for small $N$ the diffusive contributions \eqref{eq:diffusion_flux_sigma_propagator} gain in importance.
	In the case $N = 1$, there is no advection at all and the dynamics of the fluid $u ( t, x )$ is purely diffusive.
	Both limiting cases $N=1$ and $N\rightarrow\infty$ are discussed at length in the subsequent parts II and III in this series of publications \cite{Koenigstein:2021rxj,Steil:2021cbu} respectively.

	It is also well known \cite{LeVeque:1992,LeVeque:2002,RezzollaZanotti:2013,Ames:1992} that systems which involve non-linear advection tend to exhibit non-analytical behavior in the form of shock and rarefaction waves.
	We will return to this issue below. \\

	\paragraph{Diffusion:}
	Next, we turn to the contribution of the radial $\sigma$ mode to the RG flow. 
	We find that t enters the conservation law \eqref{eq:advection_diffusion_equation_zero-dimensional} as a non-linear diffusion flux \eqref{eq:diffusion_flux_sigma_propagator}, because it is overall of second order in spatial derivatives of $u ( t, x )$. 
	The characteristic property of diffusive processes is that they transport a quantity, in this case $u ( t, x )$, from regions where its density or concentration is high to regions where it is low \cite{LeVeque:1992,LeVeque:2002,RezzollaZanotti:2013,Ames:1992}.
	Diffusive processes are therefore usually important in regions of high gradients and smear out cusps, shocks \textit{etc.}, which might form via advection.
	Besides this, diffusive processes are generically undirected, which is also the case for \eqref{eq:diffusion_flux_sigma_propagator} and therefore also propagate the quantity $u ( t, x )$ in both directions, depending on the local gradients of $u ( t, x )$, which is especially relevant for models in their symmetry-broken phase with rather weak advection (small $N$). 
	The effective transport velocities via diffusion are usually much slower than those via advection, which is, due to the non-linearity, not necessarily true for RG flow equations.
	A famous example of a purely diffusive process is heat transport via the heat equation \cite{Cannon:1984,LeVeque:1992,LeVeque:2002}.
	The diffusion flux \eqref{eq:advection_diffusion_equation_zero-dimensional} can indeed be formulated as a non-linear time-dependent realization of the heat equation.
	By performing the spatial derivative in the advection-diffusion equation \eqref{eq:advection_diffusion_equation_zero-dimensional} for the purely diffusive ($N = 1$) case, one finds
		\begin{align}
			\partial_t u ( t, x ) = \, & - \frac{\tfrac{1}{2} \, \partial_t r ( t )}{[ r ( t ) + \partial_x u ( t, x ) ]^2} \, \partial_x^2 u ( t, x ) \, ,
		\end{align}
	where
		\begin{align}
			- \frac{\tfrac{1}{2} \, \partial_t r ( t )}{[ r ( t ) + \partial_x u ( t, x ) ]^2} \, ,	\label{eq:diffusion_coefficient}
		\end{align}
	plays the role of a non-linear time dependent, strictly positive diffusion coefficient. 
	The positivity of the diffusion coefficients ensures that $u ( t, x )$ is only dispersed and never accumulates locally, \textit{i.e.}, that $u ( t, x )$ tends to equilibrate towards a linear function in space. 
	A positive diffusion coefficient also ensures stability and uniqueness of (numerical) weak solutions, see, \textit{e.g}, \cite{LeVeque:1992,LeVeque:2002,Hesthaven2007,RezzollaZanotti:2013}.
	
	Directly comparing these findings with the linear heat equation with constant diffusion coefficient, we can already qualitatively predict the behavior of the diffusion transport for the RG flow of $u ( t, x )$, as long as $N$ is small and the system is diffusion-dominated.
	At a constant RG time $t$, we find that the diffusion coefficient is much larger in regions where the gradient $\partial_x u ( t, x )$ is negative with a large absolute value, compared to regions where it is positive, because in the first case the denominator of Eq.~\eqref{eq:diffusion_coefficient} is smaller than in the second case.
	This plays a crucial role for systems that involve symmetry breaking, where $\partial_x u ( t, x )$ is negative for at least some small $|x|$, while asymptotically for $|x| \rightarrow \infty$ the sign of $\partial_x u ( t, x )$ is always positive.
	Hence, for diffusion-dominated problems in RG flow equations (small number $N$ of pions), the symmetry restoration is driven by the negative gradients $\partial_x u ( t, x )$ at small $|x|$.
	Furthermore, we find that for $t \rightarrow \infty$, the numerator of the diffusion coefficient \eqref{eq:diffusion_coefficient} tends to zero such that the diffusion stops, the system equilibrates and the dynamics freezes, even though there are still gradients in $u ( t, x )$.
	This would not happen for the linear heat equation.
	The same is true for $t = 0$, where the diffusion coefficient is suppressed by $1/\Lambda$.
	However, the $t$ dependence of the diffusion coefficient of $O(1)$ models strongly depends on the space-time dimension, which will be discussed elsewhere.\\

	\paragraph{Irreversibility and entropy production:}
	In a fluid-dynamical setting, it is very easy to understand the role of the radial $\sigma$ mode: Due to its diffusive character, it is directly responsible for the irreversibility of the RG flow and RG transformations in general. 
	Diffusion is a particular example of a dissipative process, which is irreversible and increases the entropy of the system\footnote{%
		Interestingly, Ref.~\cite{Zamolodchikov:1986gt} comes to the same conclusion arguing in reverse order:
		``Some of the information on the ultraviolet behavior of the field theory is lost under renormalization transformations with $t>0$, since in the field theory it is not legitimate to examine correlations at scales smaller than the cutoff.
		We would therefore expect that a motion of the space $Q$ [a change of the set of all couplings] under the influence of the renormalization group would become an `irreversible' process, similar to the time evolution of dissipative systems.''
		We remark that also Ref.~\cite{Zumbach:1994vg} stated that a term of second order in field space derivatives in related RG flow equations ``[$\ldots$] corresponds to a dissipation in the flow and is responsible for the semi-group property of the RG.''%
		}.
	The dissipative and irreversible character can be seen as a ``thermodynamic'' version of the irreversible Kadanoff block-spin transformations \cite{Kadanoff:1966wm,Wilson:1979qg,Delamotte:2007pf}.
	Hence, the dissipation clearly singles out the RG time $t$ as a temporal direction, because it introduces a ``thermodynamic arrow of time'' and ``thermodynamic time asymmetry'' via entropy production \cite{Lebowitz:2008}.
	This also explains why
		\begin{align}
			t = - \ln \big( \tfrac{k}{\Lambda} \big)	\vphantom{\bigg(\bigg)}	\label{eq:rg-time-scale}
		\end{align}
	(including the minus sign!) is a natural choice for a temporal coordinate in higher dimensions, see also Refs.~\cite{Hasenfratz:1985dm,Zumbach:1993zz,Zumbach:1994vg,Zumbach:1994kc,Grossi:2019urj}.
	
	Interestingly, the irreversibility and the dissipative character of the system is lost if one does not include the full field dependence of the effective potential in the flow equation, but instead uses a truncated system like the Taylor expansion \eqref{eq:example_vertex_expansion}.
	Then, the system of coupled ODEs for the vertices can theoretically be integrated in either direction in RG time, as long as it consists of a finite number of couplings\footnote{%
		In momentum space this enables an integration to higher energy scales, which corresponds to a reversion of the coarse-graining in position space.
		More generally speaking, this implies that it is possible to resolve the microphysics from the macrophysics.
		Both is physically not possible and solely an artifact of the truncation.%
	}.
	The most extreme examples are the RG flows of one single $t$ dependent coupling, \textit{e.g.}, the quartic coupling of $\phi^4$ theory or the QCD $\beta$ function \cite{Politzer:1973fx,Gross:1973id,Gross:1973ju,Gross:1974cs}, see also the textbooks \cite{ZinnJustin:2002ru,Peskin:1995ev}.
	Here the integration to both higher and smaller RG scales is possible, which is the well-known result for the universal one-loop $\beta$ function and is an artifact of the restriction (truncation) to a finite number of couplings \cite{Wilson:1979qg}.
	However, this reversibility of RG transformations is not possible for the field-dependent effective potential, which is obvious from the advection-diffusion equation \eqref{eq:advection_diffusion_equation_zero-dimensional}, where entropy increases and the information about the initial condition in the UV cannot be recovered from the IR anymore.
	
	This point of view was already shared, presented, and discussed by K.~G.~Wilson:
	In Ref.~\cite{Wilson:1979qg} he pointed out the differences between his ``coarse-graining'' version of the (F)RG, which is also applicable in highly non-perturbative regimes, and the RG flow equations used by C.~Callan, K.~Symanzik, M.~Gell-Mann, F.~Low, G.~t'Hooft, S.~Weinberg, H.~Georgi, D.~Politzer \textit{et al.} to calculate the running of a single (or small number of) coupling constants, which solely describes a system correctly in a perturbative regime.
	
	The irreversibility of the RG flow and entropy production is also directly related to the presence of discontinuities in the solution, which can arise from the advective contributions to the flow.
	As shown in Refs.~\cite{Aoki:2017rjl,Grossi:2019urj,Grossi:2021ksl,Steil:2021cbu} for the large-$N$ limit, a shock wave arises when the weak solution of the PDE is multi-valued. 
	The correct solution is usually constructed by means of the Rankine-Hugoniot condition \cite{Rankine:1870,Hugoniot:1887,LeVeque:1992,LeVeque:2002,RezzollaZanotti:2013,Ames:1992}.
	This would lead to ambiguities when one tries to invert the flow (integrating backwards in time) in the presence of a shock.
	Hence, shock formation is an irreversible process and produces entropy. In summary, these are further strong arguments why the assumption of expandability of the effective average action in terms of vertices as well as the truncation of the system should in general be considered with care.
	
	Therefore, it would be extremely interesting to explicitly construct an entropy function for the flow equation, \textit{i.e.}, a quantity that is either non-decreasing or non-increasing under the RG transformations during the RG flow (depending on the sign convention), and that is a functional of the quantity $u ( t, x )$.
	The entropy for the flow equation will be a helpful instrument to design a stable numerical scheme for generic truncations \cite{LeVeque:1992,LeVeque:2002,RezzollaZanotti:2013} and will also highlight general properties of the RG flow.
	In this context we also have to mention the recent publication \cite{Cotler:2022fze} by J. Cotler and S. Rezchikov who were able to interpret the Polchinski equation as an ``optimal transport gradient flow of a field-theoretic relative entropy'' thus establishing a firm and explicit connection between an information-theoretic entropy and (F)RG flows.
	
	Additionally, a numeric entropy (function) might provide a direct link to the $\mathcal{C}$ theorem ($\mathcal{A}$ theorem) \cite{Zamolodchikov:1986gt,Banks:1987qs,Cardy:1988cwa,Osborn:1989td,Jack:1990eb,Komargodski:2011vj,Curtright:2011qg,Rosten:2010vm}, which states that in certain QFTs there exists some positive real function $\mathcal{C} ( \{ g_i \} , t )$, which depends on all coupling constants of the QFT and which is monotonically increasing\footnote{%
		It can also be defined as a monotonically decreasing function.
		This flip of sign corresponds to the difference of the mathematicians' and physicists' definition of entropy.
		We chose to the ``thermodynamic convention'' of increasing entropy for this and subsequent publications.
	} during RG flows (transformations), while it stays constant at (critical) fixed points,
		\begin{align}
			\tfrac{\mathrm{d}}{\mathrm{d} t} \,  \mathcal{C} ( \{ g_i \} , t ) \geq 0 \, .
		\end{align}
	Here, $\{ g_i \}$ denotes the set of all (possibly infinitely many) dimensionless coupling constants.
	In contrast to previous formulations~\cite{Haagensen:1993by,Generowicz:1997he,Forte:1998dx,Codello:2013iqa,Codello:2015ana,Becker:2014pea,Becker:2016zcn}, a non-local version, which is directly linked to the numerical entropy function (similar to versions presented in Refs.~\cite{Zumbach:1993zz,Zumbach:1994vg,Rosten:2010vm} for related field-dependent flow equations), would not rely on expandability in the couplings or vertices and could naturally display the dissipative character of RG transformations, which was already described by Refs.~\cite{Zamolodchikov:1986gt,Zumbach:1994vg}.
	Fixed-point solutions of the RG flow would directly correspond to steady-state or thermal-equilibrium solutions \cite{LeVeque:1992} in the fluid-dynamical picture\footnote{%
		This actually brings up the interesting question whether previous studies about global fixed-point solutions for field-dependent flow equations, which seemed to deliver interesting results, \textit{e.g.}, Refs.~\cite{Borchardt:2015rxa,Zumbach:1994vg,Yabunaka:2018mju}, should be reanalyzed from the fluid-dynamical steady-flow perspective, especially regarding their interpretation and the spatial discretization methods \cite{LeVeque:1992}.
	}.
	A caveat at this point is that a $\mathcal{C}$ function is based on the rescaled dimensionless RG flow equations.
	Hence, also a numerical entropy should be formulated in this framework, if one seeks a direct link to a $\mathcal{C}$ function.
	The dimensionless flow equations in the LPA can be recast in terms of conservation laws, which might be a good starting point.
	
	An explicit discussion of (numerical) entropy for the zero-dimensional $O(1)$ model as well as possible links to $\mathcal{C}$ functions is discussed in great detail in part II of this series of publications~\cite{Koenigstein:2021rxj}.
	The situation for the $O(N)$ model in the limit $N\rightarrow\infty$ is discussed in App. E of part III in this series of publications~\cite{Steil:2021cbu}.
	The construction of an explicit (numerical) entropy has proven to be elusive in the case of finite $N>1$ for the $O(N)$ model \cite{Koenigstein:2021rxj,Steil:2021cbu} due to the explicit position dependences in Eqs.~\eqref{eq:conservation_law_u_phi} and \eqref{eq:conservation_law_u_rho} and the related internal source terms, \textit{cf.} Eq.~\eqref{eq:advection_zero-dimensional}.\\
	
	\paragraph{Generalizations:}
	Before we turn to the numerical implementation of the RG flow equation for $u ( t, \sigma )$, we briefly comment on the generalization of the fluid-dynamical picture to RG flow equations in higher-dimensional QFTs, systems with more (field-dependent) couplings, and RG flow equations that involve fermions.
	
	In higher-dimensional QFTs, the fluid-dynamical interpretation of the RG flow of the effective potential survives, see for example Ref.~\cite{Grossi:2019urj,Grossi:2021ksl,Koenigstein:2020Talk,Stoll:2021ori}.
	A difference is that in higher dimensions the RG time enters as the negative logarithm of the ratio of the RG scale $k$ and the UV reference scale $\Lambda$, see Eq.~\eqref{eq:rg-time-scale}, while in zero dimensions $t$ just parametrizes some mass-like scale $r ( t )$, see Eq.~\eqref{eq:exponential_regulator}. 
	Furthermore, the fluxes gain further $t$ dependent prefactors via the momentum integrals of the trace in the FRG equation.
	This leads to a different time scaling but does not affect the overall discussion. \
	The inclusion of further field-independent but scale-dependent couplings (such as a scale-dependent Yukawa coupling) adds ODEs to the advection-diffusion equation for the effective potential, which does not spoil its conservative fluid-dynamical character.
	It is currently investigated by us and collaborators~\cite{Ihssen2020,Grossi:2021ksl} whether the inclusion as well as the conservative formulation of further field-dependent couplings (such as a field- and scale-dependent wave-function renormalization $Z ( t, \vec{\varphi} \, )$ in higher-dimensional models) is possible.
	However, this analysis is beyond the scope of the present work and will be presented elsewhere.
	In any case, simply adding fermions in the LPA does not destroy the fluid-dynamical character of the RG flow equation at all: On the level of the LPA for the RG flow equation of the effective potential, the contributions from fermion loops can be interpreted as a source/sink term, which only depends on $\sigma$, \textit{i.e.}, the spatial position $x$.
	For a detailed discussion of such fermionic source/sink terms at zero and non-zero temperature and especially quark chemical potential we refer the interested reader to Ref.~\cite{Stoll:2021ori}.
	Another possible generalization concerns models with more than one invariant of the underlying symmetry group of the model and respective condensation directions in field space, see, \textit{e.g.}, Refs.~\cite{Strodthoff:2011tz,Mitter:2013fxa,Rennecke:2016tkm,Lakaschus:2020caq,Fukushima:2010ji,Fejos:2020lli}.
	Here, the fluid-dynamical framework should still be applicable.
	However, a suitable identification of a complete basis of field-space directions with ``spatial directions'' of the fluid-dynamical problem and a clear separation of the single contributions into advection, diffusion, and source terms might be challenging and calls for future investigations -- especially when it comes to an actual numerical implementation.
	For first attempts of generalizing our findings to a quark-meson-diquark model, we refer to Ref.~\cite{Lakaschus:2021ewd}.

	Summarizing we find that the fluid-dynamical interpretation of flow equations has tremendous benefits, because it allows for a rather intuitive understanding of the dynamics of the system. Furthermore, it allows for a novel, physically intuitive interpretation of the RG flow and provides an understanding of its irreversibility.
	Finally, it opens up the opportunity to employ extremely powerful numerical tools from computational fluid dynamics, which are discussed in the next two sections.
	Phenomena discussed within this section are observed, quantified, and visualized in Sec.~\ref{sec:results}, in recent and parallel works \cite{Grossi:2019urj,Grossi:2021ksl,Koenigstein:2021rxj}, and in parallel and upcoming publications \cite{Koenigstein:2021,Stoll:2021ori,Steil:partIV}.\\

\subsection{Finite-volume method}
\label{subsec:finite_volume_method}

	In this section we discuss numerical solution schemes for advection-diffusion equations\footnote{%
		Oftentimes, such equations are also referred to as ``convection-diffusion equations''.
		The semantically correct term is nevertheless ``advection-diffusion equation'' because ``convection'' includes also diffusive processes besides the transport by bulk motion (advection), see also Ref.~\cite{LeVeque:1992}.%
	} 
	with source terms of the generic type
	\begin{widetext}
		\begin{align}
				\partial_t u ( t, x ) + \tfrac{\mathrm{d}}{\mathrm{d} x} \, F [ t, x, u ( t, x ) ] = \tfrac{\mathrm{d}}{\mathrm{d} x} \, Q [ t, x, u ( t, x ), \partial_x u ( t, x ) ] + S [ t, x, u ( t, x ) ] \, . \label{eq:FVcdEq}
		\end{align}
	\end{widetext}
	Whether $S$ acts as a source or sink in the dynamics of $u(t,x)$ depends on its explicit form.
	Nevertheless we will refer to $S$ as source term for convenience for the scope of this paper.
	In the following, we occasionally suppress the $t$ and $x$ dependences of $u$, $F$, $Q$, and $S$ for the sake of simplicity.
	Equation \eqref{eq:FVcdEq} is a partial differential equation describing the evolution of the conserved quantity $u \equiv u ( t, x )$ in one spatial ($x$) direction and one temporal ($t$) direction. 
	Depending on the problem at hand these two directions are not necessarily identical with physical spatial and temporal dimensions of reality, but for the following discussion we denote them as such.
	The function $F [ u ] \equiv F [ t, x, u ( t, x ) ]$ is a (non-linear) advection flux, $ Q [ u, \partial_x u ] \equiv Q [ t, x, u ( t, x) , \partial_x u ( t, x ) ]$ is a (non-linear) diffusion(dissipation) flux and $S [ u ] \equiv S [ t, x, u ( t, x) ]$ is a source term.
	The concepts discussed in the following apply directly to systems of $M$ conserved quantities $u \rightarrow \{ u_1 , \ldots , u_M \}$ and can be generalized beyond one spatial dimension to $d + 1$ dimensional space-time $( x, t ) \rightarrow ( \vec{x}, t ) = ( x_1, \ldots, x_d, t )$.
	Equations or systems similar or even identical to Eq.~\eqref{eq:FVcdEq} are often referred to as conservation laws and appear in many areas of the natural sciences, engineering, and economics.
	They are extensively studied in the field of computational fluid dynamics.
	
	Consider the advection-diffusion equation \eqref{eq:FVcdEq} with specified $F$, $Q$, and $S$ in a finite computational domain $\Omega = \mathcal{V} \times [ t_0, t_N ]$, where $\mathcal{V} \subset \mathbb{R}^1$ denotes the spatial volume, with an initial condition $u( t_0, x )$ and Dirichlet (Neumann) boundary conditions specifying $( \partial_x ) u ( t, x ) |_{x \in \partial \mathcal{V}}$.
	The question is how to evolve the initial condition in time from $t_0$ to $t_N>t_0$ to acquire a solution $u ( t_N, x )$ respecting the specified boundary conditions.
	For most problems of the type~\eqref{eq:FVcdEq} an analytic solution is not known or is even known to be nonexistent.
	Strategies for finding numerical (weak) solutions are required.
	Numerical schemes in the broad class of so-called finite-volume (FV) methods are very popular for the numerical solution of PDEs describing the conservation or balance of quantities.
	Alternative high-resolution shock-capturing (HRSC) schemes in modern computational fluid dynamics are among others finite-difference schemes including flux limiters and numerical viscosity or finite-element methods.
	
	The concept that all numerical FV methods share is a discretization of the computational domain into space-time control volumes $\mathcal{V}_j \times [ t^n, t^{n+1} ]$, where the set of spatial control volumes $\mathcal{V}_j$ covers the spatial computational domain $\mathcal{V}$.
	Integrating Eq.~\eqref{eq:FVcdEq} over such a control volume centered at $x$, using the divergence theorem (Gauss-Ostrogradsky theorem) on the fluxes and introducing the sliding cell average
		\begin{align}
			\bar{u} ( t, x ) \equiv \frac{1}{| \mathcal{V}_j |} \int_{\mathcal{V}_j} \mathrm{d} \xi \, u( \xi, t ) \, ,
		\end{align}
	where $\mathcal{V}_j = \{ \xi : | \xi - x | \leq \Delta x/2 \}$ we arrive at an equivalent integral form of Eq.~\eqref{eq:FVcdEq},
	\begin{widetext}
		\begin{align}
			& \bar{u} ( t^{n+1}, x ) =	\vphantom{\bigg(\bigg)}	\label{eq:FVintEq}
			\\
			= \, & \bar{u} ( t^n, x ) - \tfrac{1}{\Delta x} \bigg( \int_{t^n}^{t^{n+1}} \mathrm{d} \tau \, F \big[ \tau, x + \tfrac{\Delta x}{2}, u \big( \tau, x + \tfrac{\Delta x}{2} \big) \big] - \int_{t^n}^{t^{n+1}} \mathrm{d} \tau \, F \big[ \tau, x - \tfrac{\Delta x}{2}, u \big( \tau, x - \tfrac{\Delta x}{2} \big) \big] \bigg) + \ldots \, .	\vphantom{\bigg(\bigg)}	\nonumber
		\end{align}
	\end{widetext}
	Here, the ellipsis denotes the corresponding integrals for the diffusion and the source term.
	The solution of Eq.~\eqref{eq:FVintEq} presents the central challenge for an explicit FV scheme.

	A central aspect of each practical FV scheme is an appropriate and informed choice of the space-time control volumes which, depending on the scheme and problem at hand, might change during the time evolution. 
	Given a set of control volumes and a corresponding set of cell averages $\bar{u} ( t^n, x_j ) \equiv \bar{u}^n_j$ the time evolution to ${t^{n+1}\equiv t^n+\Delta t}$ requires the solution of the Riemann problems \cite{Godunov,LeVeque:1992,LeVeque:2002,Hesthaven2007,RezzollaZanotti:2013} at each cell interface.
	Part of these problems are the fluxes through the cell boundaries.
	The computation of those fluxes requires a reconstruction of the values of $u$ on the cell interfaces located at $x_{j + \frac{1}{2}}$, which we denote as $u_{j + \frac{1}{2}}^n$, from the given set of cell averages $\bar{u}_j^n$.
	This is usually done by means of a polynomial approximation respecting the given cell averages of the neighboring cells.
	The order of the chosen approximation is one of the parameters contributing to the overall spatial order (of the error) of the scheme at hand.
	
	Given the cell averages $\bar{u}_j^n$ and fluxes through the cell interfaces at $t=t^n$ it remains to solve the Riemann problems at the cell interfaces.
	The solution of the Riemann problem amounts to the exact evaluation of the flux integrals on the right-hand side of Eq.~\eqref{eq:FVintEq}.
	Depending on the complexity of the underlying conservation equation an exact solution of the Riemann problems at the cell boundaries might be either impossible or unfeasible.
	Most explicit FV schemes, especially those for general advection-diffusion equations, either use approximate Riemann solvers (\textit{e.g.}, the Roe \cite{ROE1981357} or the HLLE \cite{HLLE1,HLLE2} solver) or do not require Riemann solvers at all (\textit{e.g.}, the KT \cite{KTO2-0} scheme).
	For a pedagogic introduction into the broad field of FV methods and HRSC schemes in general we refer the interested reader to Refs.~\cite{Ames:1992,LeVeque:1992,LeVeque:2002,Hesthaven2007,RezzollaZanotti:2013} and references therein.
	
	In the following section we will introduce a particular FV scheme, which we have chosen for the numerical solution of the RG flow equations because of its flexibility, efficiency, and relative simplicity.

\subsection{Kurganov-Tadmor (KT) central scheme}
\label{subsec:kt_scheme}
	
	In this section we will summarize the central scheme presented in Ref.~\cite{KTO2-0} by A.~Kurganov and E.~Tadmor, which we will refer to in the following as KT scheme.
	The KT scheme can be implemented and applied as a black-box solver for systems of the type of Eq.~\eqref{eq:FVcdEq}.
	Apart from the PDE with its initial and boundary conditions the only additional information about the PDE required for its solution using the KT scheme is $\big| \frac{\partial F}{\partial u}\big|$ or the spectral radius of the Jacobian of $F[u]$ when considering systems of conserved quantities, see Eq.~\eqref{eq:FVajp12} and the related discussion.
	The scheme does not require a Riemann solver of any kind and as such does not rely on a characteristic decomposition of the advection flux.
	
	The KT scheme provides a direct method for evaluating the flux integrals on the right-hand side of Eq.~\eqref{eq:FVintEq}.
	The main focus lies on the treatment and implementation of the flux integrals for the advection flux $F[u]$, which is kind of a natural approach, because the advection flux determines the characteristic velocities in an advection-diffusion equation.
	The diffusion and source terms are treated separately and will be discussed at the end of this subsection.

	The KT scheme admits a meaningful ${t^{n+1}-t^n\equiv\Delta t\rightarrow 0}$ limit in the context of Eq.~\eqref{eq:FVintEq} and is thus an improvement on it predecessor the Nessyahu-Tadmor (NT) scheme \cite{NT} with which it shares its piecewise-linear MUSCL (Monotonic Upstream-centered Scheme for Conservation Laws) reconstruction \cite{MUSCL}.
	We will focus on the KT scheme in its so called semi-discrete from -- in the limit $\Delta t\rightarrow 0$ -- which involves only an explicit spatial discretization.
	The KT scheme is formally second-order accurate in the spatial direction and as such an improved version of the first-order accurate Lax–Friedrichs (LxF) scheme \cite{LxF1,LxF2}.
	A semi-discrete form reduces the PDEs \eqref{eq:FVcdEq} or equivalently \eqref{eq:FVintEq} to a set of coupled ODEs, which can be solved by a large class of general-purpose ODE solvers.
	This is especially useful when working on stiff problems or PDE systems coupled to additional ODEs.
	We will proceed with the introduction of quantities involved in the semi-discrete form \eqref{eq:FVKTO2} of the KT scheme.
	The following quantities are especially relevant for the numerical advection flux~\eqref{eq:definition_h_kt_scheme}.\\
	
	Consider a set of volume averages $\bar{u}_j^n$ at $t^n$ based on an equidistant\footnote{%
		The generalization of the KT scheme to non-uniform grids is on a conceptual level straightforward and especially useful for higher-dimensional extensions and for adaptive or moving mesh variants, see, \textit{e.g.}, Ref.~\cite{KTmovingMesh}.
		Its implementation is of course much more challenging and not needed in this work. 
		However, in the context of FRG flow equations this might be relevant for models with multiple condensate directions, see, \textit{e.g.}, Refs.~\cite{Strodthoff:2011tz,Mitter:2013fxa,Rennecke:2016tkm,Lakaschus:2020caq,Fukushima:2010ji,Fejos:2020lli}.
	} grid of volume cells $\mathcal{V}_j \equiv [ x_{j - \frac{1}{2}}, x_{j + \frac{1}{2}} ]$, with $\Delta x = x_{j + \frac{1}{2}} - x_{j - \frac{1}{2}}$.
	
	The time evolution of the averages $\bar{u}_j^n$ at $t^n$ to averages at $\bar{u}_j^{n+1}$ at $t^{n+1}$ on the same volume grid is a three-step process:
	\begin{enumerate}
		\item	The piecewise-linear MUSCL reconstruction is computed from the cell averages:
			\begin{align}
				& \tilde{u} ( t^n, x ) =	\vphantom{\bigg(\sum_j\bigg)}\label{eq:FVmuscl}
				\\
				= \, & \sum_j \big\{ \bar{u}_j^n + (\partial_x u )_j^n \, ( x - x_j ) \big\} \openone_{[ x_{j - \frac{1}{2}}, x_{j + \frac{1}{2}} ]} \, ,	\vphantom{\bigg(\sum_j\bigg)}	\nonumber
			\end{align}
		where the sum runs over all volume cells.
		The reconstruction step is needed to gain access to the function values $\tilde{u} ( t^n, x )$.
		The reconstruction uses approximations to the exact derivatives $( \partial_x u )_j^n$ by employing a scalar \textit{total variation-diminishing} (TVD) reconstruction \cite{LeVeque:1992,LeVeque:2002,HARTEN1983357},
			\begin{align}
				( \partial_x u )_j^n = \frac{\bar{u}_{j+1}^n - \bar{u}_{j}^n}{\Delta x} \, \phi \bigg( \frac{\bar{u}_{j}^n - \bar{u}_{j-1}^n}{\bar{u}_{j+1}^n - \bar{u}_{j}^n} \bigg) \, ,	\label{eq:FVuxjn}
			\end{align}
		with a TVD limiter $\phi(r)$. An overview of TVD flux limiters can be found, \textit{e.g.}, on the webpage \cite{wikiFluxLimiter}, in Refs.~\cite{LeVeque:1992,LeVeque:2002}, or in Sec.~9.3.1 of Ref.~\cite{RezzollaZanotti:2013}.
		Here, we follow  Ref.~\cite{KTO2-0} and use the so-called \textit{minmod} limiter \cite{MinModRoe}\footnote{%
			We also implemented and tested other flux limiters in the context of FRG flow equations, which however did not influence our numerical results very much.
			An optimization of the choice of flux limiters with regard to the overall runtime could be part of future work.%
		},
			\begin{align}
				\phi ( r ) = \, & \max[ 0, \min( 1, r )] \, .	\label{eq:FVminmod}
			\end{align}
		The limiter $\phi$ is used in Eq.~\eqref{eq:FVuxjn} to limit the slopes during the reconstruction process.
		This is crucial to prevent spurious oscillations around discontinuities, \textit{e.g.}, shocks, in high-resolution schemes like the KT scheme.
		The KT scheme can also be used with higher-order reconstruction schemes\footnote{%
			Examples for such improvements are the use of the third-order central weighted essentiallynon-oscillatory (C-WENO) reconstruction \cite{WENO-C,WENO-C2} outlined in Ref.~\cite{KTO3-0}, the fifth-order WENO scheme (WENO5) \cite{WENO2,WENO5} employed in Ref.~\cite{KTO5-0}, or the fifth-order monotonicity-preserving (MP5) reconstruction \cite{MP5} used in Ref.~\cite{KT-MP5}.
			WENO schemes were first introduced in Ref.~\cite{WENO} and use polynomial reconstructions based on smoothness indicators.
			In regions where the solution is estimated to be smooth high-order polynomials are used, while in regions with large gradients lower-order polynomials are employed in order to ensure an \textit{essentially non-oscillatory} \cite{WENO-C} interpolation and a TVD reconstruction.
			The MP5 scheme was first introduced in Ref.~\cite{MP5} and uses a limiting procedure together with a fourth-order polynomial reconstruction.
			The MP5 reconstruction does not require smoothness indicators and performs favorable when compared to WENO in terms of speed and accuracy according to Ref.~\cite{GRHydro}.%
		} to increase the spatial accuracy of the scheme, which is not needed for this work.
		
		When using a piecewise-constant or -linear reconstruction the cell averages $\bar{u}_j^n$ coincide with the midpoint values $u_j^n$.
		While we employ a piecewise-linear reconstruction, we still maintain the distinction between averages and midpoint values for the sake of clarity.
		
		\item	The time step from $t^{n}$ to $t^{n+1}$ is performed by computing the flux integrals on the right-hand side of Eq.~\eqref{eq:FVintEq} using the reconstruction $\tilde{u} ( t^n, x )$ from Eq.~\eqref{eq:FVmuscl} and carefully chosen control volumes discussed below. In the limit ${t^{n+1}-t^n\equiv\Delta t\rightarrow 0}$ only the expressions for $a_{j + \frac{1}{2}}^n$, $u_{j + \frac{1}{2}}^{n,-}$, and $u_{j + \frac{1}{2}}^{n,-}$ from Eqs.~\eqref{eq:FVajp12}, \eqref{eq:FVumjp12}, and \eqref{eq:FVupjp12} respectively are relevant for the semi-discrete KT scheme. The other quantities discussed for this second step of the KT scheme are however necessary to understand the underlying algorithm.
		
		At each cell interface $x_{j + \frac{1}{2}}$ the respective local speed of propagation $a_{j + \frac{1}{2}}^n$ is estimated using
			\begin{align}
				& a_{j + \frac{1}{2}}^n \equiv	\vphantom{\bigg(\bigg)}	\label{eq:FVajp12}
				\\
				\equiv \, & \max \bigg\{ \bigg| \frac{\partial F}{\partial u} \Big[ u_{j + \frac{1}{2}}^{n,+} \Big] \bigg| , \bigg| \frac{\partial F}{\partial u} \Big[ u_{j + \frac{1}{2}}^{n,-} \Big] \bigg| \bigg\} \, ,	\nonumber
			\end{align}
		with the left and right intermediate values $u_{j + \frac{1}{2}}^{n, \mp}$ of $\tilde{u} ( t^n, x )$ at the cell interface $x_{j + \frac{1}{2}}$:
			\begin{align}
				u_{j + \frac{1}{2}}^{n,-} = \, & \bar{u}_j^n + \tfrac{\Delta x}{2} \, ( \partial_x u )_j^n \, ,	\vphantom{\bigg(\bigg)} \label{eq:FVumjp12}
				\\
				u_{j + \frac{1}{2}}^{n,+} = \, & \bar{u}_{j+1}^n -\tfrac{\Delta x}{2}  \, ( \partial_x u )_{j+1}^n \, .	\vphantom{\bigg(\bigg)} \label{eq:FVupjp12}
			\end{align}
		For multi-valued conserved quantities $u = \{ u_1, \ldots, u_M \}$ the KT scheme of Ref.~\cite{KTO2-0} is limited\footnote{%
			The KT scheme of Ref.~\cite{KTO2-0} can be improved using refined estimates for the local speed of propagation. 
			In Ref.~\cite{KTO2-1} the KT scheme is refined by employing left- and right-sided local speeds $a_{j + \frac{1}{2}}^{n,\mp}$, which are extracted from the eigenvalue spectrum of the Jacobian $\frac{\partial F}{\partial u}$. 
			A further improvement in terms of estimates of local speeds of propagation engineered for non-convex hyperbolic (systems of) conservation laws is presented in Ref.~\cite{KTO5-0} using further information about the eigensystem of the Jacobian $\frac{\partial F}{\partial u}$. 
			When an explicit evaluation of the Jacobian is impossible or unfeasible numerical approximations can be employed \cite{KTO2-0,LiuTadmore2000,Jiang97non-oscillatorycentral}.
			For the numerical applications in this paper, the rather simple estimate of Eq.~\eqref{eq:FVajp12} has proven to be sufficient.
			For specific computations in part III in this series of publications \cite{Steil:2021cbu} we additionally employed the improved scheme of Ref.~\cite{KTO2-1}.%
		} to hyperbolic advection fluxes signaled by a non-degenerate eigenvalue spectrum $\lambda_1 < \ldots < \lambda_M$ of the Jacobian~$ \frac{\partial F}{\partial u} $ for all $x$, $t$, and $u$, since in this case the local speed of propagation is computed using the spectral radius $\rho ( M ) \equiv \max_i | \lambda_i ( M ) |$ in Eq.~\eqref{eq:FVajp12}.

		Using the estimated local speed of propagation, a space-time control volume ${[x_{j + \frac{1}{2}, l}^n , x_{j + \frac{1}{2}, r}^n ] \times [ t^n, t^n + \Delta t ]}$ around each cell interface $x_{j + \frac{1}{2}}$ is chosen.
		The spatial extent corresponds to the domain which is causally affected by information propagating with the local velocities away from the cell interface at $x_{j + \frac{1}{2}}$.
		The flux integrals of Eq.~\eqref{eq:FVintEq} are performed on these space-time control volumes separately using the midpoint rule to approximate the flux integrals and leading to averages $\bar{\omega}_{j}^{n + 1}$ and $\bar{\omega}_{j+\frac{1}{2}}^{n + 1}$ based on the new intermediate spatial grid spanned by the points
			\begin{align}
				x_{j + \frac{1}{2}, l}^n = \, & x_{j + \frac{1}{2}}-a_{j + \frac{1}{2}}^n\Delta t \, ,	\vphantom{\bigg(\bigg)}
				\\
				x_{j + \frac{1}{2}, r}^n = \, & x_{j + \frac{1}{2}} + a_{j + \frac{1}{2}}^n\Delta t \, .	\vphantom{\bigg(\bigg)}
			\end{align}
		In the regions $[ x_{j - \frac{1}{2}, r}^n, x_{j + \frac{1}{2}, l}^n ]$ the solutions underlying the computed averages $\bar{\omega}_{j}^{n + 1}$ are smooth while the solutions underlying the computed averages $\bar{\omega}_{j+\frac{1}{2}}^{n + 1}$ are non-smooth based on the regions $[ x_{j + \frac{1}{2}, l}^n, x_{j + \frac{1}{2}, r}^n ]$. Details of this step can be found in Ref.~\cite{KTO2-0}.
		
		\item	A MUSCL-type piecewise-linear reconstruction based on $\bar{\omega}_{j + \frac{1}{2}}^{n + 1}$ and $\bar{\omega}_{j}^{n + 1}$ is used to project these averages back onto the original uniform grid spanned by $x_{j + \frac{1}{2}}$.
		This results in a fully discrete second-order central scheme, see Eq.~(3.9) of Ref.~\cite{KTO2-0}, which gives an algebraic expression for $\bar{u}_j^{n + 1}$ in terms of the averages
			\begin{align}
				\{ \bar{u}_{j - 2}^n, \bar{u}_{j - 1}^n, \bar{u}_{j}^n, \bar{u}_{j + 1}^n, u_{j + 2}^n \}	\label{eq:kt_stencil}
			\end{align}
		and $\{ a_{j \pm \frac{1}{2}}^n \}$.
		A pictographic representation of the multi-step evolution procedure with the involved quantities and grids can be found in Fig.~3.2 of Ref.~\cite{KTO2-0}.
		The numerical viscosity of this second-order scheme is $\mathcal{O} ( \Delta x^3 )$ and does not depend on $\Delta t$, which is an improvement when compared to the $\Delta t$ dependent numerical viscosities $\mathcal{O} ( \Delta x^2/\Delta t )$ and $\mathcal{O} ( \Delta x^4/\Delta t )$ of the LxF and NT schemes, respectively~\cite{KTO2-0}.
	\end{enumerate}
	
	The $\Delta t$ independent numerical viscosity allows for a controlled limit $\Delta t \rightarrow 0$, resulting in a reduction to a practical semi-discrete scheme in the conservative form \cite{KTO2-0}, which can be implemented straightforwardly:
		\begin{align}
			\partial_t \bar{u}_j = - \frac{H_{j + \frac{1}{2}} - H_{j - \frac{1}{2}}}{\Delta x} + \ldots \, ,	\label{eq:FVKTO2H}
		\end{align}
	where the ellipsis denotes the diffusion and source fluxes.
	The numerical advection fluxes $H_{j + \frac{1}{2}}$ are given by
		\begin{align}
			H_{j + \frac{1}{2}} \equiv \, & \frac{F \big[ t, x_{j + \frac{1}{2}}, u_{j + \frac{1}{2}}^+ \big] + F \big[ t, x_{j + \frac{1}{2}}, u_{j + \frac{1}{2}}^- \big]}{2} -	\vphantom{\Bigg(\Bigg)}	\label{eq:definition_h_kt_scheme}
			\\
			& - a_{j + \frac{1}{2}} \, \frac{u_{j + \frac{1}{2}}^{+} - u_{j + \frac{1}{2}}^{-}}{2} \, .	\vphantom{\Bigg(\Bigg)}	\nonumber
		\end{align}
	This semi-discrete scheme is second-order accurate in $\Delta x$ and can be used in conjunction with various ODE time-step algorithms (in this work, we use \texttt{Mathematica}'s \textit{NDSolve} \cite{Mathematica:12.1}). 
	The KT scheme for a position-independent advection flux is conservative, meaning detailed balance at the cell interfaces is maintained.
	It is also total variation diminishing/non-increasing (TVD/TVNI) \cite{HARTEN1983357,LeVeque:1992,LeVeque:2002}\footnote{%
		This property is important for the resolution of shocks and discontinuities.
		A TVD scheme does not produce spurious oscillations around discontinuities.
		For a TVD scheme ${\sum_j \big| \bar{u}_{j + 1}^{n + 1} - \bar{u}_{j}^{n + 1} \big| \leq \sum_j \big|\bar{u}_{j + 1}^n - \bar{u}_{j}^n \big|}$ holds between all time steps, which is why the more concise expression is total variation non-increasing, \textit{cf.} Sec.~9.2.2 of Ref.\ \cite{RezzollaZanotti:2013}.
		The total variation ${\mathrm{TV}[\{\bar{u}_{j}\}]=\sum_j \big|\bar{u}_{j + 1} - \bar{u}_{j} \big|}$ is a discrete measure for the arc-length of $u(t,x)$.
		In the subsequent parts II and III in this series of publications \cite{Koenigstein:2021rxj,Steil:2021cbu} we show that the TV can serve as a (numerical) entropy for the zero-dimensional $O(N)$ model with $N=1$ and $N\rightarrow\infty$ and might provide a link to the formulation of the $\mathcal{C}$ function in truncated flow equations \cite{Zamolodchikov:1986gt}.%
		} when used with appropriate flux limiters like the minmod limiter \eqref{eq:FVminmod}.\\
	
	So far we only considered the advection term $\partial_x F[u]$ in the discussion of the KT scheme.
	The explicit treatment of source and diffusion fluxes in the semi-discrete KT scheme is as follows:
	\begin{enumerate}
		\item If the source term is independent of $u$ and of type $S ( t, x ) = \partial_x s ( t, x)$, it can be integrated directly and contributes $( s_{j + \frac{1}{2}} - s_{j - \frac{1}{2}} )/\Delta x$ with $s_{j + \frac{1}{2} } \equiv s ( t, x_{j + \frac{1}{2}} )$ to the right-hand side of Eq.~\eqref{eq:FVKTO2H}.
		Otherwise, if $S [ u ] \equiv S [ t, x, u ( t, x ) ]$ suitable approximations, like $S [ t, x_j, \bar{u}_j (t) ]$ may be used.
		For a detailed discussion source/sink terms arising in the FRG treatment of higher-dimensional QFTs we refer the interested reader to Ref.~\cite{Stoll:2021ori} and especially App. E of the aforementioned publication.

		\item	When considering a non-linear diffusion flux $Q [ u, \partial_x u ]$ Eq.~\eqref{eq:FVcdEq} is a potentially strongly degenerate parabolic equation (system) admitting potentially non-smooth solutions.
		In the KT scheme the hyperbolic and parabolic part of the PDE \eqref{eq:FVcdEq} are treated simultaneously based on the strict splitting between $F$ and $Q$.
		Kurganov and Tadmor \cite{KTO2-0} presented a discretization of the diffusion flux based on a kind of central-difference approximation,
		\begin{align}
			 P_{j + \frac{1}{2}} = \, & \tfrac{1}{2}\, Q \big[ t, x_j, \bar{u}_j, \tfrac{\bar{u}_{j + 1} - \bar{u}_j}{\Delta x} \big] +\vphantom{\bigg(\bigg)}	\label{eq:kt_original_diffusion}
			\\
			&  + \tfrac{1}{2}\, Q \big[ t, x_{j + 1}, \bar{u}_{j + 1}, \tfrac{\bar{u}_{j + 1} - \bar{u}_j}{\Delta x} \big] \, .	\vphantom{\bigg(\bigg)}	\nonumber
		\end{align}
		An alternative second-order discretization like the one put forward in App.~B of Ref.~\cite{Chertock2005} can also be successfully employed: 
		If the diffusion flux can be written as
		\begin{align}
			\qquad Q [ t, x, u, \partial_x u ] &= \varphi [ t, x, u ] \cdot \tilde{Q}[ t, x, \partial_x u ] \, .	\vphantom{\bigg(\bigg)}
		\end{align}
		then
		\begin{align}
			\qquad P_{j + \frac{1}{2}} = \, & \varphi [ t, x_{j + \frac{1}{2}}, \tfrac{1}{2} \big( u_{j + \frac{1}{2}}^{+} + u_{j + \frac{1}{2}}^{-} \big) ]\cdot \vphantom{\bigg(\bigg)}\label{eq:kt_chertock_diffusion}\\
			& \quad \cdot \tilde{Q} \big[ t, x_{j + \frac{1}{2}}, \tfrac{\bar{u}_{j + 1} - \bar{u}_j}{\Delta x} \big] \, ,\notag
		\end{align}
		which can be applied for the RG flow equation~\eqref{eq:conservation_law_u_phi} under consideration with $\varphi [ t, x, u ]=1$ and $\tilde{Q}[ t, x, \partial_x u ]=Q[ t, \partial_x u ]$, since $Q$ has no explicit position dependence when using the formulation in $\sigma$, \textit{cf.} Eq.~\eqref{eq:diffusion_flux_sigma_propagator}.
		Both approaches (using Eq.~\eqref{eq:kt_original_diffusion} or \eqref{eq:kt_chertock_diffusion} for the numerical diffusion flux $P_{j + \frac{1}{2}}$) were successfully employed in the context of this work for the diffusion flux \eqref{eq:diffusion_flux_sigma_propagator} without visible differences in quality.
		Still, all numerical results in Sec.~\ref{sec:results} are obtained using the formulation \eqref{eq:kt_original_diffusion}.
		For improved KT-type schemes employing higher-order reconstructions (like, \textit{e.g.}, C-WENO/WENO5 or MP5) higher-order discretizations for the diffusion flux are used like the fourth-order one put forward in Eqs.~(4.9) and (4.10) of Ref.~\cite{KTO3-0}.
	\end{enumerate}
	
	 The full semi-discrete KT scheme including advection, diffusion, and source flux is given by
		\begin{align}
			& \partial_t \bar{u}_j = \vphantom{\bigg(}\label{eq:FVKTO2}
			\\
			= \, & - \frac{H_{j + \frac{1}{2}} - H_{j - \frac{1}{2}}}{\Delta x} 
			+ \frac{P_{j + \frac{1}{2}} - P_{j - \frac{1}{2}}}{\Delta x} 
			+ \mathrm{source} \, . \vphantom{\bigg(} \nonumber
		\end{align}
	
	Specific spatial boundary conditions $( \partial_x ) u ( t, x ) |_{x \in \partial \mathcal{V}}$ manifest themselves in the KT scheme in the choice of volume averages for so-called \textit{ghost cells}.
	For the evaluation of the fluxes for the first two ($\bar{u}_0$ and $\bar{u}_1$) and last two ($\bar{u}_{n-2}$ and $\bar{u}_{n-1}$) cell averages ghost cells outside the computational domain are required when considering $n$ volume cells, see, \textit{e.g.}, Refs.~\cite{LeVeque:1992,LeVeque:2002} for a detailed discussion.
	The ghost cells are $u_{-2}$, $u_{-1}$, $u_{n}$, and $u_{n + 1}$ centered at $x_{-2}$, $x_{-1}$, $x_{n}$, and $x_{n+1}$, respectively.
	Depending on the problem and computational domain at hand a multitude of different boundary conditions can be implemented by an appropriate choice or reconstruction of these averages for the ghost cells.
	The specific boundary conditions for the RG flow equations in this work are discussed at length in Sub.Sec.~\ref{subsec:boundary_conditions_finite_volume} below.\\
	
	At this point we have to remark that the original KT numerical scheme presented in Ref.~\cite{KTO2-0} was constructed for position- and time-independent advection and diffusion fluxes.
	Since we employ the KT scheme in its semi-discrete form a resolution of potentially highly complicated and non-linear dynamics in $t$ is possible and ultimately outsourced to the ODE solver.
	The spatial discretization of the advection term~\eqref{eq:advection_flux_pion_propagator} in the semi-discrete KT scheme seems to be able to cope with the explicit position dependence in $F[t,x,u]$ when evaluating the numerical fluxes at the appropriate cell interfaces, \textit{cf.} Eq.~\eqref{eq:definition_h_kt_scheme}.
	In the scope of this paper we could not trace any practical problems back to the explicit position- and time-dependence of the advection and diffusion fluxes.
	The comparisons in 	Sec.~\ref{sec:results} between results obtained from a direct computation of correlation functions using the generating functional \eqref{eq:ON_expectation_value} and the results computed using RG flow equations via the KT scheme (with $t$ and $x$ dependent fluxes) can be seen as hard tests for both -- the FRG methodology as well as the (slightly modified) KT scheme -- depending on the respective perspective. In total, the precision of our results for the non-trivial test cases gives us some confidence that our approach is generically justified and the KT scheme is suitable for our purpose.
	As mentioned in the paragraph concerning entropy and irreversibility of Sub.Sub.Sec.~\ref{subsec:advection_diffusion_equation} the explicit position dependences in both RG flow Eqs.~\eqref{eq:conservation_law_u_rho} and \eqref{eq:conservation_law_u_phi} prevents a direct usage of the total variation TV as entropy functional for the zero-dimensional $O(N)$ model at finite $N>1$.
	The internal source terms in the primitive forms originating in the explicit position dependences of $F$ and $Q$ before performing the spatial derivatives lead to an increase in arc length during RG time evolution, rendering the TV useless as an entropy functional. For more details see the subsequent parts II and III in this series of publications \cite{Koenigstein:2021rxj,Steil:2021cbu}, especially App. E of part III.
	
\subsection{Boundary conditions and computational domain in FRG flow equations}
\label{subsec:boundary_conditions_finite_volume}

	In the form of the conservation law \eqref{eq:conservation_law_u_rho} or \eqref{eq:conservation_law_u_phi}, the RG flow equation \eqref{eq:flow_equation_effective_potential} is a non-linear PDE which has contributions of parabolic (diffusion terms) as well as hyperbolic type (advection terms).
	In this subsection, we specify the boundary conditions for Eq.~\eqref{eq:conservation_law_u_rho} or \eqref{eq:conservation_law_u_phi} in field space (the effective spatial $x$ direction).\\
	
	The correct implementation of field space (spatial) boundary conditions is very important for the correct numerical solution of PDEs \cite{LeVeque:1992,LeVeque:2002}.
	For (non-linear) PDEs of elliptic type, the problem of solving them is also referred to as \textit{boundary-value problem}.
	For (non-linear) PDEs of hyperbolic and parabolic type, the spatial boundary conditions are needed (in addition to the initial condition) to make finding a (weak) solution a well-defined problem.
	The latter case is also referred to as \textit{Cauchy} or \textit{initial-boundary-value problem}.
	Thus, without explicitly specifying the boundary conditions, \textit{e.g.}, of Neumann- or Dirichlet-type, as well as the initial conditions, the problem of finding a unique (weak) solution is actually ill-posed and therefore impossible to solve -- a well-known mathematical fact with particular and severe implications in, \textit{e.g.}, classical electrodynamics \cite{Jackson:1998nia}, fluid dynamics \cite{BuckleyLeverett:1942}, soliton and instanton solutions of classical field equations \cite{Rajaraman:1982is,Shifman:1994ee}, general relativity \cite{Weinberg:1972kfs,Misner:1974qy,Ryder:2009zz}, and other fields of research.
	This also holds true for the FRG. 
	However, explicit boundary conditions and especially their numerical implementation are rarely discussed the FRG literature, with ,\textit{e.g.}, Refs.~\cite{Caillol:2012zz,Pangon:2009pj,Pangon:2010uf,Codello:2013iqa} as notable exceptions.
	
	For the derivative of the effective potential $u ( t, \sigma )$, we find that the spatial boundary conditions must be imposed at $\sigma = \pm \infty$, because the field space domain of $u ( t, \sigma )$ is given by $\mathbb{R}$.
	Thus, when considering the flow equation on the non-compact domain $( - \infty, \infty )$ the problem represents a \textit{pure initial-value/Cauchy problem} \cite{Ames:1992,LeVeque:1992,LeVeque:2002} and, given the asymptotics of the flow equation and the initial condition, explicit boundary conditions at $x\rightarrow\pm\infty$ are not required.
	However, spanning a non-compact computational interval from $ - \infty$ to $ + \infty$ is practically impossible on a finite computational grid.
	A possible solution is a compactification \cite{Borchardt:2016pif} of $\mathbb{R}$ to the interval $[ - 1, + 1 ]$, via a suitable mapping $\sigma \mapsto x ( \sigma )$ usually supplemented with a mapping $u \mapsto v(u)$ rendering $v$ finite on $[ - 1, + 1 ]$.
	Another popular solution is a truncation of the computation interval at a large value $\sigma_\text{max} \sim x_\mathrm{max}$ with a suitable boundary condition \cite{Pangon:2009pj,Caillol:2012zz,Borchardt:2015rxa,Borchardt:2016pif}.
	We will return to this issue below.
	
	In any case, one of the boundaries at spatial infinity can already be replaced by a finite value by making use of the $O ( N )$ symmetry of the potential $U ( t, \vec{\varphi} \, )$ and the flow equations, which implies a $\mathbb{Z}_2$ antisymmetry of $u ( t, \sigma ) = \partial_\sigma U ( t, \sigma )$,
		\begin{align}
			&	U ( t, \sigma ) = U ( t, - \sigma ) &&	\Longleftrightarrow	&&	u ( t, \sigma ) = - u ( t, - \sigma ) \, .	\label{eq:anti-symmetry_small_u}
		\end{align}
	This reduces the spatial domain to the half-open interval $\sigma \in [ 0, + \infty )$, but now we need an additional artificial boundary condition at $\sigma = 0$, see, \textit{e.g.}, Ref.~\cite{Pangon:2009pj}.
	In previous studies, the use of the $O(N)$ symmetry was usually implemented right from the beginning by replacing the variable $\vec{\varphi}$ by the $O(N)$ invariant $\varrho = \tfrac{1}{2}\,\vec{\varphi}^{\, 2}$, whose domain is already by definition $[ 0, \infty )$.\footnote{%
		In any case, independent of the implementation of the boundary condition itself, one should make use of symmetries of the flow equations in numerical implementations.
		First of all, this leads to a reduction of the number of computational grid points in spatial direction, while keeping the spatial resolution fixed, which significantly speeds up the calculations (independently of the specific numerical method for spatial discretization).
		An additional consequence is the reduction of numerical errors: It is highly unlikely that the numerical errors are symmetric in $x$, if a symmetric interval around $x=\sigma=0$ is used.
		This might lead to an artificial breaking of the $\mathbb{Z}_2$ (anti-)symmetry by unbalanced numerical errors.
		Although these errors might be tiny and almost negligible they can be easily circumvented by exploiting the symmetries.
		Using the symmetries of a problem is a standard procedure in practical computations and of particular importance in, \textit{e.g.}, numerical fluid dynamics and numerical (general) relativity, see Refs.~\cite{Baumgarte2010Jun,Alcubierre2008,Grandclement:2007sb,Gourgoulhon:2007ue}.%
	} 
	In this case one has to define 
		\begin{align}
			u ( t, \varrho ) \equiv \partial_\varrho U ( t, \rho ) = \tfrac{1}{\sigma} \, \partial_\sigma U ( t, \sigma ) = \tfrac{1}{\sigma} \, u ( t, \sigma ) \, ,
		\end{align}
	to obtain a flow equation for $u ( t, \varrho )$ in a manifestly conservative form, see Eqs.~\eqref{eq:conservation_law_u_rho} and \eqref{eq:conservation_law_u_phi}.
	
	Before returning to the remaining boundary condition at $+ \infty$, we first consider the newly introduced artificial boundary condition at ${x = \sigma =0}$ or, correspondingly, at $ \varrho=0$.

\subsubsection{\texorpdfstring{Boundary condition at $\sigma = 0$}{sigma = 0}}
\label{subsec:phi0BC}

	At first sight it might be appealing to formulate the whole problem -- the conservation equation and the boundary condition at $\sigma = 0$ -- in the variable $\varrho$.
	However, we believe that a formulation in $\sigma$ is more suitable and easier to implement in our numerical setup.\footnote{%
		We do not claim that it is impossible to formulate well-defined discrete boundary conditions in $\varrho$ at $\varrho = 0$, as can be seen for example in Refs.~\cite{Grossi:2019urj,Grossi:2021ksl,Steil:2021cbu} for the specific case of the large-$N$ limit of the $O(N)$ model. 
		However, we were not able to provide a suitable discretization of the boundary condition at $\varrho = 0$ in the implementation of the FV method for flow equations that include diffusion via the radial $\sigma$ mode.
	}
	A key feature of (non-linear) hyperbolic/parabolic conservation equations is that their weak solutions may exhibit non-analyticities in the form of shock and rarefaction waves \textit{etc.}, which manifest themselves in the solution in cusps or discontinuities in spatial direction during the time evolution.
	These effects can develop during the time evolution even if the initial condition is smooth/analytic see, \textit{e.g.}, Refs.~\cite{KTO2-0,Chen:2001,Bateman1915,Burgers1948,LeVeque:1992,LeVeque:2002,Borchardt:2016pif}.
	As demonstrated in Refs.~\cite{Aoki:2017rjl,Wink:2020tnu,WinkHirschegg,Grossi:2019urj,Grossi:2021ksl,Ihssen2020,Stoll:2021ori} this also holds for FRG flow equations, where non-analyticities are inherent properties of the effective IR potential $U ( t_\mathrm{IR}, \sigma )$.
	These statements are also true for the point $\sigma = 0$, where $U ( t, \sigma )$ and $u ( t, \sigma )$ do not need to be analytic, see Sub.Sec.~\ref{subsec:sc4}.
	Hence, there might be a scenario where the potential $U ( t, \sigma )$, although it is symmetric in $\sigma$, has a cusp at $\sigma = 0$, which would correspond to a jump in a weak solution for $u ( t, \sigma ) = \partial_\sigma U ( t, \sigma )$ at $\sigma = 0$.
	If formulated in $\varrho$, any scenario (analytic or non-analytic at $\sigma = 0$) merely corresponds to some arbitrary value for $u ( t, \varrho ) = \partial_\varrho U ( t, \varrho )$ at $\varrho = 0$, which seems to be of great advantage, because one does not have to deal with possible discontinuities in the conserved quantity $u$.
	Furthermore, the problematic factors of $\frac{1}{\sigma}$ in the pion propagator and the advection flux \eqref{eq:advection_flux_pion_propagator}, which are diverging at $\sigma = 0$, can be avoided when formulating the flow equations in $\varrho$.
	
	Nevertheless, a problem with the variable $\varrho$ becomes apparent when turning to the discretized form of $u$ within the FV scheme presented in Sub.Secs.~\ref{subsec:finite_volume_method} and \ref{subsec:kt_scheme}: FV methods (and also other discretization schemes) usually require ghost cells at the boundaries of the computational domain, since the in- and out-flows for the $i^\text{th}$ cell are calculated from the cell averages $\bar{u}$ of its neighboring cells, \textit{cf.}\ Eq.~\eqref{eq:kt_stencil}.
	However, initially these values are not specified for the cells at the boundaries of the computational domain.
	Thus, artificial ghost cells must be introduced and the numerical values for $\bar{u}$ in these ghost cells have to be implemented by hand or reconstructed from the cells within the computational domain in accordance with the boundary conditions \cite{LeVeque:1992,LeVeque:2002}.
	In the second-order formulation of the one-dimensional KT scheme one needs two ghost cells at each of the two spatial boundaries, \textit{cf.} Eq.~\eqref{eq:kt_stencil}.
	
	However, implementing ghost cells for $u ( t, \varrho )$ at $\varrho = 0$ is conceptually difficult, because these ghost cells must be centered at negative values for $\varrho$ outside the computational domain $[ 0, \infty)$, which by definition do not exist due to the positivity of $\varrho = \tfrac{1}{2}\,\sigma^2$.
	\textit{A priori}, it is therefore not clear how numerical values $\bar{u} ( t, \varrho_i )$ should be assigned to ghost cells at negative $\varrho_i$, because symmetry arguments cannot be applied anymore.
	
	Furthermore, it is also not a feasible option to move the ghost cells to positive values of $\varrho_i$, such that the point $\varrho = 0$ is no longer part of the computational domain.\
	Namely, having ghost cells centered at small but positive $\varrho_i$ implies that one has to extrapolate the numerical values $\bar{u} ( t, \varrho_i )$ to these ghost cells and to the point $\varrho = 0$ from the other ordinary cells of the computational domain.
	However, the functional behavior of $u ( t, \varrho )$ is unknown for small $\varrho$ and is actually exactly what we want to calculate in the first place by solving the PDEs.
	Thus, any extrapolation at small $\varrho$ can only be considered an educated guess.
	It is especially dangerous, because the physical point will be part of the extrapolated ghost cells if it is located at $\varrho = 0$, which is the case for all models in their symmetric phase \cite{Stoll:2021ori}, irrespective of the dimensionality of space-time.
	Consequently, extrapolation errors at the physical point have the potential to spoil the numerical values of all $n$-point correlation functions, which are calculated at the physical point via derivatives of $u$ and contain the physics of the model.
	Even if the physical point is at finite non-zero $\varrho$ far away from the ghost cells and the boundary at $\varrho = 0$, any kind of extrapolation at small $\varrho$ leads to numerical errors, because the diffusive contributions of the radial $\sigma$ mode will propagate this information from smaller to larger $\varrho$ and hence to the physical point.
	Similar problems in formulating appropriate boundary conditions at $\varrho = 0$ also exist in other discretization schemes like finite-difference or finite-element methods.
	
	There is one exception to this conclusion: In the large-$N$ limit of the $O(N)$ model the flow equation for $u$ reduces to a pure advection equation.
	Studying the characteristic velocities, which are given by $\partial F/\partial u$, respectively, see Eq.~\eqref{eq:advection_velocity}, we find that these cannot change their sign, and information (or the conserved quantity $u$) is always propagated via advection in the direction of smaller $\varrho$ or $|\sigma|$.
	In this scenario, ghost cells can be positioned at negative $\varrho_i$ and the corresponding cell averages $\bar{u}_i$ in the ghost cells can take any numerical value since information from the ghost cells is never propagated back into the computational domain and cannot cause any errors, \textit{cf.}\ Ref.~\cite{Grossi:2019urj,Grossi:2021ksl,Steil:2021cbu}.
	Shifting the ghost cells into regions of positive $\varrho$ is still not suitable for the reasons already discussed above.\\
	
	In order to avoid all these difficulties when formulating the problem in the variable $\varrho$, we suggest a formulation in $\sigma$ and an implementation of the boundary condition at $\sigma = 0$.
	The key argument for using $\sigma$ instead of $\varrho$ is that positioning ghost cells at negative $\sigma$ poses no problem at all, since negative $\sigma$ exist in the first place. 
	Furthermore, it is clear how the cell averages $\bar{u} ( t, \sigma_i )$ in the ghost cells have to be chosen: Using the antisymmetry \eqref{eq:anti-symmetry_small_u}, one merely has to mirror the last physical cells of the computational domain at $\sigma = 0$ to the ghost cells (including a flip in sign).
	The only issue that requires careful consideration is the choice of the position of the first physical cell $x_0$ next to $\sigma = 0$: The flux term of our PDE contains factors $\frac{1}{\sigma}$ via the pion propagators, which diverge if evaluated at $\sigma = 0$.
	Therefore, we must avoid evaluating the fluxes $F [ t, x, u ( t, x ) ]$ at $x = \sigma = 0$.
	However, inspecting the KT scheme, we find that the fluxes as well as the Jacobian $ \frac{\partial F[u]}{\partial u}$ must only be evaluated at the cell boundaries $x_{j \pm \frac{1}{2}}$, \textit{cf.}\ Eqs.~\eqref{eq:FVajp12} and \eqref{eq:definition_h_kt_scheme}.
	Consequently, the natural choice for the position of the cell center $x_0$ of the first physical cell in the computational domain is at $x = \sigma = 0$, such that the in- and out-fluxes of this cell are evaluated at $x_{\pm \frac{1}{2}}$, which is not problematic.
	Incidentally, this automatically cures the problem of the possibility of non-analyticities in $u ( t, \sigma )$ at $\sigma = 0$: Even if $u ( t, \sigma )$ is discontinuous at $\sigma = 0$ we do not run into problems, because all numerical calculations are performed on the level of cell averages $\bar{u} ( t, \sigma_i )$.
	The cell average of an antisymmetric function in a cell that is centered at $\sigma = 0$ must always vanish identically, independent of all other properties of the function, see also Ref.~\cite{Pangon:2009pj,Pangon:2010uf}.
	
\begin{figure}
			\centering
			\includegraphics{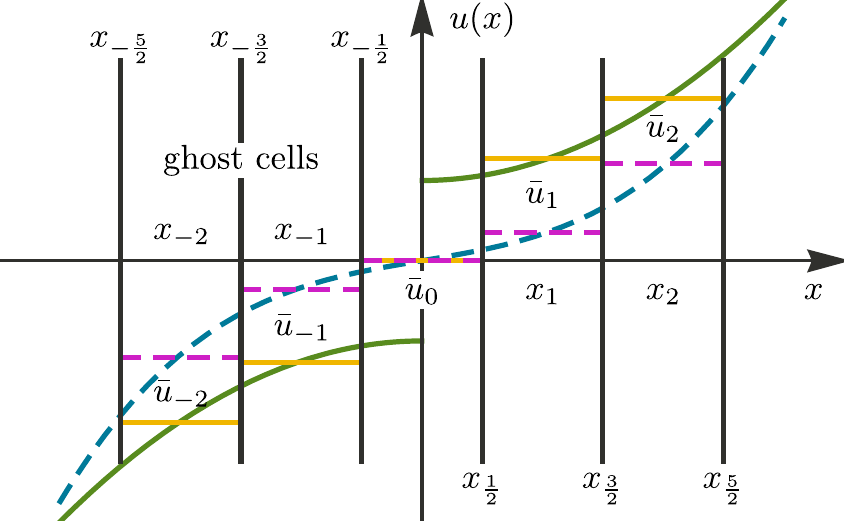}
			\caption{\label{fig:numerical_boundary_conditions_phi=0}%
				Second-order accurate FV implementation of the spatial boundary condition for $u ( t, x)$ or $\bar{u}_i (t)$, respectively, at $x = 0$. 
				We use the fact that $u ( t, x )$ is an odd function in $x$ by positioning the first computational cell $x_0$ at $x = 0$, such that the cell average is exactly zero, $\bar{u}_0 = 0$, which is true for $u (t, x)$ which are analytic (blue-dashed) as well as non-analytic (green-solid) at $x = 0$.
				The ghost-cell averages can thus be fixed by setting $\bar{u}_{-2} = - \bar{u}_2$ and $\bar{u}_{-1} = - \bar{u}_1$.
				Corresponding cell averages $\bar{u}_i$ are depicted as horizontal bars (magenta-dashed and yellow-solid).
				This boundary condition can be generalized to lower- and higher-order accurate FV schemes as well as finite-difference or finite-element schemes.%
			}
		\end{figure}

	In summary, we switch from the open computational interval $ ( - \infty, + \infty ) $ to the half-open computational interval $ [ 0, + \infty ) $ by means of the $\mathbb{Z}_2$ (anti-)symmetry using $\bar{u}_{-2} ( t ) = - \bar{u}_{2} ( t )$, $\bar{u}_{- 1} ( t ) = - \bar{u}_{1} ( t )$ for the cell averages in the ghost cells left of $x_0 = 0$ and $\bar{u}_0 ( t ) = 0$ for the cell average in the cell at $x_0$.
	This effectively corresponds to reflective boundary conditions frequently imposed in numerical fluid dynamics \cite{LeVeque:1992,LeVeque:2002}, \textit{cf.} Fig.~\ref{fig:numerical_boundary_conditions_phi=0}.
	 
\subsubsection{\texorpdfstring{Boundary condition at $\sigma \rightarrow \infty$}{sigma rightarrow infinity}}
\label{subsec:sec:BCinf} 

	Now we return to the boundary condition at $\sigma \rightarrow + \infty$. Without loss of generality we discuss the interval ${\sigma \in [ 0, + \infty )}$ since the situation in $\sigma \in (- \infty,0]$ follows from $\mathbb{Z}_2$ antisymmetry of $u(t,\sigma)$.
	
	We have already argued that there are no real boundary conditions at spatial infinity on a non-compact domain.
	The behavior of $u$ at $\sigma \rightarrow \infty$ is rather given by the asymptotics of the FRG equation, which makes the PDE an pure initial-value problem.
	The boundary condition at spatial infinity is actually fixed implicitly: As long as the initial potential $U ( t = 0, \sigma )$ is bounded from below and grows faster than $\sigma^2$ for $\sigma \rightarrow \infty$ both pion and sigma propagator tend to zero for sufficiently large $\sigma$, such that the right-hand side of the PDE \eqref{eq:flow_equation_derivative_effective_potential} vanishes during the entire FRG flow.
	In the fluid-dynamical picture this corresponds to vanishing advection and diffusion fluxes \eqref{eq:advection_flux_pion_propagator} and \eqref{eq:diffusion_flux_sigma_propagator} at $\sigma \rightarrow \infty$, which is a zero-influx boundary condition for $u ( t, \sigma )$.
	The derivative of the effective potential $u ( t, \sigma )$ is therefore fixed to its initial value $u ( t = 0, \sigma )$ at $\sigma \rightarrow \infty$.
	
	The limiting case, when the asymptotic behavior of the initial UV potential is quadratic,
		\begin{align}
			&	\lim\limits_{\sigma \rightarrow \infty} U ( t = 0, \sigma ) \sim \sigma^2 \, ,	&&	\lim\limits_{\sigma \rightarrow \infty} u ( t = 0, \sigma ) \sim \sigma \, ,
		\end{align}
	is more delicate.
	In this case, the advection and diffusion fluxes \eqref{eq:advection_flux_pion_propagator} and \eqref{eq:diffusion_flux_sigma_propagator} do not vanish for $\sigma \rightarrow \infty$ for all RG times.
	However, for small RG times $t \approx 0$, the fluxes are actually independent of $\sigma$ at large $\sigma$ due to the constant asymptotic slope of the initial condition $u ( t = 0, \sigma )$. 
	This in turn implies that the in- and out-flux for all volume cells at large $\sigma$ only depend on $t$ and must cancel exactly, such that the net flux of these cells vanishes.
	Therefore, also in this scenario $u ( t, \sigma )$ is fixed to its initial condition at $\sigma \rightarrow \infty$ not only for small $t$, but rather for all RG times $t$.
	For late RG times $t \rightarrow \infty$, the advection and diffusion fluxes \eqref{eq:advection_flux_pion_propagator} and \eqref{eq:diffusion_flux_sigma_propagator} vanish anyhow, due to the derivatives of the regulator shape functions in the numerators, \textit{i.e.}, $\partial_t r ( t ) = - \Lambda \, \mathrm{e}^{-t}$.
	In the language of fluid dynamics, initial conditions with quadratic asymptotics can therefore be interpreted as boundary conditions with time-dependent but spatially constant in-flux, \textit{cf.}\ Examples~7 and~9 in Ref.~\cite{KTO2-0}.
	
	However, both cases cannot be implemented directly on a finite computational domain and we basically have two options:
	\begin{enumerate}
		\item	We could try to map the interval $[ 0, \infty )$ to a compact interval $[0, 1]$ via a suitable map $\sigma \mapsto x ( \sigma )$.
		This also includes a suitable mapping of $u \mapsto v ( u )$ to keep the values for the conserved quantity finite on $[0, 1]$.
		This option has the advantage that the correct asymptotic behavior $u ( t, \sigma )$ can be implemented as boundary conditions for $v ( t, x )$ at $x = 1$.
		However, the same question then arises as before in the discussion of an appropriate choice of ghost cells for negative values of $\varrho$: It is highly non-trivial how the cell averages $\bar{v}_i$ should be fixed for ghost cells which no longer belong to the physical values of $x$ within the interval $[0, 1]$.
		Additionally, the two mappings would introduce at least two new numerical functional-mapping parameters.
		A suitable choice of these parameters is not obvious.
		Still, these mappings would have to ensure dense grids and high resolution around the physical point and low resolution at large field values
		All this is extremely hard to achieve.
		Therefore, we propose and favor another option.
		
		\item	The second option, which is our preferred choice, is to split the physical domain $[ 0, \infty )$ into a compact domain $[ 0, \sigma_\mathrm{max} ]$ and a non-compact domain $[ \sigma_\mathrm{max} , \infty )$.
		Here, $\sigma_\mathrm{max}$ should be chosen much larger than the physical scales of the problem and the position of the physical point, see, \textit{e.g.}, Refs.~\cite{Grossi:2019urj,Ihssen2020,Stoll:2021ori,Pangon:2009pj,Borchardt:2015rxa,Caillol:2012zz}.
		We will provide explicit tests for an appropriate choice of 	$\sigma_\mathrm{max}$ later on in Sec.~\ref{sec:results}.
		For the compact domain $[ 0, \sigma_\mathrm{max} ]$, we  keep a direct identification of the field $\sigma$ and the computational spatial variable $x$, thus $x = \sigma$.
		For higher-dimensional models this might be replaced by a linear map of $\sigma$ to a dimensionless spatial variable $x$ via appropriate rescaling with some characteristic dimensionful quantity, \textit{e.g.}, the UV cutoff $\Lambda$ or a non-vanishing condensate. 
		In any case, this allows for a direct implementation of the boundary condition at $\sigma = 0$.
		
		In the compact domain $[ 0, \sigma_\mathrm{max} ]$, we have to ensure a high spatial resolution via a sufficiently large number of cells, in order to capture all aspects of the dynamics around the physical point.
		Explicit tests to find an appropriate spatial resolution are also presented in Sec.~\ref{sec:results}.
		
		For the non-compact domain $[ \sigma_\mathrm{max}, \infty )$, instead of using a discretization scheme like the FV method, we suggest an expansion or approximation of $u ( t, \sigma )$ via polynomials or complete sets of functions with $t$ dependent expansion coefficients, which account for the asymptotic behavior of the initial condition $u ( t = 0, \sigma )$ for large $\sigma$. 
		As discussed before, it is expected that for large $\sigma$ the deviations of $u ( t, \sigma )$ from the initial condition $u ( t = 0, \sigma )$ are small during the FRG flow, such that a finite amount of expansion coefficients should be satisfactory to capture this minimal dynamics.
		
		At the point $\sigma_\mathrm{max}$, the ghost cells for the FV method in $[ 0, \sigma_\mathrm{max} ]$ can therefore be fixed via the values $u ( t, \sigma )$ from the asymptotic expansion in the non-compact interval $[ \sigma_\mathrm{max}, \infty )$.
	\end{enumerate}
	Interestingly, our numerical tests showed that, as long as $\sigma_\mathrm{max}$ is chosen sufficiently large, the fluxes at $\sigma_\mathrm{max}$ are already negligibly small.
	As a consequence, the deviation of $u ( t, \sigma )$ from the initial condition in the non-compact interval $[ \sigma_\mathrm{max}, \infty )$ is extremely small and can be ignored.
	In this case, the computational boundary conditions for the ghost cells at $\sigma_\mathrm{max}$ can be fixed via an extrapolation using the asymptotics of the initial condition.
	For extremely high spatial resolution, hence rather small $\Delta x$, even a simple linear extrapolation might be sufficient.
	
	On the other hand, choosing $\sigma_\mathrm{max}$ rather large while keeping a high spatial resolution in the compact computational domain $[ 0, \sigma_\mathrm{max} ]$ requires a large number of cells.
	However, this slows down the computations drastically.
	For problems where this issue becomes relevant, we suggest to further divide the compact domain $[ 0, \sigma_\mathrm{max} ]$ into several smaller subdomains.
	In each of these subdomains one can implement the FV method with different spatial resolution $\Delta x$ for each domain. 
	This ensures high resolution at small $\sigma$ next to the physical point and also allows to truncate the spatial interval at large $\sigma_\mathrm{max}$, while keeping a decent and manageable total number of cells \cite{Grossi:2019urj,Grossi:2021ksl}.
	An alternative approach would be switching from equally sized volume cells on a uniform grid to a non-uniform (potentially even moving/time dependent) grid, see, \textit{e.g.}, Ref.~\cite{KTmovingMesh}.
	However, in the context of FRG flow equations this might be relevant for models with multiple condensate directions, see, \textit{e.g.}, Refs.~\cite{Strodthoff:2011tz,Mitter:2013fxa,Rennecke:2016tkm,Lakaschus:2020caq}.
	In our test cases for the zero-dimensional $O(N)$ model the subdivision of the compact interval $[ 0, \sigma_\mathrm{max} ]$ or a formulation on non-uniform grids was, however, not necessary.

	We close the discussion of spatial boundary conditions with some final remarks: We have discussed that the correct implementation of spatial boundary conditions is of utmost importance.
	Otherwise there is no guarantee for the solution of a PDE and especially our FRG flow equations to converge against the correct result.
	Additionally, we remark that our specific choice and implementation might not be the best option at hand for arbitrary (higher-dimensional) models and arbitrary initial conditions within the FRG framework.
	In the current context of the zero-dimensional $O(N)$ model initial conditions without a proper large-$| \vec{\phi} \, |$ asymptotics, \textit{e.g.}, $[ 2 - \sin ( \vec{\phi}^{\, 2} )] \, \vec{\phi}^{\, 2}$ or even worse $[ 2 - | \sin( \vec{\phi}^{\, 2} ) | ] \, \vec{\phi}^{\, 2}$, and/or periodic potentials could be a very interesting topics for further research.

\section{Zero-dimensional field theory as testing ground for FRG}
\label{sec:results}

	After our general discussion of the theoretical basis for the solution of FRG flow equations, we shall discuss concrete applications in the following subsections. 
	To this end, we study the RG flow of various zero-dimensional field theories which differ by distinct initial conditions.
	Our choices for the initial conditions range from smooth potentials to extreme choices featuring non-analyticities.
	Note that such extreme choices are not only relevant to demonstrate the numerical performance and stability of our implementation but also for phenomenological reasons.
	In fact, in higher dimensions we expect non-analytic behavior to build up, \textit{e.g.}, in the IR limit, as associated with spontaneous symmetry breaking and the emergence of convexity of the effective action. 

\subsection{Test case I: Non-analytic initial condition}
\label{subsec:sc1}

\begin{figure}
			\centering
			\includegraphics{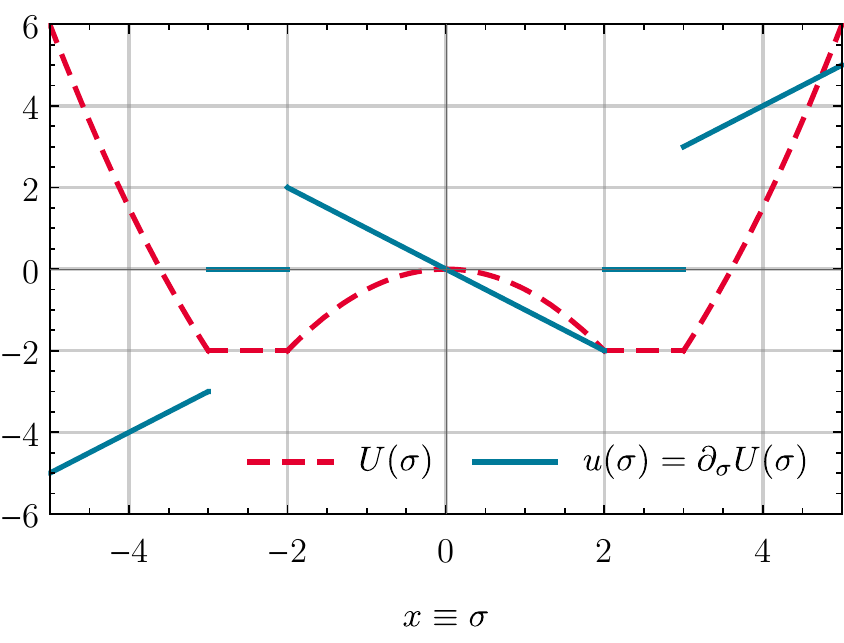}
			\caption{\label{fig:sc_i_uv_initial_condition}%
				The plot shows the UV potential $U ( \sigma )$ (red-dashed) and its first derivative $u ( \sigma ) = \partial_\sigma U ( \sigma )$ (blue, solid) of test case Eq.~\eqref{eq:testing_scenario_non-analytic_quadaratic_asymptote} evaluated on the constant background field configuration.%
			}
		\end{figure}

	Consider the following initial UV potential,
		\begin{align}
			U ( \vec{\varphi} \, ) =
			\begin{cases}
				- \tfrac{1}{2} \, \vec{\varphi}^{\, 2} \, ,			&	\text{if} \quad \varphi \leq 2 \, ,	\vphantom{\bigg(\bigg)}
				\\
				- 2 \, ,									&	\text{if} \quad 2 < \varphi \leq 3 \, ,	\vphantom{\bigg(\bigg)}
				\\
				+ \tfrac{1}{2} \, ( \vec{\varphi}^{\, 2} - 13 ) \, ,	&	\text{if} \quad 3 < \varphi \, ,	\vphantom{\bigg(\bigg)}
			\end{cases}	\label{eq:testing_scenario_non-analytic_quadaratic_asymptote}
		\end{align}
	where $\varphi = | \vec{\varphi} \, |$, \textit{cf.} Fig.~\ref{fig:sc_i_uv_initial_condition}.
	The test case is designed this way for the following reasons:
	\begin{enumerate}
		\item	All parameters of the potential $U ( \vec{\varphi} )$ are by default dimensionless and chosen to be approximately of order one, such that scales can easily be compared with each other.

		\item	The UV potential $U ( \sigma )$ has non-analytical points at $\sigma = 2$ and $\sigma = 3$, which correspond to discontinuities in its derivative $u ( \sigma )$.
		In the fluid-dynamical context such discontinuities present a Riemann problem and result in rarefaction waves.
		In QFT and 	thermodynamics such discontinuities can be associated with phase 	transitions, see App.~\ref{app:mermin_wagner}.
		The non-analytical behavior of this potential makes commonly used techniques like the FRG Taylor expansion inapplicable.
		Furthermore, naive discretizations that rely on smoothness are doomed to fail.
		One has to choose numerical schemes that can handle this numerically challenging dynamics.

		\item	The potential is initialized in the symmetry-broken phase, with infinitely many degenerate minima at $\sigma \in ( 2, \, 3]$.
		Furthermore, the UV potential is neither convex nor smooth.
		However, in the IR the $O(N)$ symmetry has to be restored and there must only be one minimum at $\sigma = 0$, due to the Coleman-Mermin-Wagner-Hohenberg theorem.
		Furthermore, for $t \rightarrow \infty$ the potential has to be convex and smooth, see App.~\ref{app:mermin_wagner}.
		This non-trivial transition and complicated dynamics from the UV to the IR is a numerically challenging test.
			
		\item	Furthermore, we choose a potential which is asymptotically quadratic in $\sigma$.
		This is to ensure that the large-$\sigma$ boundary condition for $u ( t, \sigma )$ is fully under control and errors can be excluded, see Sub.Sec.~\ref{subsec:boundary_conditions_finite_volume}. 
		This allows for a high-precision analysis of all other sources of numerical errors.
	\end{enumerate}
	The reference values for the exact IR 1PI vertex functions $\Gamma^{(2n)}$ of the $O(N)$ model are calculated numerically from the UV potential \eqref{eq:testing_scenario_non-analytic_quadaratic_asymptote} via the integral \eqref{eq:ON_expectation_value} using Eqs.~\eqref{eq:on-model_relation_2pf_phi2} --\eqref{eq:on-model_relation_6pf_phi2}.
	They are listed in Table~\ref{tab:sc_1_n_point_functions_exact} for selected $N$.
		\begin{table}[b]
			\caption{\label{tab:sc_1_n_point_functions_exact}%
				The table lists the (up to numerical-integration errors) exact results for the $\Gamma^{(2n)}$ of the $O(N)$ model with the initial UV potential \eqref{eq:testing_scenario_non-analytic_quadaratic_asymptote} for selected $N$.
				They are obtained by a high-precision one-dimensional numerical integration of the expectation values $\langle ( \vec{\phi}^{\, 2} )^n \rangle$ using \texttt{Mathematica}'s numerical integration routine  \textit{NIntegrate} \cite{Mathematica:12.1} with a \textit{PrecisionGoal} and \textit{AccuracyGoal} of $10$. 
				Here, we present the first ten digits only.%
			}
			\begin{ruledtabular}
				\begin{tabular}{l c c c}
					$N$		&	$\Gamma^{(2)}$	&	$\Gamma^{(4)}$	&	$\Gamma^{(6)}$	\\
					\colrule
					$1$		&	$0.1768130358$		&	$0.0520549107$		&	$\phantom{-}0.0865733100$ \\
					$3$		&	$0.3973542395$		&	$0.1408641479$		&	$\phantom{-}0.2249964322$ \\
					$10$	&	$0.8451440328$		&	$0.1519326337$		&	$-0.0691341063$
				\end{tabular}
			\end{ruledtabular}
		\end{table}

\subsubsection{General discussion of the FRG flow --\texorpdfstring{\\}{}advection and diffusion}

	We start our analysis with a general discussion of the FRG flow with initial condition \eqref{eq:testing_scenario_non-analytic_quadaratic_asymptote}.
	
	To this end, we fix $O(N = 3)$ to include both diffusive and advective contributions via the radial $\sigma$ mode and two pions. For $N = 3$ the number of pions is reasonably small and the (diffusive) effects of the $\sigma$ mode remain visible.
	Furthermore, we choose $[ 0, x_\mathrm{max} = 10 ]$ as the spatial computational domain with $800$ volume cells and use the KT scheme from Sub.Sec.~\ref{subsec:kt_scheme} for spatial discretization.
	The initial cell averages $\bar{u}_i ( t = 0 )$ are computed by exact averaging\footnote{%
		Using the exact averages for $\bar{u}_i(t=0)$ has proven advantageous in terms of achievable numerical precision in the IR compared to taking the mid-point values of the exact derivative of ${ \bar{u}_i ( t = 0 ) = \partial_\sigma U ( \sigma ) |_{\sigma = \sigma_i} }$ when considering non-analytic initial conditions like Eqs.~\eqref{eq:testing_scenario_non-analytic_quadaratic_asymptote} or \eqref{eq:testing_scenario_4}.%
	} 
		\begin{align}
			\bar{u}_i ( t = 0 ) = \tfrac{1}{\Delta \sigma} \, \big[ U \big( \sigma_{i + \frac{1}{2}} \big) - U \big( \sigma_{i - \frac{1}{2}} \big) \big] ,
		\end{align}
	using the UV initial condition \eqref{eq:testing_scenario_non-analytic_quadaratic_asymptote}.
	We use linear extrapolation as spatial boundary condition at $x_\mathrm{max}$.
	The UV cutoff is set to $\Lambda = 10^6$ at $t = 0$. 
	The integration is performed with \texttt{Mathematica}'s ODE solver \textit{NDSolve} \cite{Mathematica:12.1} with a \textit{PrecisionGoal} and \textit{AccuracyGoal} of 10 and stopped in the IR at $r ( t_\mathrm{IR} = 60 ) \approx 10^{-20}$ using the exponential regulator shape function \eqref{eq:exponential_regulator}.
	We find that these parameters suffice to produce decent results, as discussed in the following subsubsections.
	There, we quantitatively analyze sources and severity of possible errors.
	
\begin{figure}
			\centering
			\includegraphics{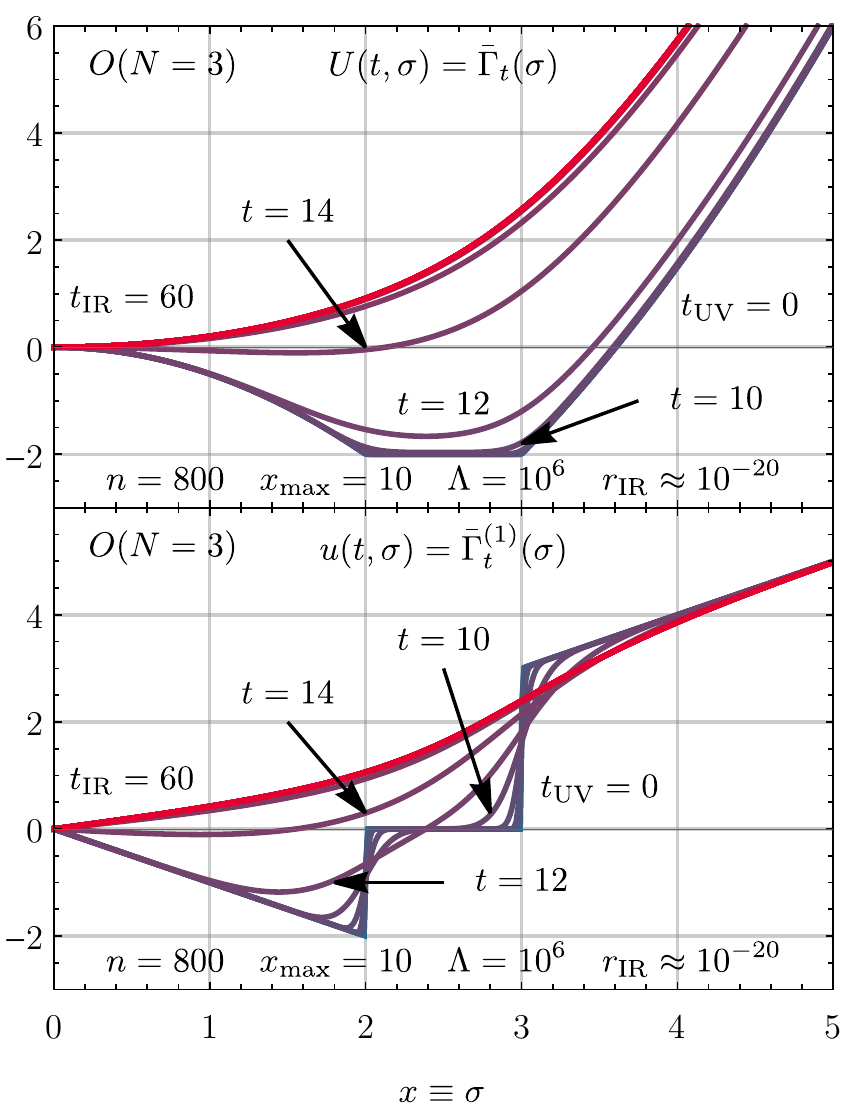}
			\caption{\label{fig:sc_i_on_3_n_800_xmax_10_lambda_1e6_tir_60_rg_flow}%
				The FRG flow of the effective potential $U ( t, \sigma )$ (upper panel) and its derivative $u ( t , \sigma ) = \partial_\sigma U ( t , \sigma )$ (lower panel) for the zero-dimensional $O ( 3 )$ model with initial condition Eq.~\eqref{eq:testing_scenario_non-analytic_quadaratic_asymptote} evaluated at $t = 0, \, 2, \, 4, \, \ldots, \, 60$ (integer values for $t$ were chosen for convenience and readability). 
				The blue curves correspond to the UV and the red curves to the IR.
				We used the exponential regulator Eq.~\eqref{eq:exponential_regulator} with UV cutoff $\Lambda = 10^6$.
				For the sake of readability, the plot does not show the region $x = 5$ to $x = 10$, because the tiny differences between $u ( t , \sigma )$ and $u ( 0 , \sigma )$ are not visible in this region and vanish for large $x = \sigma$ anyhow.
			}
		\end{figure}

\begin{figure}
			\centering
			\includegraphics{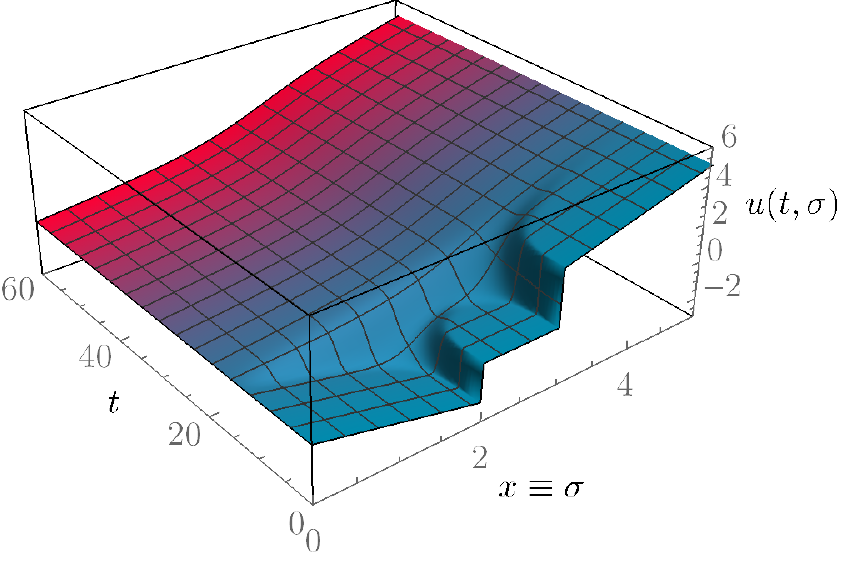}
			\caption{\label{fig:sc_i_on_3_n_800_xmax_10_lambda_1e6_tir_60_rg_flow_du_3d}%
				The RG flow of the derivative of the effective potential $u ( t , \sigma ) = \partial_\sigma U ( t , \sigma )$ for the zero-dimensional $O ( 3 )$ model with initial condition Eq.~\eqref{eq:testing_scenario_non-analytic_quadaratic_asymptote}. 
				This is a 3-dimensional rendering of the flow displayed Fig.~\ref{fig:sc_i_on_3_n_800_xmax_10_lambda_1e6_tir_60_rg_flow} (lower panel).%
			}
		\end{figure}

\begin{figure}
			\centering
			\includegraphics{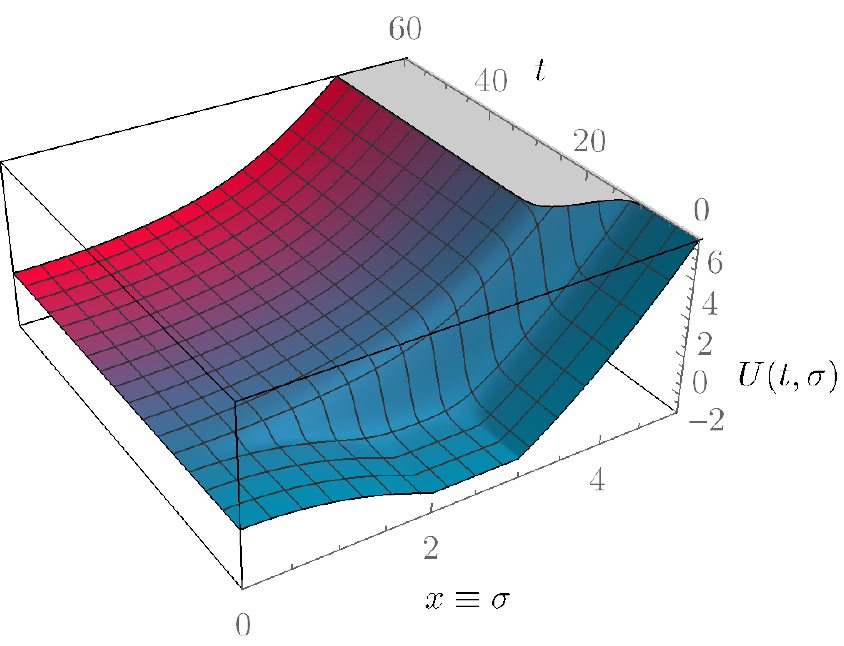}
			\caption{\label{fig:sc_i_on_3_n_800_xmax_10_lambda_1e6_tir_60_rg_flow_u_3d}%
				The RG flow of the effective potential $U (t , \sigma )$ for the zero-dimensional $O(3)$ model with initial condition Eq.~\eqref{eq:testing_scenario_non-analytic_quadaratic_asymptote}.
				This is a 3-dimensional rendering of the flow displayed Fig.~\ref{fig:sc_i_on_3_n_800_xmax_10_lambda_1e6_tir_60_rg_flow} (upper panel).%
			}
		\end{figure}

	We first provide qualitative results of our numerical methods in Figs.~\ref{fig:sc_i_on_3_n_800_xmax_10_lambda_1e6_tir_60_rg_flow}, \ref{fig:sc_i_on_3_n_800_xmax_10_lambda_1e6_tir_60_rg_flow_du_3d}, and \ref{fig:sc_i_on_3_n_800_xmax_10_lambda_1e6_tir_60_rg_flow_u_3d}, where we show the FRG flow of the effective potential $U ( t, \sigma )$ and its derivative $u ( t, \sigma ) = \partial_\sigma U ( t, \sigma )$ from the UV (blue) to the IR (red). 
	
	In the beginning, \textit{i.e.}, in the UV, the system is in the broken phase. 
	This changes only marginally until $t \approx 7$, which indicates that the UV cutoff is chosen sufficiently large.
	When $r(t)$ reaches the scales of the model at $t \gtrsim 11$ most of the dynamics takes place (symmetry restoration) and $u ( t, \sigma )$ changes rapidly and drastically until it freezes out in the IR.
	In the IR the system is in the restored phase.
	After $t \approx 26$ the potential does not change anymore, which indicates that one has reached a sufficiently small IR scale to stop the integration.
	We render this statement more precise in the following subsubsections.
	Note that the evolution in $t$ is logarithmic and corresponds to a change in scale over $25$ orders of magnitude.
	At first glance this range sounds excessive, but its necessity is explained in detail below.

	During the FRG evolution one observes that diffusive processes smear out the discontinuities at the non-analytic points at $\sigma = 2$ and $\sigma = 3$.
	We also find that the diffusion acts in both directions -- towards larger and smaller values of $\sigma$ -- as expected from our discussion in Sub.Sec.~\ref{subsec:conservative_form_of_frg_flow_equations}. 
	Nevertheless, the diffusion effects do not reach the computational boundary, which is outside the plot range at $\sigma_\mathrm{max} = 10$.
	This suggests that $\sigma_\mathrm{max} = 10$ is sufficiently large.
	Additionally, we observe a propagation of the conserved quantity $u$ towards smaller values of $\sigma$ via advection. 
	Close to the initial discontinuities these advective processes can be interpreted as rarefaction waves. 
	In a more pictorial language, the advection and diffusion ``fill up the pit'' in $u ( t, \sigma )$ at small values of $\sigma$ by moving more and more of the quantity $u$ from larger values of $\sigma$ to smaller $\sigma$ (via advection and diffusion) as well as from small negative $\sigma$ to small positive sigma (via diffusion).
	Eventually the symmetry is restored and $u ( t, \sigma )$ is smoothed towards the IR by diffusion. 
	Furthermore, the conserved quantity $u$ does not ``pile up'' at $\sigma = 0$ after symmetry restoration, because at negative $\sigma$ exactly the opposite dynamics happens, due to the $\mathbb{Z}_2$ antisymmetry of $u ( t, \sigma )$, which is encoded in the boundary condition at $\sigma = 0$, see Sub.Sec.~\ref{subsec:boundary_conditions_finite_volume}.
	The differences between advective and diffusive contributions become apparent when comparing the same system for different $N$, see below.
	
	From a numerical perspective, the KT scheme is able to handle the highly non-linear dynamics, including the non-analyticities in $u ( t, \sigma )$, without any spurious oscillatory behavior (under-/over-shooting) and allows for a stable $t$ integration to extremely small IR scales.
	
	For the sake of completeness and illustrative purposes, we also provide the RG flow of the integral of $u ( t, \sigma )$, \textit{i.e.}, the effective potential $U ( t, \sigma )$, in Figs.~\ref{fig:sc_i_on_3_n_800_xmax_10_lambda_1e6_tir_60_rg_flow} and \ref{fig:sc_i_on_3_n_800_xmax_10_lambda_1e6_tir_60_rg_flow_u_3d}.
	Here, the integration constant was set to zero\footnote{%
		$U ( t, 0 ) = 0$ is dictated by our choice of normalization for the zero-point function(s), see Eq.~\eqref{eq:normalization}.
	} and the integration was performed via Riemann summation\footnote{%
		At this point we should mention that numerical results for $U ( t, \sigma )$ via Riemann summation should be treated with great caution: Numerical errors in the cell averages $\bar{u} ( t, x_j )$ which arise from the numerical FRG flow can accumulate during integration (here summation) along $\sigma = x$.
		More precise quadrature methods should be used if precise, quantitative values for $U ( t, \sigma )$ are needed. 
		This will be discussed elsewhere \cite{Koenigstein:2021}.
	} of the discrete cell averages,
		\begin{align}
			U ( t, x_i ) = \Delta x \sum_{j = 0}^{i} \frac{\bar{u} ( t, x_j )}{( 1 + \delta_{j 0} + \delta_{j i} )} \, .	\label{eq:riemann_sum}
		\end{align}
	Figure \ref{fig:sc_i_on_3_n_800_xmax_10_lambda_1e6_tir_60_rg_flow_u_3d} perfectly illustrates the restoration of the $O(3)$ symmetry of the potential $U ( t, \sigma )$ during the FRG flow via ``vaporization'' of the infinitely many minima.
	Nevertheless, we find that it is hardly possible to intuitively understand the contributions of the radial $\sigma$ mode and the pions to the FRG flow on the level of $U ( t, \sigma )$ only.
	This complements the discussion in Sub.Sec.~\ref{subsec:conservative_form_of_frg_flow_equations} and lends support to our claim that the fluid-dynamical interpretation of the FRG flow in terms of $u ( t, \sigma )$ is superior to the canonical formulation in terms of $U ( t, \sigma )$ commonly used in the FRG literature.
	
\begin{figure}
			\centering
			\includegraphics{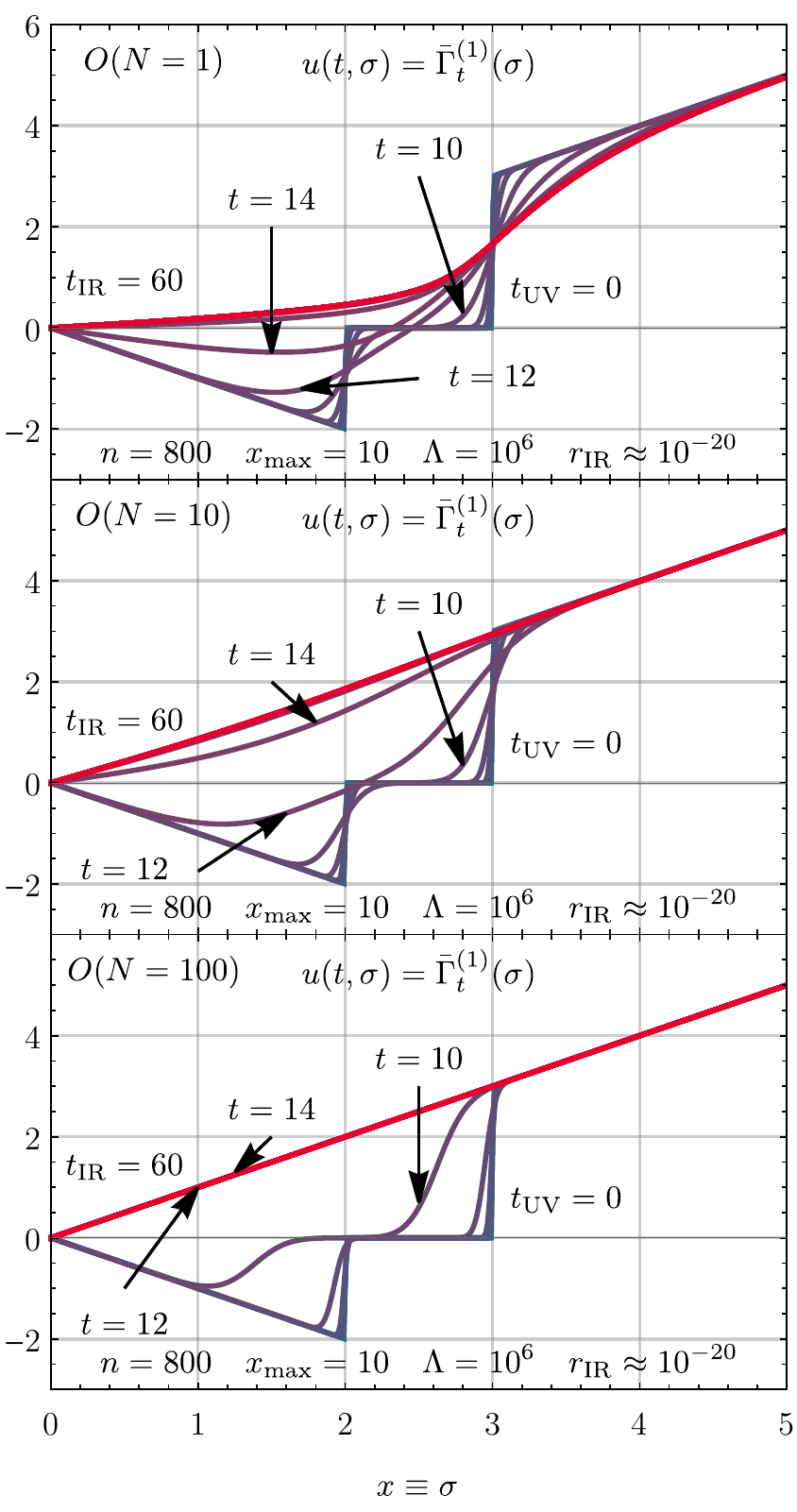}
			\caption{\label{fig:sc_i_on_1_10_100_n_800_xmax_10_lambda_1e6_tir_60_rg_flow}%
				The FRG flow of the derivative of the effective potential $u ( t , \sigma ) = \partial_\sigma U ( t, \sigma )$ for the zero-dimensional $O(N)$ model for $N = 1, \, 10, \, 100$ with initial condition Eq.~\eqref{eq:testing_scenario_non-analytic_quadaratic_asymptote}.
				All parameters are identical to those in Fig.~\ref{fig:sc_i_on_3_n_800_xmax_10_lambda_1e6_tir_60_rg_flow}.
			}
		\end{figure}

	Before discussing quantitative results and sources of (numerical) errors in RG flows, we briefly return to the interpretation of the radial $\sigma$ mode as diffusion and the interpretation of the pions as advection in the FRG flow along the field space direction.
	To this end, we discuss FRG flows for the same initial UV potential \eqref{eq:testing_scenario_non-analytic_quadaratic_asymptote} as before, but for different $N$.
	This  corresponds to a different number of pions in the flow and different advection velocities \eqref{eq:advection_velocity}.
	All other parameters remain unchanged.
	In addition to the $N = 3$ case in Fig.~\ref{fig:sc_i_on_3_n_800_xmax_10_lambda_1e6_tir_60_rg_flow}, we provide the RG flows of $u ( t, \sigma )$ for $N = 1, 10, 100$ in Fig.~\ref{fig:sc_i_on_1_10_100_n_800_xmax_10_lambda_1e6_tir_60_rg_flow}.
	
	The figure again demonstrates on a qualitative level that the sigma mode enters the FRG as diffusion, while pions enter as advection: Increasing the number of pions the problem becomes more advection-driven exhibiting more pronounced waves and faster propagation.
	This can be seen by comparing the plots at equal RG times.
	For $N = 1$, the problem reduces to the pure diffusion equation \eqref{eq:pde_gamma}, where the dynamics is slowest and the diffusion propagates the fluid from negative $\sigma$ to small positive $\sigma$ close to $\sigma = 0$.
	Furthermore, one observes stronger smearing of the discontinuities at $\sigma = 2$ and $\sigma = 3$.
	The $N = 100$ case is extremely advection-dominated\footnote{%
		We will discuss the qualitative and quantitative differences between RG flows at large and infinite $N$ in part III of this series of publications \cite{Steil:2021cbu}.%
	}, similar to the situation in large-$N$ limit of the $O(N)$ model, where the $\sigma$ mode can be completely ignored.
	In the fluid-dynamical language, this corresponds to a complete suppression of diffusion, which is clearly observed from the fast propagation of the rarefaction waves and almost negligible smoothing of the discontinuities at $\sigma = 2$ and $\sigma = 3$.

\subsubsection{Tests of the spatial resolution \texorpdfstring{$\Delta x$}{Delta x}}

	This subsubsection is dedicated to quantifying numerical errors in the FRG flow arising from the finite spatial resolution $\Delta x$ of the cells in the KT scheme.
	Any spatial discretization comes with a discretization error.
	The KT scheme, which is used throughout this paper, is of second-order accuracy $\Delta x$.
	Therefore, the numerical errors arising from the spatial discretization should scale with $\Delta x^2$ when $\Delta x$ is decreased.

	As test values (observables) we use the modulus of the relative errors of the 1PI $n$-point vertex functions $\Gamma^{(2n)}$ for $n = 1, 2, 3$,
		\begin{align}
			\bigg| \frac{\Gamma^{(2n)}_\mathrm{KT}}{\Gamma^{(2n)}} - 1 \bigg| \, ,	\label{eq:relative_errors_gamma_2n}
		\end{align}
	where $\Gamma^{(2n)}_\mathrm{KT}$  is calculated from the FRG (via the KT scheme) and $\Gamma^{(2n)}$ from the (functional) integral, see Table~\ref{tab:sc_1_n_point_functions_exact}.
	In order to determine how much of the relative numerical error~\eqref{eq:relative_errors_gamma_2n} is directly related to the spatial discretization, we have to optimize all other parameters of the computation in order to reduce other sources of errors.
	We basically choose the same parameter set and UV initial condition \eqref{eq:testing_scenario_non-analytic_quadaratic_asymptote} which was also used for the qualitative analysis in the previous subsection and only vary the number of cells $n$ to change the resolution $\Delta x$. 
	We keep $\Lambda = 10^6$, $x_\mathrm{max} = 10$ and $t_\mathrm{IR} = 60$, which turn out to be decent choices as will become clear in the following. 
	Without loss of generality we keep $N=3$.
	
	Before we embark on our discussion, we remark that the spatial-discretization error enters the relative errors~\eqref{eq:relative_errors_gamma_2n} in a twofold way: 
	\begin{enumerate}
		\item	There is the discretization error which comes from the KT scheme during the FRG time integration.
		This error shows up directly in the IR values $u ( t_\mathrm{IR}, x_i )$ and should scale according to $\Delta x^2$ for the chosen second-order KT scheme.

		\item	There is a discretization error which is related to the extraction of the 1PI $n$-point vertex functions from the discrete $\bar{u} ( t_\mathrm{IR}, x_i )$. 
		They have to be calculated at the physical point (the minimum at $x = \sigma = 0$) via numerical differentiation, which also comes with a discretization error.
		The latter is also related to the spatial resolution $\Delta x$.
		In general (especially in higher-dimensional models) it is not clear whether these numerical derivatives at the physical point are always well-defined.
		We argued before that $u ( t, \sigma )$ might exhibit non-analytical behavior also at the physical point in the IR, \textit{cf.}\ Refs.~\cite{Grossi:2019urj,Grossi:2021ksl,Borchardt:2016pif,Stoll:2021ori}, such that a naive numerical differentiation is not allowed in general.
		This will be discussed and analyzed in detail elsewhere \cite{Koenigstein:2021}.
		In the special case of zero-dimensional QFTs, we prove in App.~\ref{app:mermin_wagner} that the IR effective action and the IR potential $U ( t \rightarrow \infty, \vec{\varphi}\, )$ are smooth, which also translates to $u ( t \rightarrow \infty, \sigma )$, such that numerical differentiation (\textit{e.g.}, via finite-difference approximations) is well-defined.
	\end{enumerate}
				
	However, even though finite-difference formulas are reliable approximations for derivatives of a smooth function and have a well-defined truncation-error scaling in powers of $\Delta x$, there remains a well-known subtlety: While decreasing the resolution $\Delta x$, one eventually will reach a point where the error starts increasing again contrary to the formal truncation-error scaling. 
	This is related to the ill-conditioned nature of finite-difference formulas and to explicit rounding errors in floating-point arithmetic, which increase the error of the numerical derivative below a certain $\Delta x$, see, \textit{e.g.}, Chap.~5.7 of Ref.~\cite{PresTeukVettFlan92}.
	Related spurious cancellations occur if the discrete data of a smooth function include some sort of noise.
	In our case this ``noise'' is directly related to the spatial-discretization errors from the KT scheme and the errors from RG time integration using numerical ODE solvers.
	These errors might be tiny, but can easily inflate the errors of the numerical derivatives, especially for higher-order derivatives.
	
	In conclusion, while decreasing $\Delta x$ we expect that long before the relative errors from the KT scheme start increasing again (because the KT scheme begins operating close to machine precision or because the error of the time stepping becomes relevant) the relative errors \eqref{eq:relative_errors_gamma_2n} start increasing due to the numerical differentiation of slightly ``noisy data''. 
	This phenomenon is expected to set in at larger $\Delta x$ for approximations for higher-order derivatives and long before the formal error scaling of the KT scheme is no longer valid.

\begin{figure}
			\centering
			\includegraphics{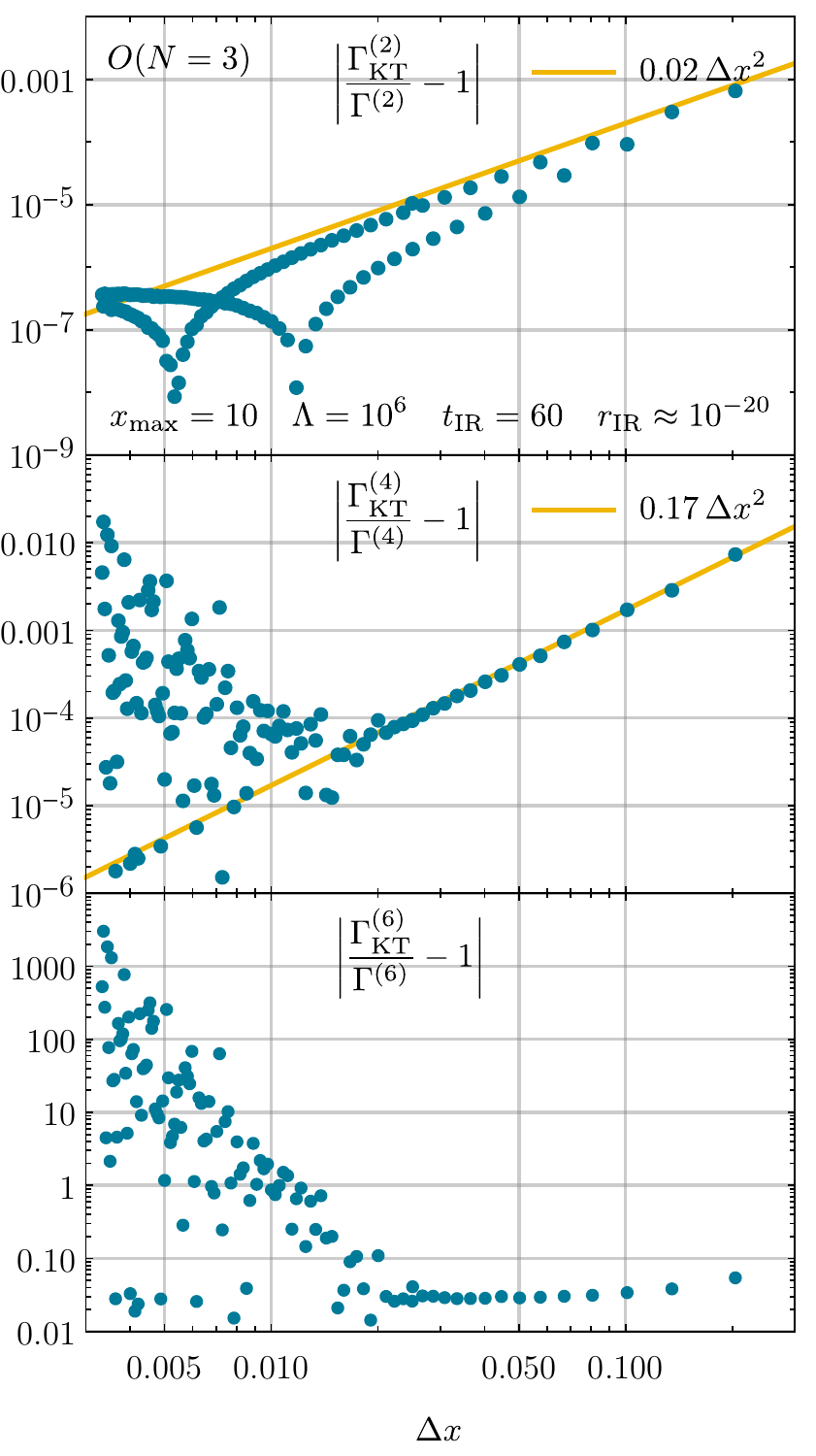}
			\caption{\label{fig:sc_i_on_3_xmax_10_lambda_1e6_tir_60_deltax_scaling}%
				The relative error of the numerical results (blue dots) from the KT scheme for the 1PI $n$-point vertex functions $\Gamma^{(2n)}$ for $n = 1, 2, 3$ as a function of $\Delta x$ with initial potential \eqref{eq:testing_scenario_non-analytic_quadaratic_asymptote}.
				The numerical derivatives at $\sigma = 0$ of $u(t_\mathrm{IR}=60, \sigma)$ were calculated via the second-order accurate central schemes \eqref{eq:derivative_1_central_error_2}, \eqref{eq:derivative_3_central_error_2}, and \eqref{eq:derivative_5_central_error_2}.
				The plot was produced with $x_\mathrm{max} = 10$, but could have been calculated for any sufficiently large $x_\mathrm{max}$.
				We used the exponential regulator \eqref{eq:exponential_regulator} with UV cutoff $\Lambda = 10^6$.
				The yellow straight lines are for optical guidance.%
			}
		\end{figure}

	Our explicit results for the scaling of the relative errors with decreasing spatial resolution are presented in Fig.~\ref{fig:sc_i_on_3_xmax_10_lambda_1e6_tir_60_deltax_scaling}, where we show the relative errors \eqref{eq:relative_errors_gamma_2n} for the \hbox{two-}, \hbox{four-}, and six-point functions in a double-logarithmic plot over $\Delta x$.
	For $\Gamma^{(2)}$ and $\Gamma^{(4)}$ we find that the corresponding relative errors scale with $\Delta x^2$ (or even slightly better) as expected from the error scaling of the KT scheme as well as the error scaling of the finite-difference stencils \eqref{eq:derivative_1_central_error_2} and \eqref{eq:derivative_3_central_error_2}.
	We observe two groups of points for $\Gamma^{(2)}$ (upper panel of Fig.~\ref{fig:sc_i_on_3_xmax_10_lambda_1e6_tir_60_deltax_scaling}), which are related to discretization errors of the discontinuous initial condition \eqref{eq:testing_scenario_non-analytic_quadaratic_asymptote} at $x=2$ and $x=3$. 
	The error scaling of $0.02\, \Delta x^2$ for $\Gamma^{(2)}$ is a conservative estimate for the observed errors, which are actually lower for $\Delta x >0.005$. 
	For $\Delta x <0.005$ we observe deviations from the conservative estimate for the error scaling of $\Gamma^{(2)}$ related to other error sources.
	In the middle panel of Fig.~\ref{fig:sc_i_on_3_xmax_10_lambda_1e6_tir_60_deltax_scaling}, we clearly see that there is an optimal minimal $\Delta x \approx 0.02$ where the correct formal scaling of the numerical derivative breaks down and the relative error of $\Gamma^{(4)}$ increases again for smaller $\Delta x$.
	We can be sure that this breakdown of the error scaling is related to the numerical differentiation and not the KT scheme because we observe scaling with at least $\Delta x^2$ for $\Gamma^{(2)}$ in the upper panel of Fig.~\ref{fig:sc_i_on_3_xmax_10_lambda_1e6_tir_60_deltax_scaling} well below $\Delta x \approx 0.02$.
	This is expected for lower-order numerical derivatives.
	Furthermore, we find that for $\Gamma^{(6)}$ (lower panel of Fig.~\ref{fig:sc_i_on_3_xmax_10_lambda_1e6_tir_60_deltax_scaling}) the order of the numerical derivative is already too large, such that the theoretical error scaling of the KT scheme cannot be seen at all and is completely obscured by the errors from the numerical differentiation of $\bar{u} ( t_\mathrm{IR}, x_i )$.
	
	We conclude that the KT scheme is perfectly suited for the spatial discretization of the RG flow equation for $u ( t, \sigma )$ and shows correct scaling with decreasing spatial resolution $\Delta x$. 
	This is also confirmed by tests with different initial conditions, see below.
	
	In addition, we actually found that a more severe problem is the correct extraction of physical observables from the IR values $\bar{u} ( t_\mathrm{IR}, x_i )$, which are usually related to derivatives of $u ( t_\mathrm{IR}, \sigma )$.
	We further conclude that this problem is expected to be worse in higher dimensions, were the IR potential is no longer guaranteed to be smooth.
	We therefore suggest to search for better ways of calculating those derivatives as well as for careful analysis tools for numerically calculated 1PI $n$-point vertex functions in the vicinity of non-analyticities in general.
	However, this is beyond the scope of the present work.
	
	We remark that our numerical findings indicate that -- independent of the specific numerical discretization scheme -- the number of grid points or expansion coefficients \textit{etc.}\ might have been chosen too small in previous studies to obtain a decent resolution.
	However, other works, \textit{cf.} Refs.~\cite{Pelaez:2015nsa,Caillol:2012zz,Pangon:2009pj,Pangon:2010uf,Borchardt:2015rxa,Borchardt:2016pif}, which also discuss the limitations of their numerical schemes in detail, have used a rather large number of discretization points -- in some cases to compensate the demand for continuity of the specific scheme.
	
	In the next subsections we will mostly use a spatial resolution of 
		\begin{align}
			\Delta x = \frac{x_\mathrm{max}}{n - 1} \simeq 0.025 \, ,
		\end{align}
	where we can trust the results for the two- and four-point functions.
	The relative errors for the six-point function will only be plotted for the sake of completeness, but cannot be included in any reasonable quantitative analysis of other sources of (numerical) errors in RG flow equations although they are still at an acceptably small level.

\subsubsection{Tests of the size of the computational domain}
\label{subsec:test_size_computational_domain}

	In this subsection, we discuss the influence of the size of the computational domain $[ 0, \sigma_\mathrm{max} ]$ on the relative errors of the IR observables \eqref{eq:relative_errors_gamma_2n}.
	As discussed in Sub.Sec.~\ref{subsec:boundary_conditions_finite_volume}, we expect that, if the spatial boundary conditions are not implemented with great caution and the computational domain is too small, one cannot trust the results from the numerical integration of the RG flow.
	If the computational domain is too small, we expect large errors, because the boundary conditions at $\sigma_\mathrm{max}$ are no longer valid due to wrong extrapolation to the ghost cells and consequently wrongly estimated in-flux.
	
	In the case with UV initial condition \eqref{eq:testing_scenario_non-analytic_quadaratic_asymptote}, the boundary condition at $\sigma_\mathrm{max}$ is implemented as a linear extrapolation of $u ( t, \sigma )$ to the two ghost cells of the KT scheme to mimic the asymptotic behavior of $u ( t, \sigma )$. 
	As long as $\sigma_\mathrm{max}$ is sufficiently large, we expect only tiny deviations of $u ( t, \sigma )$ from its initial UV value $u ( t_\mathrm{UV} = 0, \sigma )$ next to $\sigma_\mathrm{max}$.
	However, if $\sigma_\mathrm{max}$ is too small and approaches to the model scales, we expect the diffusion effects to reach the boundary of the computational domain, such that a linear extrapolation is no longer a good approximation in order to determine the spatial boundary condition.
	
\begin{figure}
			\centering
			\includegraphics{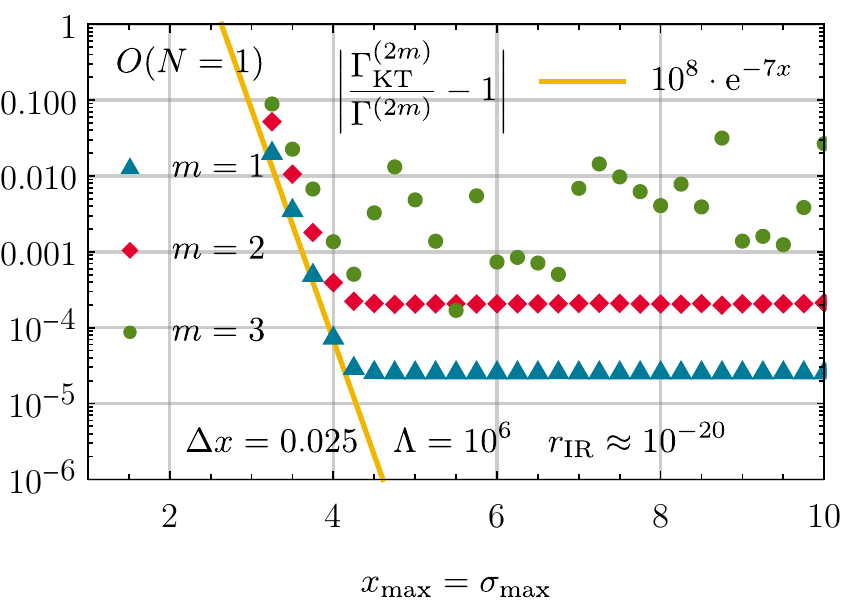}
			\caption{\label{fig:sc_i_on_1_deltax_25e-3_lambda_1e6_tir_60_errors_xmax}%
				The relative error for $\Gamma^{(2m)}$ for $m = 1, 2, 3$ for the UV potential \eqref{eq:testing_scenario_non-analytic_quadaratic_asymptote} of the $O ( 1 )$ model as a function of $x_\mathrm{max}$, while keeping the cell size constant, $\Delta x = 0.025$. $\Gamma^{(2m)}$ are computed from the discrete values of the derivative of the IR potential $u ( t_\mathrm{IR} = 60, \sigma )$ via the second-order accurate central finite-difference stencils \eqref{eq:derivative_1_central_error_2}, \eqref{eq:derivative_3_central_error_2}, and \eqref{eq:derivative_5_central_error_2} at $\sigma = 0$.
				We use the exponential regulator \eqref{eq:exponential_regulator} with UV cutoff $\Lambda = 10^6$.
				The yellow straight line is for optical guidance.%
			}
		\end{figure}

\begin{figure}
			\centering
			\includegraphics{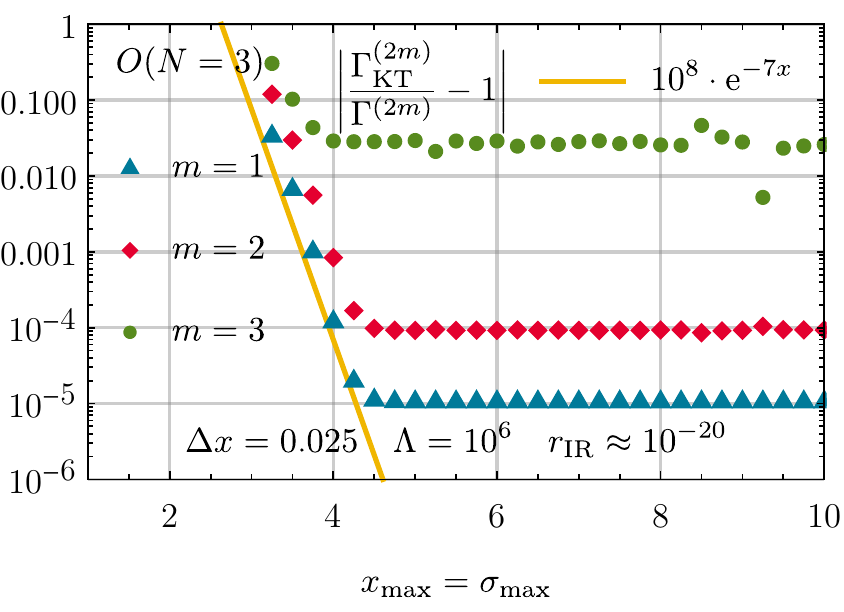}
			\caption{\label{fig:sc_i_on_3_deltax_25e-3_lambda_1e6_tir_60_errors_xmax}%
				The relative error for $\Gamma^{(2m)}$ for $m = 1, 2, 3$ for the UV potential \eqref{eq:testing_scenario_non-analytic_quadaratic_asymptote} of the $O ( 3 )$ model as a function of $x_\mathrm{max}$. 
				All parameters are identical to Fig.~\ref{fig:sc_i_on_1_deltax_25e-3_lambda_1e6_tir_60_errors_xmax}.
			}
		\end{figure}

	To this end, we test the scaling of the relative errors \eqref{eq:relative_errors_gamma_2n} with decreasing computational domain size $x_\mathrm{max} = \sigma_\mathrm{max}$ for $N = 1$ (purely diffusive) and $N = 3$. 
	The results and (numerical) parameters are shown in Figs.~\ref{fig:sc_i_on_1_deltax_25e-3_lambda_1e6_tir_60_errors_xmax} and \ref{fig:sc_i_on_3_deltax_25e-3_lambda_1e6_tir_60_errors_xmax}.
	In both cases we find that the relative errors are independent of $\sigma_\mathrm{max}$ for sufficiently large $\sigma_\mathrm{max}$.
	However, if the spatial cutoff $\sigma_\mathrm{max}$ is approaching the model scales (here the discontinuity in $u ( t_\mathrm{UV} = 0, \sigma )$ at $\sigma = 3$, see Fig.~\ref{fig:sc_i_uv_initial_condition}) the relative errors for $\Gamma^{(2)}$ and $\Gamma^{(4)}$ start rising exponentially.
	
	Contrary to our expectations, the results for $N = 1$ and $N = 3$ are very similar and the exponential rise of the relative errors sets in at a similar $\sigma_\mathrm{max}$.
	We expected that for the purely diffusive scenario with $N = 1$, the diffusion effects arising from the large gradients at $\sigma = 3$ might have more time to reach and influence the shape of $u ( t, \sigma )$ at larger values of $\sigma$, which does not seem to be the case.
	Our employed monitors for numerical errors -- the 1PI $n$-point vertex functions in the IR computed at $\sigma=0$ and $t=0$ -- are rather intensive to such changes.
	A possible explanation is the fact that errors from the boundary at $\sigma_\mathrm{max}$ propagate into the computational domain at a finite speed, which is rather low in the purely diffusive case and in general small at large~$\sigma$, and thus do not influence the physical point at $t=0$ and $\sigma=0$. 

	Nevertheless, we conclude from Figs.~\ref{fig:sc_i_on_1_deltax_25e-3_lambda_1e6_tir_60_errors_xmax} and \ref{fig:sc_i_on_3_deltax_25e-3_lambda_1e6_tir_60_errors_xmax} that it is extremely important to use sufficiently large computational domains to minimize numerical errors in field-dependent FRG flows. 
	This implies that $\sigma_\mathrm{max}$ should be chosen much larger than all relevant scales of the model.
	
	From our findings, it is therefore expected that choosing a large $\sigma_\mathrm{max}$ might even gain in importance in higher-dimensional models, where the physical point may be located at a non-trivial minimum in the IR (\textit{e.g.}, the quark-meson model in its broken phase \cite{Ihssen2020}): The 1PI $n$-point vertex functions are calculated at this physical point. 
	If the physical point is closer to the boundary of the computational domain the relative errors for observables might even be larger than for our zero-dimensional model where the physical point moves towards $\sigma = 0$ during the FRG flow. 
	In terms of errors originating from the boundary at $\sigma_\mathrm{max}$, the physical point at $\sigma = 0$ is ideal since it has the largest spatial and -- in a sense causal, due to the finite speed of propagation -- distance to $\sigma_\mathrm{max}$.
	
	Lastly, we have to warn that there is no panacea for the construction of a sufficiently large computational domain and the choice of $\sigma_\mathrm{max}$ has to be adjusted to the specific model and specific initial condition under consideration. 
	For some problems even more involved approaches (like using several computational grids of different resolution $\Delta x$) might be needed or are at least highly advantageous \cite{Grossi:2019urj,Grossi:2021ksl}.
	In any case one has to check that the IR results do not depend on the size of the computational domain (even if exact reference values for observables are unknown), \textit{cf.}\ Refs.~\cite{Pangon:2009pj,Caillol:2012zz,Stoll:2021ori}.
	This can be done by fixing appropriate values for the spatial resolution $\Delta x$ as well as for all other (numerical) parameters and successively increasing $\sigma_\mathrm{max}$ until the IR observables do not change anymore.

\subsubsection{Tests of the UV and IR scales}\label{subsec:RGconsistency}

	We now turn to a long-standing discussion in the FRG community, namely the question: How do we have to choose the initial UV and numerical IR cutoff scale for the calculation of the 
RG flow for a specific model? 

	A common argument is based on the energy scales of a given model. The UV initial condition is fixed at UV cutoffs $\Lambda$ that are close to the largest energy scale of the model.
	Higher $\Lambda$ are excluded by arguing that at higher energy scales other physical degrees of freedom (\textit{e.g.}, other interaction channels or new particles) are relevant and the model at hand is only valid within a certain energy regime.
	On the other hand, the IR cutoff $k_\mathrm{IR}$ scale is oftentimes fixed by arguing that if it decreases below the lowest energy scale of the model, the FRG flow is effectively ``frozen in'' and the effective potential no longer changes anyway.
	A relatively low UV initial scale and a high IR cutoff lead to rather short flow times of only $t_\mathrm{UV} - t_\mathrm{IR} \approx 3 - 4$.
	
	Another approach, which is sometimes employed in conjunction with the first strategy, is guided by the principle of ``numerical stability'' of the RG flow, where cutoffs are chosen in a certain way to ``improve performance and stability'' during the numerical RG time integration.
	In turn, in Refs.~\cite{Pelaez:2015nsa,Caillol:2012zz,Pangon:2009pj,Pangon:2010uf,Borchardt:2016pif} relatively small IR cutoff scales are reached due to the use of numerical stable schemes or the control of stability.
	Careful extrapolations into the deep IR like the ones discussed in, \textit{e.g.}, Refs.~\cite{Pelaez:2015nsa,Grossi:2019urj,Grossi:2021ksl} are another possibility to achieve low IR cutoffs.
	Note that, for theories in $d>0$ dimensions, numerical integration into the (deep) IR becomes very demanding due to multiple reasons, see also Refs.~\cite{Pelaez:2015nsa,Grossi:2019urj,Grossi:2021ksl,Stoll:2021ori}.
	This is probably the main reason why often too large numerical IR cutoffs are used.
	
	In general, however, there is a well-defined strategy for the choice of the UV cutoff scale, which is known as \textit{RG consistency}.
	It states that the full effective action $\Gamma [ \Phi ]$ in the IR must be independent of the UV initial scale \cite{Braun:2018svj},
		\begin{align}
			\Lambda \, \frac{\mathrm{d} \Gamma [ \Phi ]}{\mathrm{d} \Lambda} = 0 \, .	\label{eq:rg_consistency}
		\end{align}
	In this framework, one should consider effective models as ``stand-alone theories'',  which have their own functional integral and their own well-defined expectation values.
	This framework is consistent with our introduction to zero-dimensional QFTs and FRG in Sec.~\ref{sec:frg_introduction} and implies that the UV cutoff scale $\Lambda$ has to be much larger than all scales in the model.
	(In higher-dimensional QFTs the ``scales of the model'' are defined via couplings of dimension energy as well as external parameters like temperature or chemical potential.)
	In this sense, a high initial UV scale is necessary to include all fluctuations. 
	It was already demonstrated in Ref.~\cite{Braun:2018svj} that if the initial UV scale $\Lambda$ is chosen too small and too close to the model scales or external scales, physical results are spoiled drastically by slightly varying $\Lambda$ and Eq.~\eqref{eq:rg_consistency} is not fulfilled anymore, \textit{cf.} Refs.~\cite{Braun:2003ii,Herbst:2013ufa,Springer2017,PhysRevD.87.076004} for related discussions in the context of low-energy effective theories of QCD.
	
	A lower limit for $\Lambda$ arises from the fact that for a given initial condition $U(t=0,\sigma)$ and at all $\sigma$
		\begin{align}
			\Lambda + \tfrac{1}{\sigma} \, \partial_\sigma U( t = 0, \sigma ) > \, & 0 \, ,	\vphantom{\bigg(\bigg)}	\label{eq:LambdaMin1}
			\\
			\Lambda + \partial_\sigma^2 U( t = 0, \sigma ) > \, & 0 \, ,	\vphantom{\bigg(\bigg)}	\label{eq:LambdaMin2}
		\end{align}
	must hold to have a non-singular flow equation~\eqref{eq:conservation_law_u_phi}. 
	This is discussed, \textit{e.g.}, in Refs.~\cite{Pelaez:2015nsa,Schaefer:2001cn} and represents a minimal requirement for $\Lambda$ when considering a given initial condition $U(t=0,\sigma)$. However, guaranteeing the inequalities \eqref{eq:LambdaMin1} and \eqref{eq:LambdaMin2} does by itself not guarantee RG consistency in the sense of Eq.~\eqref{eq:rg_consistency}.
	
	For higher-dimensional QFTs it is actually complicated to quantify the relative error of observables from violations of Eq.~\eqref{eq:rg_consistency}, because ``exact'' reference values, \textit{e.g.}, by numerical calculation of expectation values from the functional integral, are rarely known, especially for low-energy effective models. 
	In higher dimensions one can only quantify the changes of observables while varying $\Lambda$ and search for a sufficiently large $\Lambda$, where Eq.~\eqref{eq:rg_consistency} holds.
	In zero-dimensional QFT this is different, because we can directly calculate the relative errors for observables like 1PI $n$-point vertex functions, \textit{cf.} Eq.~\eqref{eq:relative_errors_gamma_2n}, for different values of $\Lambda$.
	
\begin{figure}
			\centering
			\includegraphics{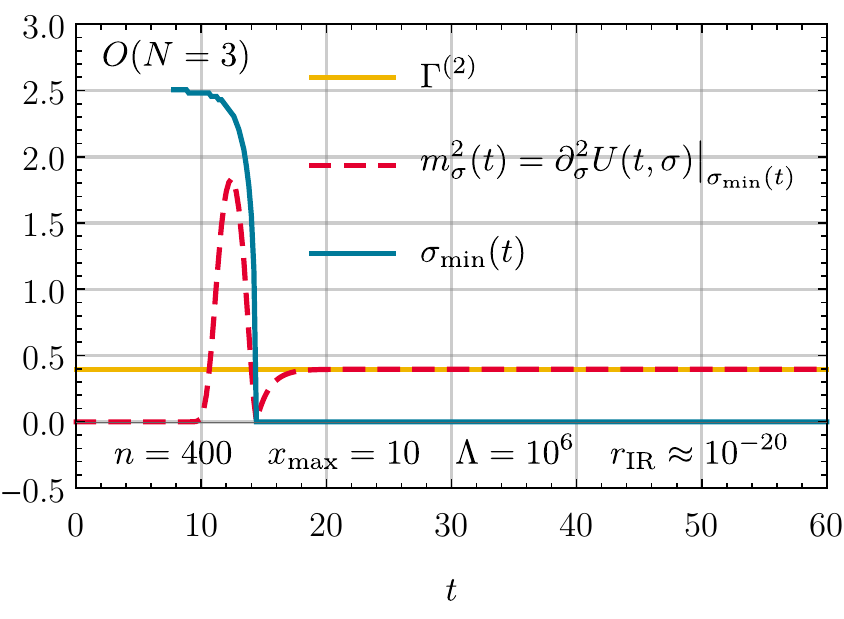}
			\caption{\label{fig:sc_i_on_3_n_400_xmax_10_lambda_1e6_tir_60_mass_minimum}%
				The RG flow of the minimum $\sigma_\mathrm{min} ( t )$ ({blue}) of the effective potential $U ( t, \sigma )$ as well as the FRG flow of the curvature mass $m_\sigma^2 ( t )$ of the $\sigma$ mode ({red-dashed}) evaluated on the equations of motion~\eqref{eq:definition_j_t} (at the flowing minimum) during the RG flow.
				The {blue} curve sets in after a unique minimum at $\pm \sigma_\mathrm{min} ( t )$ has formed.
				As UV initial condition we use Eq.~\eqref{eq:testing_scenario_non-analytic_quadaratic_asymptote}. 
				We used the exponential regulator \eqref{eq:exponential_regulator} with UV cutoff $\Lambda = 10^6$.
				The curvature mass $m_\sigma^2 ( t )$ was extracted from $u ( t, \sigma )$ via Eq.~\eqref{eq:derivative_1_forward_error_2} at the moving $\sigma_\mathrm{min} ( t )$.
				The horizontal ({yellow}) line denotes the exact IR result for $\Gamma^{(2)}$ at $\sigma = 0$, which must agree with $m_\sigma^2$ in the IR, where $\sigma_\mathrm{min} ( t ) = 0$.%
			}
		\end{figure}

	Similar arguments apply to the IR cutoff, where the numerical integration of the RG flow is stopped.
	Here, one must clearly state that the full effective average action $\Gamma [ \Phi ]$ in the IR is unambiguously defined via the limit ${t \rightarrow \infty \Longleftrightarrow r(t) \rightarrow 0}$ of $\bar{\Gamma}_t [ \Phi ]$, \textit{cf.}\ Eq.~\eqref{eq:scale_dependent_effective_average_action}.
	In practice, a direct integration to $t \rightarrow \infty$ is numerically impossible, which implies that one has at least to make sure that the numerical RG time integration is stopped no earlier than when all observables of interest do not change anymore, or one has to systematically extrapolate to $t \rightarrow \infty$, see, \textit{e.g.}, Refs.~\cite{Grossi:2019urj,Grossi:2021ksl}.
	It is worth mentioning that, depending on the specific observable, these ``freeze-out scales'' can be extremely different, see Fig.~\ref{fig:sc_i_on_3_n_400_xmax_10_lambda_1e6_tir_60_mass_minimum}.\\
	
\begin{figure}
			\centering
			\includegraphics{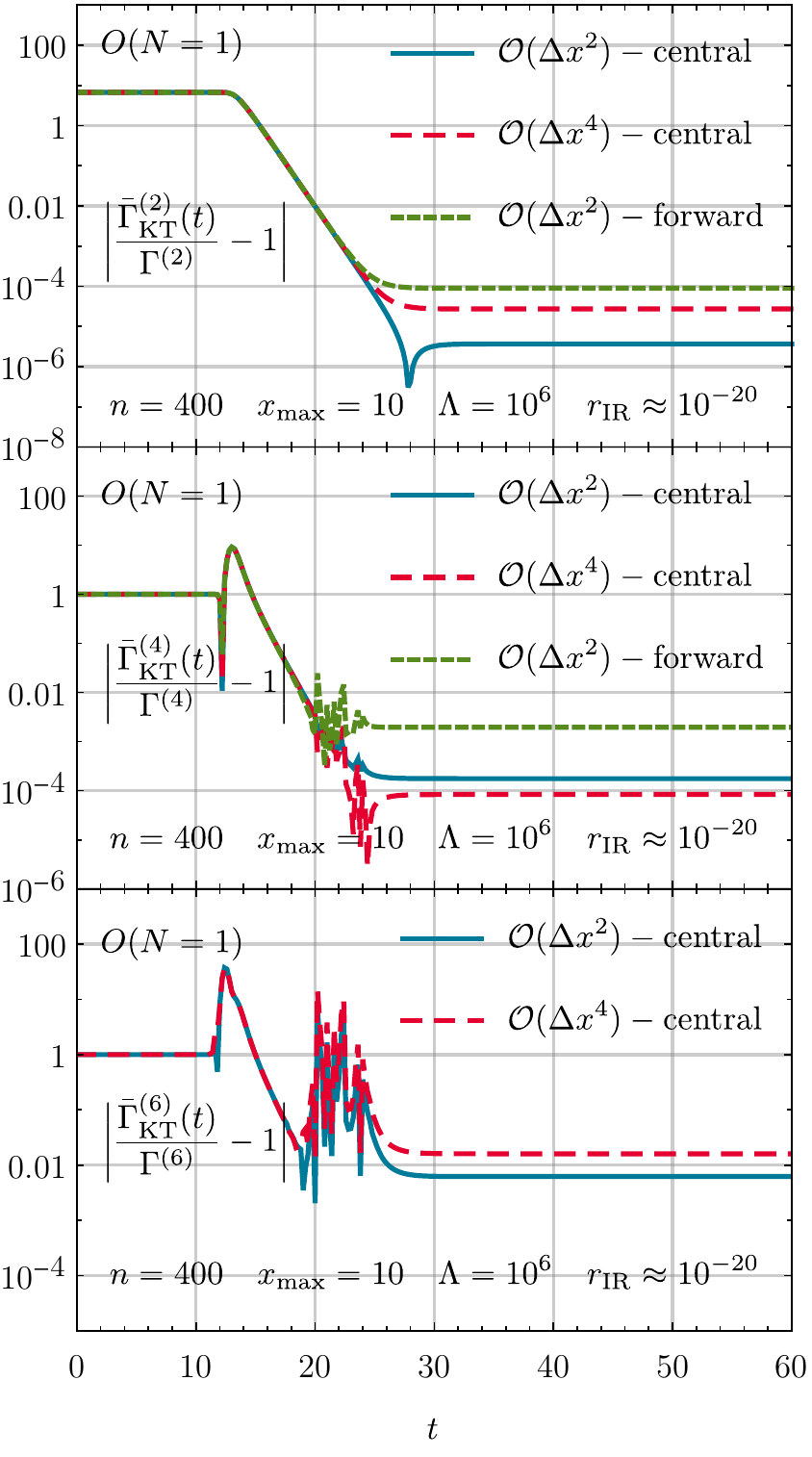}
			\caption{\label{fig:sc_i_on_1_n_400_xmax_10_lambda_1e6_tir_60_flow_errors}%
				The relative error for $\Gamma^{(2m)}$, for $m = 1, 2, 3$, calculated with the KT scheme as a function of the RG time $t$ for the $O(1)$ model.
				The initial UV potential is given by Eq.~\eqref{eq:testing_scenario_non-analytic_quadaratic_asymptote}. 
				We use the exponential regulator \eqref{eq:exponential_regulator} with UV cutoff $\Lambda = 10^6$. 
				The computational grid has 400 cells and $\sigma_\mathrm{max} = x_\mathrm{max} = 10$. $\Gamma^{(2m)}$ are extracted from $u ( t_\mathrm{IR} = 60, \sigma )$ via the finite-difference stencils \eqref{eq:derivative_1_central_error_2} -- \eqref{eq:derivative_5_central_error_4}.%
			}
		\end{figure}

\begin{figure}
			\centering
			\includegraphics{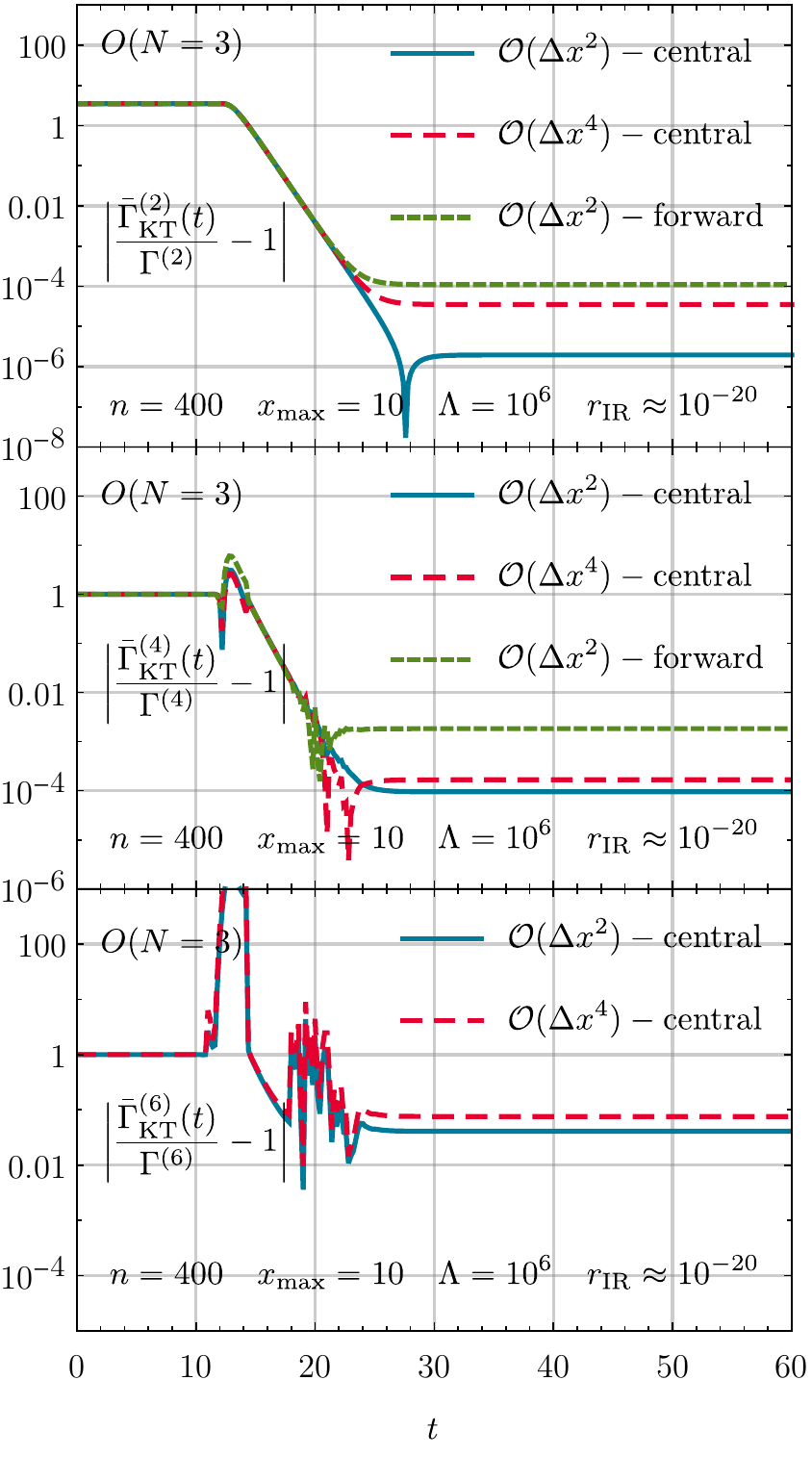}
			\caption{\label{fig:sc_i_on_3_n_400_xmax_10_lambda_1e6_tir_60_flow_errors}%
				The relative error for $\Gamma^{(2m)}$, for $m = 1, 2, 3$, 	calculated with the KT scheme as a function of the RG time $t$ for the $O(3)$ model. 
				The initial condition and all other parameters are identical to those of Fig.~\ref{fig:sc_i_on_1_n_400_xmax_10_lambda_1e6_tir_60_flow_errors}.%
			}
		\end{figure}

\begin{figure}
			\centering
			\includegraphics{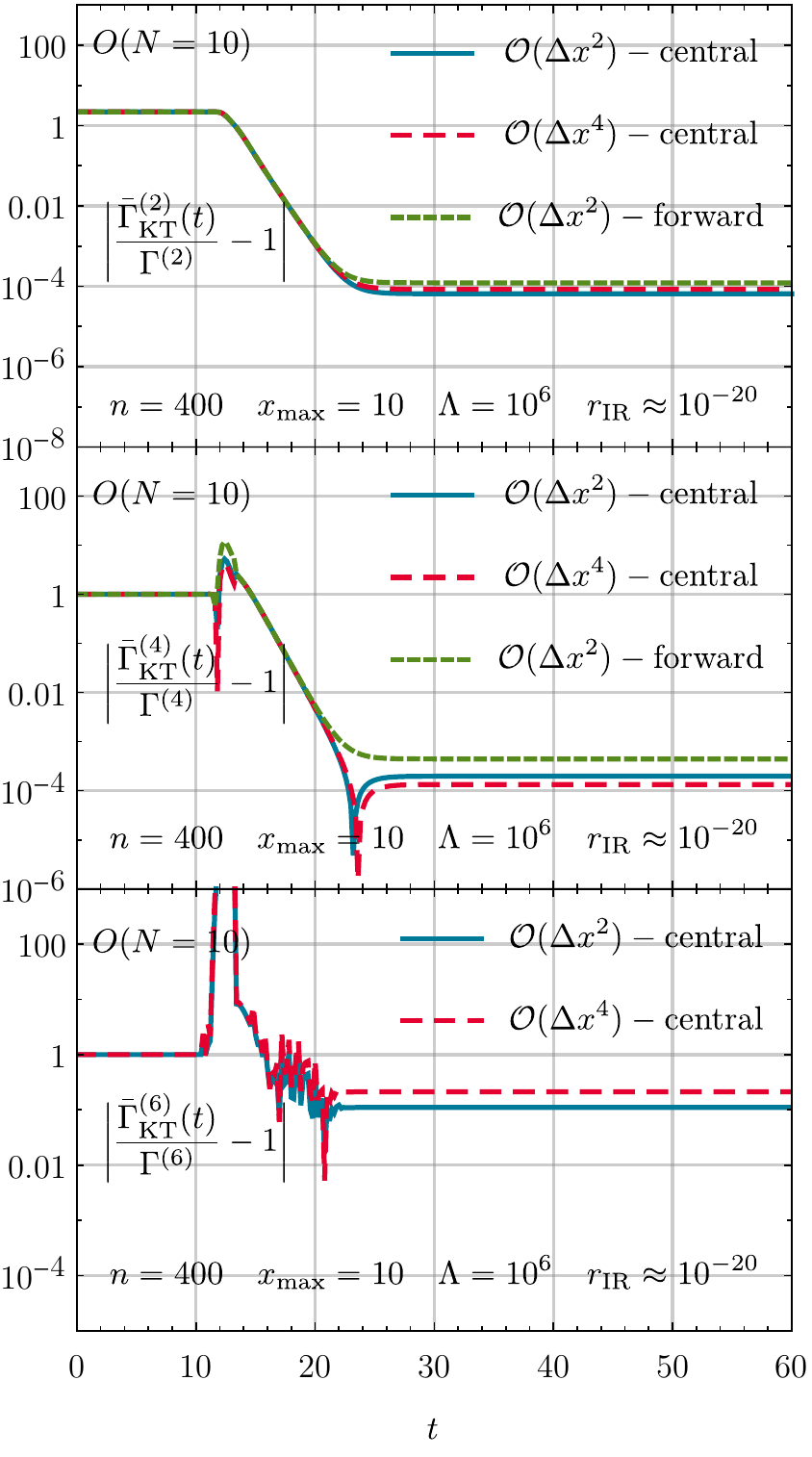}
			\caption{\label{fig:sc_i_on_10_n_400_xmax_10_lambda_1e6_tir_60_flow_errors}%
				The relative error for $\Gamma^{(2m)}$, for $m = 1, 2, 3$, calculated with the KT scheme as a function of the RG time $t$ for the $O(10)$ model.
				The initial condition and all other parameters are identical to those of Fig.~\ref{fig:sc_i_on_1_n_400_xmax_10_lambda_1e6_tir_60_flow_errors}.%
			}
		\end{figure}

	In the following, we will therefore explicitly explore the influence of UV  and IR cutoff scales on the relative errors \eqref{eq:relative_errors_gamma_2n} for the $\Gamma^{(2n)}$.
	We start our discussion by providing results for the relative errors \eqref{eq:relative_errors_gamma_2n} depending on the RG time $t$ for different $N$ of $O(N)$ and UV initial condition \eqref{eq:testing_scenario_non-analytic_quadaratic_asymptote}.
	In Figs.~\ref{fig:sc_i_on_1_n_400_xmax_10_lambda_1e6_tir_60_flow_errors}, \ref{fig:sc_i_on_3_n_400_xmax_10_lambda_1e6_tir_60_flow_errors}, and \ref{fig:sc_i_on_10_n_400_xmax_10_lambda_1e6_tir_60_flow_errors} we plot the relative errors of $\Gamma^{(2n)}$ for $n = 1, 2, 3$ for $N = 1, 3, 10$, which are all extracted via various finite-difference stencils from $u ( t, \sigma )$ at the physical point $\sigma = 0$ and different $t$ during the FRG flow.
	All (numerical) parameters are mentioned in the figures or the respective captions.

	For all three figures (\textit{i.e.}, for $N = 1, 3, 10$) and independent of the choice of discretization of the numerical derivatives, we observe plateaus for the relative errors for $\Gamma^{(2n)}$ at the beginning and the end of the FRG time evolution.
	The plateau at small $t$ corresponds to the UV regime and indicates that the UV cutoff is chosen sufficiently large because no fluctuations are present at the IR physical point until $r(t)$ reaches the scales of the model.
	RG consistency \eqref{eq:rg_consistency}, hence UV-cutoff independence should therefore be fulfilled, as long as we initialize our RG flow at some RG scale which is at the far left of this plateau.
	Such a plateau at small $t$ is a sufficient condition for RG consistency but not a necessary one, because quantum fluctuations could already work at positions in field space away from the IR physical point and only influence higher-order correlation functions. 
	We will quantify this within the next paragraphs. 
	In the plots various finite-difference stencils with distinct error scaling in $\Delta x$ are used to demonstrate that the plateaus are independent of other sources of errors, like spatial discretization errors\footnote{%
		Incidentally, Figs.~\ref{fig:sc_i_on_1_n_400_xmax_10_lambda_1e6_tir_60_flow_errors}, \ref{fig:sc_i_on_3_n_400_xmax_10_lambda_1e6_tir_60_flow_errors}, and \ref{fig:sc_i_on_10_n_400_xmax_10_lambda_1e6_tir_60_flow_errors} also underline our statement that the spatial discretization errors deriving from the numerical differentiation of $u ( t, \sigma )$ are much more severe than the discretization errors of the KT scheme. Otherwise, the curves for the various finite-difference stencils would coincide in the IR.%
	}.
	
	For intermediate $t$, we observe strong dynamics and fast changes in the relative errors for the $\Gamma^{(2n)}$. The actual values of the relative errors for intermediate $t$ is irrelevant for the current discussion on UV and IR scales.
	
	The plateau at late RG times $t$ corresponds to the IR scale of the theory and indicates that the physical observables are frozen and do not change anymore, such that the numerical time integration can be stopped.
	As expected, we find that the explicit IR scale strongly depends on the choice of $N$, thus the number of pions and the amount of advection. The smaller $N$ and the more diffusive the system, the longer it takes to reach the IR\footnote{%
		This is a well-known observation from all kinds of fluid-dynamical systems.
		It takes much longer to reach a (thermal) equilibrium state via diffusion only than by including advective processes.
	}: For $N = 10$ the freeze-out already sets in at $t \approx 26$, while for $N = 1$ one has to wait until $t \approx 30$ to find that the dynamics ends.
	This is a difference of two orders of magnitude in the RG scale.
	In general, our toy-model tests indicate that rather small IR scales are needed to actually reach the regime where the observables are frozen. 
	Still, for $N = 10$, $r(t \approx 26) \approx 5 \cdot 10^{-6}$, \textit{i.e.}, the IR regime begins six orders of magnitude below the model scales.
	
	This observation might also partially translate to higher-dimensional models, meaning that commonly used IR cutoffs might be systematically chosen too large, such that predictive power is lost. Nevertheless, we expect this problem to be the less severe the higher the space-time dimensionality of a model under consideration, because of the larger phase-space (momentum suppression)\footnote{%
		The smaller the space-time dimension of a model, the more important are long-range interactions -- quantum fluctuations at small RG scales $k$ -- for the macroscopic observables, which is of course most extreme for $d = 0$. Furthermore, field space interactions tend to become more important for a small number of space-time dimensions.%
	}.
	
	Furthermore, we observe from Fig.~\ref{fig:sc_i_on_3_n_400_xmax_10_lambda_1e6_tir_60_mass_minimum} as well as Figs.\ \ref{fig:sc_i_on_1_n_400_xmax_10_lambda_1e6_tir_60_flow_errors}, \ref{fig:sc_i_on_3_n_400_xmax_10_lambda_1e6_tir_60_flow_errors}, and \ref{fig:sc_i_on_10_n_400_xmax_10_lambda_1e6_tir_60_flow_errors} that the freeze-out scale is slightly different for different observables, because higher 1PI $n$-point vertex functions seem to be more sensitive to tiny changes in $u ( t, \sigma )$.
	In particular, we observe from Fig.~\ref{fig:sc_i_on_3_n_400_xmax_10_lambda_1e6_tir_60_mass_minimum} that the minimum $\sigma_\mathrm{min}$ is already frozen at $t \approx 14$, while the curvature mass $m_\sigma^2$ still changes drastically after $t \approx 14$ over several orders of magnitude in RG scale.
	This is especially interesting for higher-dimensional models: Oftentimes the freeze-out of the minimum is considered a suitable IR scale to stop the FRG flow, which is definitely not justified, since the derivatives of the potential at the physical point are usually still changing.
	Using the changing rates of the curvature mass instead of the position of the minimum as a monitor for the dynamic range -- viable numerical IR cutoffs -- has proven crucial in the FRG study \cite{Stoll:2021ori} of the Gross-Neveu(-Yukawa) model by several authors and their collaborators.
	
	Next, we explicitly quantify the relative errors of $\Gamma^{(2n)}$, which derive from too small UV cutoffs $\Lambda$ and the violation of RG consistency \eqref{eq:rg_consistency}.
	To this end, we plot the relative errors \eqref{eq:relative_errors_gamma_2n} as a function of the UV cutoff $\Lambda$, while keeping the IR cutoff scale fixed at extremely small $r ( t_\mathrm{IR}) = 10^{-20}$.
		\begin{figure}
			\centering
			\includegraphics{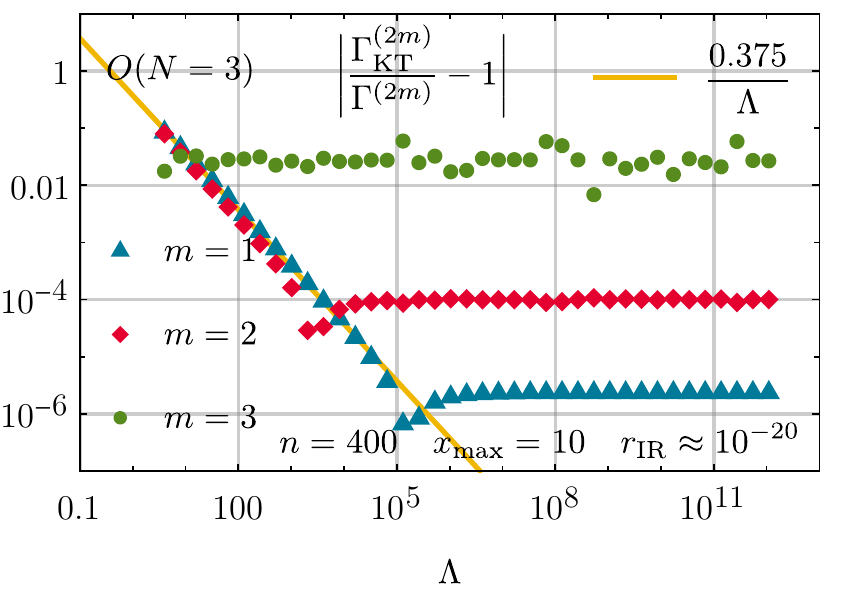}
			\caption{\label{fig:sc_i_on_3_n_400_xmax_10_rir_10e-20_cutoff_test}%
				The relative error for $\Gamma^{(2m)}$ for $m = 1, 2, 3$ from the KT scheme as a function of the UV cutoff scale $\Lambda$ for the initial potential \eqref{eq:testing_scenario_non-analytic_quadaratic_asymptote}.
				We use the exponential regulator \eqref{eq:exponential_regulator} and keep the IR cutoff scale constant at $r ( t_\mathrm{IR} ) = 10^{-20}$ for all runs.
				Furthermore, for all data points the computational grid size is fixed at $\sigma_\mathrm{max} = x_\mathrm{max} = 10$ and the number of volume cells is set to $n = 400$.
				$\Gamma^{(2m)}$ are calculated from $u ( t_\mathrm{IR} = 60, \sigma )$ via the approximations \eqref{eq:derivative_1_central_error_2}, \eqref{eq:derivative_3_central_error_2}, and \eqref{eq:derivative_5_central_error_2} for the numerical derivative. The yellow straight line is for optical guidance.%
			}
		\end{figure}
		
	In Fig.~\ref{fig:sc_i_on_3_n_400_xmax_10_rir_10e-20_cutoff_test} we observe that the IR observables become independent of $\Lambda$ at rather large $\Lambda \approx 10^6$.
	This is several orders of magnitude above the model scales, contrary to what is often used in FRG studies in higher dimensions.
	If the UV cutoff is chosen too small, we find that the relative errors of $\Gamma^{(2n)}$ grow proportional to $\frac{1}{\Lambda}$, as estimated in Eq.~\eqref{eq:error_scaling_uv_cutoff}.
	Surprisingly, it turns out that the RG-consistency condition \eqref{eq:rg_consistency} is already violated at rather large UV cutoff scales $\Lambda \approx 10^5$ and is only fulfilled for $\Lambda \gtrsim 10^5$. 
	We conclude that it requires great care when specifying the UV scale in a FRG calculation.
	
\begin{figure}
			\centering
			\includegraphics{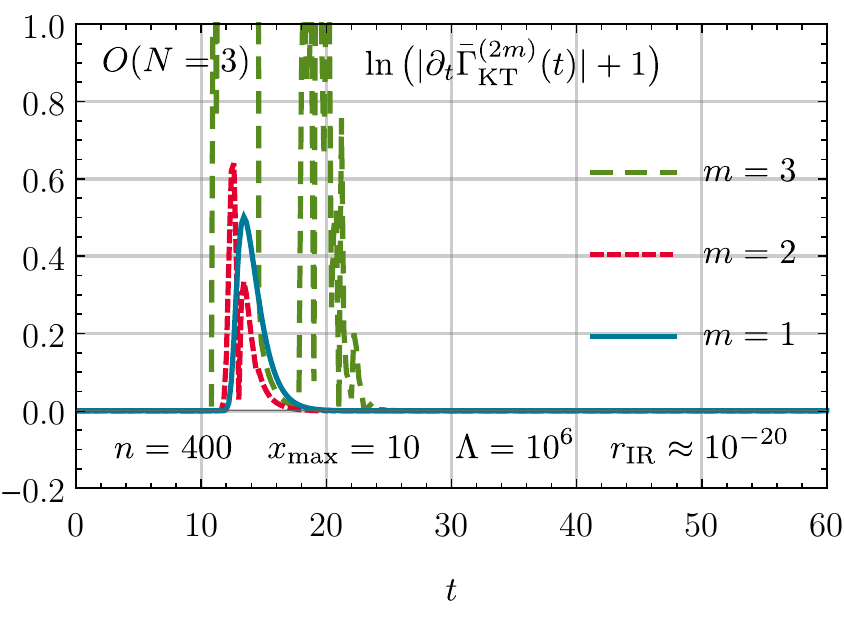}
			\caption{\label{fig:sc_i_on_3_n_400_xmax_10_lambda_1e6_tir_60_changing_rates}%
				The rate of change in $t$ of $\bar{\Gamma}^{(2m)} ( t )$ at the IR minimum $\sigma = 0$ for $n = 1, 2, 3$ during the RG flow.
				This rate of change is defined as the numerical RG time derivative $\partial_t \bar{\Gamma}^{(2m)} ( t )$ over the RG time.
				$\partial_t \bar{\Gamma}^{(2m)} ( t )$ are calculated via a finite-difference approximation ${[\bar{\Gamma}^{(2m)} (t) - \bar{\Gamma}^{(2m)} (t - \Delta t) ]/\Delta t}$, where $\Delta t = 0.2$.
				$\bar{\Gamma}^{(2m)} ( t )$ are obtained via numerical derivatives \eqref{eq:derivative_1_central_error_2}, \eqref{eq:derivative_3_central_error_2}, and \eqref{eq:derivative_5_central_error_2} of $u (t, \sigma)$ at $x = \sigma = 0$.
				For convenience, we added $1$ and took the logarithm to pronounce the regions of high changing rates of the observables $\bar{\Gamma}^{(2m)} ( t )$ and to identify the freeze-out plateau, where they have to vanish.
				We used the exponential regulator \eqref{eq:exponential_regulator} with UV cutoff $\Lambda = 10^6$.%
			}
		\end{figure}

	Before we close this discussion, we provide a natural measure to estimate the correct UV and IR scales of a model or theory, even if there are no exact reference values for observables that can be used for comparison with the FRG results.
	To this end, we plot in Fig.~\ref{fig:sc_i_on_3_n_400_xmax_10_lambda_1e6_tir_60_changing_rates} the shifted logarithm of the changing rates of the $\bar{\Gamma}^{(2n)} ( t )$ at the IR minimum $\sigma = 0$ over RG time $t$.
	These quantities have to vanish in the UV and the IR, when the relative errors \eqref{eq:relative_errors_gamma_2n} are not changing.
	
	A similar investigation can be done for any other model or theory and can be used as an indication to ensure sufficiently large UV and sufficiently small IR cutoffs: A first estimate may be obtained by choosing $\Lambda$ and $t_\mathrm{IR}$ in a way that the plateaus (or scaling regimes) in figures similar to Fig.~\ref{fig:sc_i_on_3_n_400_xmax_10_lambda_1e6_tir_60_changing_rates} are of approximately equal RG time duration than the time interval in which the actual dynamics takes place. 
	In the absence of an explicit and accessible error estimate rates of change are a cheap and simple tool to study the UV and IR limits of RG time evolution, \textit{cf.}\ Ref.~\cite{Stoll:2021ori}.

\subsection{Test case II: \texorpdfstring{$\phi^4$}{phi4} theory}

\begin{figure}
			\centering
			\includegraphics{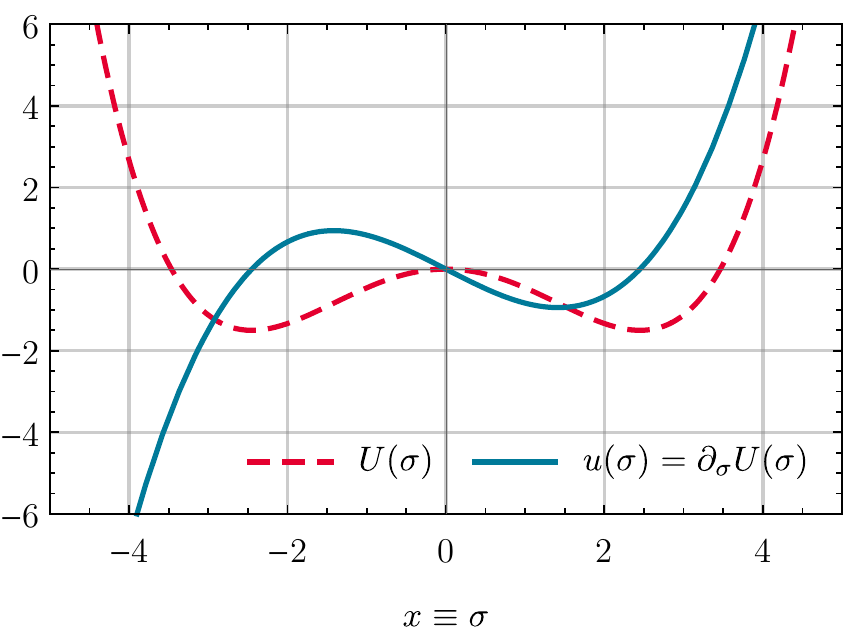}
			\caption{\label{fig:sc_ii_n_uv_initial_condition}%
				The UV potential $U ( \sigma )$ (red-dashed) and its first derivative $u ( \sigma ) = \partial_\sigma U ( \sigma )$ (blue-solid) of our test case \eqref{eq:testing_scenario_phi4} evaluated on the constant background field configuration.
			}
		\end{figure}

	The second test case is a zero-dimensional version of $\phi^4$ theory with a \textit{``Mexican hat''}-type potential well-known from standard textbook discussions of spontaneous symmetry breaking \cite{Goldstone:1961eq,Goldstone:1962es}.
	The initial UV potential reads
		\begin{align}
			U ( \vec{\varphi} \, ) = - \tfrac{1}{2} \, \vec{\varphi}^{\, 2} + \tfrac{1}{4!} \, ( \vec{\varphi}^{\, 2} )^2 \, .	\label{eq:testing_scenario_phi4}
		\end{align}
	The corresponding initial condition for the FRG flow is illustrated in Fig.~\ref{fig:sc_ii_n_uv_initial_condition}.
	The reference values for the exact IR 1PI vertex functions $\Gamma^{(2n)}$ of the $O(N)$ model \eqref{eq:on-model_relation_2pf_phi2} -- \eqref{eq:on-model_relation_6pf_phi2} are calculated numerically from the UV potential \eqref{eq:testing_scenario_phi4} and are listed for selected values of $N$ in Table~\ref{tab:sc_2_n_point_functions_exact}, which also includes reference values at $N = 4$ for the analogous potential with positive mass term $+ \tfrac{1}{2} \vec{\varphi}^{\, 2}$.
	
	In the following subsubsections, we discuss results obtained with the KT scheme and numerical results computed using the Taylor (vertex) expansion outlined in Sub.Sub.Sec.~\ref{subsubsec:vertex_expansion}.
		\begin{table}[b]
			\caption{\label{tab:sc_2_n_point_functions_exact}%
				The exact results for $\Gamma^{(2n)}$ for the $O(N)$ model with the initial UV potential \eqref{eq:testing_scenario_phi4} for $N = 1$ and $N = 4$ as well as $N=4$ with a positive mass term.
				The results are obtained by a high-precision one-dimensional numerical integration of the expectation values $\langle ( \vec{\phi}^{\, 2} )^n \rangle$ using \texttt{Mathematica}'s numerical integration routine \textit{NIntegrate} \cite{Mathematica:12.1} with a \textit{PrecisionGoal} and \textit{AccuracyGoal} of $10$.
				Here, we present the first ten digits only.
			}
			\begin{ruledtabular}
				\begin{tabular}{l c c c}
					$N$		&	$\Gamma^{(2)}$	&	$\Gamma^{(4)}$	&	$\Gamma^{(6)}$	\\
					\colrule
					$1$		&	$0.1995098930$		&	$0.0622583604$		&	$0.1077442107$ \\
					$4$		&	$0.5064440744$		&	$0.1824153741$		&	$0.2802880035$ \\
					$4$\footnote{Using $U( \vec{\varphi} \, ) = + \tfrac{1}{2} \, \vec{\varphi}^{\, 2} + \tfrac{1}{4!} \, ( \vec{\varphi}^{\, 2} )^2$, thus Eq.~\eqref{eq:testing_scenario_phi4} with a positive mass term.}		&	$1.5809246562$		&	$0.6118483669$		&	$0.5686310448$
				\end{tabular}
			\end{ruledtabular}
		\end{table}

\subsubsection{Results obtained using the KT scheme}

	In this subsubsection we will discuss selected numerical results of the application of the KT scheme for the analytic initial condition \eqref{eq:testing_scenario_phi4}.
	We have performed the full set of numerical tests discussed in Sub.Sec.~\ref{subsec:sc1} for this initial condition and found results supporting the general statements made there.
	For the sake of brevity, we will therefore not repeat the complete discussion of that section.\\

\begin{figure}
			\centering
			\includegraphics{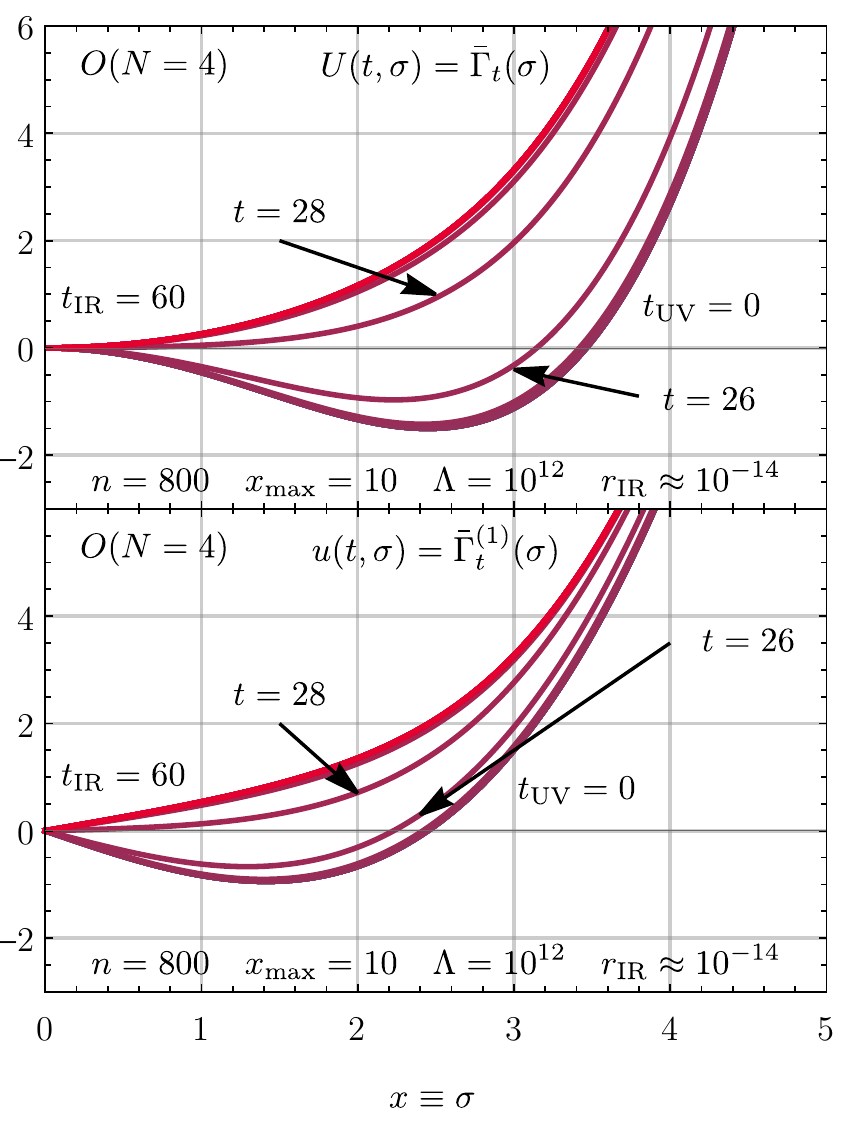}
			\caption{\label{fig:sc_ii_n_on_4_n_800_xmax_10_lambda_1e12_tir_60_rg_flow}%
				The FRG flow of the effective potential $U ( t, \sigma )$ (upper panel) and its derivative $u ( t , \sigma ) = \partial_\sigma U ( t , \sigma )$ (lower panel) for the zero-dimensional $O ( 4 )$ model with initial condition Eq.~\eqref{eq:testing_scenario_phi4}, evaluated at $t = 0, \, 2, \, 4, \, \ldots, \, 60$ (integer values for $t$ were chosen for convenience and readability). 
				The (overlapping) {blue} and {violet} curves correspond to the UV and the {red} curves to the IR.
				We used the exponential regulator \eqref{eq:exponential_regulator} with UV cutoff $\Lambda = 10^{12}$.
				The plot does not show the region $x = 5$ to $x = 10$, because the tiny differences between $u ( t, \sigma )$ and $u ( t_\mathrm{UV}, \sigma )$ are not visible in this region and vanish for large $x = \sigma$ anyhow.%
			}
		\end{figure}

	\paragraph*{UV and IR scales:}
	In Fig.~\ref{fig:sc_ii_n_on_4_n_800_xmax_10_lambda_1e12_tir_60_rg_flow} we present the RG flow of the derivative of the effective potential $u ( t, \sigma )$ from the UV ({blue}) to the IR ({red}).
	For the smooth initial condition -- in the absence of large gradients -- the highly non-linear advection and diffusion contribute almost an equal amount to the dynamics.
	Between $t \approx 25$ and $t \approx 30$ we observe significant changes in the shape of the potential: the non-trivial minimum moves towards $\sigma = 0$ and vanishes at $t \approx 28$ resulting in a convex potential with a trivial minimum at $\sigma = 0$ as expected and required.
	At small and large $t$ outside the apparent dynamic range between $t \approx 25$ and $t \approx 30$ we observe only very marginal changes in Fig.~\ref{fig:sc_ii_n_on_4_n_800_xmax_10_lambda_1e12_tir_60_rg_flow}.

\begin{figure}
			\centering
			\includegraphics{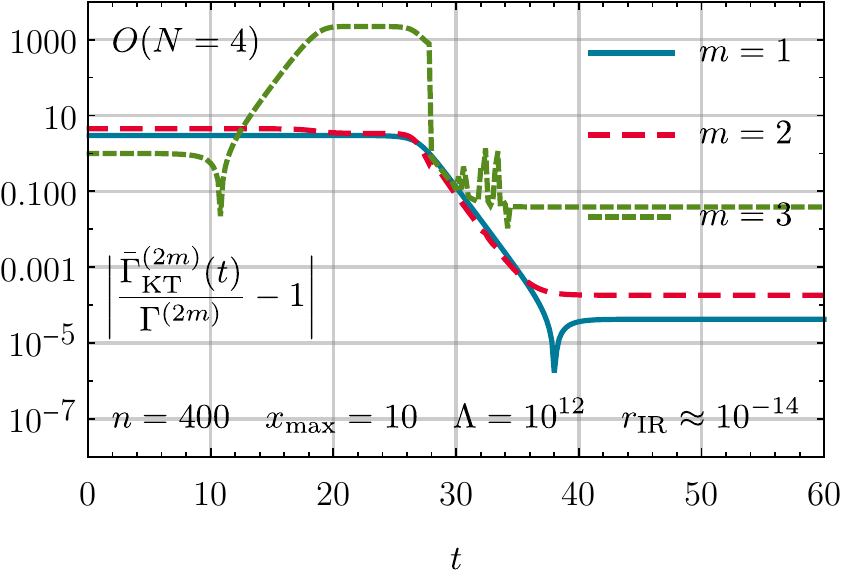}
			\caption{\label{fig:sc_ii_n_on_4_n_400_xmax_10_lambda_1e12_tir_60_flow_errors}%
				The relative error for $\Gamma^{(2m)}$, for $m = 1, 2, 3$, calculated with the KT scheme as a function of the RG time $t$ for the $O(4)$ model.
				The initial UV potential is given by Eq.~\eqref{eq:testing_scenario_phi4}.
				We use the exponential regulator \eqref{eq:exponential_regulator} with UV cutoff $\Lambda = 10^{12}$.
				The computational grid has 400 cells and $\sigma_\mathrm{max} = x_\mathrm{max} = 10$.
				$\Gamma^{(2m)}$ are extracted from $u ( t_\mathrm{IR} = 60, \sigma )$ via the finite-difference stencils \eqref{eq:derivative_1_central_error_2}, \eqref{eq:derivative_3_central_error_2}, and \eqref{eq:derivative_5_central_error_2}.%
			}
		\end{figure}

	A close inspection of the relative errors for the first three non-vanishing $n$-point vertex functions in Fig.~\ref{fig:sc_ii_n_on_4_n_400_xmax_10_lambda_1e12_tir_60_flow_errors} reveals that actually the relevant dynamics sets in much earlier at $t \approx 10$ and the values for the $n$-point vertex functions freeze out at late times around $t \approx 40$, which is due to the diffusion close to $\sigma = 0$.
	On the level of $u ( t, \sigma )$ these subtle changes in the $n$-point vertex functions cannot be observed by a simple visual inspection of Fig.~\ref{fig:sc_ii_n_on_4_n_800_xmax_10_lambda_1e12_tir_60_rg_flow}.
	
	The plateaus in the UV (at small $t$) and the IR (at large $t$) support the choice of $\Lambda=10^{12}$ and $t_\mathrm{IR}=60$ to be valid initial UV and IR cutoff scales in terms of RG consistency. 
	The present initial UV scale is larger when compared to $\Lambda=10^{6}$ used for most computations involving the non-analytic potential considered in the previous section. 
	($\Lambda=10^{6}$ corresponds to $t \approx 14$ in the present case, which is already in the dynamic region in Fig.~\ref{fig:sc_ii_n_on_4_n_400_xmax_10_lambda_1e12_tir_60_flow_errors}.)
	Hence, the inclusion of a quartic interaction term in Eq.~\eqref{eq:testing_scenario_phi4} seems to require higher initial UV scales to ensure RG consistency.
	This supports the  statements made in Sub.Sub.Sec.~\ref{subsec:RGconsistency}: RG consistency and UV and IR scales have to be re-evaluated when changing the initial condition in the UV or the model under consideration, since characteristic internal scales then also change.\\

\begin{figure}
			\centering
			\includegraphics[width=\linewidth]{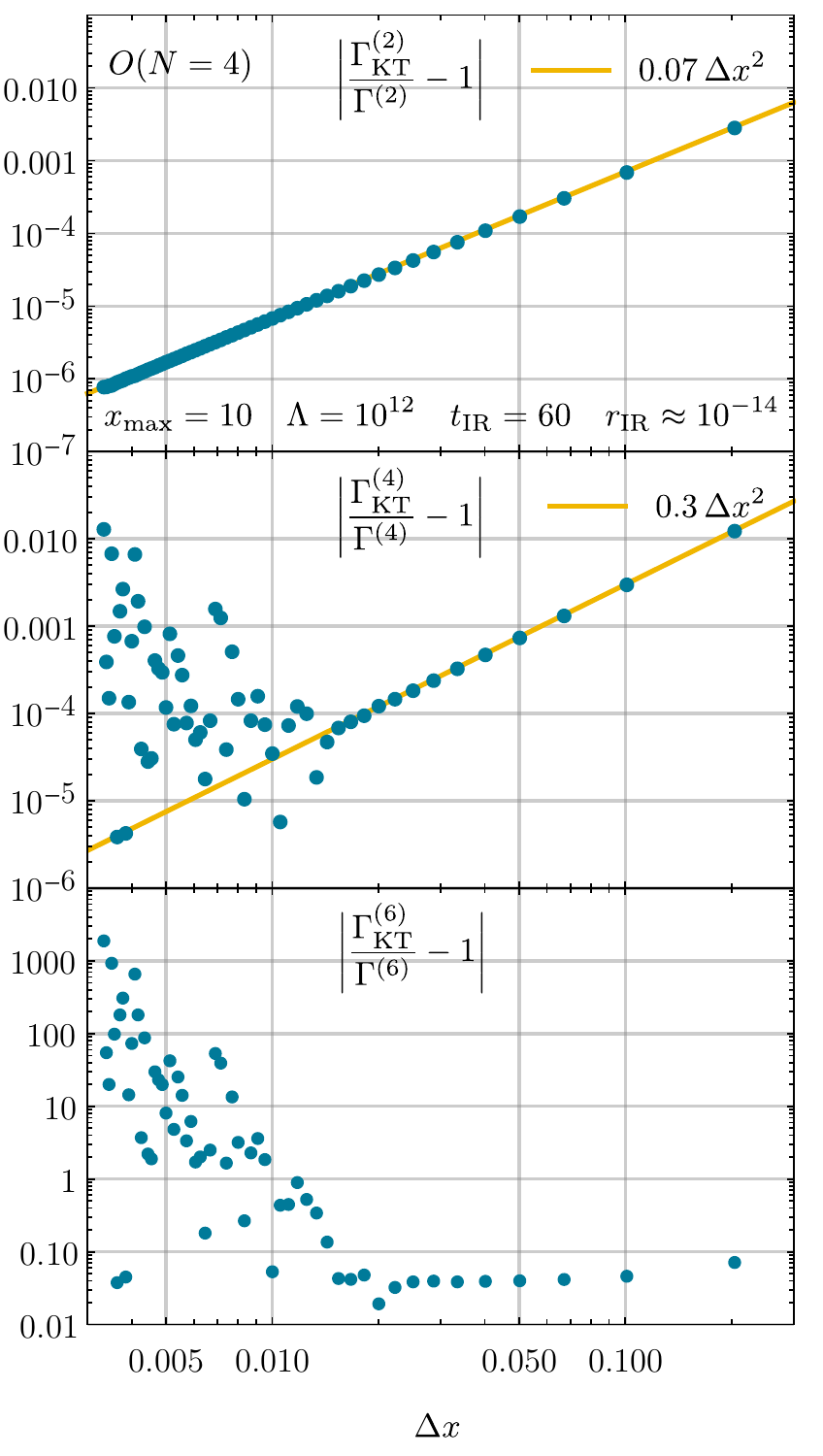}
			\caption{\label{fig:fig:sc_ii_n_on_4_xmax_10_lambda_1e12_tir_60_deltax_scaling}%
			The relative error as a function of the cell size $\Delta x$ for the numerical results (blue dots) from the KT scheme for the coefficients $\Gamma^{(2n)}$ for $n = 1, 2, 3$ with initial potential \eqref{eq:testing_scenario_phi4}.
			The numerical derivatives at $\sigma = 0$ of $u ( t_\mathrm{IR} = 60, \sigma )$ were calculated via the second-order accurate central schemes \eqref{eq:derivative_1_central_error_2}, \eqref{eq:derivative_3_central_error_2}, and \eqref{eq:derivative_5_central_error_2}.
			Here, $x_\mathrm{max} = 10$, but we could have used any sufficiently large $x_\mathrm{max}$.
			We used the exponential regulator \eqref{eq:exponential_regulator} with UV cutoff $\Lambda = 10^{12}$.
			The yellow straight lines are for optical guidance.%
			}
		\end{figure}

\begin{figure}
			\centering
			\includegraphics{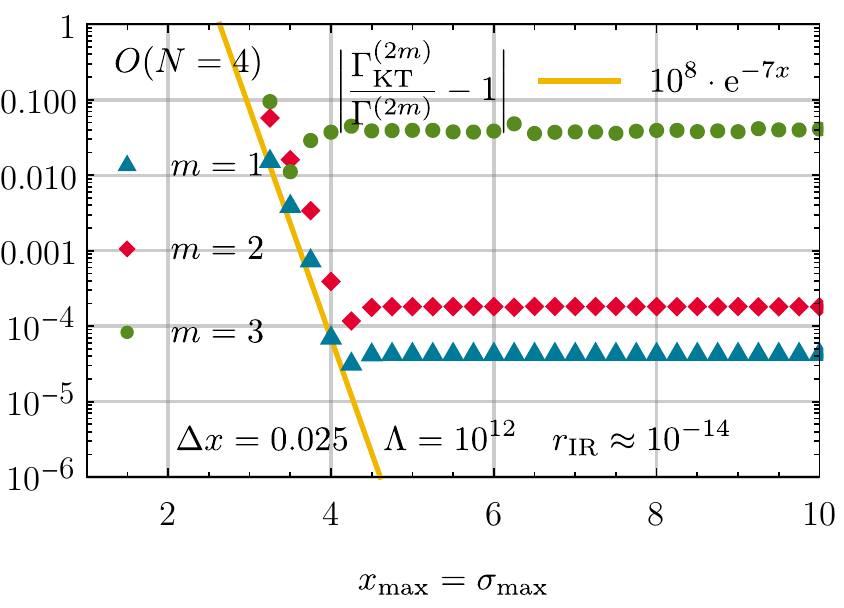}
			\caption{\label{fig:sc_ii_n_on_4_deltax_25e-3_lambda_1e12_tir_60_errors_xmax}%
				The relative error for $\Gamma^{(2m)}$ for $m = 1, 2, 3$ for the UV potential \eqref{eq:testing_scenario_phi4} of the $O ( 4 )$ model as a function of $x_\mathrm{max}$, keeping the cell size $\Delta x = 0.025$ constant.
				$\Gamma^{(2m)}$ are computed from the discrete values of the derivative of the IR potential $u ( t_\mathrm{IR} = 60, \sigma )$ via the second-order accurate central finite-difference stencils \eqref{eq:derivative_1_central_error_2}, \eqref{eq:derivative_3_central_error_2}, and \eqref{eq:derivative_5_central_error_2} at $\sigma = 0$.
				We used the exponential regulator \eqref{eq:exponential_regulator} with UV cutoff $\Lambda = 10^{12}$.
				The yellow straight line is for optical guidance.%
			}
		\end{figure}

	\paragraph*{Computational grid and domain size:}
	We conclude this subsubsection with a brief discussion regarding the computational grid and domain size.
	The relative error for the first three non-vanishing $n$-point vertex functions is shown as a function of the cell size $\Delta x$ in Fig.~\ref{fig:fig:sc_ii_n_on_4_xmax_10_lambda_1e12_tir_60_deltax_scaling}.
	For the two-point function we recover a perfect error scaling with $\Delta x^2$ down to extremely small $\Delta x$.
	The last data point in Fig.~\ref{fig:fig:sc_ii_n_on_4_xmax_10_lambda_1e12_tir_60_deltax_scaling} is at ${\Delta x \approx 3.3 \cdot 10^{-3}}$ corresponding to $n=3000$ cells.
	For the two-point function the rounding errors of the employed finite-difference extraction \eqref{eq:derivative_1_central_error_2} for $\Gamma^{(2)}$ and the finite precision of the ODE integrator (\textit{NDSolve} from \texttt{Mathematica} \cite{Mathematica:12.1} with a \textit{PrecisionGoal} and \textit{AccuracyGoal} of 10) seem to be small for all depicted $\Delta x$ in this scenario. 
	A comparison with the present perfect error scaling for $\Gamma^{(2)}$ supports the comments made about discretization errors for the discontinuous initial condition \eqref{eq:testing_scenario_non-analytic_quadaratic_asymptote} in Fig.~\ref{fig:sc_i_on_3_xmax_10_lambda_1e6_tir_60_deltax_scaling}.
	For the higher-order $n$-point vertex functions $\Gamma^{(4)}$ and $\Gamma^{(6)}$, however, we find that rounding errors related to the finite-difference extractions \eqref{eq:derivative_3_central_error_2} and \eqref{eq:derivative_5_central_error_2} limit the achievable precision.
	Again, we identify $\Delta x \approx 0.025$ as an optimal cell size for the extraction of $\Gamma^{(4)}$ and $\Gamma^{(6)}$ but note that typical relative errors for $\Gamma^{(6)}$ are rather high at $\approx 4 \%$ around $\Delta x \approx 0.025$.
	In Fig.~\ref{fig:sc_ii_n_on_4_deltax_25e-3_lambda_1e12_tir_60_errors_xmax}, we study the effect of the size of the computational domain $x_\mathrm{max}$ on the achievable relative errors for $\Gamma^{(2)}$, $\Gamma^{(4)}$, and $\Gamma^{(6)}$ at a constant $\Delta x =0.025$. 
	One major difference between the $\phi^4$ potential \eqref{eq:testing_scenario_phi4} studied in this section and the non-analytic potential \eqref{eq:testing_scenario_non-analytic_quadaratic_asymptote} of the previous subsection is their asymptotic behavior for large $\sigma$.
	For large $\sigma$ the leading-order term of the $\phi^4$ potential is -- as the name suggests -- quartic while the non-analytic potential of the previous section grows only $\sim \sigma^2$.
	In terms of the conserved quantity $u = \partial_\sigma U$ one might expect problems when using a linear extrapolation for the ghost cells at large $\sigma$ as discussed in Sub.Sub.Sec.~\ref{subsec:sec:BCinf} with a potential where $u$ grows $\sim \sigma^3$ for large $\sigma$.
	For the non-analytic initial condition \eqref{eq:testing_scenario_phi4} we avoided this possible source of error by construction.
	However, considering the results plotted in Fig.~\ref{fig:sc_ii_n_on_4_deltax_25e-3_lambda_1e12_tir_60_errors_xmax} together with the perfect error scaling displayed in the previous Fig.~\ref{fig:fig:sc_ii_n_on_4_xmax_10_lambda_1e12_tir_60_deltax_scaling}, we conclude that a linear extrapolation is not problematic even in the case of cubic asymptotics for $u$.
	This might be again in part related to the large spatial distance between the physical minimum in the IR and the upper boundary of the grid.
	For $x_\mathrm{max}\gtrsim 5$ we find a complete insensitivity of the relative errors on the interval size.

\subsubsection{Results obtained using the FRG Taylor (vertex) expansion}
\label{subsubsec:vertex_expansion_phi4}

	In this subsubsection we confront the theoretical results and concerns stated in Sub.Sub.Sec.~\ref{subsubsec:vertex_expansion} with respect to the Taylor expansion around the fixed expansion point $\vec{\varphi} = 0$ with the exact results for the zero-dimensional $O(N)$ model.
	The $\phi^4$ potential of Eq.~\eqref{eq:testing_scenario_phi4} is the, in terms of initial conditions, simplest UV potential with a non-trivial minimum.
	At the end of this subsection we will briefly discuss the vertex expansion for the $\phi^4$ potential with positive mass term and therefore a scenario without a non-trivial minimum, which has to be considered the simplest non-trivial UV initial condition in the context of the vertex expansion for the zero-dimensional $O(N)$ model.\\
	
	\paragraph*{RG flow of the $n$-point vertex functions:}
	In the following we integrate the ODE system \eqref{eq:example_vertex_expansion} truncated at $m = 2 n_\mathrm{trunc}$ with the initial condition
		\begin{align}
			&	\bar{\Gamma}^{(2)} ( 0 ) = - 1 \, ,	&&	\bar{\Gamma}^{(4)} ( 0 ) = + 1 \, ,	\vphantom{\bigg(\bigg)}
			\\
			&	\bar{\Gamma}^{(2n)} ( 0 ) = 0 \, , \quad \forall n > 2 \, ,	\vphantom{\bigg(\bigg)}	\nonumber
		\end{align}
	corresponding to the potential \eqref{eq:testing_scenario_phi4} numerically up to $t_\mathrm{IR}=60$ employing the exponential regulator \eqref{eq:exponential_regulator} with $\Lambda=10^{12}$ and using the same ODE solver \textit{NDSolve} from \texttt{Mathematica} \cite{Mathematica:12.1} with a \textit{PrecisionGoal} and \textit{AccuracyGoal} of 10 as before.
	Using the $n$-point vertex functions at the physical minimum as the flow variables makes an additional extraction procedure (like finite differences) obviously obsolete.
	The $n$-point correlation functions in the IR can be directly obtained from the values $\bar{\Gamma}^{(2n)} ( t_\mathrm{IR} ) = \Gamma^{(2n)}$.
		
\begin{figure}
			\centering
			\includegraphics{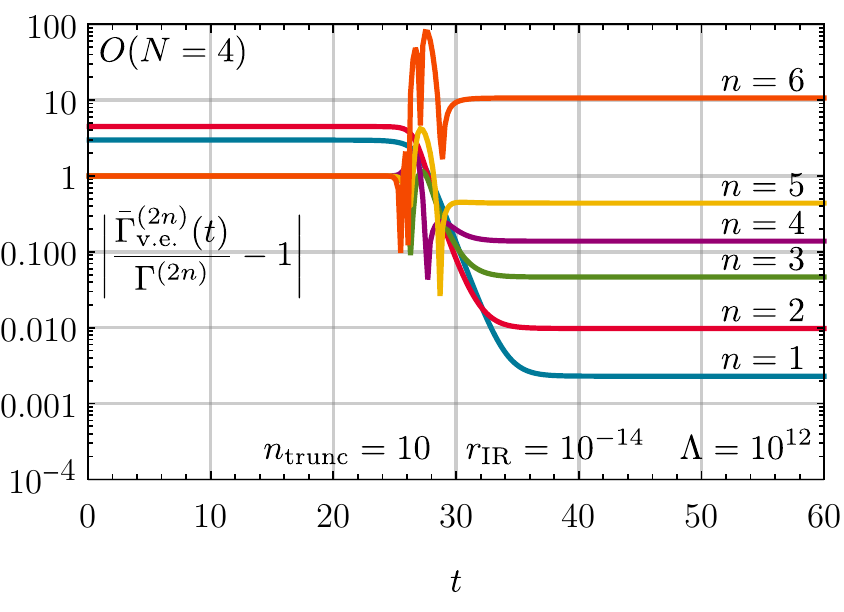}
			\caption{\label{fig:sc_ii_n_on_4_lambda_1e12_tir_60_ntrunc_10_vertex_exp_flow}%
				The relative errors for $\Gamma^{(2n)}$ as a function of the RG time $t$ for $n \in \{ 1, \ldots, 6 \}$ for the $O (4)$ model.
				$\Gamma^{(2n)}$ were calculated via the FRG flow of the FRG Taylor (vertex) expansion with truncation order $m = 2 n_\mathrm{trunc} = 20$ using the exponential regulator \eqref{eq:exponential_regulator}. 
				As initial condition we use the UV potential \eqref{eq:testing_scenario_phi4}.
			}
		\end{figure}

	In Fig.~\ref{fig:sc_ii_n_on_4_lambda_1e12_tir_60_ntrunc_10_vertex_exp_flow} we show the flow of the relative deviations for the first six non-vanishing $n$-point vertex functions towards the IR using $m = 2 n_\mathrm{trunc} = 20$ vertices in the expansion for the $O(4)$ model.
	We can identify a dynamic range between $t \approx 24$ and $t \approx 38$ in which the vertices vary significantly and change their signs before they reach their respective IR values.
	This range is substantially smaller than the dynamic range observed when solving the full PDE \eqref{eq:conservation_law_u_phi} using the KT scheme, see Fig.~\ref{fig:sc_ii_n_on_4_n_400_xmax_10_lambda_1e12_tir_60_flow_errors}.
	In the IR, the errors range from $2.3 \cdot 10^{-3}$ for $\Gamma^{(2)}$ to $1.1 \cdot 10^{1}$ for $\Gamma^{(12)}$.
	However, the strict hierarchy observed in Fig.~\ref{fig:sc_ii_n_on_4_lambda_1e12_tir_60_ntrunc_10_vertex_exp_flow} for $n = 1, \ldots, 6$ is not a general feature of the vertex expansion for this model.
	Using different $n_\mathrm{trunc}$ or including higher-order vertices changes this hierarchy.\\
		
\begin{figure}
			\centering
			\includegraphics{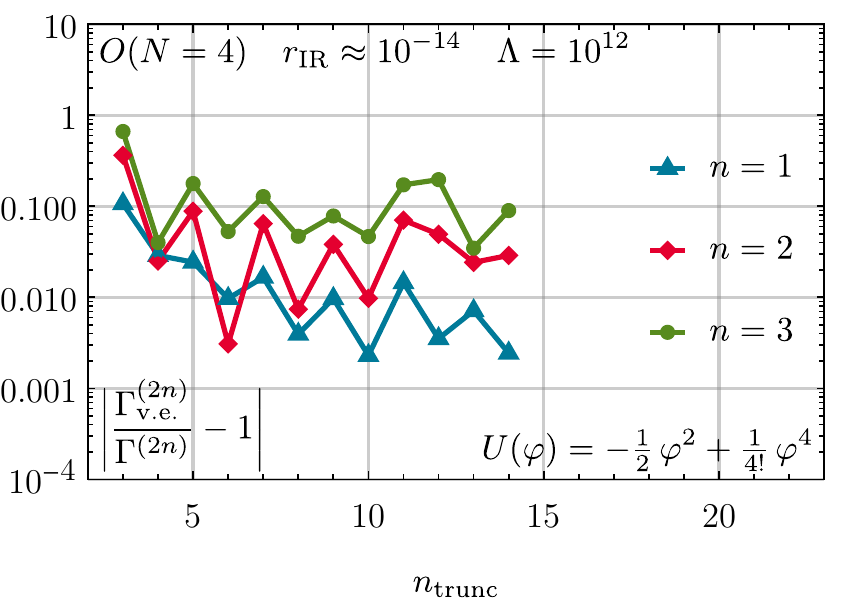}
			\caption{\label{fig:sc_ii_n_on_4_lambda_1e12_tir_60_vertex_exp_error}%
				The relative errors for $\Gamma^{(2n)}$ in the IR for $n = 1, 2, 3$ for the $O (4)$ model, calculated via the FRG flow of the FRG Taylor (vertex) expansion to order $m = 2 n_\mathrm{trunc}$ with $n_\mathrm{trunc} \in \{ 3, \ldots , 14 \}$ using the exponential regulator \eqref{eq:exponential_regulator}.
				As initial condition we use the UV potential \eqref{eq:testing_scenario_phi4}.
				The discrete results for integer $n_\mathrm{trunc}$ are connected by straight lines to improve readability and for a better trend analysis.%
			}
		\end{figure}

	\paragraph*{Truncation error:}
	The truncation error for the $O(4)$ model is discussed at hand of Fig.~\ref{fig:sc_ii_n_on_4_lambda_1e12_tir_60_vertex_exp_error}, where we show the relative error for $\Gamma^{(2)}$, $\Gamma^{(4)}$, and $\Gamma^{(6)}$ for the vertex expansion using different truncation orders $m=2n_\mathrm{trunc}$ between $n_\mathrm{trunc}=3$ and $n_\mathrm{trunc}=14$. 
	Beyond $n_\mathrm{trunc}=10$ the relative error for the $n$-point vertex function no longer decreases and we observe in general rather strong oscillations using different $n_\mathrm{trunc}$.
	The errors for the two and four-point function are with $2.3 \cdot 10^{-3}$ and $9.8 \cdot 10^{-3}$ larger than the errors ($4.2 \cdot 10^{-5}$ and $1.8 \cdot 10^{-4}$ respectively) obtained in the KT scheme, see,  \textit{e.g.}, Fig.~\ref{fig:sc_ii_n_on_4_deltax_25e-3_lambda_1e12_tir_60_errors_xmax}. 
	The relative error for the six-point function is with $4.7 \cdot 10^{-2}$ comparable to the $3.7 \cdot 10^{-2}$ error obtained in the KT scheme.
	While the extraction of higher-order $n$-point functions beyond $n = 6$ is in general possible in the vertex expansion, their relative errors grow overall rapidly with increasing $n$.
		
	For the initial condition \eqref{eq:testing_scenario_phi4} we do not observe any meaningful error scaling in orders of $n_\mathrm{trunc}$.
	Furthermore a numerical solution at and beyond $n_\mathrm{trunc}=15$ has proven impossible with the current set-up.
	At $n_\mathrm{trunc}=15$ an ODE integration to the IR at $r ( t_\mathrm{IR} = 60 ) \approx 10^{-14}$ is impossible due to an instability of the ODE system occurring at $t \approx 30$ where all coefficients $\Gamma^{(2n)}(t)$ with $n>1$ start diverging.
	This divergence seems to be driven by $\Gamma^{(30)}(t)$.
	The ODE system is in general poorly conditioned since $\Gamma^{(2n)}(t)$ for different $n$ vary vastly over multiple orders of magnitude.
	The instability at  $t \approx 30$ cannot be overcome by increasing the numerical precision of the employed ODE integrator (\textit{NDSolve} from \texttt{Mathematica} \cite{Mathematica:12.1}) and seems to be an inherent problem of the ODE systems with $n_\mathrm{trunc}\geq15$.
		
\begin{figure}
			\centering
			\includegraphics{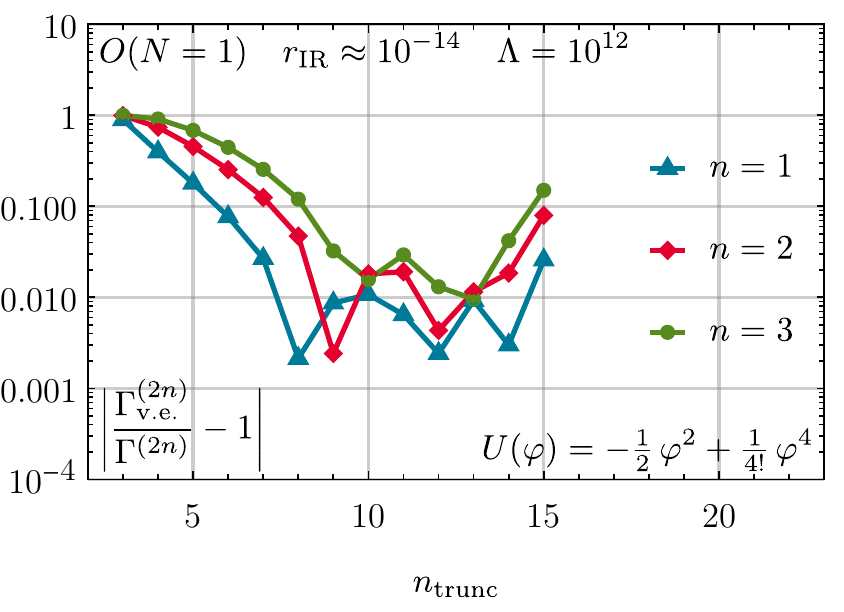}
			\caption{\label{fig:sc_ii_n_on_1_lambda_1e12_tir_60_vertex_exp_error}%
				The relative errors for $\Gamma^{(2n)}$ in the IR for $n = 1, 2, 3$ for the $O (1)$ model, calculated via the FRG flow of the FRG Taylor (vertex) expansion to order $m = 2 n_\mathrm{trunc}$ with $n_\mathrm{trunc} \in \{ 3, \ldots , 15 \}$ using the exponential regulator \eqref{eq:exponential_regulator}.
				As initial condition we use the UV potential \eqref{eq:testing_scenario_phi4}.
				The discrete results for integer $n_\mathrm{trunc}$ are connected by straight lines to improve readability and for a better trend analysis.%
			}
		\end{figure}

	The vertex expansion for $\Gamma^{(2n)}(t)$, with a fixed expansion point at $\vec{\varphi} = 0$, for the zero-dimensional $O ( 4 )$ model and the simple initial condition \eqref{eq:testing_scenario_phi4} with its non-trivial global minimum in the UV is severely limited in its performance.
	The absence of a proper error scaling in orders of $n_\mathrm{trunc}$ and the instability of the ODE system beyond $n_\mathrm{trunc}=14$ support the conceptual reservations presented in Sub.Sub.Sec.~\ref{subsubsec:vertex_expansion}.
	It seems that the expansion around $\vec{\varphi} = 0$ is either incapable of capturing the dynamics driven by the non-trivial minima located at $| \vec{\varphi} \, | = \sqrt{6}$ in the UV or the desired solution might be non-analytic in $\vec{\varphi} = 0$.
	The situation does not improve when considering the same initial condition in the purely diffusive $O(1)$ model.
	In Fig.~\ref{fig:sc_ii_n_on_1_lambda_1e12_tir_60_vertex_exp_error} we display relative errors for the first three non-vanishing $\Gamma^{(2n)}$ as a function of $n_\mathrm{trunc}$ for the initial condition \eqref{eq:testing_scenario_phi4} in the $O(1)$ model.
	The overall errors are even worse when compared to the $O(4)$ results discussed previously.
	The ODE integration becomes impossible at $n_\mathrm{trunc} = 16$ where we encounter an instability at $t \approx 31$.\\
	
\begin{figure}
			\centering
			\includegraphics{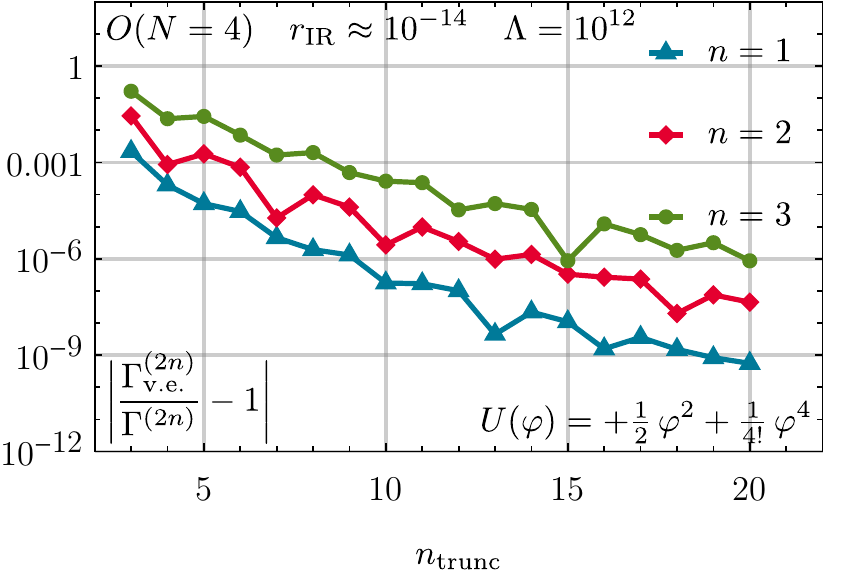}
			\caption{\label{fig:sc_ii_p_on_4_lambda_1e12_tir_60_vertex_exp_error}%
				The relative errors for $\Gamma^{(2n)}$ in the IR for $n = 1, 2, 3$ and for the $O (4)$ model, calculated via the FRG flow of the FRG Taylor (vertex) expansion to order $m = 2 n_\mathrm{trunc}$ with $n_\mathrm{trunc} \in \{ 3, \ldots , 20 \}$ using the exponential regulator \eqref{eq:exponential_regulator}.
				As initial condition we use the UV potential \eqref{eq:testing_scenario_phi4} with positive instead of negative mass term.
				The discrete results for integer $n_\mathrm{trunc}$ are connected by straight lines to improve readability and for a better trend analysis.
			}
		\end{figure}

	\paragraph*{$\phi^4$ potential with positive mass term:}
	We continue our discussion of the FRG Taylor (vertex) expansion by considering the modified initial condition $+ \tfrac{1}{2} \, \vec{\varphi}^{\, 2} + \tfrac{1}{4!} \, ( \vec{\varphi}^{\,2} )^2$ with a positive mass term and therefore without a non-trivial minimum.
	In the context of zero-dimensional $O(N)$ models this initial condition is in the family of UV potentials discussed qualitatively at length and to some extent even quantitatively in Refs.~\cite{Keitel:2011pn,Rosa:2016czs,Moroz:2011thesis}.
	In Fig.~\ref{fig:sc_ii_p_on_4_lambda_1e12_tir_60_vertex_exp_error} we show relative errors for the first three non-vanishing $\Gamma^{(2n)}$ as a function of $n_\mathrm{trunc}$ for this initial condition for the $O(4)$ model.
	These results where obtained using \texttt{Mathematica}'s \textit{NDSolve} \cite{Mathematica:12.1} with an increased \textit{PrecisionGoal} and \textit{AccuracyGoal} of 12, which became necessary for a proper truncation-error scaling beyond $n_\mathrm{trunc}=15$ for the two-point function.
	In Fig.~\ref{fig:sc_ii_p_on_4_lambda_1e12_tir_60_vertex_exp_error} we observe a truncation-error scaling following power laws in $n_\mathrm{trunc}$ with approximately $n_\mathrm{trunc}^{-8.2}$, $n_\mathrm{trunc}^{-7.6}$, and $n_\mathrm{trunc}^{-7.3}$ for the two-point, four-point, and six-point function, respectively.
	For this initial condition the expansion point $\vec{\varphi} = 0$ is located at the global minimum of the potential and the potential is also convex for all $t$.
	The dynamics of the FRG flow is rather unspectacular for this potential, see Fig.\ 13 of Ref.~\cite{Keitel:2011pn} or Fig.\ 4 of Ref.~\cite{Koenigstein:2021rxj} for a visualization. For the two- and four-point functions, the numerical results at $n_\mathrm{trunc}=3$ ($\Leftrightarrow m=6$) have already acceptable relative errors of $\approx 2.2 \cdot 10^{-3}$ and $\approx 2.8 \cdot 10^{-2}$, respectively, which was observed and discussed in Ref.~\cite{Keitel:2011pn}, where results for the vertex expansions were presented only up to $n_\mathrm{trunc}=3$.
		
	The vertex expansion outperforms the KT scheme in this setting in terms of relative errors.
	The performance and practical applicability of the vertex expansion seem to depend strongly on the initial condition under consideration.
	We will discuss another analytic initial condition for the vertex expansion briefly in the next Sub.Sec.~\ref{subsec:sc3}.\\

	\paragraph*{Numerical irreversibility:}
	Before we conclude this subsubsection we will briefly comment on the irreversibility of RG flows when employing the FRG Taylor (vertex) expansion.
	We discussed in subsection \ref{subsec:advection_diffusion_equation} that the projection onto a finite set of couplings underlying the FRG Taylor (vertex) expansion theoretically allows for an unphysical reversibility of the RG flow.
	The ODE systems for the running couplings of the FRG Taylor (vertex) expansion in principle allow for an integration both in positive and negative RG time direction.
	Thus an unphysical resolution of micro physics from macro physics -- an inversion of the underlying RG transformations connecting them -- is possible when considering a finite set of couplings $\{\bar{\Gamma}^{(2n)}(t)\}$.
	
	We performed practical test with the $\phi^4$ theory discussed in this subsubsection.
	For the $\phi^4$ theory with positive mass term discussed in the previous paragraph a complete inversion of the RG flow (from $t=60$ back to $t=0$ using $\Lambda=10^{12}$) is numerically possible for systems with $n_\mathrm{trunc}<7$ for $N=1$.
	For larger systems the strong oscillations of the higher-order couplings prevent a numerical integration back to the UV.
	The ODE system becomes numerically unstable when approaching $t\approx 24$ from above.
	The recovery of the exact UV initial condition is very good for small $n_\mathrm{trunc}$ but deteriorates when approaching $n_\mathrm{trunc}=7$.
	For the $\phi^4$ theory with positive mass term this situations remains qualitatively unchanged for higher $N>1$.
	
	For the $\phi^4$ theory with negative mass term an inversion of the RG flow from the IR to the UV is numerically impossible.
	We were not able to find a $n_\mathrm{trunc}$ and $N$ in heuristic tests which allowed for a numerical inversion of the RG flow from $t=60$ back to $t=0$ using $\Lambda=10^{12}$.
	The dynamics related to the vaporization of the non-trivial minimum seems to prevent a numerical inversion.
	In our heuristic tests it has proven impossible to form back the non-trivial minimum when approaching the UV from the IR.
	This is a rather interesting observation which might warrant a detailed investigation of the ODE systems involved in the FRG Taylor (vertex) expansion.
	Further investigations in higher-dimensional models might be interesting.\\

	\paragraph*{Concluding remarks:}
	In this subsubsection we have discussed numerical results for the vertex expansion in $\bar{\Gamma}^{(2n)}(t)$ around $\vec{\varphi} = 0 $ with the quartic potential \eqref{eq:testing_scenario_phi4} with positive and negative mass terms $\pm \tfrac{1}{2} \, \vec{\varphi}^{\, 2}$.
	The numerical performance in terms of achievable relative errors for the $n$-point vertex functions in the IR is rather poor for the potential with the negative mass term and very good for the potential with the positive mass term.
	In both situations the initial conditions are analytic but for a negative mass term the corresponding non-trivial minimum has significant implications on the dynamics of the FRG flow. The vertex expansion with a fixed expansion point at $\vec{\varphi} = 0$ is not able to capture the relevant dynamics.
	The zero-dimensional $O(N)$ model has proven very challenging for the vertex expansion.
	It should be noted that in this paper we discussed the simplest possible vertex-expansion scheme.
	Other versions of the FRG Taylor (vertex) expansion including a moving expansion point or a rescaling of the expansion coefficients might improve the performance of the expansion scheme in certain cases, \textit{cf.} Refs.~\cite{Litim:2002cf,Schaefer:2001cn,Rennecke:2015lur}.
	Implementing and testing different approaches to the vertex expansion for zero-dimensional $O(N)$ models would certainly be an interesting topic for further studies.
		
	The conceptual reservations about the application of a Taylor (vertex) expansion to the advection-diffusion equations considered in this work still stand.
	In scenarios with RG flows driven by an interplay of advection and diffusion around non-trivial minima and/or large gradients of the conserved quantity $u$ the vertex expansion is inevitably doomed to fail.
	It is not possible to capture the dynamics of such equations reliably with the simple vertex expansion discussed here.
	A numerical inversion of the RG flow is also impossible in those scenarios.
		
	It should also be noted that the absence of momenta and therefore phase-space suppression in zero dimensions leads to a strong coupling in field space for the potential. 
	This might be a possible explanation for our observations in $d=0$. For ``real'' QFTs in $d > 0$ dimensions the vertex expansion is and was applied very successfully in various contexts, see, \textit{e.g.}, Refs.~\cite{Benitez:2009xg,Benitez:2011xx} for scalar field theories and Refs.~\cite{Mitter:2014wpa,Cyrol:2017ewj,Cyrol:2017qkl} for gauge theories.
	In $d > 0$, there is indeed a significant difference between an expansion in vertices carrying and capturing the non-trivial momentum dependences and Taylor expansions of interaction potentials.
	Capturing and tracking the momentum dependence with a suitable ansatz is very important in the study of FRG flows in $d>0$ and in certain models a proper resolution of the dynamics in momentum space may be even more important than a full-fledged treatment in field space.

\subsection{Test case III: \texorpdfstring{$\phi^6$}{phi6} potential}
\label{subsec:sc3}

\begin{figure}
			\centering
			\includegraphics{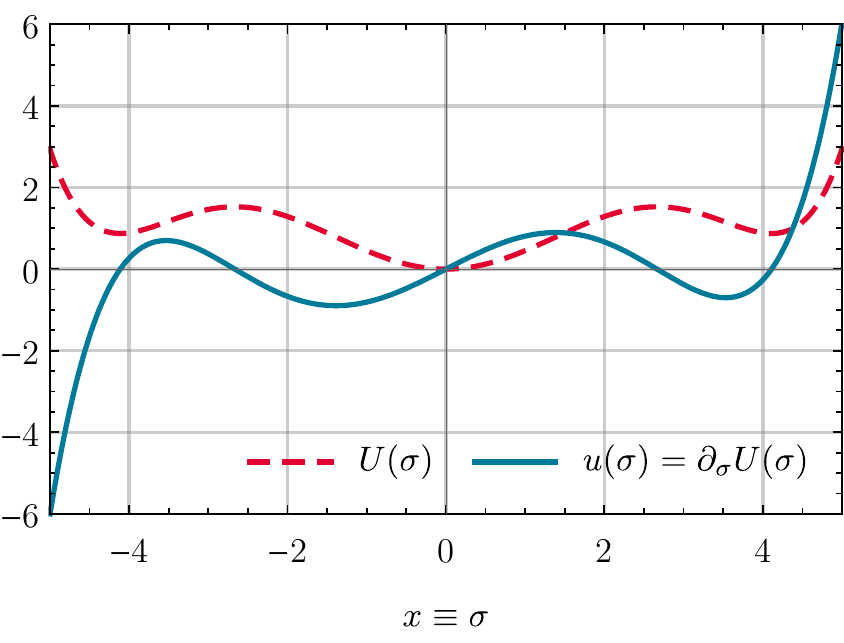}
			\caption{\label{fig:sc_iii_uv_initial_condition}%
				The UV potential $U ( \sigma )$ ({red-dashed}) and its first derivative $u ( \sigma ) = \partial_\sigma U ( \sigma )$ ({blue-solid}) of our test case Eq.~\eqref{eq:testing_scenario_phi6} evaluated on the constant background field configuration.%
			}
		\end{figure}

	For the third test case we consider the potential
		\begin{align}
			U ( \vec{\varphi} \, ) = \tfrac{1}{2} \, \vec{\varphi}^{\, 2} - \tfrac{1}{20} \, ( \vec{\varphi}^{\, 2} )^2 + \tfrac{1}{6!} \, ( \vec{\varphi}^{\, 2} )^3\, .	\label{eq:testing_scenario_phi6}
		\end{align}
	This potential includes terms up to $( \vec{\varphi}^{\, 2} )^3$ and has two local minima and one local maximum and is therefore not convex.
	The global minimum is located at $\vec{\varphi} = 0$ and the potential and its derivative (evaluated on the constant field configuration $\sigma$) are depicted in Fig.~\ref{fig:sc_iii_uv_initial_condition}.
	For the sake of completeness we included reference values for the first three non-vanishing $n$-point vertex functions in Table~\ref{tab:sc_3_n_point_functions_exact} for the $O(4)$ model.
	
	We have again performed the full set of numerical tests of Sub.Sec.~\ref{subsec:sc1} and found results supporting the general statements made in that subsection.
	For the sake of brevity we will not repeat the complete discussion of Sub.Sec.~\ref{subsec:sc1} but instead focus again on selected results.\\
		
		\begin{table}[b]
			\caption{\label{tab:sc_3_n_point_functions_exact}%
				The exact results for $\Gamma^{(2n)}$ for the $O(4)$ model with the initial UV potential \eqref{eq:testing_scenario_phi6}, obtained by a high-precision one-dimensional numerical integration of the expectation values $\langle ( \vec{\phi}^{\, 2} )^n \rangle$ using \texttt{Mathematica}'s numerical integration routine \textit{NIntegrate} \cite{Mathematica:12.1} with a \textit{PrecisionGoal} and \textit{AccuracyGoal} of $10$.
				Here, we present the first ten digits only.
			}
			\begin{ruledtabular}
				\begin{tabular}{l c c c}
					$N$		&	$\Gamma^{(2)}$	&	$\Gamma^{(4)}$	&	$\Gamma^{(6)}$	\\
					\colrule
					$4$		&	$0.2503331837$		&	$0.0481313248$		&	$0.0432822719$
				\end{tabular}
			\end{ruledtabular}
		\end{table}
		
\begin{figure}
			\centering
			\includegraphics{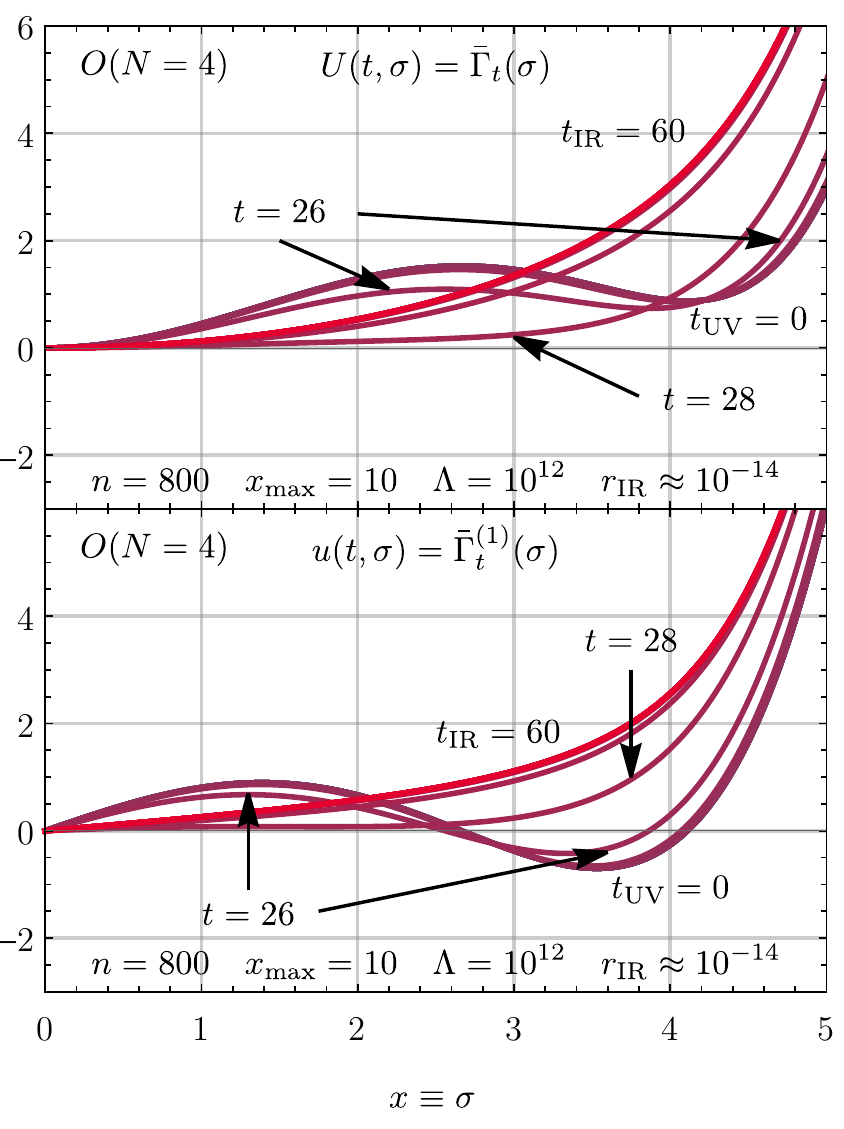}
			\caption{\label{fig:sc_iii_on_4_n_800_xmax_10_lambda_1e12_tir_60_rg_flow}%
				The FRG flow of the effective potential $U ( t, \sigma )$ (upper panel) and its derivative $u ( t , \sigma ) = \partial_\sigma U ( t , \sigma )$ (lower panel) for the zero-dimensional $O ( 4 )$ model with initial condition \eqref{eq:testing_scenario_phi6}, evaluated at $t = 0, \, 2, \, 4, \, \ldots, \, 60$ (integer values of $t$ were chosen for convenience and readability).
				The (overlapping) {blue} and {violet} curves correspond to the UV and the {red} curves to the IR.
				We used the exponential regulator \eqref{eq:exponential_regulator} with UV cutoff $\Lambda = 10^{12}$.
				For the sake of readability, the plot does not show the region $x = 5$ to $x = 10$, because the tiny differences between $u ( t, \sigma )$ and $u ( t_\mathrm{UV}, \sigma )$ are not visible in this region and vanish for large $x = \sigma$ anyhow.%
			}
		\end{figure}

\begin{figure}
			\centering
			\includegraphics{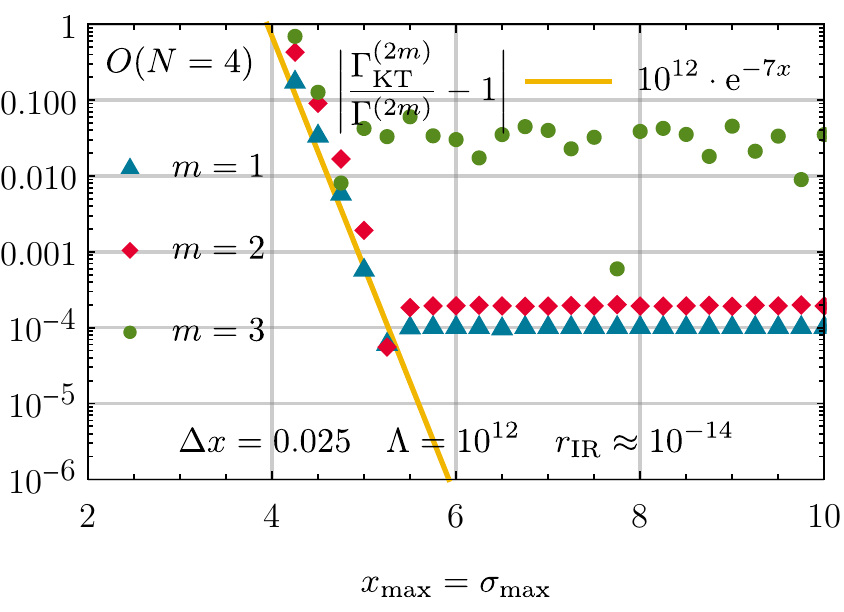}
			\caption{\label{fig:sc_iii_on_4_deltax_25e-3_lambda_1e12_tir_60_errors_xmax}%
				The relative error for $\Gamma^{(2m)}$ for $m = 1, 2, 3$, for the $O ( 4 )$ model using the UV potential \eqref{eq:testing_scenario_phi6}, as a function of the size of the computational interval $x_\mathrm{max}$. The cell size is $\Delta x = 0.025$. $\Gamma^{(2m)}$ are computed from the discrete values of the derivative of the IR potential $u ( t_\mathrm{IR} = 60, \sigma )$ via the second-order accurate central finite-difference stencils \eqref{eq:derivative_1_central_error_2}, \eqref{eq:derivative_3_central_error_2}, and \eqref{eq:derivative_5_central_error_2} at $\sigma = 0$.
				We used the exponential regulator \eqref{eq:exponential_regulator} with UV cutoff $\Lambda = 10^{12}$.%
			}
		\end{figure}

	Figure~\ref{fig:sc_iii_on_4_n_800_xmax_10_lambda_1e12_tir_60_rg_flow} shows the RG flow with the initial condition \eqref{eq:testing_scenario_phi6} for the $O(4)$ model computed with the KT scheme again using \texttt{Mathematica}'s \textit{NDSolve} \cite{Mathematica:12.1} with \textit{PrecisionGoal} and \textit{AccuracyGoal} of 10 for the RG time evolution.
	Both non-trivial local extrema fade away during RG time evolution towards the IR.
	At $t \approx 28$ the potential $U ( t, \sigma )$ becomes convex as $u ( t, \sigma )$ turns strictly positive for $\sigma > 0$.
	We again observe that the linear extrapolation used at the right boundary $x_\mathrm{max}$ of the computational domain seems surprisingly efficient even for an initial condition with quintic asymptotics.
	Studying Fig.~\ref{fig:sc_iii_on_4_deltax_25e-3_lambda_1e12_tir_60_errors_xmax} we observe that the relative errors in the IR become independent of the size of the computational domain for $x_\mathrm{max}\gtrsim 6$.
		
	We were not able to evolve the ODE system of the vertex expansion with the current initial condition to the IR for any setup at all\footnote{%
		We thank J.~Eser for discussions on this issue and a cross check using his FRG Taylor expansion code \cite{Divotgey:2019xea,Cichutek:2020bli,Eser:2018jqo,Eser:2019pvd}, which reproduced our findings.
	}. 
	Independent of $n_\mathrm{trunc}$ and ODE integrator (\texttt{Mathematica}'s \textit{NDSolve} \cite{Mathematica:12.1}) settings we encounter a numerical instability of the ODE systems at around $t \approx 28$ preventing a complete integration to the IR.
	The expansion coefficients $\bar{\Gamma}^{(2n)}(t)$ simply diverge at $t \approx 28$.
	From Fig.~\ref{fig:sc_iii_on_4_n_800_xmax_10_lambda_1e12_tir_60_rg_flow} we deduce that this is approximately the RG-time point at which the non-trivial extrema vanish and the potential turns convex.
	The precise underlying dynamics generated by the full PDE and resolved by the KT scheme cannot be captured by the vertex expansion (at least not in our set-up). 
	However, also switching to a set-up with a $t$ dependent expansion point will not cure this problem, because the expansion point (the global minimum) does not move for this initial potential.
	This again supports the claims made in Sub.Sub.Sec.~\ref{subsubsec:vertex_expansion_phi4} about the very limited practical applicability of the vertex expansion to the $O(N)$ model in zero dimensions.
		
	The instability of the solution of the coupled system of ODEs can be understood \textit{a posteriori} considering that at that time the potential may generate a non-analyticity.
	Inevitably, due to the non-analyticity of the potential, Wilbraham-Gibbs oscillations \cite{Wilbraham:1848,Gibbs:1898,Gibbs:1899} arise in the Taylor expansion, making the expansion scheme unstable \cite{boyd2001chebyshev}\footnote{%
		This phenomenon is also observed and discussed in detail in the context of Fourier expansions of periodic potentials in the FRG in Sec.~2.2.2 of Ref.~\cite{Pangon:2010uf}.
	}.
	The appearance of a non-analytic behavior is also understood via a rise of entropy \cite{Koenigstein:2021rxj,Steil:2021cbu}.
		
	However, a vertex expansion for a convex sextic potential including only positive coefficients in the UV is possible, similar to $\phi^4$ theory with a positive mass term discussed at the end of the previous Sub.Sub.Sec.~\ref{subsubsec:vertex_expansion_phi4}.
	A numerical inversion of the RG flow is again possible for systems with a small number of couplings.
	It seems that only convex, analytic UV initial conditions and the resulting rather simple FRG flows can be treated with a vertex expansion in $\bar{\Gamma}^{(2n)}(t)$ around $\vec{\varphi} = 0$ in the zero-dimensional $O(N)$ model. 

\subsection{Test case IV: the \texorpdfstring{$\sigma = 0$}{sigma = 0} boundary}
\label{subsec:sc4}

\begin{figure}
		\centering
		\includegraphics{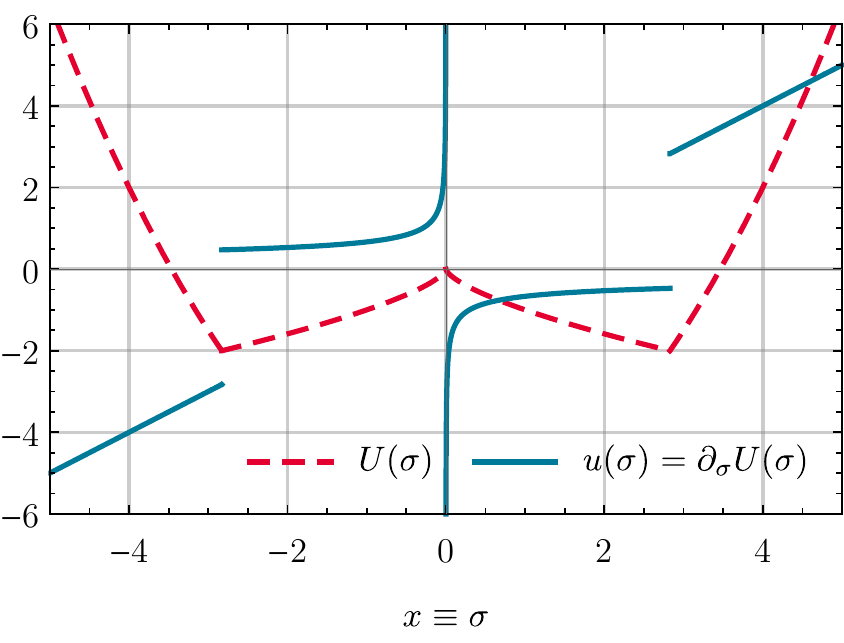}
		\caption{\label{fig:sc_iv_uv_initial_condition}%
			The UV potential $U ( \sigma )$ ({red-dashed}) and its first derivative $u ( \sigma ) = \partial_\sigma U ( \sigma )$ ({blue-solid}) of the test case \eqref{eq:testing_scenario_4} evaluated on the constant background field configuration.%
			}
		\end{figure}

	The last test case is again a non-analytic and discontinuous potential,
		\begin{align}
			U ( \vec{\varphi} \, ) =
			\begin{cases}
				- ( \vec{\varphi}^{\, 2} )^{\tfrac{1}{3}} \, ,			&	\text{if} \quad \varphi \leq \sqrt{8} \, ,	\vphantom{\bigg(\bigg)}
				\\
				\tfrac{1}{2} \, \vec{\varphi}^{\, 2} - 6 \, ,							&	\text{if} \quad \varphi > \sqrt{8} \, ,	\vphantom{\bigg(\bigg)}
			\end{cases}	\label{eq:testing_scenario_4}
		\end{align}
	where $\varphi = | \vec{\varphi} \, |$.
	The numerically challenging features are the cusp at $\varphi = 0$ as well as a non-trivial minimum at the kink at $\varphi = \sqrt{8}$.
	As displayed in Fig.~\ref{fig:sc_iv_uv_initial_condition} (evaluated on the constant field configuration), the cusp at $\sigma = 0$ in $U$ translates to a pole in $u = \partial_\sigma U$.
	This scenario was engineered as an extreme test case for the boundary condition at $\sigma = 0$ discussed at length in Sub.Sub.Sec.~\ref{subsec:phi0BC}.
	
	We have again performed the full set of numerical tests of Sub.Sec.~\ref{subsec:sc1} and found results supporting the general statements made in Sub.Sec.~\ref{subsec:sc1}.
	For the sake of brevity we will not repeat the complete discussion of Sub.Sec.~\ref{subsec:sc1} but instead focus again on selected results.

\begin{figure}
			\centering
			\includegraphics{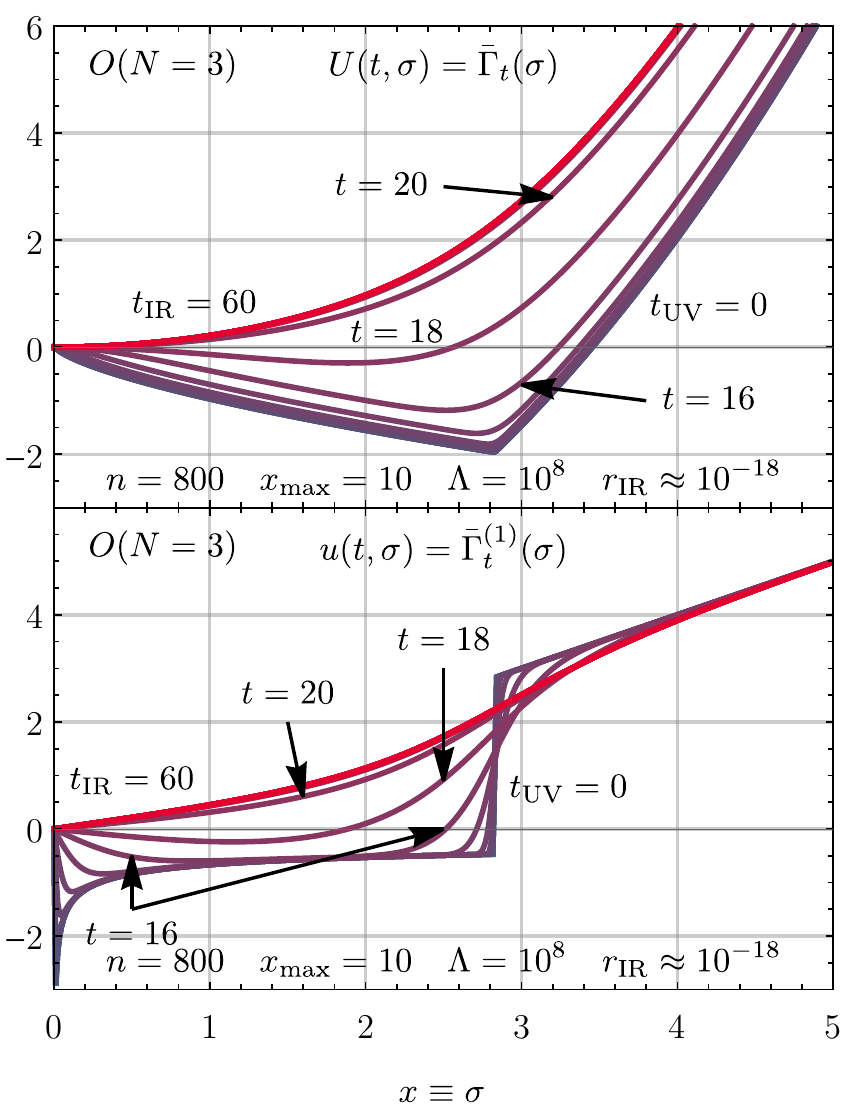}
			\caption{\label{fig:sc_iv_on_3_n_800_xmax_10_lambda_1e8_tir_60_rg_flow}%
				The FRG flow of the effective potential $U ( t, \sigma )$ (upper panel) and its derivative $u ( t , \sigma ) = \partial_\sigma U ( t , \sigma )$ (lower panel) for the zero-dimensional $O (3 )$ model with initial condition \eqref{eq:testing_scenario_4} evaluated at $t = 0, \, 2, \, 4, \, \ldots, \, 60$ (integer values for $t$ were chosen for convenience and readability).
				The {blue} curves correspond to the UV and the {red} curves to the IR. 
				We used the exponential regulator \eqref{eq:exponential_regulator} with UV cutoff $\Lambda = 10^8$.
				For the sake of readability, the plot does not show the region $x = 5$ to $x = 10$, because the tiny differences between $u ( t, \sigma )$ and $u ( t_\mathrm{UV}, \sigma )$ are not visible in this region and vanish for large $x = \sigma$ anyhow.%
			}
		\end{figure}

\begin{figure}
			\centering
			\includegraphics{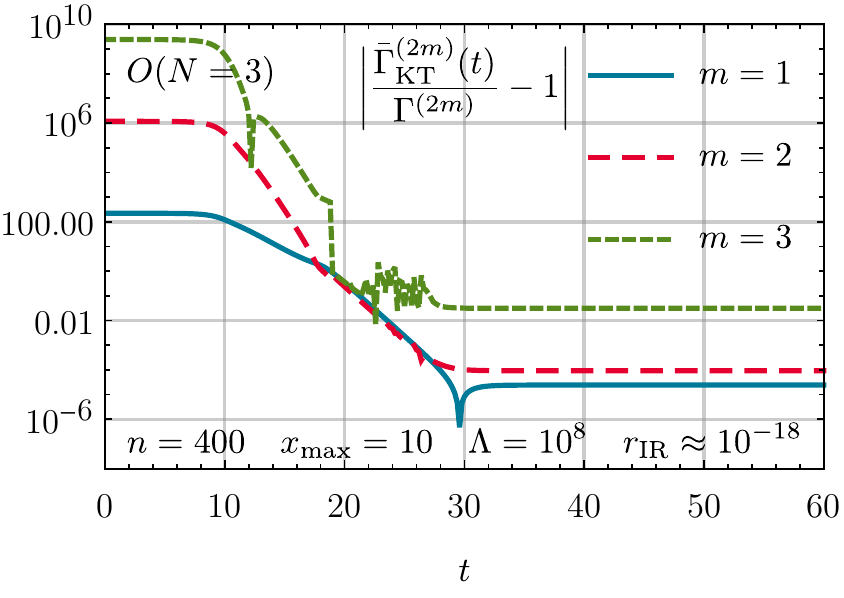}
			\caption{\label{fig:sc_iv_on_3_n_400_xmax_10_lambda_1e8_tir_60_flow_errors}%
				The relative error for $\Gamma^{(2m)}$, for $m = 1, 2$, calculated with the KT scheme as a function of the RG time $t$ for the $O(3)$ model.
				The initial UV potential is given by Eq.\ \eqref{eq:testing_scenario_4}.
				We use the exponential regulator \eqref{eq:exponential_regulator} with UV cutoff $\Lambda = 10^8$.
				The computational grid has 400 cells and $\sigma_\mathrm{max} = x_\mathrm{max} = 10$. $\Gamma^{(2m)}$ are extracted from $u ( t_\mathrm{IR} = 60, \sigma )$ via the finite-difference stencils \eqref{eq:derivative_1_central_error_2}, \eqref{eq:derivative_3_central_error_2}, and \eqref{eq:derivative_5_central_error_2}.
			}
		\end{figure}

	Figure~\ref{fig:sc_iv_on_3_n_800_xmax_10_lambda_1e8_tir_60_rg_flow} depicts the RG flow for the $O(3)$ model computed with the KT scheme for the UV initial condition \eqref{eq:testing_scenario_4}.
	Figure~\ref{fig:sc_iv_on_3_n_400_xmax_10_lambda_1e8_tir_60_flow_errors} displays the flow of the first three non-vanishing $n$-point vertex functions.
	With our implementation of the KT scheme using \texttt{Mathematica}'s \textit{NDSolve} \cite{Mathematica:12.1} with a \textit{PrecisionGoal} and \textit{AccuracyGoal} of 10 we are able to compute precise solutions, where the achievable precision for $\Gamma^{(4)}$ and $\Gamma^{(6)}$ is, as discussed in the previous sections, limited by the finite-difference rounding errors.
	The discretization-error scaling shows the same peculiarities as the case of Sub.Sec.~\ref{subsec:sc1} due to the discontinuities in the initial conditions.
	The corresponding  reference values for the $O(3)$ model are listed in Table~\ref{tab:sc_4_n_point_functions_exact}.
	The dynamics during the FRG flow is dominated by the pole at $\sigma = 0$ and the discontinuity at $\sigma = \sqrt{8}$ in $u$.
	The diffusion smears out the discontinuity and advection transports it towards $\sigma = 0$ ``filling up the well'' at $\sigma = 0$.
	Considering the corresponding values for $u$ for $\sigma < 0$ using the anti-symmetry of $u$ the boundary at $\sigma = 0$ can be seen as a point where waves of opposite amplitude annihilate.
	
	Only the carefully engineered boundary condition at $\sigma = 0$ together with corresponding ghost cells allows for practical computations with the present initial condition. The pole at $\sigma = 0$ presents no problem in practical computations because the boundary condition at $\sigma = 0$ makes use of the anti-symmetry of $u ( t, \sigma )$. 
	The first cell containing the pole is centered at $\sigma = 0$ and due to the anti-symmetry, the corresponding cell average $\bar{u}_0(t)$ vanishes by construction.
	Enforcing $\bar{u}_0(t)=0$ and for the two ghost cells $\bar{u}_{-2}(t)=-\bar{u}_{2}(t)$ and $\bar{u}_{-1}(t)=-\bar{u}_{1}(t)$ at each time step allows for a stable and accurate RG time evolution even for such extreme initial conditions like the one of Eq.~\eqref{eq:testing_scenario_4}.
	
	Treating this initial condition using a formulation in the invariant $\varrho = \tfrac{1}{2} \, \sigma^2$ with some naive boundary conditions without strict mathematical justification is hazardous, because $u ( t, \varrho ) = \partial_\varrho U ( t, \varrho )$ diverges as $\varrho^{-2/3}$ as $\varrho \rightarrow 0$.
	As mentioned in Sub.Sub.Sec.~\ref{subsec:phi0BC}, it is unclear to us how to deal with the $\varrho = 0$ boundary especially in a case like the one discussed in this subsection.
	
		\begin{table}[b]
			\caption{\label{tab:sc_4_n_point_functions_exact}%
				The exact results for $\Gamma^{(2n)}$ for the $O(3)$ model with the initial UV potential \eqref{eq:testing_scenario_4}, obtained by a high-precision one-dimensional numerical integration of the expectation values $\langle ( \vec{\phi}^{\, 2} )^n \rangle$ using \texttt{Mathematica}'s numerical integration routine \textit{NIntegrate} \cite{Mathematica:12.1}  with a \textit{PrecisionGoal} and \textit{AccuracyGoal} of $10$.
				Here, we present the first ten digits only.
			}
			\begin{ruledtabular}
				\begin{tabular}{l c c c}
					$N$		&	$\Gamma^{(2)}$	&	$\Gamma^{(4)}$	&	$\Gamma^{(6)}$	\\
					\colrule
					$3$		&	$0.4216739793$		&	$0.1535593029$		&	$0.2492523147$
				\end{tabular}
			\end{ruledtabular}
		\end{table}

\begin{figure}
			\centering
			\includegraphics{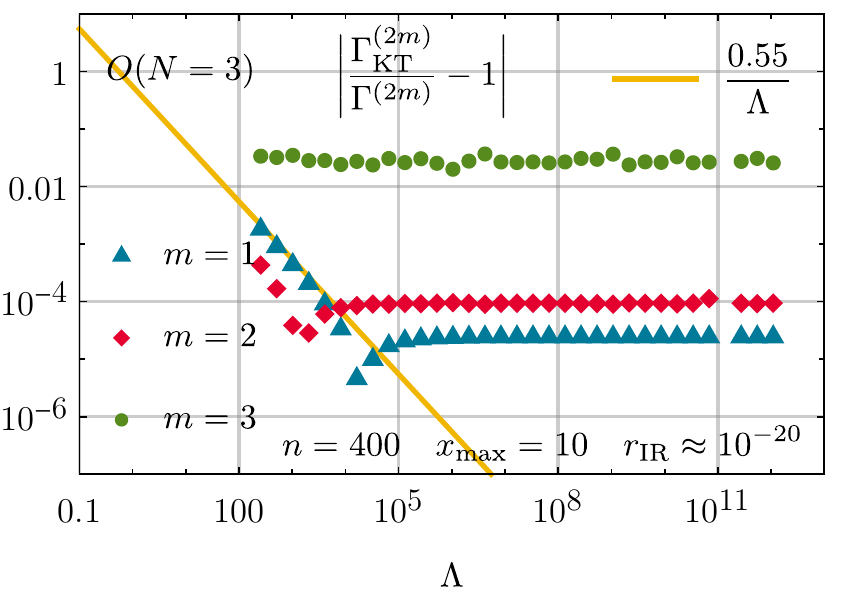}
			\caption{\label{fig:sc_iv_on_3_n_400_xmax_10_rir_10e-20_cutoff_test}%
				The relative error for $\Gamma^{(2m)}$ for $m = 1, 2, 3$ from the KT scheme as a function of the UV cutoff scale $\Lambda$, calculated for the initial potential \eqref{eq:testing_scenario_4}.
				We use the exponential regulator \eqref{eq:exponential_regulator} and keep the IR cutoff scale constant at $r(t_\mathrm{IR}) = 10^{-15}$ for all runs.
				Furthermore, for all data points the computational grid size is $\sigma_\mathrm{max} = x_\mathrm{max} = 10$ and the number of cells is $n = 400$. $\Gamma^{(2m)}$ are calculated from $u ( t_\mathrm{IR} = 60, \sigma )$ via the approximations \eqref{eq:derivative_1_central_error_2}, \eqref{eq:derivative_3_central_error_2}, and \eqref{eq:derivative_5_central_error_2} for the numerical derivative.
				The straight yellow line is for optical guidance.%
			}
		\end{figure}

	We conclude this subsection with a short discussion of RG consistency.
	The plateaus in Fig.~\ref{fig:sc_iv_on_3_n_400_xmax_10_lambda_1e8_tir_60_flow_errors} in the UV (at small $t$) and the IR (at large $t$) are again a strong indication for appropriately chosen UV and IR scales.
	From Fig.~\ref{fig:sc_iv_on_3_n_400_xmax_10_rir_10e-20_cutoff_test}, showing the initial UV-scale dependence of $\Gamma^{(2)}$, $\Gamma^{(4)}$, and $\Gamma^{(6)}$, one observes that, even in the presence of the pole at $\sigma = 0$ in ${u ( t = 0, \sigma )}$, an initial UV scale of $\Lambda=10^8$ is sufficient to realize RG consistency. 
	Arguably even $\Lambda=10^6$ -- the scale used in Sub.Sec.~\ref{subsec:sc1} -- would suffice, suggesting that in the current case the scale is primarily set by the discontinuity and linear asymptotics at and beyond $\sigma = \sqrt{8}$, which both are also present (with very similar values) in the initial condition \eqref{eq:testing_scenario_non-analytic_quadaratic_asymptote} of Sub.Sec.~\ref{subsec:sc1}.
	
	However, decreasing $\Delta x$ would lead to larger numerical gradients for the initial condition at $\sigma = 0$ due to the discretization of the pole in $u$, which in turn implies that $\Lambda$ has to be simultaneously increased in order to keep the propagators \eqref{eq:advection_flux_pion_propagator} and \eqref{eq:diffusion_flux_sigma_propagator} dominated by $\Lambda$ in the UV.
	
	Also, if the cusp at $\sigma = 0$ in the initial UV potential $U ( t = 0, \sigma )$ in Fig.~\ref{fig:sc_iv_uv_initial_condition} pointed downwards and $u ( 0, x )$ had negative gradients on both sides of the corresponding pole, it would formally be extremely hard to guarantee the inequalities \eqref{eq:LambdaMin1} and \eqref{eq:LambdaMin2} and to have a non-singular flow equation in the UV, because the giant negative gradients would not be restricted to the cell at $\sigma = 0$.
	In a discretized version with non-zero $\Delta x$ a calculation is still possible, as long as $\Lambda$ is chosen extremely large, much larger than the huge, but finite negative gradient of $u$.\footnote{%
		Similar effects are expected in FRG flows of higher-dimensional models with non-zero chemical potential at zero temperature. The chemical potential enters field space as a shock wave in field space with infinite negative slope in $u$ at positive $\sigma$. This will be discussed elsewhere \cite{Koenigstein:2021,Stoll:2021ori}.%
	}	
	Hence, RG consistency is not only a physical requirement, but also sets strict limits on the choice of numerical parameters, respectively.
		
\section{Conclusions and outlook}
\label{sec:conclusions}

	In the present work we have discussed the connection between (F)RG flow equations and conservation laws of fluid dynamics and studied implications of this connection.
	
	In Sec.~\ref{sec:frg_introduction} we have provided an introduction to the FRG in zero space-time dimensions.
	In this admittedly rather academic limit QFTs are exactly solvable in terms of ordinary one-dimensional integrals.
	However, their treatment in the framework of the FRG is not principally different from that of their higher-dimensional counterparts, which allows to understand many features of this approach in a simpler quantum field-theoretical setting, as well as check its validity against exact results.
	In particular, in $d = 0$ space-time dimensions the solution of the FRG flow equation (Wetterich equation)~\eqref{eq:wetterich_general} directly as a PDE requires no additional truncation, such as the gradient or vertex expansion.
	In Sec.~\ref{sec:zero-dimensional_o(n)-model} we have then discussed the application of the FRG method to the zero-dimensional $O(N)$ model.
	The corresponding RG flow equation \eqref{eq:flow_equation_effective_potential} is a highly non-linear PDE and thus provides an interesting and challenging application for algorithms that solve such equations.
	We are currently working on an extension of the $O(N$) model including fermions (Grassmann numbers) as well as related truncation strategies in the context of local approximations \cite{Steil:partIV}.
	
	In Sec.~\ref{sec:frg-flow_equations_and_numerical_fluid_dynamics} we have discussed the formulation of particular RG flow equations -- among them the flow equation~\eqref{eq:conservation_law_u_phi} of the O($N$) model -- in terms of conservation laws well-known in the context of fluid dynamics.
	This allows to employ existing, highly developed techniques to solve these kind of PDEs.
	The flow equations in conservative form discussed in this paper have the form of non-linear advection-diffusion equations.
	The identification of advective and diffusive contributions, related to the contributions from the pions and the radial $\sigma$ mode, respectively, allows for a simple, appealing, and physically intuitive understanding of RG flows, \textit{i.e.}, the RG time evolution from the UV to the IR, in terms of classical, time-honored fluid-dynamical concepts.
	Using the conservative formulation, it is possible to make a connection between FRG solutions in the IR to steady-state and/or thermal-equilibrium solutions.
	In this context, the diffusive character as well as the possibility of shock and rarefaction wave formation and interaction during the RG time evolution are direct manifestations of the irreversibility of RG flows and the corresponding RG transformations, which can be quantified using the concept of numerical entropy.
	We have also briefly discussed possible connections between irreversibility, numerical entropy, and the so-called $\mathcal{C}$ theorem.
	A more detailed analysis of this aspect will be presented in part II of this series of publications~\cite{Koenigstein:2021rxj}.
	
	We have discussed proper boundary conditions for the RG flow equation of the $O(N)$ model on compact intervals.
	We have then used a finite-volume method, the so-called Kurganov--Tadmor (KT) scheme, for the explicit numerical solution of RG flow equations in conservative form.
	A formulation in terms of the constant background field $\sigma$ instead of the corresponding $O(N)$ invariant $\varrho = \tfrac{1}{2} \, \sigma^2$ has proven advantageous with respect to implementing the boundary conditions for the flow equation of the $O(N)$ model in conservative form when considering a compact computational domain.
	
	In Sec.~\ref{sec:results} we have studied a set of test cases realizing various UV initial conditions for the zero-dimensional $O(N)$ model.
	We have demonstrated the applicability of the KT scheme for RG flow equations by comparing the results against exact solutions for the $n$-point vertex functions of the $O(N)$ model as obtained from a direct integration of the partition function.
	We have performed several precision tests by quantifying discretization and boundary effects.
	We have also discussed the advective and diffusive contributions to the RG flow on a qualitative level by varying the number of scalars $N$ in the $O(N)$ model.
	For large $N$, the system becomes advection-dominated~\cite{Grossi:2019urj,Grossi:2021ksl,Steil:2021cbu} since the pionic modes dominate the flow equation, whereas for small $N$ the diffusive contribution of the radial $\sigma$ mode becomes the dominant (in the case $N = 1$ even the only) driving force.
	The study of discontinuous initial conditions in this context highlights the capability of the KT scheme for the study of highly non-perturbative phenomena and phase transitions in the FRG framework.
	In all test cases, we do not observe a violation of the Coleman-Mermin-Wagner-Hohenberg theorem, \textit{i.e.}, we find that there is no spontaneous symmetry breaking in $d = 0$ in the IR limit. In a parallel publication, we continue this discussion and analyze systematic differences between the large-$N$ and infinite-$N$ within the zero-dimensional $O(N)$ model \cite{Steil:2021cbu}.
	
	Discussing the FRG Taylor (vertex) expansion as a possible truncation scheme for the Wetterich equation in the context of zero-dimensional models, we have observed that the absence of momentum suppression in $d=0$ leads to an extremely strong coupling in field space.
	In turn, this greatly limits the applicability of the FRG Taylor (vertex) expansion for the zero-dimensional models discussed in this paper.
	These findings are supported by and directly related to our novel findings regarding the irreversibility of the RG flow, if strictly analyzed as a dissipative system.
	This partially generalizes to higher-dimensional systems.
	
	We have also performed quantitative studies of the dependence on the value of the IR cutoff $r_\mathrm{IR}$ which has to be chosen in explicit (numerical) solutions of the flow equation.
	Moreover, we have discussed RG consistency, which is related to the initial UV scale $\Lambda$ for a given initial action $\mathcal{S}$.
	We find that computations in the FRG framework require sufficiently low IR cutoffs and sufficiently large initial UV scales in order to recover the exact $n$-point vertex functions.
	As demonstrated by our results, the explicit values for $k_\mathrm{IR}$ and $\Lambda$ depend on the initial action under consideration.
	
	Apart from further studies~\cite{Koenigstein:2021rxj,Steil:2021cbu,Steil:partIV} in $d = 0$ space-time dimensions, it will be very interesting to see the implications of our present work for studies of higher-dimensional QFTs.
	In the context of FRG studies of theories in $d > 0$ space-time dimensions, the current discussions and results for $d=0$ are highly relevant and in large parts directly applicable in the context of the LPA of the Wetterich equation.
	Studies of the $O(N)$ model using a conservative formulation of the LPA flow equation in $d=3$ have been performed in the large-$N$ limit by two of us in Ref.~\cite{Grossi:2019urj}.
	In a recent publication~\cite{Grossi:2021ksl} (co-authored by two of us), the possibility of conservative formulations beyond the LPA truncation has been discussed and further studies using this novel framework for selected theories in ${d = 2}$ and ${d = 4}$ are in preparation \cite{Koenigstein:2021,Stoll:2021ori}.
	We strongly believe that leveraging the vast existing knowledge for conservation laws in studies using the FRG and its (truncated) flow equations is very promising and the research in this direction should be continued and extended.

\begin{acknowledgments}
	The authors thank L.~Kurth, J.~Stoll, and N.~Zorbach for valuable comments on the first draft of the manuscript.
	The authors thank J.~Eser, F.~Murgana, J.~Stoll, and N.~Zorbach for valuable discussions and crosschecking selected numerical results.
	A.K.\ thanks J.A.~Fotakis and L.R.~Weih for enlightening discussions on boundary conditions and numerical implementations in the context of numerical fluid dynamics.
	We further thank F.~Divotgey, F.~Giacosa, F.~Ihssen, P.~Lakaschus, C.~Niehof, K.~Otto, J.M.~Pawlowski, S.~Rechenberger, B.-J.~Schaefer, A.~Sciarra, L.~von Smekal, R.-A.~Tripolt, J.~Wambach, C.~Wetterich for valuable discussions.
	
	J.B., M.B., A.K., D.H.R., and M.J.S.\ acknowledge support from the \textit{Deutsche Forschungsgemeinschaft} (DFG, German Research Foundation) through the Collaborative Research Center TransRegio CRC-TR 211 ``Strong-interaction matter under extreme conditions'' -- project number 315477589 -- TRR 211.
	
	A.K.\ acknowledges support from the \textit{Friedrich-Naumann-Foundation for Freedom}.
	
	A.K.\ and M.J.S.\ acknowledge support from the \textit{Giersch Foundation} and the \textit{Helmholtz Graduate School for Hadron and Ion Research}.
	
	J.B.\ acknowledges support from the DFG under Grant No.\ BR 4005/4-1 and BR 4005/6-1 (Heisenberg program).
	
	E.G.\ is supported by the U.S.\ Department of Energy, Office of Science, Office of Nuclear Physics, grant No.\ DE-FG-02-08ER41450.
	
	N.W.\ is supported by the \textit{Deutsche Forschungsgemeinschaft} (DFG, German Research Foundation) under Germany's Excellence Strategy EXC 2181/1 - 390900948 (the Heidelberg STRUCTURES Excellence Cluster) and under the Collaborative Research Center SFB 1225 (ISOQUANT) and by the BMBF grant 05P18VHFCA.\\
	
	All numerical numerical results as well as all figures in this work were obtained and designed using \texttt{Mathematica} \cite{Mathematica:12.1} including the following \textit{ResourceFunction}(s) from the \textit{Wolfram Function Repository}:
	\textit{PlotGrid} \cite{Lang:plotgrid}, \textit{PolygonMarker} \cite{Popkov:polygonmarker}, and \textit{MaTeXInstall} \cite{Horvat:matex}.
	The ``Feynman'' diagrams in Eqs.~\eqref{eq:flow_equation_effective_potential} and \eqref{eq:flow_equation_derivative_effective_potential} were generated via \texttt{Axodraw Version 2} \cite{Collins:2016aya}.\\
\end{acknowledgments}

\appendix

\section{Numerical derivatives}

	In Sec.~\ref{sec:results} we need to extract the 1PI vertex functions $\Gamma^{(2n)}_{\varphi_{i_1} \cdots \varphi_{i_{2n}}}$ at the physical point $\sigma = 0$ from the IR results of the RG flows (or respectively the coefficients $\Gamma^{(2n)}$, which contain the same information).
	To this end, we compute numerical derivatives of the discrete values of the derivative of the effective potential $u ( t_\mathrm{IR}, \sigma ) = \partial_\sigma U ( t_\mathrm{IR}, \sigma )$, which were calculated via the FV method.
	In this work, the following finite-difference approximations \cite{Fornberg1988,abramowitz+stegun} are used,
\begin{widetext}
		\begin{align}
			f^{(1)}_{i , \mathrm{central}} = \, & \frac{- f_{i - 1} + f_{i + 1}}{2 \, \Delta x} + \mathcal{O} ( \Delta x^2 ) \, ,	\vphantom{\bigg(\bigg)}	\label{eq:derivative_1_central_error_2}
			\\
			f^{(1)}_{i , \mathrm{central}} = \, & \frac{f_{i - 2} - 8 \, f_{i - 1} + 8 \, f_{i + 1} - f_{i + 2}}{12 \, \Delta x} + \mathcal{O} ( \Delta x^4 ) \, ,	\vphantom{\bigg(\bigg)}	\label{eq:derivative_1_central_error_4}
			\\
			f^{(1)}_{i , \mathrm{forward}}  = \, & \frac{- 3 \, f_i + 4 \, f_{i + 1} - f_{i + 2}}{2 \, \Delta x} + \mathcal{O} ( \Delta x^2 ) \, ,	\vphantom{\bigg(\bigg)}	\label{eq:derivative_1_forward_error_2}
			\\
			f^{(3)}_{i , \mathrm{central}} = \, & \frac{- f_{i - 2} + 2 \, f_{i - 1} - 2 \, f_{i + 1} + f_{i + 2}}{2 \, \Delta x^3} + \mathcal{O} ( \Delta x^2 ) \, ,	\vphantom{\bigg(\bigg)}	\label{eq:derivative_3_central_error_2}
			\\
			f^{(3)}_{i , \mathrm{central}} = \, & \frac{f_{i - 3} - 8 \, f_{i - 2} + 13 \, f_{i - 1} - 13 \, f_{i + 1} + 8 \,  f_{i + 2} - f_{i + 3}}{8 \, \Delta x^3} + \mathcal{O} ( \Delta x^4 ) \, ,	\vphantom{\bigg(\bigg)}	\label{eq:derivative_3_central_error_4}
			\\
			f^{(3)}_{i , \mathrm{forward}} = \, & \frac{- 5 \, f_{i} + 18 \, f_{i + 1} - 24 \, f_{i + 2} + 14 \, f_{i + 3} - 3 \, f_{i + 4} }{2 \, \Delta x^3} + \mathcal{O} ( \Delta x^2 ) \, ,	\vphantom{\bigg(\bigg)}	\label{eq:derivative_3_central_error_1}
			\\
			f^{(5)}_{i , \mathrm{central}} = \, & \frac{- f_{i - 3} + 4 \, f_{i - 2} - 5 \, f_{i - 1} + 5 \, f_{i + 1} - 4 \,  f_{i + 2} + f_{i + 3}}{2 \, \Delta x^5} + \mathcal{O} ( \Delta x^2 ) \, ,	\vphantom{\bigg(\bigg)}	\label{eq:derivative_5_central_error_2}
			\\
			f^{(5)}_{i , \mathrm{central}} = \, & \frac{f_{i - 4} - 9 \, f_{i - 3} + 26 \, f_{i - 2} - 29 \, f_{i - 1} + 29 \, f_{i + 1} - 26 \,  f_{i + 2} + 9 \, f_{i + 3} - f_{i + 4} }{6 \, \Delta x^5} + \mathcal{O} ( \Delta x^4 ) \, ,	\vphantom{\bigg(\bigg)}	\label{eq:derivative_5_central_error_4}
		\end{align}
\end{widetext}
	where $f_i = f ( x_i )$, $f^{(n)}$ denotes the $n^\text{th}$ derivative and ``central'' and ``forward'' stand for central/forward stencil approximations.
	The scaling order $m$ of the error is indicated by $\mathcal{O}(\Delta x^m)$.
	In our numerical implementation, the central-scheme approximations are further simplified by exploiting the anti-symmetry property $u ( t, - \sigma ) = - u ( t, \sigma )$ of the derivative of the effective potential.
	In consequence, the central stencils are effectively forward stencils.
	Furthermore, at the same order of accuracy, the ``anti-symmetrized'' central stencils need one point less than the actual forward stencils of same error order of accuracy.
	In Figs.\ \ref{fig:sc_i_on_1_n_400_xmax_10_lambda_1e6_tir_60_flow_errors}, \ref{fig:sc_i_on_3_n_400_xmax_10_lambda_1e6_tir_60_flow_errors}, and \ref{fig:sc_i_on_10_n_400_xmax_10_lambda_1e6_tir_60_flow_errors} we find that this property singles out the central stencils as the most favorable choice, because the accumulation of errors in the derivative stencil, which originally derive from the numerical solution of the flow equation, can be reduced this way, by including as few points as possible in the numerical derivative approximations.
	
	We stress that the use of low-order finite-difference approximations to the derivative is only justified because the effective IR potential $U ( t_\mathrm{IR}, \sigma )$ has to be smooth, which is discussed at length in App.~\ref{app:mermin_wagner}.
	For higher-dimensional models, the use of finite-difference approximations to extract information from the IR effective potential $U ( t_\mathrm{IR}, \sigma )$ might not always be justified due to the possibility of non-analyticities in the vicinity of the physical point, where the 1PI $n$-point vertex functions have to be calculated. Further investigation is needed.

\section{Coleman-Mermin-Wagner-Hohenberg theorem in zero dimensions:
Absence of spontaneous symmetry breaking and of phase transitions}
\label{app:mermin_wagner}

	In this appendix we comment on spontaneous symmetry breaking, phase transitions, and the Coleman-Mermin-Wagner-Hohenberg theorem for the special case of zero space-time dimensions.
	The discussion follows partially Refs.~\cite{Moroz:2011thesis,Wipf:2013vp} and is presented here for the sake of completeness.

\subsubsection{Ehrenfest classification of phase transitions}

\begin{figure}
			\centering
			\includegraphics{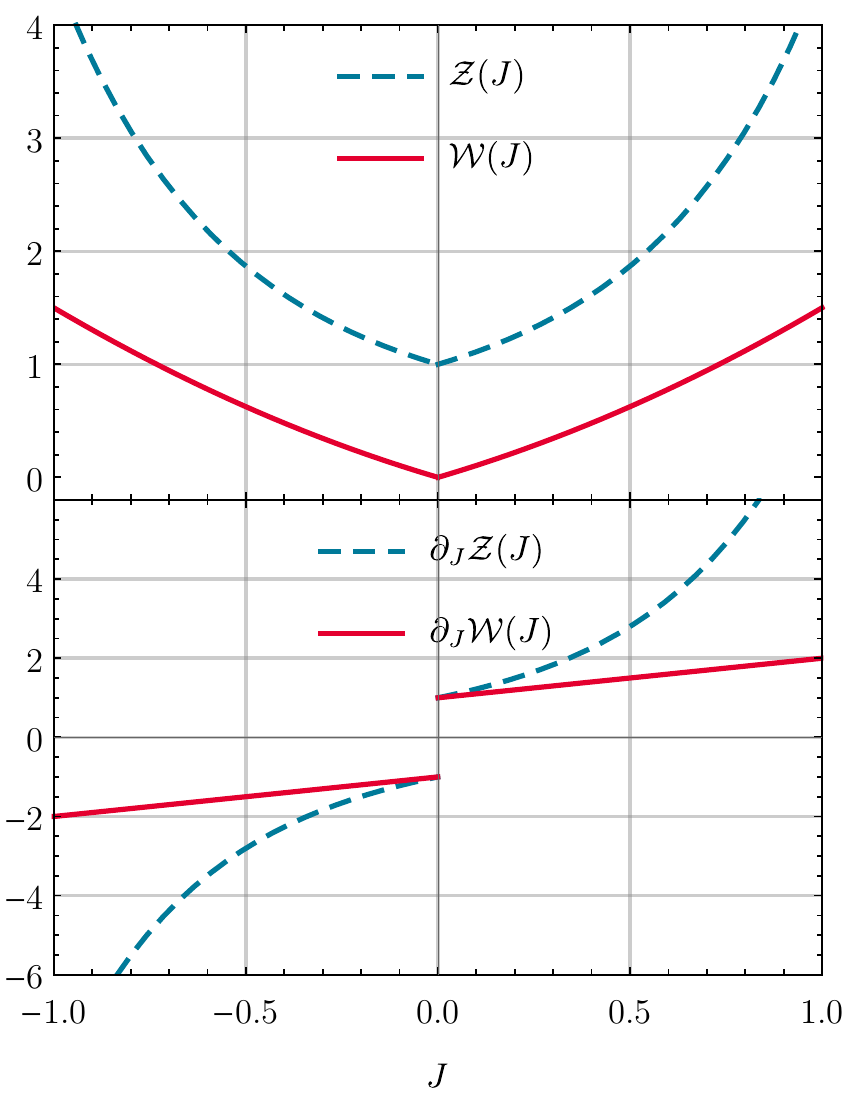}
			\caption{\label{fig:z_w_j_first_order}%
				A hypothetical realization of a first-order phase transition as a discontinuity in the first derivative of the thermodynamic potential $\mathcal{W} ( J )$.
				The functional form of $\mathcal{W}(J)$, which leads to a discontinuity in the first derivative of $\partial_J \mathcal{W}(J)$, was chosen as 
				\begin{minipage}{\linewidth}
					\begin{align}
						\mathcal{W}(J) = \sqrt{J^2} + \tfrac{1}{2} \, J^2 \, ,\vphantom{\Big)}
					\end{align}
				\end{minipage}
				which cannot correspond to a real scenario for zero-dimensional models.
			}
		\end{figure}

\begin{figure}
			\centering
			\includegraphics{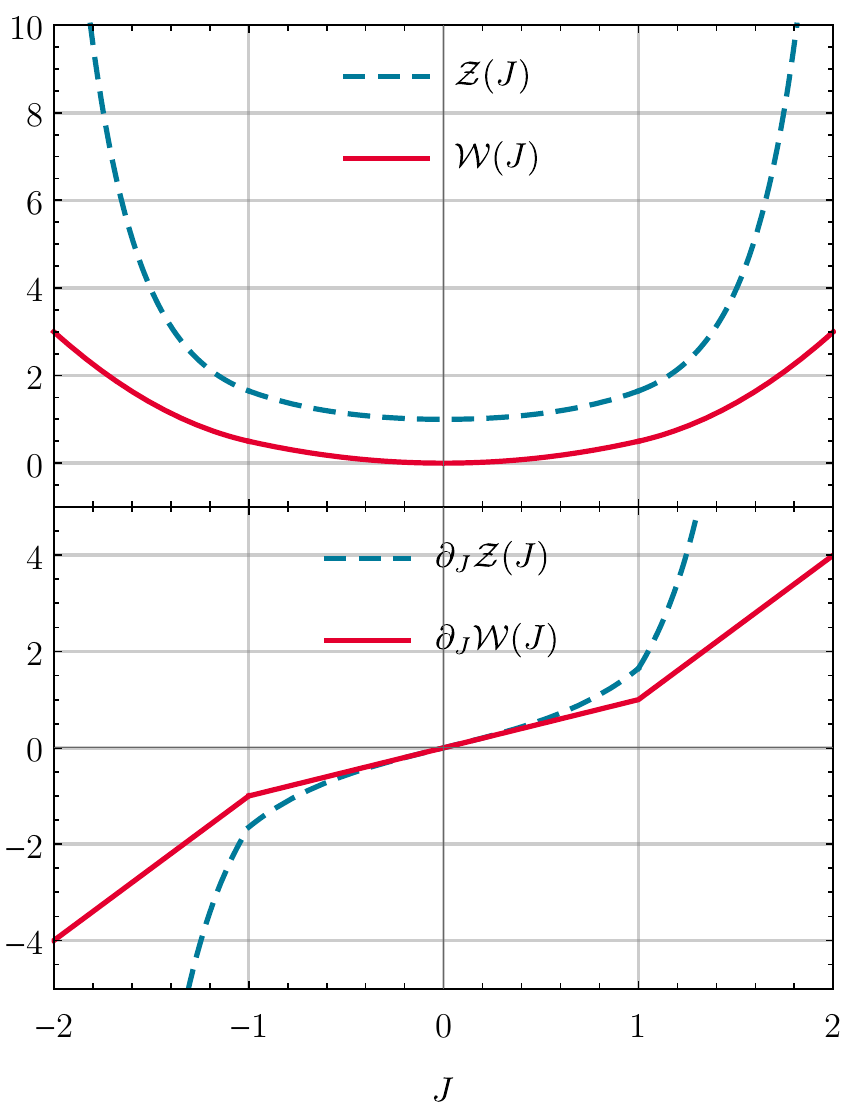}
			\caption{\label{fig:z_w_j_second_order}%
				A hypothetical realization of a second-order phase transition as a discontinuity in the second derivative of the thermodynamic potential $\mathcal{W}(J)$.
				The functional form of $\mathcal{W}(J)$, which leads to cusps in the first derivative of $\partial_J \mathcal{W} ( J )$ and hence to discontinuities in the second derivative $\partial_J^2 \mathcal{W} ( J )$, was chosen as
				\begin{minipage}{\linewidth}
					\begin{align}
						\mathcal{W} ( J ) =
						\begin{cases}
							\tfrac{3}{2} \, J^2 + 2 \, J + 1 \, ,	&	\text{if} \quad J \leq - 1 \, ,\vphantom{\Big)}
							\\
							\tfrac{1}{2} \, J^2 \, ,				&	\text{if} \quad - 1 < J < + 1 \, , \vphantom{\Big)}
							\\
							\tfrac{3}{2} \, J^2 - 2 \, J + 1 \, ,	&	\text{if} \quad + 1 \leq J \, , \vphantom{\Big)}
						\end{cases}
					\end{align}
				\end{minipage}
				which cannot correspond to a real scenario for zero-dimensional models.
			}
		\end{figure}

	Although often designated as outdated, the Ehrenfest classification of phase transitions \cite{Ehrenfest1933} is an illustrative start for our discussion.
	In thermodynamics, the Ehrenfest classification of phase transitions is based on the Helmholtz or Gibbs free energy, $F$ or $G$, which both are thermodynamic potentials.
	All phase transitions are defined as discontinuities in the derivatives of such thermodynamic potentials with respect to\ an intensive thermodynamic variable like the temperature \cite{GreinerNeiseStoecker}.
	A phase transition of order $n$ is a discontinuity in the $n^\text{th}$ derivative, while all derivatives of lower order must stay continuous.
	Meanwhile, in statistical mechanics the Helmholtz free energy $F$ is given by (or proportional to) the logarithm of the partition function or, if generalized to thermal QFT, the logarithm of the functional integral.
	If we apply this to the zero-dimensional QFT of the scalar field from our introductory Sec.~\ref{sec:frg_introduction}, we find
		\begin{align}
			F(J) = G(J) \equiv \ln \mathcal{Z}(J) = \mathcal{W}(J) \, ,	\label{eq:free_energy}
		\end{align}
	where we have used Eqs.~\eqref{eq:partition_function} and \eqref{eq:schwinger_functional}.
	Because quantities like temperature, pressure, \textit{etc.}\ do not exist in zero dimensions, the external source $J$ is the only ``thermodynamical state variable'' in the Helmholtz free energy $F(J)$, which is consequently also identical to the Gibbs free energy $G(J)$.%
	\footnote{%
		Although the notion of intensive and extensive quantities seems to be pointless in zero dimensions, the external source field $J$ can be associated with the zero-dimensional analogue of an intensive thermodynamic state variable, because it plays a similar role as an external magnetic field in higher-dimensional $O(N)$ models.
	}
	According to Ehrenfest a phase transition of $n^\text{th}$ order would therefore be associated with a discontinuity in the $n^\text{th}$ derivative of $\mathcal{W}(J)$ or equivalently in the $n^\text{th}$ derivative of $\mathcal{Z}(J)$ with respect to $J \in \mathbb{R}$.
	In order to illustrate how phase transitions are realized in $\mathcal{Z}(J)$ and $\mathcal{W}(J)$ while changing $J$, we construct two completely artificial examples for $\mathcal{W}(J)$, which correspond to a first and second-order phase transition, see Figs.\ \ref{fig:z_w_j_first_order} and \ref{fig:z_w_j_second_order} respectively.
	
	However, it can be shown (see below) that $\mathcal{Z}(J) \in C^\infty$ in zero dimensions, which also implies that the other functions in Eq.~\eqref{eq:free_energy} are infinitely often continuously differentiable, thus smooth functions of $J \in \mathbb{R}$.
	This holds as long as $\mathcal{S}(\phi)$ is bounded from below and grows faster than $| \phi |$ for $| \phi | \rightarrow \infty$, which are the conditions that were already discussed in Sec.Sub.~\ref{subsec:partition_function} in order to have well-defined expectation values \eqref{eq:expectation_value_1}.
	The proof is as follows \cite{Moroz:2011thesis}:
	
	The function $\mathcal{Z}(J)$, defined via Eq.~\eqref{eq:partition_function}, is called smooth (or $\in C^\infty$) if for all $n \in \mathbb{N}$ the left- and right-derivatives with respect to $J$ coincide at any $J \in \mathbb{R}$,
		\begin{align}
			\lim\limits_{\epsilon \rightarrow 0} \, \bigg[ \frac{\mathrm{d}^n \mathcal{Z} (J)}{\mathrm{d} J^n} \bigg|_{J + \epsilon} - \frac{\mathrm{d}^n \mathcal{Z} (J)}{\mathrm{d} J^n} \bigg|_{J - \epsilon} \bigg] = 0 \, .
		\end{align}
	Using Eq.\ \eqref{eq:partition_function} this can be checked explicitly
		\begin{align}
			& \lim\limits_{\epsilon \rightarrow 0} \, \bigg[ \frac{\mathrm{d}^n \mathcal{Z} (J)}{\mathrm{d} J^n} \bigg|_{J + \epsilon} - \frac{\mathrm{d}^n \mathcal{Z} (J)}{\mathrm{d} J^n} \bigg|_{J - \epsilon} \bigg] =	\vphantom{\bigg(\bigg)}	\label{eq:proof_smoothness_z}
			\\
			= \, & \, \lim\limits_{\epsilon \rightarrow 0} \, \mathcal{N} \int_{-\infty}^{\infty} \mathrm{d} \phi \, \phi^n \, \mathrm{e}^{ - \mathcal{S}(\phi) + J \, \phi} \, \big( \mathrm{e}^{\epsilon \, \phi} - \mathrm{e}^{ - \epsilon \, \phi} \big) =	\vphantom{\bigg(\bigg)}	\nonumber
			\\
			= \, & \, \lim\limits_{\epsilon \rightarrow 0} \, \bigg[ 2 \, \epsilon \, \mathcal{N} \int_{-\infty}^{\infty} \mathrm{d} \phi \, \phi^{n+1} \, \mathrm{e}^{ - \mathcal{S}(\phi) + J \, \phi} + \mathcal{O}(\epsilon^2) \bigg] = 0 \, .	\vphantom{\bigg(\bigg)}	\nonumber
		\end{align}
	Here we expanded the exponentials $\mathrm{e}^{\pm \epsilon \, \phi}$ for small $\epsilon$ and used that the remaining $\phi$ integral is finite due to the constraints for $\mathcal{S}(\phi)$. 
	
	The assumption that $\mathcal{S}(\phi)$ is bounded from below and grows faster than $| \phi |$ for $| \phi | \rightarrow \infty$ implying well defined and finite expectation values might in fact be strong enough to guarantee not only $\mathcal{Z}(J) \in C^\infty$ but also analyticity of $\mathcal{Z}(J)$. 
	A formal proof of the latter should be possible along the lines of the proof sketch for smoothness~\eqref{eq:proof_smoothness_z}. Proving complex differentiability, thus holomorphicity and thus ultimately analyticity of $Z(J)$ should be possible using the bounded/dominated convergence theorem which formalizes the requirements on $\mathcal{S}(\phi)$ \cite{Niehof:2021}.
	The analyticity of $\mathcal{Z}(J)$ has no direct relevance (apart from the implication of $\mathcal{Z}(J) \in C^\infty$ of course) for the discussion of phase transitions but it is relevant for the FRG Taylor (vertex) expansion discussed in Sub.Sub.Secs.~\ref{subsubsec:vertex_expansion} and \ref{subsubsec:vertex_expansion_phi4}.
	
\begin{figure}
			\centering
			\includegraphics{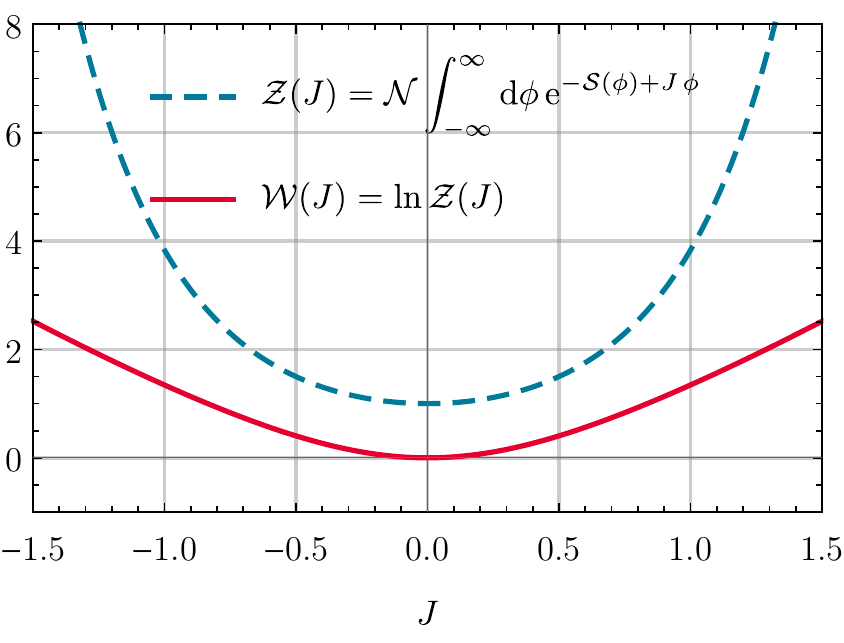}
			\caption{\label{fig:w_z_j_c_infinity}%
				The functions $\mathcal{Z}(J)$ and $\mathcal{W}(J)$ for the non-analytic action \eqref{eq:example_non-analytic_action}, obtained by numerical integration of Eq.~\eqref{eq:partition_function}.
				The plot can be extended to arbitrary $J$, where $\mathcal{Z} ( J )$ and $\mathcal{W} ( J )$ stay convex and differentiable.
				It can be seen by visual inspection that $\mathcal{Z} ( J )$ does not contain any kinks or discontinuities, which to some extent ``visualizes'' the proof \eqref{eq:proof_smoothness_z} for non-analytic actions $\mathcal{S} [ \phi ]$.
			}
		\end{figure}

	We have shown explicitly that $\mathcal{Z}(J) \in C^\infty$, which implies that there cannot be any phase transition according to the Ehrenfest classification in our zero-dimensional one-boson system, and Figs.\ \ref{fig:z_w_j_first_order} and \ref{fig:z_w_j_second_order} cannot be realized in actual zero-dimensional scenarios in the IR.
	This also generalizes to zero-dimensional QFTs with more elaborate field content, but not to higher-dimensional systems.
	Of course, $\mathcal{Z}(J) \in C^\infty$ holds for any choice of $\mathcal{S}[\phi]$ that fulfills the minimal requirements mentioned above, which also includes non-analytic actions like Eqs.~\eqref{eq:example_non-analytic_action}, \eqref{eq:testing_scenario_non-analytic_quadaratic_asymptote}, and \eqref{eq:testing_scenario_4}.
	This seems at first sight surprising.
	Because we are working in zero dimensions, we can however even visualize this counter-intuitive result of the proof and simply plot $\mathcal{Z}(J)$ and $\mathcal{W}(J)$ by numerical evaluation of the $\phi$ integral for arbitrary $J$.
	This is done in Fig.~\ref{fig:w_z_j_c_infinity}, which shows $\mathcal{Z} ( J )$ and $\mathcal{W} ( J )$ for the non-analytic action \eqref{eq:example_non-analytic_action}.
	
	However, the FRG formalism is based on the effective action $\Gamma [ \varphi ]$, which is defined via the Legendre transformation \eqref{eq:gamma_legendre} of the Schwinger functional $\mathcal{W}(J)$.
	From a thermodynamic point of view, this Legendre transformation corresponds to a change from one thermodynamic potential to another one, by switching from one intensive thermodynamic state variable to its corresponding extensive counterpart or vice versa.
	The transformation of $\mathcal{W} ( J )$ to $\Gamma ( \varphi )$ is usually associated with the transformation from the free energy $F$ to the grand canonical potential $\Omega$ in statistical  mechanics and thermal QFT \cite{Peskin:1995ev}, which is directly related to the pressure of the system.
	Here, however, we are merely interested in how the absence of phase transitions manifests itself in $\Gamma ( \varphi )$ (and $\bar{\Gamma}_t (\varphi)$ during the FRG flow) after the Legendre transformation.
	
\begin{figure}
			\centering
			\includegraphics{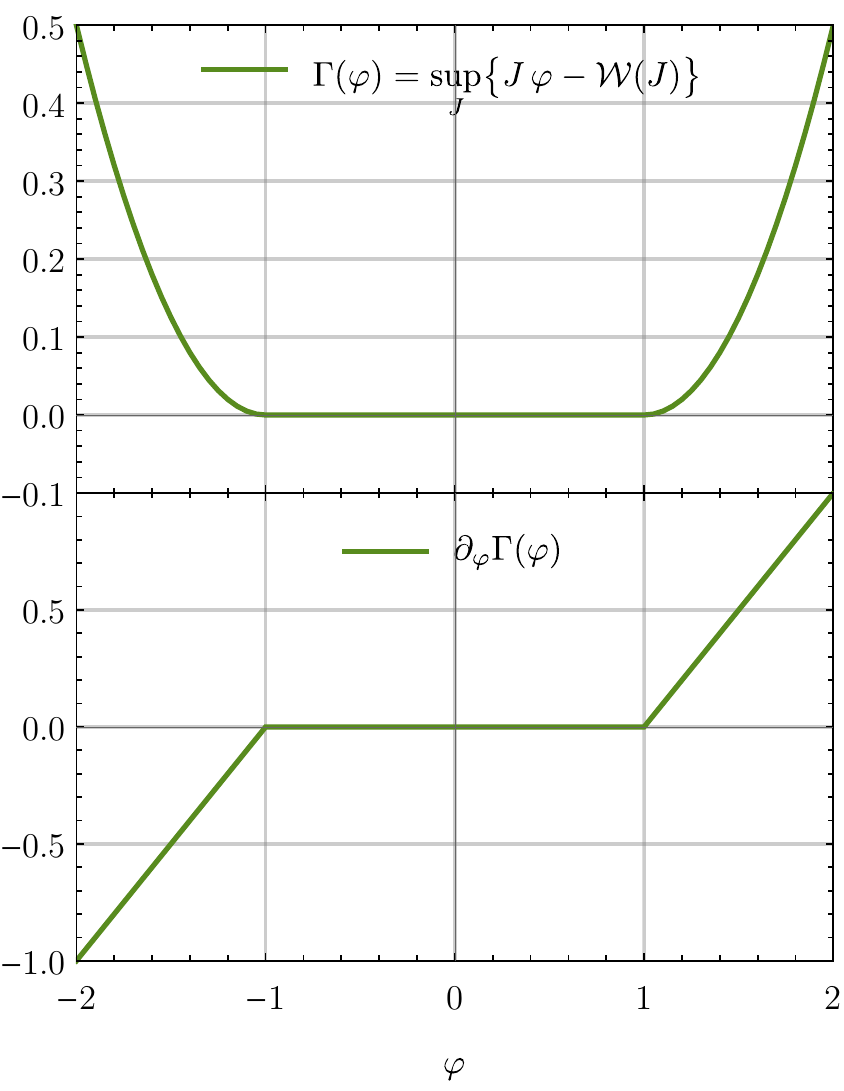}
			\caption{\label{fig:gamma_phi_first_order}%
				The figure shows how a hypothetical first-order phase transition is realized on the level of $\Gamma ( \varphi )$.
				The upper panel shows the Legendre transform of $\mathcal{W} ( J )$ from Fig.~\ref{fig:z_w_j_first_order}.
				The lower panel illustrates the first derivative of $\Gamma ( \varphi )$.
			}
		\end{figure}

\begin{figure}
			\centering
			\includegraphics{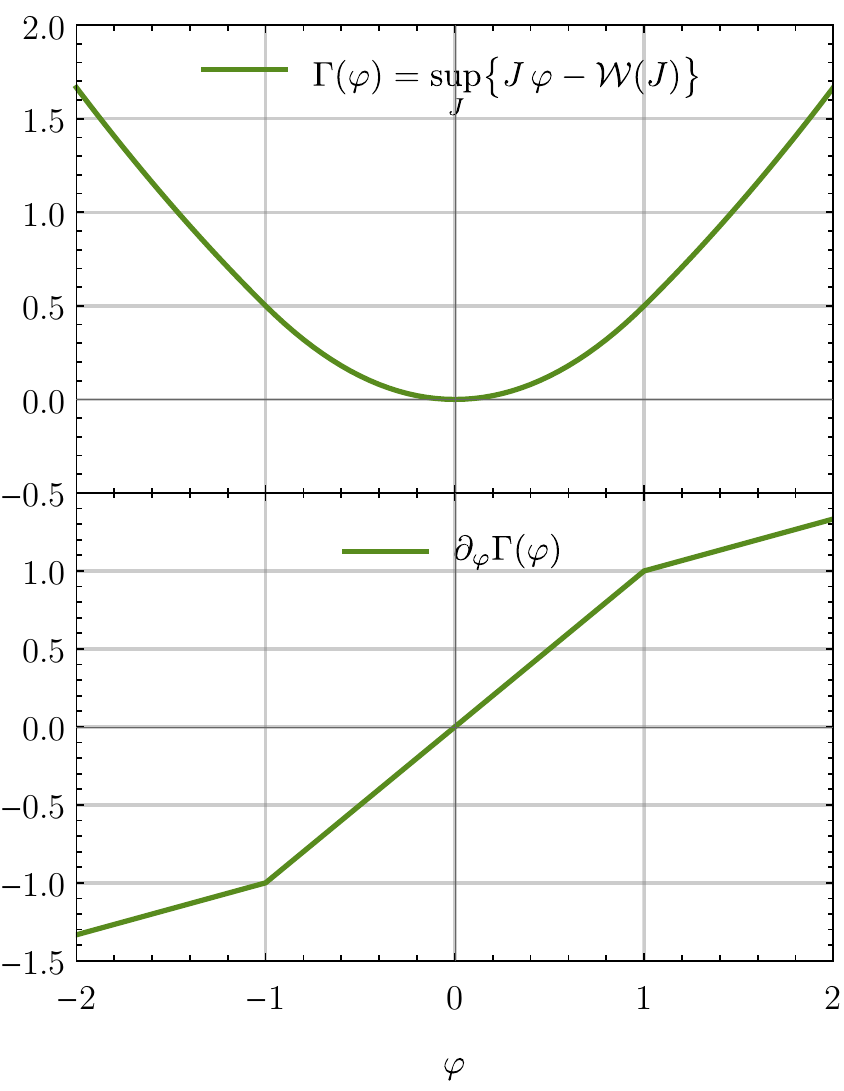}
			\caption{\label{fig:gamma_phi_second_order}%
				The figure shoes how a hypothetical second-order phase transition is realized on the level of $\Gamma ( \varphi )$.
				The upper panel shows the Legendre transform of $\mathcal{W}(J)$ from Fig.~\ref{fig:z_w_j_second_order}.
				The lower panel illustrates the first derivative of $\Gamma ( \varphi )$.
			}
		\end{figure}

	From thermodynamics and our discussion after Eq.~\eqref{eq:jacobian_w} we know that all functions in Eq.~\eqref{eq:free_energy} have to be convex.
	Furthermore, we also discussed that the convexity translates from $\mathcal{W}(J)$ to $\Gamma ( \varphi )$ via the Legendre transformation \eqref{eq:gamma_legendre}, or, on the level of the thermodynamic potentials, from $F$ to $\Omega$, respectively.
	Consequently, a first-order phase transition according to Ehrenfest, which would correspond to a cusp in the functions \eqref{eq:free_energy}, would translate to a plateau and non-analytic behavior in $\Gamma [ \varphi ]$, see Fig.~\ref{fig:gamma_phi_first_order} and Refs.~\cite{Fujimoto:1982tc,Wipf:2013vp}.
	A second-order phase transition translates to cusps in the first derivative of $\Gamma ( \varphi )$ and discontinuities in its second derivative, see Fig.~\ref{fig:gamma_phi_second_order}.
		
	However, from our above discussion we know that $\mathcal{W}(J)$ is smooth in zero dimensions and that this property also translates to $\Gamma ( \varphi )$ via the Legendre transformation \eqref{eq:gamma_legendre}.
	Thus $\Gamma ( \varphi )$ must also be smooth and convex, which again implies the absence of phase transitions according to the Ehrenfest classification of phase transitions in zero dimensions, but this  time on the level of $\Gamma ( \varphi )$ in the IR.
	Thus, the full quantum effective action $\Gamma ( \varphi )$ in the IR for real zero-dimensional systems cannot behave like the functions in Figs.\ \ref{fig:gamma_phi_first_order} and \ref{fig:gamma_phi_second_order}.
	
\begin{figure}
			\centering
			\includegraphics{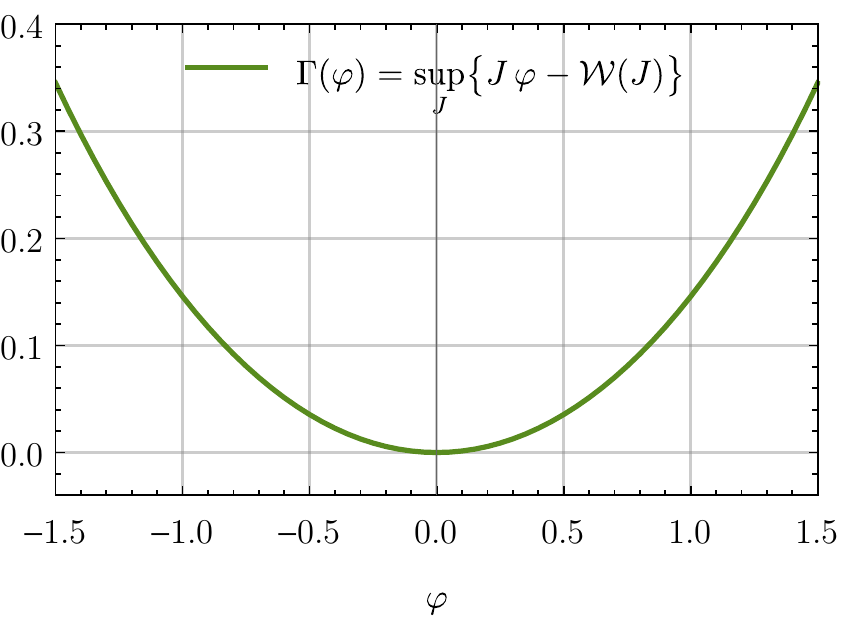}
			\caption{\label{fig:gamma_phi_c_infinity}%
				The full quantum effective action $\Gamma ( \varphi )$ in the IR corresponding to the ultraviolet classical action \eqref{eq:example_non-analytic_action}.
				The plot was obtained via a numerical Legendre transformation of $\mathcal{W} ( J )$ from from Fig.~\ref{fig:w_z_j_c_infinity}.
			}
		\end{figure}

	For the sake of completeness, in Fig.~\ref{fig:gamma_phi_c_infinity} we also provide a plot of $\Gamma ( \varphi )$ for the action \eqref{eq:example_non-analytic_action}, which was obtained via a Legendre transformation of $\mathcal{W}(J)$ from Fig.~\ref{fig:w_z_j_c_infinity}.
	Solving the Wetterich equation \eqref{eq:pde_gamma} with initial condition \eqref{eq:example_non-analytic_action} yields the same result for $\Gamma ( \varphi )$.
	For the effective average action $\bar{\Gamma}_t [ \varphi ]$, things are, however, more involved.
	We will return to this issue at the end of this appendix.

\bookmarksetupnext{level=2}
\subsubsection{Landau's theory of phase transitions}
\label{subsubsec:Landau_theory}

	Landau's theory of phase transitions \cite{Landau:1937obd} is based on the symmetries of a system.%
	\footnote{%
		Additionally, Landau theory assumes expandability of the free energy in field space.
	}
	A Landau phase transition is associated with the spontaneous breaking or restoration of a symmetry.
	This is usually associated with the (dis)appearance of an order parameter while an intensive external thermodynamic state variable is changed.
	
	In usual Landau theory in higher-dimensional systems, the order parameter is a quantity which characterizes the state of the system.
	A common example for an order parameter is a condensate.
	Usually, a non-vanishing condensate signals symmetry breaking, because it is associated with a non-trivial ground state (vacuum) of a thermodynamic potential, which, in turn, breaks the full symmetry of the system.
	
	Sometimes also the canonically conjugate extensive thermodynamic state variable of an intensive external thermodynamic state variable is used as an order parameter \cite{GreinerNeiseStoecker}.
	A typical example is the magnetization of a ferro-magnet.
	Here, however, one would not vary the magnetic field and look for spontaneous symmetry breaking, because considering non-zero magnetic field already corresponds to an explicit breaking of the symmetry.
	Taking a thermodynamic state variable as an order parameter to study spontaneous symmetry breaking implies that one should vary an unrelated intensive state variable, \textit{e.g.}, the temperature.
	
	Zero-dimensional systems are special, because the usual thermodynamic state variables like temperature, pressure, \textit{etc.}\ do not exist.
	The only external ``thermodynamic state variable'' is the source field $J$.
	Thus, in our zero-dimensional toy model from Sec.Sub.~\ref{subsec:partition_function}, the order parameter can only be associated with the mean field $\varphi ( J )$, which is the canonically conjugate of the external source $J$ and is furthermore identical to the expectation value $\langle \phi \rangle_J$ in the presence of the source,
		\begin{align}
			\varphi ( J ) = \langle \phi \rangle_J = \partial_J \mathcal{W} (J) \, .
		\end{align}
	This corresponds to the second scenario, where the order parameter is a thermodynamic variable itself.\\
	
	If we assume the classical action $\mathcal{S} [ \varphi ]$ and the integral measure of the partition function \eqref{eq:partition_function} to be symmetric under $\mathbb{Z}_2$ transformations $\phi \mapsto - \phi$, we clearly see that the presence of a fixed non-zero source $J \neq 0$ breaks the symmetry of the system and the mean field $\varphi ( J ) = \langle \phi \rangle_J$ will not vanish for finite $J$.
	
	For $J = 0$, however, the expectation value $\varphi ( J ) = \langle \phi \rangle_J$ must vanish.
	This seems to signal a phase transition according to the Landau theory of phase transitions, if $\varphi ( J )$ is considered to be the order parameter.
	On the other hand, according to the Ehrenfest classification, there are no phase transitions in zero dimensions, which sounds like a contradiction.
	The solution is the following:
	
	It is the precise definition of the order parameter and the distinction between spontaneous and explicit symmetry breaking in zero dimensions.
	Taking non-zero values for the external source field $J$ actually corresponds to explicit symmetry breaking and not spontaneous symmetry breaking within the Landau classification.
	
	On the other hand, we have just seen that ${\mathcal{Z} ( J ) \in C^\infty}$ and ${\mathcal{W} ( J ) \in C^\infty}$, which implies that the transition from $\varphi ( J ) = 0$ to $\varphi ( J ) \neq 0$ by increasing $J$ is smooth, which corresponds to a ``crossover transition'' in the Ehrenfest classification via explicit symmetry breaking.
	
	However, the main issue is that we should rather consider the condensate, which is defined as the minimum of $\Gamma ( \varphi )$, as the order parameter of Landau phase transitions to really compare with the Ehrenfest classification.
	In zero dimensions, due to the convexity and smoothness of $\Gamma ( \varphi )$, this order parameter must always vanish, because the minimum of the IR effective potential is always at $\varphi = 0$.
	In turn, the Ehrenfest and the Landau classification of phase transitions coincide and both predict the absence of phase transitions and spontaneous symmetry breaking in the zero-dimensional model under consideration.

\bookmarksetupnext{level=2}
\subsubsection{The Coleman-Mermin-Wagner-Hohenberg theorem}

	The above findings can also be interpreted as a special case of the Coleman-Mermin-Wagner-Hohenberg theorem \cite{Mermin:1966,Hohenberg:1967,Coleman:1973ci}, which states that for systems of dimension $d \leq 2$ and sufficiently short-range interactions there cannot be spontaneous breaking of continuous symmetries at non-vanishing temperature.
	Of course, zero-dimensional models do not include temperature.
	Still, all interactions are ``short-range'', because the whole system is only defined in a single point.
	
	Furthermore, we can think of a zero-dimensional system in terms of the high-temperature limit of a one-dimensional system that only includes a compact temperature direction and no spatial directions \cite{Moroz:2011thesis} -- thus a model on a circle with ``circumference'' $\beta = \frac{2\pi}{T}$.
	Sending $T \rightarrow \infty$, the circle ``shrinks to a point'', hence to zero dimensions.
	In discrete momentum space, only the bosonic Matsubara zero modes survive \cite{Braaten:1995cm}.
	All remaining quantities can formally be rescaled with the inverse temperature $\beta \equiv 1/T$ and one ends up with a zero-dimensional bosonic system, via this \textit{high-temperature dimensional reduction}.
	In this sense the Coleman-Mermin-Wagner-Hohenberg theorem applies and is confirmed for the special case of zero dimensions by our previous discussion, see also the discussion in Ref.~\cite{Fister:2015eca}.

\bookmarksetupnext{level=2}
\subsubsection{Phase transitions during the RG flow}

	In this work, we make use of the absence of spontaneous symmetry breaking in zero dimensions to test our numerical methods.
	These methods have to deliver results which are in agreement with the discussion of this appendix to be trustworthy.
	
	Interestingly, also the scale-dependent generating functionals $\mathcal{Z} ( t, J )$, $\mathcal{W} ( t, J )$, and $\Gamma ( t, \varphi )$ are smooth functions for any $t$:
	Introducing the artificial mass term $\frac{1}{2} \, r ( t ) \, \phi^2$ in Eq.~\eqref{eq:scale_dependent_z} does not violate any of the requirements of the proof \eqref{eq:proof_smoothness_z} in this appendix.
	$\mathcal{W}_t [ J ]$ is still simply defined via the logarithm of $\mathcal{Z}_t [ J ]$, see Eq.~\eqref{eq:scale_dependent_w}, which translates the proof from $\mathcal{Z}_t [ J ]$ to $\mathcal{W}_t [ J ]$.
	Furthermore, also the scale-dependent Legendre-transformation \eqref{eq:scale_dependent_effective_action} is well defined and the relations among the $n$-point correlation functions in Sub.Sec.~\ref{subsec:n_pt_functions} hold for any $t$, such that $\Gamma_t [ \varphi ]$ is smooth.
	
	In actual FRG calculations, however, we are working with the scale-dependent effective average action $\bar{\Gamma}_t [ \varphi ]$, defined via Eq.~\eqref{eq:scale_dependent_effective_average_action}.
	In general, this should not affect the smoothness property, because $\Delta \mathcal{S}_t [ \varphi ]$ is also $C^\infty$ with respect to $\varphi$.
	Still, a consequence is that $\bar{\Gamma}_t [ \varphi ]$ does -- in contrast to $\Gamma_t [ \varphi ]$ -- not need to be convex for all $t$ and only turns convex for $t \rightarrow \infty$, when $\Gamma [ \varphi ]$, $\Gamma_t [ \varphi ]$, and $\bar{\Gamma}_t [ \varphi ]$ coincide.
	This can easily be seen from Eq.~\eqref{eq:scale_dependent_effective_average_action}, where $\Delta \mathcal{S}_t [ \varphi ]$ enters with a negative sign as a huge negative mass term for small $t$.
	Hence, on the level of the effective average action $\bar{\Gamma}_t [ \varphi ]$, we can easily find Landau-type phase transitions during the FRG flow, by initializing $\bar{\Gamma}_t [ \varphi ]$ with a classical action $\mathcal{S} [ \varphi ]$ that includes a non-trivial minimum.
	During the FRG flow, this condensate must be vaporized by the bosonic fluctuations, such that convexity is restored in the IR.
	
	Additionally, we can initialize the FRG flow with non-analytic classical actions $\mathcal{S} [ \varphi ]$ in the UV due to Eq.~\eqref{eq:initial_condition_gamma}.
	This, however, might contradict previous statements and violates the smoothness of $\bar{\Gamma}_t [ \varphi ]$ at $t = 0$.
	This subtle issue is resolved as follows.
	It is the ``approximate'' sign in Eq.~\eqref{eq:initial_condition_gamma} that tells us to initialize the FRG flow with non-analytic initial conditions in the UV, if $\mathcal{S}[ \phi ]$ is non-analytic.
	Such discontinuities in the UV can be interpreted as Ehrenfest type phase transitions when varying $J$.
	The initialization with a non-analytic initial condition is valid as long as the UV cutoff $\Lambda$ is chosen sufficiently large such that the regulator insertion still approximates a delta distribution in the UV.
	Nevertheless, it is still an approximation that violates smoothness, but it is necessary for practical calculations.
	Interestingly, the FRG equation \eqref{eq:pde_gamma} seems to immediately cure this tiny ``inconsistency'' at small $t$, due to its diffusive character which smears out the non-analyticities (at least in zero-dimensions).
	Anyhow, the FRG flow must drive $\bar{\Gamma}_t [ \varphi ]$ to be smooth and convex in the IR in zero dimensions.
	The in this context pathological infinite $N$ limit is an exception \cite{Tetradis:1995br,Litim:1995ex,DAttanasio:1997yph}, which we will discuss in detail in part III of this series of publications \cite{Steil:2021cbu}. 
	
	All in all, this provides us with perfectly suited test cases, where we can explicitly check this challenging dynamics (the restoration of symmetry and smoothness on the level of $\bar{\Gamma}_{t > 0} [ \varphi ]$) during the FRG flow numerically.

\bibliography{zero_dim_part_1}  

\begin{thebibliography}{248}%
\makeatletter
\providecommand \@ifxundefined [1]{%
 \@ifx{#1\undefined}
}%
\providecommand \@ifnum [1]{%
 \ifnum #1\expandafter \@firstoftwo
 \else \expandafter \@secondoftwo
 \fi
}%
\providecommand \@ifx [1]{%
 \ifx #1\expandafter \@firstoftwo
 \else \expandafter \@secondoftwo
 \fi
}%
\providecommand \natexlab [1]{#1}%
\providecommand \enquote  [1]{``#1''}%
\providecommand \bibnamefont  [1]{#1}%
\providecommand \bibfnamefont [1]{#1}%
\providecommand \citenamefont [1]{#1}%
\providecommand \href@noop [0]{\@secondoftwo}%
\providecommand \href [0]{\begingroup \@sanitize@url \@href}%
\providecommand \@href[1]{\@@startlink{#1}\@@href}%
\providecommand \@@href[1]{\endgroup#1\@@endlink}%
\providecommand \@sanitize@url [0]{\catcode `\\12\catcode `\$12\catcode
  `\&12\catcode `\#12\catcode `\^12\catcode `\_12\catcode `\%12\relax}%
\providecommand \@@startlink[1]{}%
\providecommand \@@endlink[0]{}%
\providecommand \url  [0]{\begingroup\@sanitize@url \@url }%
\providecommand \@url [1]{\endgroup\@href {#1}{\urlprefix }}%
\providecommand \urlprefix  [0]{URL }%
\providecommand \Eprint [0]{\href }%
\providecommand \doibase [0]{https://doi.org/}%
\providecommand \selectlanguage [0]{\@gobble}%
\providecommand \bibinfo  [0]{\@secondoftwo}%
\providecommand \bibfield  [0]{\@secondoftwo}%
\providecommand \translation [1]{[#1]}%
\providecommand \BibitemOpen [0]{}%
\providecommand \bibitemStop [0]{}%
\providecommand \bibitemNoStop [0]{.\EOS\space}%
\providecommand \EOS [0]{\spacefactor3000\relax}%
\providecommand \BibitemShut  [1]{\csname bibitem#1\endcsname}%
\let\auto@bib@innerbib\@empty
\bibitem [{\citenamefont {Philipsen}(2013)}]{Philipsen:2012nu}%
  \BibitemOpen
  \bibfield  {author} {\bibinfo {author} {\bibfnamefont {O.}~\bibnamefont
  {Philipsen}},\ }\bibfield  {title} {\bibinfo {title} {{The QCD equation of
  state from the lattice}},\ }\href
  {https://doi.org/10.1016/j.ppnp.2012.09.003} {\bibfield  {journal} {\bibinfo
  {journal} {Prog. Part. Nucl. Phys.}\ }\textbf {\bibinfo {volume} {70}},\
  \bibinfo {pages} {55} (\bibinfo {year} {2013})},\ \Eprint
  {https://arxiv.org/abs/1207.5999} {arXiv:1207.5999 [hep-lat]} \BibitemShut
  {NoStop}%
\bibitem [{\citenamefont {Ding}\ \emph {et~al.}(2015)\citenamefont {Ding},
  \citenamefont {Karsch},\ and\ \citenamefont {Mukherjee}}]{Ding:2015ona}%
  \BibitemOpen
  \bibfield  {author} {\bibinfo {author} {\bibfnamefont {H.-T.}\ \bibnamefont
  {Ding}}, \bibinfo {author} {\bibfnamefont {F.}~\bibnamefont {Karsch}},\ and\
  \bibinfo {author} {\bibfnamefont {S.}~\bibnamefont {Mukherjee}},\ }\bibfield
  {title} {\bibinfo {title} {{Thermodynamics of strong-interaction matter from
  Lattice QCD}},\ }\href {https://doi.org/10.1142/S0218301315300076} {\bibfield
   {journal} {\bibinfo  {journal} {Int. J. Mod. Phys. E}\ }\textbf {\bibinfo
  {volume} {24}},\ \bibinfo {pages} {1530007} (\bibinfo {year} {2015})},\
  \Eprint {https://arxiv.org/abs/1504.05274} {arXiv:1504.05274 [hep-lat]}
  \BibitemShut {NoStop}%
\bibitem [{\citenamefont {Guenther}\ \emph {et~al.}(2018)\citenamefont
  {Guenther}, \citenamefont {Bellwied}, \citenamefont {Borsanyi}, \citenamefont
  {Fodor}, \citenamefont {Katz}, \citenamefont {Pasztor}, \citenamefont
  {Ratti},\ and\ \citenamefont {Szabo}}]{Guenther:2017grd}%
  \BibitemOpen
  \bibfield  {author} {\bibinfo {author} {\bibfnamefont {J.~N.}\ \bibnamefont
  {Guenther}}, \bibinfo {author} {\bibfnamefont {R.}~\bibnamefont {Bellwied}},
  \bibinfo {author} {\bibfnamefont {S.}~\bibnamefont {Borsanyi}}, \bibinfo
  {author} {\bibfnamefont {Z.}~\bibnamefont {Fodor}}, \bibinfo {author}
  {\bibfnamefont {S.~D.}\ \bibnamefont {Katz}}, \bibinfo {author}
  {\bibfnamefont {A.}~\bibnamefont {Pasztor}}, \bibinfo {author} {\bibfnamefont
  {C.}~\bibnamefont {Ratti}},\ and\ \bibinfo {author} {\bibfnamefont
  {K.}~\bibnamefont {Szabo}},\ }\bibfield  {title} {\bibinfo {title} {{Recent
  lattice QCD results at non-zero baryon densities}},\ }\href
  {https://doi.org/10.22323/1.311.0032} {\bibfield  {journal} {\bibinfo
  {journal} {PoS}\ }\textbf {\bibinfo {volume} {CPOD2017}},\ \bibinfo {pages}
  {032} (\bibinfo {year} {2018})}\BibitemShut {NoStop}%
\bibitem [{\citenamefont {Attanasio}\ \emph {et~al.}(2020)\citenamefont
  {Attanasio}, \citenamefont {J\"ager},\ and\ \citenamefont
  {Ziegler}}]{Attanasio:2020spv}%
  \BibitemOpen
  \bibfield  {author} {\bibinfo {author} {\bibfnamefont {F.}~\bibnamefont
  {Attanasio}}, \bibinfo {author} {\bibfnamefont {B.}~\bibnamefont {J\"ager}},\
  and\ \bibinfo {author} {\bibfnamefont {F.~P.~G.}\ \bibnamefont {Ziegler}},\
  }\bibfield  {title} {\bibinfo {title} {{Complex Langevin simulations and the
  QCD phase diagram: Recent developments}},\ }\href
  {https://doi.org/10.1140/epja/s10050-020-00256-z} {\bibfield  {journal}
  {\bibinfo  {journal} {Eur. Phys. J. A}\ }\textbf {\bibinfo {volume} {56}},\
  \bibinfo {pages} {251} (\bibinfo {year} {2020})},\ \Eprint
  {https://arxiv.org/abs/2006.00476} {arXiv:2006.00476 [hep-lat]} \BibitemShut
  {NoStop}%
\bibitem [{\citenamefont {Scherer}(2003)}]{Scherer:2002tk}%
  \BibitemOpen
  \bibfield  {author} {\bibinfo {author} {\bibfnamefont {S.}~\bibnamefont
  {Scherer}},\ }\bibfield  {title} {\bibinfo {title} {{Introduction to Chiral
  Perturbation Theory}},\ }\href {https://doi.org/10.1007/0-306-47916-8_2}
  {\bibfield  {journal} {\bibinfo  {journal} {Adv. Nucl. Phys.}\ }\textbf
  {\bibinfo {volume} {27}},\ \bibinfo {pages} {277} (\bibinfo {year} {2003})},\
  \Eprint {https://arxiv.org/abs/hep-ph/0210398} {arXiv:hep-ph/0210398}
  \BibitemShut {NoStop}%
\bibitem [{\citenamefont {Epelbaum}\ \emph {et~al.}(2009)\citenamefont
  {Epelbaum}, \citenamefont {Hammer},\ and\ \citenamefont
  {Meissner}}]{Epelbaum:2008ga}%
  \BibitemOpen
  \bibfield  {author} {\bibinfo {author} {\bibfnamefont {E.}~\bibnamefont
  {Epelbaum}}, \bibinfo {author} {\bibfnamefont {H.-W.}\ \bibnamefont
  {Hammer}},\ and\ \bibinfo {author} {\bibfnamefont {U.-G.}\ \bibnamefont
  {Meissner}},\ }\bibfield  {title} {\bibinfo {title} {{Modern Theory of
  Nuclear Forces}},\ }\href {https://doi.org/10.1103/RevModPhys.81.1773}
  {\bibfield  {journal} {\bibinfo  {journal} {Rev. Mod. Phys.}\ }\textbf
  {\bibinfo {volume} {81}},\ \bibinfo {pages} {1773} (\bibinfo {year}
  {2009})},\ \Eprint {https://arxiv.org/abs/0811.1338} {arXiv:0811.1338
  [nucl-th]} \BibitemShut {NoStop}%
\bibitem [{\citenamefont {'t~Hooft}(1974)}]{tHooft:1973alw}%
  \BibitemOpen
  \bibfield  {author} {\bibinfo {author} {\bibfnamefont {G.}~\bibnamefont
  {'t~Hooft}},\ }\bibfield  {title} {\bibinfo {title} {{A planar diagram theory
  for strong interactions}},\ }\href
  {https://doi.org/10.1016/0550-3213(74)90154-0} {\bibfield  {journal}
  {\bibinfo  {journal} {Nucl. Phys. B}\ }\textbf {\bibinfo {volume} {72}},\
  \bibinfo {pages} {461} (\bibinfo {year} {1974})}\BibitemShut {NoStop}%
\bibitem [{\citenamefont {Witten}(1979)}]{Witten:1979kh}%
  \BibitemOpen
  \bibfield  {author} {\bibinfo {author} {\bibfnamefont {E.}~\bibnamefont
  {Witten}},\ }\bibfield  {title} {\bibinfo {title} {{Baryons in the $1/N$
  expansion}},\ }\href {https://doi.org/10.1016/0550-3213(79)90232-3}
  {\bibfield  {journal} {\bibinfo  {journal} {Nucl. Phys. B}\ }\textbf
  {\bibinfo {volume} {160}},\ \bibinfo {pages} {57} (\bibinfo {year}
  {1979})}\BibitemShut {NoStop}%
\bibitem [{\citenamefont {Moshe}\ and\ \citenamefont
  {Zinn-Justin}(2003)}]{Moshe:2003xn}%
  \BibitemOpen
  \bibfield  {author} {\bibinfo {author} {\bibfnamefont {M.}~\bibnamefont
  {Moshe}}\ and\ \bibinfo {author} {\bibfnamefont {J.}~\bibnamefont
  {Zinn-Justin}},\ }\bibfield  {title} {\bibinfo {title} {{Quantum field theory
  in the large N limit: A Review}},\ }\href
  {https://doi.org/10.1016/S0370-1573(03)00263-1} {\bibfield  {journal}
  {\bibinfo  {journal} {Phys. Rept.}\ }\textbf {\bibinfo {volume} {385}},\
  \bibinfo {pages} {69} (\bibinfo {year} {2003})},\ \Eprint
  {https://arxiv.org/abs/hep-th/0306133} {arXiv:hep-th/0306133} \BibitemShut
  {NoStop}%
\bibitem [{\citenamefont {Maldacena}(1999)}]{Maldacena:1997re}%
  \BibitemOpen
  \bibfield  {author} {\bibinfo {author} {\bibfnamefont {J.~M.}\ \bibnamefont
  {Maldacena}},\ }\bibfield  {title} {\bibinfo {title} {{The large $N$ limit of
  superconformal field theories and supergravity}},\ }\href
  {https://doi.org/10.1023/A:1026654312961} {\bibfield  {journal} {\bibinfo
  {journal} {Int. J. Theor. Phys.}\ }\textbf {\bibinfo {volume} {38}},\
  \bibinfo {pages} {1113} (\bibinfo {year} {1999})},\ \Eprint
  {https://arxiv.org/abs/hep-th/9711200} {arXiv:hep-th/9711200} \BibitemShut
  {NoStop}%
\bibitem [{\citenamefont {Witten}(1998)}]{Witten:1998qj}%
  \BibitemOpen
  \bibfield  {author} {\bibinfo {author} {\bibfnamefont {E.}~\bibnamefont
  {Witten}},\ }\bibfield  {title} {\bibinfo {title} {{Anti-de Sitter space and
  holography}},\ }\href {https://doi.org/10.4310/ATMP.1998.v2.n2.a2} {\bibfield
   {journal} {\bibinfo  {journal} {Adv. Theor. Math. Phys.}\ }\textbf {\bibinfo
  {volume} {2}},\ \bibinfo {pages} {253} (\bibinfo {year} {1998})},\ \Eprint
  {https://arxiv.org/abs/hep-th/9802150} {arXiv:hep-th/9802150} \BibitemShut
  {NoStop}%
\bibitem [{\citenamefont {Fischer}(2019)}]{Fischer:2018sdj}%
  \BibitemOpen
  \bibfield  {author} {\bibinfo {author} {\bibfnamefont {C.~S.}\ \bibnamefont
  {Fischer}},\ }\bibfield  {title} {\bibinfo {title} {{QCD at finite
  temperature and chemical potential from Dyson\textendash{}Schwinger
  equations}},\ }\href {https://doi.org/10.1016/j.ppnp.2019.01.002} {\bibfield
  {journal} {\bibinfo  {journal} {Prog. Part. Nucl. Phys.}\ }\textbf {\bibinfo
  {volume} {105}},\ \bibinfo {pages} {1} (\bibinfo {year} {2019})},\ \Eprint
  {https://arxiv.org/abs/1810.12938} {arXiv:1810.12938 [hep-ph]} \BibitemShut
  {NoStop}%
\bibitem [{\citenamefont {Dupuis}\ \emph {et~al.}(2021)\citenamefont {Dupuis},
  \citenamefont {Canet}, \citenamefont {Eichhorn}, \citenamefont {Metzner},
  \citenamefont {Pawlowski}, \citenamefont {Tissier},\ and\ \citenamefont
  {Wschebor}}]{Dupuis:2020fhh}%
  \BibitemOpen
  \bibfield  {author} {\bibinfo {author} {\bibfnamefont {N.}~\bibnamefont
  {Dupuis}}, \bibinfo {author} {\bibfnamefont {L.}~\bibnamefont {Canet}},
  \bibinfo {author} {\bibfnamefont {A.}~\bibnamefont {Eichhorn}}, \bibinfo
  {author} {\bibfnamefont {W.}~\bibnamefont {Metzner}}, \bibinfo {author}
  {\bibfnamefont {J.~M.}\ \bibnamefont {Pawlowski}}, \bibinfo {author}
  {\bibfnamefont {M.}~\bibnamefont {Tissier}},\ and\ \bibinfo {author}
  {\bibfnamefont {N.}~\bibnamefont {Wschebor}},\ }\bibfield  {title} {\bibinfo
  {title} {{The nonperturbative functional renormalization group and its
  applications}},\ }\href {https://doi.org/10.1016/j.physrep.2021.01.001}
  {\bibfield  {journal} {\bibinfo  {journal} {Phys. Rept.}\ }\textbf {\bibinfo
  {volume} {910}},\ \bibinfo {pages} {1} (\bibinfo {year} {2021})},\ \Eprint
  {https://arxiv.org/abs/2006.04853} {arXiv:2006.04853 [cond-mat.stat-mech]}
  \BibitemShut {NoStop}%
\bibitem [{\citenamefont {Cichutek}\ \emph {et~al.}(2020)\citenamefont
  {Cichutek}, \citenamefont {Divotgey},\ and\ \citenamefont
  {Eser}}]{Cichutek:2020bli}%
  \BibitemOpen
  \bibfield  {author} {\bibinfo {author} {\bibfnamefont {N.}~\bibnamefont
  {Cichutek}}, \bibinfo {author} {\bibfnamefont {F.}~\bibnamefont {Divotgey}},\
  and\ \bibinfo {author} {\bibfnamefont {J.}~\bibnamefont {Eser}},\ }\bibfield
  {title} {\bibinfo {title} {{Fluctuation-induced higher-derivative couplings
  and infrared dynamics of the quark-meson-diquark model}},\ }\href
  {https://doi.org/10.1103/PhysRevD.102.034030} {\bibfield  {journal} {\bibinfo
   {journal} {Phys. Rev. D}\ }\textbf {\bibinfo {volume} {102}},\ \bibinfo
  {pages} {034030} (\bibinfo {year} {2020})},\ \Eprint
  {https://arxiv.org/abs/2006.12473} {arXiv:2006.12473 [hep-ph]} \BibitemShut
  {NoStop}%
\bibitem [{\citenamefont {Papp}\ \emph {et~al.}(2000)\citenamefont {Papp},
  \citenamefont {Schaefer}, \citenamefont {Pirner},\ and\ \citenamefont
  {Wambach}}]{Papp:1999he}%
  \BibitemOpen
  \bibfield  {author} {\bibinfo {author} {\bibfnamefont {G.}~\bibnamefont
  {Papp}}, \bibinfo {author} {\bibfnamefont {B.-J.}\ \bibnamefont {Schaefer}},
  \bibinfo {author} {\bibfnamefont {H.~J.}\ \bibnamefont {Pirner}},\ and\
  \bibinfo {author} {\bibfnamefont {J.}~\bibnamefont {Wambach}},\ }\bibfield
  {title} {\bibinfo {title} {{On the convergence of the expansion of
  renormalization group flow equation}},\ }\href
  {https://doi.org/10.1103/PhysRevD.61.096002} {\bibfield  {journal} {\bibinfo
  {journal} {Phys. Rev. D}\ }\textbf {\bibinfo {volume} {61}},\ \bibinfo
  {pages} {096002} (\bibinfo {year} {2000})},\ \Eprint
  {https://arxiv.org/abs/hep-ph/9909246} {arXiv:hep-ph/9909246} \BibitemShut
  {NoStop}%
\bibitem [{\citenamefont {Pawlowski}\ and\ \citenamefont
  {Rennecke}(2014)}]{Pawlowski:2014zaa}%
  \BibitemOpen
  \bibfield  {author} {\bibinfo {author} {\bibfnamefont {J.~M.}\ \bibnamefont
  {Pawlowski}}\ and\ \bibinfo {author} {\bibfnamefont {F.}~\bibnamefont
  {Rennecke}},\ }\bibfield  {title} {\bibinfo {title} {{Higher order
  quark-mesonic scattering processes and the phase structure of QCD}},\ }\href
  {https://doi.org/10.1103/PhysRevD.90.076002} {\bibfield  {journal} {\bibinfo
  {journal} {Phys. Rev. D}\ }\textbf {\bibinfo {volume} {90}},\ \bibinfo
  {pages} {076002} (\bibinfo {year} {2014})},\ \Eprint
  {https://arxiv.org/abs/1403.1179} {arXiv:1403.1179 [hep-ph]} \BibitemShut
  {NoStop}%
\bibitem [{\citenamefont {Fischer}\ and\ \citenamefont
  {Pawlowski}(2007)}]{Fischer:2006vf}%
  \BibitemOpen
  \bibfield  {author} {\bibinfo {author} {\bibfnamefont {C.~S.}\ \bibnamefont
  {Fischer}}\ and\ \bibinfo {author} {\bibfnamefont {J.~M.}\ \bibnamefont
  {Pawlowski}},\ }\bibfield  {title} {\bibinfo {title} {{Uniqueness of infrared
  asymptotics in Landau gauge Yang-Mills theory}},\ }\href
  {https://doi.org/10.1103/PhysRevD.75.025012} {\bibfield  {journal} {\bibinfo
  {journal} {Phys. Rev. D}\ }\textbf {\bibinfo {volume} {75}},\ \bibinfo
  {pages} {025012} (\bibinfo {year} {2007})},\ \Eprint
  {https://arxiv.org/abs/hep-th/0609009} {arXiv:hep-th/0609009} \BibitemShut
  {NoStop}%
\bibitem [{\citenamefont {Eser}\ \emph {et~al.}(2018)\citenamefont {Eser},
  \citenamefont {Divotgey}, \citenamefont {Mitter},\ and\ \citenamefont
  {Rischke}}]{Eser:2018jqo}%
  \BibitemOpen
  \bibfield  {author} {\bibinfo {author} {\bibfnamefont {J.}~\bibnamefont
  {Eser}}, \bibinfo {author} {\bibfnamefont {F.}~\bibnamefont {Divotgey}},
  \bibinfo {author} {\bibfnamefont {M.}~\bibnamefont {Mitter}},\ and\ \bibinfo
  {author} {\bibfnamefont {D.~H.}\ \bibnamefont {Rischke}},\ }\bibfield
  {title} {\bibinfo {title} {{Low-energy limit of the $O(4)$ quark-meson model
  from the functional renormalization group approach}},\ }\href
  {https://doi.org/10.1103/PhysRevD.98.014024} {\bibfield  {journal} {\bibinfo
  {journal} {Phys. Rev. D}\ }\textbf {\bibinfo {volume} {98}},\ \bibinfo
  {pages} {014024} (\bibinfo {year} {2018})},\ \Eprint
  {https://arxiv.org/abs/1804.01787} {arXiv:1804.01787 [hep-ph]} \BibitemShut
  {NoStop}%
\bibitem [{\citenamefont {Divotgey}\ \emph {et~al.}(2019)\citenamefont
  {Divotgey}, \citenamefont {Eser},\ and\ \citenamefont
  {Mitter}}]{Divotgey:2019xea}%
  \BibitemOpen
  \bibfield  {author} {\bibinfo {author} {\bibfnamefont {F.}~\bibnamefont
  {Divotgey}}, \bibinfo {author} {\bibfnamefont {J.}~\bibnamefont {Eser}},\
  and\ \bibinfo {author} {\bibfnamefont {M.}~\bibnamefont {Mitter}},\
  }\bibfield  {title} {\bibinfo {title} {{Dynamical generation of low-energy
  couplings from quark-meson fluctuations}},\ }\href
  {https://doi.org/10.1103/PhysRevD.99.054023} {\bibfield  {journal} {\bibinfo
  {journal} {Phys. Rev. D}\ }\textbf {\bibinfo {volume} {99}},\ \bibinfo
  {pages} {054023} (\bibinfo {year} {2019})},\ \Eprint
  {https://arxiv.org/abs/1901.02472} {arXiv:1901.02472 [hep-ph]} \BibitemShut
  {NoStop}%
\bibitem [{\citenamefont {Eser}\ \emph {et~al.}(2019)\citenamefont {Eser},
  \citenamefont {Divotgey},\ and\ \citenamefont {Mitter}}]{Eser:2019pvd}%
  \BibitemOpen
  \bibfield  {author} {\bibinfo {author} {\bibfnamefont {J.}~\bibnamefont
  {Eser}}, \bibinfo {author} {\bibfnamefont {F.}~\bibnamefont {Divotgey}},\
  and\ \bibinfo {author} {\bibfnamefont {M.}~\bibnamefont {Mitter}},\
  }\bibfield  {title} {\bibinfo {title} {{Low-energy limit of the $O(4)$
  quark-meson model}},\ }\href {https://doi.org/10.22323/1.317.0060} {\bibfield
   {journal} {\bibinfo  {journal} {PoS}\ }\textbf {\bibinfo {volume}
  {CD2018}},\ \bibinfo {pages} {060} (\bibinfo {year} {2019})},\ \Eprint
  {https://arxiv.org/abs/1902.04804} {arXiv:1902.04804 [hep-ph]} \BibitemShut
  {NoStop}%
\bibitem [{\citenamefont {Eser}\ and\ \citenamefont
  {Blaizot}(2021)}]{Eser:2021ivo}%
  \BibitemOpen
  \bibfield  {author} {\bibinfo {author} {\bibfnamefont {J.}~\bibnamefont
  {Eser}}\ and\ \bibinfo {author} {\bibfnamefont {J.-P.}\ \bibnamefont
  {Blaizot}},\ }\href@noop {} {\bibinfo {title} {{S-wave pion-pion scattering
  lengths from nucleon-meson fluctuations}}} (\bibinfo {year} {2021}),\ \Eprint
  {https://arxiv.org/abs/2112.14579} {arXiv:2112.14579 [hep-ph]} \BibitemShut
  {NoStop}%
\bibitem [{\citenamefont {Bonanno}\ and\ \citenamefont
  {Lacagnina}(2004)}]{Bonanno:2004pq}%
  \BibitemOpen
  \bibfield  {author} {\bibinfo {author} {\bibfnamefont {A.~M.}\ \bibnamefont
  {Bonanno}}\ and\ \bibinfo {author} {\bibfnamefont {G.}~\bibnamefont
  {Lacagnina}},\ }\bibfield  {title} {\bibinfo {title} {{Spontaneous symmetry
  breaking and proper time flow equations}},\ }\href
  {https://doi.org/10.1016/j.nuclphysb.2004.06.003} {\bibfield  {journal}
  {\bibinfo  {journal} {Nucl. Phys. B}\ }\textbf {\bibinfo {volume} {693}},\
  \bibinfo {pages} {36} (\bibinfo {year} {2004})},\ \Eprint
  {https://arxiv.org/abs/hep-th/0403176} {arXiv:hep-th/0403176} \BibitemShut
  {NoStop}%
\bibitem [{\citenamefont {Mark\'o}\ and\ \citenamefont
  {Sz\'ep}(2019)}]{Marko:2019rsl}%
  \BibitemOpen
  \bibfield  {author} {\bibinfo {author} {\bibfnamefont {G.}~\bibnamefont
  {Mark\'o}}\ and\ \bibinfo {author} {\bibfnamefont {Z.}~\bibnamefont
  {Sz\'ep}},\ }\bibfield  {title} {\bibinfo {title} {{$O(4)$ $\phi^4$ model as
  an effective light meson theory: A lattice-continuum comparison}},\ }\href
  {https://doi.org/10.1103/PhysRevD.100.056017} {\bibfield  {journal} {\bibinfo
   {journal} {Phys. Rev. D}\ }\textbf {\bibinfo {volume} {100}},\ \bibinfo
  {pages} {056017} (\bibinfo {year} {2019})},\ \Eprint
  {https://arxiv.org/abs/1907.07985} {arXiv:1907.07985 [hep-ph]} \BibitemShut
  {NoStop}%
\bibitem [{\citenamefont {Cyrol}\ \emph
  {et~al.}(2018{\natexlab{a}})\citenamefont {Cyrol}, \citenamefont {Mitter},
  \citenamefont {Pawlowski},\ and\ \citenamefont {Strodthoff}}]{Cyrol:2017ewj}%
  \BibitemOpen
  \bibfield  {author} {\bibinfo {author} {\bibfnamefont {A.~K.}\ \bibnamefont
  {Cyrol}}, \bibinfo {author} {\bibfnamefont {M.}~\bibnamefont {Mitter}},
  \bibinfo {author} {\bibfnamefont {J.~M.}\ \bibnamefont {Pawlowski}},\ and\
  \bibinfo {author} {\bibfnamefont {N.}~\bibnamefont {Strodthoff}},\ }\bibfield
   {title} {\bibinfo {title} {{Nonperturbative quark, gluon, and meson
  correlators of unquenched QCD}},\ }\href
  {https://doi.org/10.1103/PhysRevD.97.054006} {\bibfield  {journal} {\bibinfo
  {journal} {Phys. Rev. D}\ }\textbf {\bibinfo {volume} {97}},\ \bibinfo
  {pages} {054006} (\bibinfo {year} {2018}{\natexlab{a}})},\ \Eprint
  {https://arxiv.org/abs/1706.06326} {arXiv:1706.06326 [hep-ph]} \BibitemShut
  {NoStop}%
\bibitem [{\citenamefont {Cyrol}\ \emph {et~al.}(2016)\citenamefont {Cyrol},
  \citenamefont {Fister}, \citenamefont {Mitter}, \citenamefont {Pawlowski},\
  and\ \citenamefont {Strodthoff}}]{Cyrol:2016tym}%
  \BibitemOpen
  \bibfield  {author} {\bibinfo {author} {\bibfnamefont {A.~K.}\ \bibnamefont
  {Cyrol}}, \bibinfo {author} {\bibfnamefont {L.}~\bibnamefont {Fister}},
  \bibinfo {author} {\bibfnamefont {M.}~\bibnamefont {Mitter}}, \bibinfo
  {author} {\bibfnamefont {J.~M.}\ \bibnamefont {Pawlowski}},\ and\ \bibinfo
  {author} {\bibfnamefont {N.}~\bibnamefont {Strodthoff}},\ }\bibfield  {title}
  {\bibinfo {title} {{Landau gauge Yang-Mills correlation functions}},\ }\href
  {https://doi.org/10.1103/PhysRevD.94.054005} {\bibfield  {journal} {\bibinfo
  {journal} {Phys. Rev. D}\ }\textbf {\bibinfo {volume} {94}},\ \bibinfo
  {pages} {054005} (\bibinfo {year} {2016})},\ \Eprint
  {https://arxiv.org/abs/1605.01856} {arXiv:1605.01856 [hep-ph]} \BibitemShut
  {NoStop}%
\bibitem [{\citenamefont {Grossi}\ and\ \citenamefont
  {Wink}(2019)}]{Grossi:2019urj}%
  \BibitemOpen
  \bibfield  {author} {\bibinfo {author} {\bibfnamefont {E.}~\bibnamefont
  {Grossi}}\ and\ \bibinfo {author} {\bibfnamefont {N.}~\bibnamefont {Wink}},\
  }\href@noop {} {\bibinfo {title} {{Resolving phase transitions with
  Discontinuous Galerkin methods}}} (\bibinfo {year} {2019}),\ \Eprint
  {https://arxiv.org/abs/1903.09503} {arXiv:1903.09503 [hep-th]} \BibitemShut
  {NoStop}%
\bibitem [{\citenamefont {Grossi}\ \emph {et~al.}(2021)\citenamefont {Grossi},
  \citenamefont {Ihssen}, \citenamefont {Pawlowski},\ and\ \citenamefont
  {Wink}}]{Grossi:2021ksl}%
  \BibitemOpen
  \bibfield  {author} {\bibinfo {author} {\bibfnamefont {E.}~\bibnamefont
  {Grossi}}, \bibinfo {author} {\bibfnamefont {F.~J.}\ \bibnamefont {Ihssen}},
  \bibinfo {author} {\bibfnamefont {J.~M.}\ \bibnamefont {Pawlowski}},\ and\
  \bibinfo {author} {\bibfnamefont {N.}~\bibnamefont {Wink}},\ }\bibfield
  {title} {\bibinfo {title} {{Shocks and quark-meson scatterings at large
  density}},\ }\href {https://doi.org/10.1103/PhysRevD.104.016028} {\bibfield
  {journal} {\bibinfo  {journal} {Phys. Rev. D}\ }\textbf {\bibinfo {volume}
  {104}},\ \bibinfo {pages} {016028} (\bibinfo {year} {2021})},\ \Eprint
  {https://arxiv.org/abs/2102.01602} {arXiv:2102.01602 [hep-ph]} \BibitemShut
  {NoStop}%
\bibitem [{\citenamefont {Tetradis}\ and\ \citenamefont
  {Litim}(1996)}]{Tetradis:1995br}%
  \BibitemOpen
  \bibfield  {author} {\bibinfo {author} {\bibfnamefont {N.}~\bibnamefont
  {Tetradis}}\ and\ \bibinfo {author} {\bibfnamefont {D.~F.}\ \bibnamefont
  {Litim}},\ }\bibfield  {title} {\bibinfo {title} {{Analytical solutions of
  exact renormalization group equations}},\ }\href
  {https://doi.org/10.1016/0550-3213(95)00642-7} {\bibfield  {journal}
  {\bibinfo  {journal} {Nucl. Phys. B}\ }\textbf {\bibinfo {volume} {464}},\
  \bibinfo {pages} {492} (\bibinfo {year} {1996})},\ \Eprint
  {https://arxiv.org/abs/hep-th/9512073} {arXiv:hep-th/9512073} \BibitemShut
  {NoStop}%
\bibitem [{\citenamefont {Litim}\ and\ \citenamefont
  {Tetradis}(1995)}]{Litim:1995ex}%
  \BibitemOpen
  \bibfield  {author} {\bibinfo {author} {\bibfnamefont {D.}~\bibnamefont
  {Litim}}\ and\ \bibinfo {author} {\bibfnamefont {N.}~\bibnamefont
  {Tetradis}},\ }\href@noop {} {\bibinfo {title} {{Approximate solutions of
  exact renormalization group equations}}} (\bibinfo {year} {1995}),\ \Eprint
  {https://arxiv.org/abs/hep-th/9501042} {arXiv:hep-th/9501042} \BibitemShut
  {NoStop}%
\bibitem [{\citenamefont {D'Attanasio}\ and\ \citenamefont
  {Morris}(1997)}]{DAttanasio:1997yph}%
  \BibitemOpen
  \bibfield  {author} {\bibinfo {author} {\bibfnamefont {M.}~\bibnamefont
  {D'Attanasio}}\ and\ \bibinfo {author} {\bibfnamefont {T.~R.}\ \bibnamefont
  {Morris}},\ }\bibfield  {title} {\bibinfo {title} {{Large N and the
  renormalization group}},\ }\href
  {https://doi.org/10.1016/S0370-2693(97)00866-6} {\bibfield  {journal}
  {\bibinfo  {journal} {Phys. Lett. B}\ }\textbf {\bibinfo {volume} {409}},\
  \bibinfo {pages} {363} (\bibinfo {year} {1997})},\ \Eprint
  {https://arxiv.org/abs/hep-th/9704094} {arXiv:hep-th/9704094} \BibitemShut
  {NoStop}%
\bibitem [{\citenamefont {Steil}\ and\ \citenamefont
  {Koenigstein}(2022)}]{Steil:2021cbu}%
  \BibitemOpen
  \bibfield  {author} {\bibinfo {author} {\bibfnamefont {M.~J.}\ \bibnamefont
  {Steil}}\ and\ \bibinfo {author} {\bibfnamefont {A.}~\bibnamefont
  {Koenigstein}},\ }\bibfield  {title} {\bibinfo {title} {{Numerical fluid
  dynamics for FRG flow equations: Zero-dimensional QFTs as numerical test
  cases. III. Shock and rarefaction waves in RG flows reveal limitations of the
  $N\ensuremath{\rightarrow}\ensuremath{\infty}$ limit in $O(N)$-type
  models}},\ }\href {https://doi.org/10.1103/PhysRevD.106.065014} {\bibfield
  {journal} {\bibinfo  {journal} {Phys. Rev. D}\ }\textbf {\bibinfo {volume}
  {106}},\ \bibinfo {pages} {065014} (\bibinfo {year} {2022})},\ \Eprint
  {https://arxiv.org/abs/2108.04037} {arXiv:2108.04037 [cond-mat.stat-mech]}
  \BibitemShut {NoStop}%
\bibitem [{\citenamefont {Stoll}\ \emph {et~al.}(2021)\citenamefont {Stoll},
  \citenamefont {Zorbach}, \citenamefont {Koenigstein}, \citenamefont {Steil},\
  and\ \citenamefont {Rechenberger}}]{Stoll:2021ori}%
  \BibitemOpen
  \bibfield  {author} {\bibinfo {author} {\bibfnamefont {J.}~\bibnamefont
  {Stoll}}, \bibinfo {author} {\bibfnamefont {N.}~\bibnamefont {Zorbach}},
  \bibinfo {author} {\bibfnamefont {A.}~\bibnamefont {Koenigstein}}, \bibinfo
  {author} {\bibfnamefont {M.~J.}\ \bibnamefont {Steil}},\ and\ \bibinfo
  {author} {\bibfnamefont {S.}~\bibnamefont {Rechenberger}},\ }\href@noop {}
  {\bibinfo {title} {{Bosonic fluctuations in the $( 1 + 1 )$-dimensional
  Gross-Neveu(-Yukawa) model at varying $\mu$ and $T$ and finite $N$}}}
  (\bibinfo {year} {2021}),\ \Eprint {https://arxiv.org/abs/2108.10616}
  {arXiv:2108.10616 [hep-ph]} \BibitemShut {NoStop}%
\bibitem [{\citenamefont {Koenigstein}\ \emph {et~al.}(2022)\citenamefont
  {Koenigstein}, \citenamefont {Steil}, \citenamefont {Wink}, \citenamefont
  {Grossi},\ and\ \citenamefont {Braun}}]{Koenigstein:2021rxj}%
  \BibitemOpen
  \bibfield  {author} {\bibinfo {author} {\bibfnamefont {A.}~\bibnamefont
  {Koenigstein}}, \bibinfo {author} {\bibfnamefont {M.~J.}\ \bibnamefont
  {Steil}}, \bibinfo {author} {\bibfnamefont {N.}~\bibnamefont {Wink}},
  \bibinfo {author} {\bibfnamefont {E.}~\bibnamefont {Grossi}},\ and\ \bibinfo
  {author} {\bibfnamefont {J.}~\bibnamefont {Braun}},\ }\bibfield  {title}
  {\bibinfo {title} {{Numerical fluid dynamics for FRG flow equations:
  Zero-dimensional QFTs as numerical test cases. II. Entropy production and
  irreversibility of RG flows}},\ }\href
  {https://doi.org/10.1103/PhysRevD.106.065013} {\bibfield  {journal} {\bibinfo
   {journal} {Phys. Rev. D}\ }\textbf {\bibinfo {volume} {106}},\ \bibinfo
  {pages} {065013} (\bibinfo {year} {2022})},\ \Eprint
  {https://arxiv.org/abs/2108.10085} {arXiv:2108.10085 [cond-mat.stat-mech]}
  \BibitemShut {NoStop}%
\bibitem [{\citenamefont {Steil}\ \emph {et~al.}()\citenamefont {Steil},
  \citenamefont {Koenigstein},\ and\ \citenamefont {Braun}}]{Steil:partIV}%
  \BibitemOpen
  \bibfield  {author} {\bibinfo {author} {\bibfnamefont {M.~J.}\ \bibnamefont
  {Steil}}, \bibinfo {author} {\bibfnamefont {A.}~\bibnamefont {Koenigstein}},\
  and\ \bibinfo {author} {\bibfnamefont {J.}~\bibnamefont {Braun}},\
  }\href@noop {} {\bibinfo {title} {{Numerical fluid dynamics for FRG flow
  equations: Zero-dimensional QFTs as numerical test cases - Part IV: A
  fermion-boson model}}},\ \bibinfo {howpublished} {in preparation}\BibitemShut
  {NoStop}%
\bibitem [{\citenamefont {Wetterich}(1993{\natexlab{a}})}]{Wetterich:1992yh}%
  \BibitemOpen
  \bibfield  {author} {\bibinfo {author} {\bibfnamefont {C.}~\bibnamefont
  {Wetterich}},\ }\bibfield  {title} {\bibinfo {title} {{Exact evolution
  equation for the effective potential}},\ }\href
  {https://doi.org/10.1016/0370-2693(93)90726-X} {\bibfield  {journal}
  {\bibinfo  {journal} {Phys. Lett. B}\ }\textbf {\bibinfo {volume} {301}},\
  \bibinfo {pages} {90} (\bibinfo {year} {1993}{\natexlab{a}})},\ \Eprint
  {https://arxiv.org/abs/1710.05815} {arXiv:1710.05815 [hep-th]} \BibitemShut
  {NoStop}%
\bibitem [{\citenamefont {Ellwanger}(1994)}]{Ellwanger:1993mw}%
  \BibitemOpen
  \bibfield  {author} {\bibinfo {author} {\bibfnamefont {U.}~\bibnamefont
  {Ellwanger}},\ }\bibfield  {title} {\bibinfo {title} {{Flow equations for $N$
  point functions and bound states}},\ }\href
  {https://doi.org/10.1007/BF01555911} {\bibfield  {journal} {\bibinfo
  {journal} {Z. Phys. C}\ }\textbf {\bibinfo {volume} {62}},\ \bibinfo {pages}
  {503} (\bibinfo {year} {1994})},\ \Eprint
  {https://arxiv.org/abs/hep-ph/9308260} {arXiv:hep-ph/9308260} \BibitemShut
  {NoStop}%
\bibitem [{\citenamefont {Morris}(1994)}]{Morris:1993qb}%
  \BibitemOpen
  \bibfield  {author} {\bibinfo {author} {\bibfnamefont {T.~R.}\ \bibnamefont
  {Morris}},\ }\bibfield  {title} {\bibinfo {title} {{The Exact Renormalization
  Group and approximate solutions}},\ }\href
  {https://doi.org/10.1142/S0217751X94000972} {\bibfield  {journal} {\bibinfo
  {journal} {Int. J. Mod. Phys. A}\ }\textbf {\bibinfo {volume} {09}},\
  \bibinfo {pages} {2411} (\bibinfo {year} {1994})},\ \Eprint
  {https://arxiv.org/abs/hep-ph/9308265} {arXiv:hep-ph/9308265} \BibitemShut
  {NoStop}%
\bibitem [{\citenamefont {Weinberg}(1996)}]{Weinberg:1996kr}%
  \BibitemOpen
  \bibfield  {author} {\bibinfo {author} {\bibfnamefont {S.}~\bibnamefont
  {Weinberg}},\ }\href {https://doi.org/10.1017/CBO9781139644174} {\emph
  {\bibinfo {title} {{The quantum theory of fields: Modern applications}}}},\
  Vol.~\bibinfo {volume} {2}\ (\bibinfo  {publisher} {Cambridge University
  Press},\ \bibinfo {address} {Cambridge, England, UK},\ \bibinfo {year}
  {1996})\BibitemShut {NoStop}%
\bibitem [{\citenamefont {Peskin}\ and\ \citenamefont
  {Schroeder}(1995)}]{Peskin:1995ev}%
  \BibitemOpen
  \bibfield  {author} {\bibinfo {author} {\bibfnamefont {M.~E.}\ \bibnamefont
  {Peskin}}\ and\ \bibinfo {author} {\bibfnamefont {D.~V.}\ \bibnamefont
  {Schroeder}},\ }\href@noop {} {\emph {\bibinfo {title} {{An introduction to
  quantum field theory}}}}\ (\bibinfo  {publisher} {Addison-Wesley},\ \bibinfo
  {address} {Reading, USA},\ \bibinfo {year} {1995})\BibitemShut {NoStop}%
\bibitem [{\citenamefont {Zinn-Justin}(2002)}]{ZinnJustin:2002ru}%
  \BibitemOpen
  \bibfield  {author} {\bibinfo {author} {\bibfnamefont {J.}~\bibnamefont
  {Zinn-Justin}},\ }\href
  {https://doi.org/10.1093/acprof:oso/9780198509233.001.0001} {\emph {\bibinfo
  {title} {{Quantum field theory and critical phenomena}}}},\ \bibinfo
  {edition} {4th}\ ed.,\ \bibinfo {series} {Int. Ser. Monogr. Phys.}, Vol.\
  \bibinfo {volume} {113}\ (\bibinfo  {publisher} {Oxford University Press},\
  \bibinfo {year} {2002})\ pp.\ \bibinfo {pages} {1--1054},\ \bibinfo {note} {a
  Clarendon Press Publication}\BibitemShut {NoStop}%
\bibitem [{\citenamefont {Kleinert}(2004)}]{Kleinert:2004ev}%
  \BibitemOpen
  \bibfield  {author} {\bibinfo {author} {\bibfnamefont {H.}~\bibnamefont
  {Kleinert}},\ }\href {https://doi.org/10.1142/7305} {\emph {\bibinfo {title}
  {{Path integrals in quantum mechanics, statistics, polymer physics, and
  financial markets}}}},\ \bibinfo {edition} {5th}\ ed.\ (\bibinfo  {publisher}
  {World Scientific},\ \bibinfo {year} {2004})\ p.\ \bibinfo {pages}
  {1624}\BibitemShut {NoStop}%
\bibitem [{\citenamefont {Press}\ \emph
  {et~al.}(1992{\natexlab{a}})\citenamefont {Press}, \citenamefont {Teukolsky},
  \citenamefont {Vetterling},\ and\ \citenamefont {Flannery}}]{Press:1992zz}%
  \BibitemOpen
  \bibfield  {author} {\bibinfo {author} {\bibfnamefont {W.~H.}\ \bibnamefont
  {Press}}, \bibinfo {author} {\bibfnamefont {S.~A.}\ \bibnamefont
  {Teukolsky}}, \bibinfo {author} {\bibfnamefont {W.~T.}\ \bibnamefont
  {Vetterling}},\ and\ \bibinfo {author} {\bibfnamefont {B.~P.}\ \bibnamefont
  {Flannery}},\ }\href@noop {} {\emph {\bibinfo {title} {{Numerical recipes in
  FORTRAN: The art of scientific computing}}}},\ \bibinfo {edition} {2nd}\ ed.\
  (\bibinfo  {publisher} {Cambridge University Press},\ \bibinfo {address}
  {Cambridge, USA},\ \bibinfo {year} {1992})\BibitemShut {NoStop}%
\bibitem [{\citenamefont {Press}\ \emph
  {et~al.}(1992{\natexlab{b}})\citenamefont {Press}, \citenamefont {Teukolsky},
  \citenamefont {Vetterling},\ and\ \citenamefont
  {Flannery}}]{PresTeukVettFlan92}%
  \BibitemOpen
  \bibfield  {author} {\bibinfo {author} {\bibfnamefont {W.~H.}\ \bibnamefont
  {Press}}, \bibinfo {author} {\bibfnamefont {S.~A.}\ \bibnamefont
  {Teukolsky}}, \bibinfo {author} {\bibfnamefont {W.~T.}\ \bibnamefont
  {Vetterling}},\ and\ \bibinfo {author} {\bibfnamefont {B.~P.}\ \bibnamefont
  {Flannery}},\ }\href@noop {} {\emph {\bibinfo {title} {{Numerical Recipes in
  C: The art of scientific computing}}}},\ \bibinfo {edition} {2nd}\ ed.\
  (\bibinfo  {publisher} {Cambridge University Press},\ \bibinfo {address}
  {Cambridge, USA},\ \bibinfo {year} {1992})\BibitemShut {NoStop}%
\bibitem [{\citenamefont {Keitel}\ and\ \citenamefont
  {Bartosch}(2012)}]{Keitel:2011pn}%
  \BibitemOpen
  \bibfield  {author} {\bibinfo {author} {\bibfnamefont {J.}~\bibnamefont
  {Keitel}}\ and\ \bibinfo {author} {\bibfnamefont {L.}~\bibnamefont
  {Bartosch}},\ }\bibfield  {title} {\bibinfo {title} {{The zero-dimensional
  $O(N)$ vector model as a benchmark for perturbation theory, the large-$N$
  expansion and the Functional Renormalization Group}},\ }\href
  {https://doi.org/10.1088/1751-8113/45/10/105401} {\bibfield  {journal}
  {\bibinfo  {journal} {J. Phys.}\ }\textbf {\bibinfo {volume} {A45}},\
  \bibinfo {pages} {105401} (\bibinfo {year} {2012})},\ \Eprint
  {https://arxiv.org/abs/1109.3013} {arXiv:1109.3013 [cond-mat.stat-mech]}
  \BibitemShut {NoStop}%
\bibitem [{\citenamefont {Bessis}\ \emph {et~al.}(1980)\citenamefont {Bessis},
  \citenamefont {Itzykson},\ and\ \citenamefont {Zuber}}]{Bessis:1980ss}%
  \BibitemOpen
  \bibfield  {author} {\bibinfo {author} {\bibfnamefont {D.}~\bibnamefont
  {Bessis}}, \bibinfo {author} {\bibfnamefont {C.}~\bibnamefont {Itzykson}},\
  and\ \bibinfo {author} {\bibfnamefont {J.-B.}\ \bibnamefont {Zuber}},\
  }\bibfield  {title} {\bibinfo {title} {{Quantum field theory techniques in
  graphical enumeration}},\ }\href
  {https://doi.org/10.1016/0196-8858(80)90008-1} {\bibfield  {journal}
  {\bibinfo  {journal} {Adv. Appl. Math.}\ }\textbf {\bibinfo {volume} {1}},\
  \bibinfo {pages} {109} (\bibinfo {year} {1980})}\BibitemShut {NoStop}%
\bibitem [{\citenamefont {Zinn-Justin}(1998)}]{Zinn-Justin:1998hwu}%
  \BibitemOpen
  \bibfield  {author} {\bibinfo {author} {\bibfnamefont {J.}~\bibnamefont
  {Zinn-Justin}},\ }\bibfield  {title} {\bibinfo {title} {{Vector models in the
  large $N$ limit: A Few applications}},\ }in\ \href@noop {} {\emph {\bibinfo
  {booktitle} {{11th Taiwan Spring School on Particles and Fields}}}}\
  (\bibinfo {year} {1998})\ \Eprint {https://arxiv.org/abs/hep-th/9810198}
  {arXiv:hep-th/9810198} \BibitemShut {NoStop}%
\bibitem [{\citenamefont {Di~Vecchia}\ \emph {et~al.}(1991)\citenamefont
  {Di~Vecchia}, \citenamefont {Kato},\ and\ \citenamefont
  {Ohta}}]{DiVecchia:1990ce}%
  \BibitemOpen
  \bibfield  {author} {\bibinfo {author} {\bibfnamefont {P.}~\bibnamefont
  {Di~Vecchia}}, \bibinfo {author} {\bibfnamefont {M.}~\bibnamefont {Kato}},\
  and\ \bibinfo {author} {\bibfnamefont {N.}~\bibnamefont {Ohta}},\ }\bibfield
  {title} {\bibinfo {title} {{Double scaling limit in $O(N)$ vector models}},\
  }\href {https://doi.org/10.1016/0550-3213(91)90478-G} {\bibfield  {journal}
  {\bibinfo  {journal} {Nucl. Phys. B}\ }\textbf {\bibinfo {volume} {357}},\
  \bibinfo {pages} {495} (\bibinfo {year} {1991})}\BibitemShut {NoStop}%
\bibitem [{\citenamefont {Hikami}\ and\ \citenamefont
  {Brezin}(1979)}]{Hikami:1978ya}%
  \BibitemOpen
  \bibfield  {author} {\bibinfo {author} {\bibfnamefont {S.}~\bibnamefont
  {Hikami}}\ and\ \bibinfo {author} {\bibfnamefont {E.}~\bibnamefont
  {Brezin}},\ }\bibfield  {title} {\bibinfo {title} {{Large order behavior of
  the $1/N$ expansion in zero and one dimensions}},\ }\href
  {https://doi.org/10.1088/0305-4470/12/6/006} {\bibfield  {journal} {\bibinfo
  {journal} {J. Phys. A}\ }\textbf {\bibinfo {volume} {12}},\ \bibinfo {pages}
  {759} (\bibinfo {year} {1979})}\BibitemShut {NoStop}%
\bibitem [{\citenamefont {Nishigaki}\ and\ \citenamefont
  {Yoneya}(1991)}]{Nishigaki:1990sk}%
  \BibitemOpen
  \bibfield  {author} {\bibinfo {author} {\bibfnamefont {S.}~\bibnamefont
  {Nishigaki}}\ and\ \bibinfo {author} {\bibfnamefont {T.}~\bibnamefont
  {Yoneya}},\ }\bibfield  {title} {\bibinfo {title} {{A nonperturbative theory
  of randomly branching chains}},\ }\href
  {https://doi.org/10.1016/0550-3213(91)90215-J} {\bibfield  {journal}
  {\bibinfo  {journal} {Nucl. Phys. B}\ }\textbf {\bibinfo {volume} {348}},\
  \bibinfo {pages} {787} (\bibinfo {year} {1991})}\BibitemShut {NoStop}%
\bibitem [{\citenamefont {Schelstraete}\ and\ \citenamefont
  {Verschelde}(1994)}]{Schelstraete:1994sc}%
  \BibitemOpen
  \bibfield  {author} {\bibinfo {author} {\bibfnamefont {S.}~\bibnamefont
  {Schelstraete}}\ and\ \bibinfo {author} {\bibfnamefont {H.}~\bibnamefont
  {Verschelde}},\ }\bibfield  {title} {\bibinfo {title} {{Large $N$ limit of
  $O(N)$ vector models}},\ }\href
  {https://doi.org/10.1016/0370-2693(94)90855-9} {\bibfield  {journal}
  {\bibinfo  {journal} {Phys. Lett. B}\ }\textbf {\bibinfo {volume} {332}},\
  \bibinfo {pages} {36} (\bibinfo {year} {1994})},\ \Eprint
  {https://arxiv.org/abs/hep-th/9405158} {arXiv:hep-th/9405158} \BibitemShut
  {NoStop}%
\bibitem [{\citenamefont {Pawlowski}(2013)}]{Pawlowski:talk}%
  \BibitemOpen
  \bibfield  {author} {\bibinfo {author} {\bibfnamefont {J.~M.}\ \bibnamefont
  {Pawlowski}},\ }\href
  {https://www.thphys.uni-heidelberg.de/~pawlowski/NPgauge12/bonus/idea.pdf}
  {\bibinfo {title} {{Solving integrals with flow equations}}},\ \bibinfo
  {howpublished} {{Slides for the lecture \textit{Non-perturbative aspects of
  gauge theories} winter term 2012/2013}} (\bibinfo {year} {2013}),\ \bibinfo
  {note} {[Online; accessed 2020.10.29]}\BibitemShut {NoStop}%
\bibitem [{\citenamefont {Moroz}(2011)}]{Moroz:2011thesis}%
  \BibitemOpen
  \bibfield  {author} {\bibinfo {author} {\bibfnamefont {S.}~\bibnamefont
  {Moroz}},\ }\emph {\bibinfo {title} {{Few-body physics with functional
  renormalization}}},\ \href {https://doi.org/10.11588/heidok.00011602}
  {\bibinfo {type} {Phd thesis}},\ \bibinfo  {school} {University of
  Heidelberg} (\bibinfo {year} {2011})\BibitemShut {NoStop}%
\bibitem [{\citenamefont {Flörchinger}(2010)}]{Fl_rchinger_2010}%
  \BibitemOpen
  \bibfield  {author} {\bibinfo {author} {\bibfnamefont {S.}~\bibnamefont
  {Flörchinger}},\ }\bibfield  {title} {\bibinfo {title} {{Functional
  Renormalization and Ultracold Quantum Gases}},\ }\bibfield  {journal}
  {\bibinfo  {journal} {Springer Theses}\ }\href
  {https://doi.org/10.1007/978-3-642-14113-3} {10.1007/978-3-642-14113-3}
  (\bibinfo {year} {2010})\BibitemShut {NoStop}%
\bibitem [{\citenamefont {Skinner}(2018)}]{SkinnerScript}%
  \BibitemOpen
  \bibfield  {author} {\bibinfo {author} {\bibfnamefont {D.}~\bibnamefont
  {Skinner}},\ }\href {http://www.damtp.cam.ac.uk/user/dbs26/AQFT.html}
  {\bibinfo {title} {{Lecture notes: Quantum Field Theory II}}} (\bibinfo
  {year} {2018}),\ \bibinfo {note} {[Online; accessed 2021.01.12]}\BibitemShut
  {NoStop}%
\bibitem [{\citenamefont {Strocchi}(2013)}]{Strocchi:2013awa}%
  \BibitemOpen
  \bibfield  {author} {\bibinfo {author} {\bibfnamefont {F.}~\bibnamefont
  {Strocchi}},\ }\href
  {https://doi.org/10.1093/acprof:oso/9780199671571.001.0001} {\emph {\bibinfo
  {title} {{An introduction to non-perturbative foundations of quantum field
  theory}}}},\ Vol.\ \bibinfo {volume} {158}\ (\bibinfo  {publisher} {Oxford
  University Press, Oxford},\ \bibinfo {year} {2013})\BibitemShut {NoStop}%
\bibitem [{\citenamefont {Kemler}\ and\ \citenamefont
  {Braun}(2013)}]{Kemler:2013yka}%
  \BibitemOpen
  \bibfield  {author} {\bibinfo {author} {\bibfnamefont {S.}~\bibnamefont
  {Kemler}}\ and\ \bibinfo {author} {\bibfnamefont {J.}~\bibnamefont {Braun}},\
  }\bibfield  {title} {\bibinfo {title} {{Towards a Renormalization Group
  approach to density functional theory -- general formalism and case
  studies}},\ }\href {https://doi.org/10.1088/0954-3899/40/8/085105} {\bibfield
   {journal} {\bibinfo  {journal} {J. Phys. G}\ }\textbf {\bibinfo {volume}
  {40}},\ \bibinfo {pages} {085105} (\bibinfo {year} {2013})},\ \Eprint
  {https://arxiv.org/abs/1304.1161} {arXiv:1304.1161 [nucl-th]} \BibitemShut
  {NoStop}%
\bibitem [{\citenamefont {Rentrop}\ \emph {et~al.}(2015)\citenamefont
  {Rentrop}, \citenamefont {Jakobs},\ and\ \citenamefont
  {Meden}}]{Rentrop_2015}%
  \BibitemOpen
  \bibfield  {author} {\bibinfo {author} {\bibfnamefont {J.~F.}\ \bibnamefont
  {Rentrop}}, \bibinfo {author} {\bibfnamefont {S.~G.}\ \bibnamefont
  {Jakobs}},\ and\ \bibinfo {author} {\bibfnamefont {V.}~\bibnamefont
  {Meden}},\ }\bibfield  {title} {\bibinfo {title} {{Two-particle irreducible
  Functional Renormalization Group schemes -- a comparative study}},\ }\href
  {https://doi.org/10.1088/1751-8113/48/14/145002} {\bibfield  {journal}
  {\bibinfo  {journal} {Journal of Physics A: Mathematical and Theoretical}\
  }\textbf {\bibinfo {volume} {48}},\ \bibinfo {pages} {145002} (\bibinfo
  {year} {2015})}\BibitemShut {NoStop}%
\bibitem [{\citenamefont {Rosa}\ \emph {et~al.}(2016)\citenamefont {Rosa},
  \citenamefont {Farias},\ and\ \citenamefont {Ramos}}]{Rosa:2016czs}%
  \BibitemOpen
  \bibfield  {author} {\bibinfo {author} {\bibfnamefont {D.~S.}\ \bibnamefont
  {Rosa}}, \bibinfo {author} {\bibfnamefont {R.~L.~S.}\ \bibnamefont
  {Farias}},\ and\ \bibinfo {author} {\bibfnamefont {R.~O.}\ \bibnamefont
  {Ramos}},\ }\bibfield  {title} {\bibinfo {title} {{Reliability of the
  optimized perturbation theory in the 0-dimensional $O(N)$ scalar field
  model}},\ }\href {https://doi.org/10.1016/j.physa.2016.07.067} {\bibfield
  {journal} {\bibinfo  {journal} {Physica A}\ }\textbf {\bibinfo {volume}
  {464}},\ \bibinfo {pages} {11} (\bibinfo {year} {2016})},\ \Eprint
  {https://arxiv.org/abs/1604.00537} {arXiv:1604.00537 [hep-ph]} \BibitemShut
  {NoStop}%
\bibitem [{\citenamefont {Liang}\ \emph {et~al.}(2018)\citenamefont {Liang},
  \citenamefont {Niu},\ and\ \citenamefont {Hatsuda}}]{Liang:2017whg}%
  \BibitemOpen
  \bibfield  {author} {\bibinfo {author} {\bibfnamefont {H.}~\bibnamefont
  {Liang}}, \bibinfo {author} {\bibfnamefont {Y.}~\bibnamefont {Niu}},\ and\
  \bibinfo {author} {\bibfnamefont {T.}~\bibnamefont {Hatsuda}},\ }\bibfield
  {title} {\bibinfo {title} {{Functional Renormalization Group and Kohn-Sham
  scheme in density functional theory}},\ }\href
  {https://doi.org/10.1016/j.physletb.2018.02.034} {\bibfield  {journal}
  {\bibinfo  {journal} {Phys. Lett. B}\ }\textbf {\bibinfo {volume} {779}},\
  \bibinfo {pages} {436} (\bibinfo {year} {2018})},\ \Eprint
  {https://arxiv.org/abs/1710.00650} {arXiv:1710.00650 [cond-mat.str-el]}
  \BibitemShut {NoStop}%
\bibitem [{\citenamefont {Millington}\ and\ \citenamefont
  {Saffin}(2019)}]{Millington:2019nkw}%
  \BibitemOpen
  \bibfield  {author} {\bibinfo {author} {\bibfnamefont {P.}~\bibnamefont
  {Millington}}\ and\ \bibinfo {author} {\bibfnamefont {P.~M.}\ \bibnamefont
  {Saffin}},\ }\bibfield  {title} {\bibinfo {title} {{Visualising quantum
  effective action calculations in zero dimensions}},\ }\href
  {https://doi.org/10.1088/1751-8121/ab37e6} {\bibfield  {journal} {\bibinfo
  {journal} {J. Phys. A}\ }\textbf {\bibinfo {volume} {52}},\ \bibinfo {pages}
  {405401} (\bibinfo {year} {2019})},\ \Eprint
  {https://arxiv.org/abs/1905.09674} {arXiv:1905.09674 [hep-th]} \BibitemShut
  {NoStop}%
\bibitem [{\citenamefont {Alexander}\ \emph {et~al.}(2019)\citenamefont
  {Alexander}, \citenamefont {Millington}, \citenamefont {Nursey},\ and\
  \citenamefont {Saffin}}]{Alexander:2019cgw}%
  \BibitemOpen
  \bibfield  {author} {\bibinfo {author} {\bibfnamefont {E.}~\bibnamefont
  {Alexander}}, \bibinfo {author} {\bibfnamefont {P.}~\bibnamefont
  {Millington}}, \bibinfo {author} {\bibfnamefont {J.}~\bibnamefont {Nursey}},\
  and\ \bibinfo {author} {\bibfnamefont {P.~M.}\ \bibnamefont {Saffin}},\
  }\bibfield  {title} {\bibinfo {title} {{Alternative flow equation for the
  functional renormalization group}},\ }\href
  {https://doi.org/10.1103/PhysRevD.104.069906} {\bibfield  {journal} {\bibinfo
   {journal} {Phys. Rev. D}\ }\textbf {\bibinfo {volume} {100}},\ \bibinfo
  {pages} {101702} (\bibinfo {year} {2019})},\ \Eprint
  {https://arxiv.org/abs/1907.06503} {arXiv:1907.06503 [hep-th]} \BibitemShut
  {NoStop}%
\bibitem [{\citenamefont {Catalano}(2019)}]{Catalano:2019}%
  \BibitemOpen
  \bibfield  {author} {\bibinfo {author} {\bibfnamefont {A.~G.}\ \bibnamefont
  {Catalano}},\ }\emph {\bibinfo {title} {{Application of renormalization group
  techniques to the solution of integrals and Schrödinger eigenvalue
  equations}}},\ \href {http://webthesis.biblio.polito.it/id/eprint/18673}
  {\bibinfo {type} {master's thesis}},\ \bibinfo  {school} {Politecnico di
  Torino} (\bibinfo {year} {2019})\BibitemShut {NoStop}%
\bibitem [{\citenamefont {Millington}(2020)}]{Millington:2020Talk}%
  \BibitemOpen
  \bibfield  {author} {\bibinfo {author} {\bibfnamefont {P.}~\bibnamefont
  {Millington}},\ }\href
  {https://www2.yukawa.kyoto-u.ac.jp/~erg2020/ERG2020_slides/Millington.pdf}
  {\bibinfo {title} {{An alternative flow equation from the regulator-sourced
  $2$PI effective action}}},\ \bibinfo {howpublished} {{Talk at the 10th
  International Conference on Exact Renormalization Group 2020 (ERG2020)}}
  (\bibinfo {year} {2020}),\ \bibinfo {note} {[Online; accessed
  2021.01.12]}\BibitemShut {NoStop}%
\bibitem [{\citenamefont {Millington}\ and\ \citenamefont
  {Saffin}(2021)}]{Millington:2021ftp}%
  \BibitemOpen
  \bibfield  {author} {\bibinfo {author} {\bibfnamefont {P.}~\bibnamefont
  {Millington}}\ and\ \bibinfo {author} {\bibfnamefont {P.~M.}\ \bibnamefont
  {Saffin}},\ }\bibfield  {title} {\bibinfo {title} {{Benchmarking
  regulator-sourced 2PI and average 1PI flow equations in zero dimensions}},\
  }\href {https://doi.org/10.1088/1751-8121/ac2d00} {\bibfield  {journal}
  {\bibinfo  {journal} {J. Phys. A}\ }\textbf {\bibinfo {volume} {54}},\
  \bibinfo {pages} {465401} (\bibinfo {year} {2021})},\ \Eprint
  {https://arxiv.org/abs/2107.12914} {arXiv:2107.12914 [hep-th]} \BibitemShut
  {NoStop}%
\bibitem [{\citenamefont {Kades}\ \emph {et~al.}(2021)\citenamefont {Kades},
  \citenamefont {G\"arttner}, \citenamefont {Gasenzer},\ and\ \citenamefont
  {Pawlowski}}]{Kades:2021hir}%
  \BibitemOpen
  \bibfield  {author} {\bibinfo {author} {\bibfnamefont {L.}~\bibnamefont
  {Kades}}, \bibinfo {author} {\bibfnamefont {M.}~\bibnamefont {G\"arttner}},
  \bibinfo {author} {\bibfnamefont {T.}~\bibnamefont {Gasenzer}},\ and\
  \bibinfo {author} {\bibfnamefont {J.~M.}\ \bibnamefont {Pawlowski}},\
  }\href@noop {} {\bibinfo {title} {{Towards sampling complex actions}}}
  (\bibinfo {year} {2021}),\ \Eprint {https://arxiv.org/abs/2106.09367}
  {arXiv:2106.09367 [hep-lat]} \BibitemShut {NoStop}%
\bibitem [{\citenamefont {Fraboulet}(2021)}]{Fraboulet:2021amf}%
  \BibitemOpen
  \bibfield  {author} {\bibinfo {author} {\bibfnamefont {K.}~\bibnamefont
  {Fraboulet}},\ }\emph {\bibinfo {title} {{Path-integral approaches to
  strongly-coupled quantum many-body systems}}},\ \href@noop {} {Ph.D.
  thesis},\ \bibinfo  {school} {LPMC, Palaiseau, IJCLab, Orsay} (\bibinfo
  {year} {2021})\BibitemShut {NoStop}%
\bibitem [{\citenamefont {Mermin}\ and\ \citenamefont
  {Wagner}(1966)}]{Mermin:1966}%
  \BibitemOpen
  \bibfield  {author} {\bibinfo {author} {\bibfnamefont {N.~D.}\ \bibnamefont
  {Mermin}}\ and\ \bibinfo {author} {\bibfnamefont {H.}~\bibnamefont
  {Wagner}},\ }\bibfield  {title} {\bibinfo {title} {{Absence of ferromagnetism
  or antiferromagnetism in one- or two-dimensional isotropic Heisenberg
  models}},\ }\href {https://doi.org/10.1103/PhysRevLett.17.1133} {\bibfield
  {journal} {\bibinfo  {journal} {Phys. Rev. Lett.}\ }\textbf {\bibinfo
  {volume} {17}},\ \bibinfo {pages} {1133} (\bibinfo {year}
  {1966})}\BibitemShut {NoStop}%
\bibitem [{\citenamefont {Hohenberg}(1967)}]{Hohenberg:1967}%
  \BibitemOpen
  \bibfield  {author} {\bibinfo {author} {\bibfnamefont {P.~C.}\ \bibnamefont
  {Hohenberg}},\ }\bibfield  {title} {\bibinfo {title} {{Existence of
  long-range order in one and two dimensions}},\ }\href
  {https://doi.org/10.1103/PhysRev.158.383} {\bibfield  {journal} {\bibinfo
  {journal} {Phys. Rev.}\ }\textbf {\bibinfo {volume} {158}},\ \bibinfo {pages}
  {383} (\bibinfo {year} {1967})}\BibitemShut {NoStop}%
\bibitem [{\citenamefont {Coleman}(1973)}]{Coleman:1973ci}%
  \BibitemOpen
  \bibfield  {author} {\bibinfo {author} {\bibfnamefont {S.~R.}\ \bibnamefont
  {Coleman}},\ }\bibfield  {title} {\bibinfo {title} {{There are no Goldstone
  bosons in two-dimensions}},\ }\href {https://doi.org/10.1007/BF01646487}
  {\bibfield  {journal} {\bibinfo  {journal} {Commun. Math. Phys.}\ }\textbf
  {\bibinfo {volume} {31}},\ \bibinfo {pages} {259} (\bibinfo {year}
  {1973})}\BibitemShut {NoStop}%
\bibitem [{\citenamefont {Litim}(2000)}]{Litim:2000ci}%
  \BibitemOpen
  \bibfield  {author} {\bibinfo {author} {\bibfnamefont {D.~F.}\ \bibnamefont
  {Litim}},\ }\bibfield  {title} {\bibinfo {title} {{Optimization of the exact
  renormalization group}},\ }\href
  {https://doi.org/10.1016/S0370-2693(00)00748-6} {\bibfield  {journal}
  {\bibinfo  {journal} {Phys. Lett. B}\ }\textbf {\bibinfo {volume} {486}},\
  \bibinfo {pages} {92} (\bibinfo {year} {2000})},\ \Eprint
  {https://arxiv.org/abs/hep-th/0005245} {arXiv:hep-th/0005245} \BibitemShut
  {NoStop}%
\bibitem [{\citenamefont {Pawlowski}\ \emph {et~al.}(2017)\citenamefont
  {Pawlowski}, \citenamefont {Scherer}, \citenamefont {Schmidt},\ and\
  \citenamefont {Wetzel}}]{Pawlowski:2015mlf}%
  \BibitemOpen
  \bibfield  {author} {\bibinfo {author} {\bibfnamefont {J.~M.}\ \bibnamefont
  {Pawlowski}}, \bibinfo {author} {\bibfnamefont {M.~M.}\ \bibnamefont
  {Scherer}}, \bibinfo {author} {\bibfnamefont {R.}~\bibnamefont {Schmidt}},\
  and\ \bibinfo {author} {\bibfnamefont {S.~J.}\ \bibnamefont {Wetzel}},\
  }\bibfield  {title} {\bibinfo {title} {{Physics and the choice of regulators
  in functional renormalisation group flows}},\ }\href
  {https://doi.org/10.1016/j.aop.2017.06.017} {\bibfield  {journal} {\bibinfo
  {journal} {Annals Phys.}\ }\textbf {\bibinfo {volume} {384}},\ \bibinfo
  {pages} {165} (\bibinfo {year} {2017})},\ \Eprint
  {https://arxiv.org/abs/1512.03598} {arXiv:1512.03598 [hep-th]} \BibitemShut
  {NoStop}%
\bibitem [{\citenamefont {Braun}\ \emph
  {et~al.}(2021{\natexlab{a}})\citenamefont {Braun}, \citenamefont
  {D\"ornfeld}, \citenamefont {Schallmo},\ and\ \citenamefont
  {T\"opfel}}]{Braun:2020bhy}%
  \BibitemOpen
  \bibfield  {author} {\bibinfo {author} {\bibfnamefont {J.}~\bibnamefont
  {Braun}}, \bibinfo {author} {\bibfnamefont {T.}~\bibnamefont {D\"ornfeld}},
  \bibinfo {author} {\bibfnamefont {B.}~\bibnamefont {Schallmo}},\ and\
  \bibinfo {author} {\bibfnamefont {S.}~\bibnamefont {T\"opfel}},\ }\bibfield
  {title} {\bibinfo {title} {{Renormalization group studies of dense
  relativistic systems}},\ }\href {https://doi.org/10.1103/PhysRevD.104.096002}
  {\bibfield  {journal} {\bibinfo  {journal} {Phys. Rev. D}\ }\textbf {\bibinfo
  {volume} {104}},\ \bibinfo {pages} {096002} (\bibinfo {year}
  {2021}{\natexlab{a}})},\ \Eprint {https://arxiv.org/abs/2008.05978}
  {arXiv:2008.05978 [hep-ph]} \BibitemShut {NoStop}%
\bibitem [{\citenamefont {Pawlowski}(2007)}]{Pawlowski:2005xe}%
  \BibitemOpen
  \bibfield  {author} {\bibinfo {author} {\bibfnamefont {J.~M.}\ \bibnamefont
  {Pawlowski}},\ }\bibfield  {title} {\bibinfo {title} {{Aspects of the
  functional renormalisation group}},\ }\href
  {https://doi.org/10.1016/j.aop.2007.01.007} {\bibfield  {journal} {\bibinfo
  {journal} {Annals Phys.}\ }\textbf {\bibinfo {volume} {322}},\ \bibinfo
  {pages} {2831} (\bibinfo {year} {2007})},\ \Eprint
  {https://arxiv.org/abs/hep-th/0512261} {arXiv:hep-th/0512261} \BibitemShut
  {NoStop}%
\bibitem [{\citenamefont {Rosten}(2012)}]{Rosten:2010vm}%
  \BibitemOpen
  \bibfield  {author} {\bibinfo {author} {\bibfnamefont {O.~J.}\ \bibnamefont
  {Rosten}},\ }\bibfield  {title} {\bibinfo {title} {{Fundamentals of the Exact
  Renormalization Group}},\ }\href
  {https://doi.org/10.1016/j.physrep.2011.12.003} {\bibfield  {journal}
  {\bibinfo  {journal} {Phys. Rept.}\ }\textbf {\bibinfo {volume} {511}},\
  \bibinfo {pages} {177} (\bibinfo {year} {2012})},\ \Eprint
  {https://arxiv.org/abs/1003.1366} {arXiv:1003.1366 [hep-th]} \BibitemShut
  {NoStop}%
\bibitem [{\citenamefont {Osborn}\ and\ \citenamefont
  {Twigg}(2012)}]{Osborn:2011kw}%
  \BibitemOpen
  \bibfield  {author} {\bibinfo {author} {\bibfnamefont {H.}~\bibnamefont
  {Osborn}}\ and\ \bibinfo {author} {\bibfnamefont {D.~E.}\ \bibnamefont
  {Twigg}},\ }\bibfield  {title} {\bibinfo {title} {{Remarks on Exact RG
  equations}},\ }\href {https://doi.org/10.1016/j.aop.2011.10.011} {\bibfield
  {journal} {\bibinfo  {journal} {Annals Phys.}\ }\textbf {\bibinfo {volume}
  {327}},\ \bibinfo {pages} {29} (\bibinfo {year} {2012})},\ \Eprint
  {https://arxiv.org/abs/1108.5340} {arXiv:1108.5340 [hep-th]} \BibitemShut
  {NoStop}%
\bibitem [{\citenamefont {Salmhofer}(2020)}]{Salmhofer:2020Talk}%
  \BibitemOpen
  \bibfield  {author} {\bibinfo {author} {\bibfnamefont {M.}~\bibnamefont
  {Salmhofer}},\ }\href
  {https://www2.yukawa.kyoto-u.ac.jp/~erg2020/ERG2020_slides/Salmhofer.pdf}
  {\bibinfo {title} {{Rigorous Renormalization Group}}},\ \bibinfo
  {howpublished} {{Talk at the 10th International Conference on Exact
  Renormalization Group 2020 (ERG2020)}} (\bibinfo {year} {2020}),\ \bibinfo
  {note} {[Online; accessed 2021.01.12]}\BibitemShut {NoStop}%
\bibitem [{\citenamefont {Cannon}(1984)}]{Cannon:1984}%
  \BibitemOpen
  \bibfield  {author} {\bibinfo {author} {\bibfnamefont {J.~R.}\ \bibnamefont
  {Cannon}},\ }\href {https://doi.org/10.1017/CBO9781139086967} {\emph
  {\bibinfo {title} {{The one-dimensional heat equation}}}}\ (\bibinfo
  {publisher} {Cambridge University Press},\ \bibinfo {year}
  {1984})\BibitemShut {NoStop}%
\bibitem [{\citenamefont {LeVeque}(1992)}]{LeVeque:1992}%
  \BibitemOpen
  \bibfield  {author} {\bibinfo {author} {\bibfnamefont {R.~J.}\ \bibnamefont
  {LeVeque}},\ }\href {https://doi.org/10.1007/978-3-0348-8629-1} {\emph
  {\bibinfo {title} {{Numerical methods for conservation laws}}}},\ \bibinfo
  {edition} {2nd}\ ed.\ (\bibinfo  {publisher} {Birkhäuser},\ \bibinfo
  {address} {Basel},\ \bibinfo {year} {1992})\BibitemShut {NoStop}%
\bibitem [{\citenamefont {LeVeque}(2002)}]{LeVeque:2002}%
  \BibitemOpen
  \bibfield  {author} {\bibinfo {author} {\bibfnamefont {R.~J.}\ \bibnamefont
  {LeVeque}},\ }\href {https://doi.org/10.1017/CBO9780511791253} {\emph
  {\bibinfo {title} {{Finite-volume methods for hyperbolic problems}}}},\
  Cambridge Texts in Applied Mathematics\ (\bibinfo  {publisher} {Cambridge
  University Press},\ \bibinfo {year} {2002})\BibitemShut {NoStop}%
\bibitem [{\citenamefont {Iliopoulos}\ \emph {et~al.}(1975)\citenamefont
  {Iliopoulos}, \citenamefont {Itzykson},\ and\ \citenamefont
  {Martin}}]{Iliopoulos:1974ur}%
  \BibitemOpen
  \bibfield  {author} {\bibinfo {author} {\bibfnamefont {J.}~\bibnamefont
  {Iliopoulos}}, \bibinfo {author} {\bibfnamefont {C.}~\bibnamefont
  {Itzykson}},\ and\ \bibinfo {author} {\bibfnamefont {A.}~\bibnamefont
  {Martin}},\ }\bibfield  {title} {\bibinfo {title} {{Functional methods and
  perturbation theory}},\ }\href {https://doi.org/10.1103/RevModPhys.47.165}
  {\bibfield  {journal} {\bibinfo  {journal} {Rev. Mod. Phys.}\ }\textbf
  {\bibinfo {volume} {47}},\ \bibinfo {pages} {165} (\bibinfo {year}
  {1975})}\BibitemShut {NoStop}%
\bibitem [{\citenamefont {McCullagh}\ and\ \citenamefont
  {Kolassa}(2009)}]{McCullagh:2009}%
  \BibitemOpen
  \bibfield  {author} {\bibinfo {author} {\bibfnamefont {P.}~\bibnamefont
  {McCullagh}}\ and\ \bibinfo {author} {\bibfnamefont {J.}~\bibnamefont
  {Kolassa}},\ }\bibfield  {title} {\bibinfo {title} {{C}umulants},\ }\href
  {https://doi.org/10.4249/scholarpedia.4699} {\bibfield  {journal} {\bibinfo
  {journal} {Scholarpedia}\ }\textbf {\bibinfo {volume} {4}},\ \bibinfo {pages}
  {4699} (\bibinfo {year} {2009})},\ \bibinfo {note} {revision
  \#137322}\BibitemShut {NoStop}%
\bibitem [{\citenamefont {Fujimoto}\ \emph {et~al.}(1983)\citenamefont
  {Fujimoto}, \citenamefont {O'Raifeartaigh},\ and\ \citenamefont
  {Parravicini}}]{Fujimoto:1982tc}%
  \BibitemOpen
  \bibfield  {author} {\bibinfo {author} {\bibfnamefont {Y.}~\bibnamefont
  {Fujimoto}}, \bibinfo {author} {\bibfnamefont {L.}~\bibnamefont
  {O'Raifeartaigh}},\ and\ \bibinfo {author} {\bibfnamefont {G.}~\bibnamefont
  {Parravicini}},\ }\bibfield  {title} {\bibinfo {title} {{Effective potential
  for non-convex potentials}},\ }\href
  {https://doi.org/10.1016/0550-3213(83)90305-X} {\bibfield  {journal}
  {\bibinfo  {journal} {Nucl. Phys. B}\ }\textbf {\bibinfo {volume} {212}},\
  \bibinfo {pages} {268} (\bibinfo {year} {1983})}\BibitemShut {NoStop}%
\bibitem [{\citenamefont {Wipf}(2013)}]{Wipf:2013vp}%
  \BibitemOpen
  \bibfield  {author} {\bibinfo {author} {\bibfnamefont {A.}~\bibnamefont
  {Wipf}},\ }\href {https://doi.org/10.1007/978-3-642-33105-3} {\emph {\bibinfo
  {title} {{Statistical Approach to Quantum Field Theory}}}},\ \bibinfo
  {series} {Lect.Notes Phys.}\ No.\ \bibinfo {number} {864}\ (\bibinfo
  {publisher} {Springer-Verlag},\ \bibinfo {address} {Berlin, Germany},\
  \bibinfo {year} {2013})\BibitemShut {NoStop}%
\bibitem [{\citenamefont {Reuter}\ and\ \citenamefont
  {Wetterich}(1994)}]{Reuter:1993kw}%
  \BibitemOpen
  \bibfield  {author} {\bibinfo {author} {\bibfnamefont {M.}~\bibnamefont
  {Reuter}}\ and\ \bibinfo {author} {\bibfnamefont {C.}~\bibnamefont
  {Wetterich}},\ }\bibfield  {title} {\bibinfo {title} {{Effective average
  action for gauge theories and exact evolution equations}},\ }\href
  {https://doi.org/10.1016/0550-3213(94)90543-6} {\bibfield  {journal}
  {\bibinfo  {journal} {Nucl. Phys. B}\ }\textbf {\bibinfo {volume} {417}},\
  \bibinfo {pages} {181} (\bibinfo {year} {1994})}\BibitemShut {NoStop}%
\bibitem [{\citenamefont {Reuter}(1998)}]{Reuter:1996cp}%
  \BibitemOpen
  \bibfield  {author} {\bibinfo {author} {\bibfnamefont {M.}~\bibnamefont
  {Reuter}},\ }\bibfield  {title} {\bibinfo {title} {{Nonperturbative evolution
  equation for quantum gravity}},\ }\href
  {https://doi.org/10.1103/PhysRevD.57.971} {\bibfield  {journal} {\bibinfo
  {journal} {Phys. Rev. D}\ }\textbf {\bibinfo {volume} {57}},\ \bibinfo
  {pages} {971} (\bibinfo {year} {1998})},\ \Eprint
  {https://arxiv.org/abs/hep-th/9605030} {arXiv:hep-th/9605030} \BibitemShut
  {NoStop}%
\bibitem [{\citenamefont {Reuter}\ and\ \citenamefont
  {Saueressig}(2002)}]{Reuter:2001ag}%
  \BibitemOpen
  \bibfield  {author} {\bibinfo {author} {\bibfnamefont {M.}~\bibnamefont
  {Reuter}}\ and\ \bibinfo {author} {\bibfnamefont {F.}~\bibnamefont
  {Saueressig}},\ }\bibfield  {title} {\bibinfo {title} {{Renormalization group
  flow of quantum gravity in the Einstein-Hilbert truncation}},\ }\href
  {https://doi.org/10.1103/PhysRevD.65.065016} {\bibfield  {journal} {\bibinfo
  {journal} {Phys. Rev. D}\ }\textbf {\bibinfo {volume} {65}},\ \bibinfo
  {pages} {065016} (\bibinfo {year} {2002})},\ \Eprint
  {https://arxiv.org/abs/hep-th/0110054} {arXiv:hep-th/0110054} \BibitemShut
  {NoStop}%
\bibitem [{\citenamefont {Berges}\ \emph {et~al.}(2002)\citenamefont {Berges},
  \citenamefont {Tetradis},\ and\ \citenamefont {Wetterich}}]{Berges:2000ew}%
  \BibitemOpen
  \bibfield  {author} {\bibinfo {author} {\bibfnamefont {J.}~\bibnamefont
  {Berges}}, \bibinfo {author} {\bibfnamefont {N.}~\bibnamefont {Tetradis}},\
  and\ \bibinfo {author} {\bibfnamefont {C.}~\bibnamefont {Wetterich}},\
  }\bibfield  {title} {\bibinfo {title} {{Nonperturbative renormalization flow
  in quantum field theory and statistical physics}},\ }\href
  {https://doi.org/10.1016/S0370-1573(01)00098-9} {\bibfield  {journal}
  {\bibinfo  {journal} {Phys. Rept.}\ }\textbf {\bibinfo {volume} {363}},\
  \bibinfo {pages} {223} (\bibinfo {year} {2002})},\ \Eprint
  {https://arxiv.org/abs/hep-ph/0005122} {arXiv:hep-ph/0005122} \BibitemShut
  {NoStop}%
\bibitem [{\citenamefont {Kopietz}\ \emph {et~al.}(2010)\citenamefont
  {Kopietz}, \citenamefont {Bartosch},\ and\ \citenamefont
  {Sch\"utz}}]{Kopietz:2010zz}%
  \BibitemOpen
  \bibfield  {author} {\bibinfo {author} {\bibfnamefont {P.}~\bibnamefont
  {Kopietz}}, \bibinfo {author} {\bibfnamefont {L.}~\bibnamefont {Bartosch}},\
  and\ \bibinfo {author} {\bibfnamefont {F.}~\bibnamefont {Sch\"utz}},\ }\href
  {https://doi.org/10.1007/978-3-642-05094-7} {\emph {\bibinfo {title}
  {{Introduction to the Functional Renormalization Group}}}},\ \bibinfo
  {series} {Lecture Notes in Physics}, Vol.\ \bibinfo {volume} {798}\ (\bibinfo
   {publisher} {Springer-Verlag Berlin Heidelberg},\ \bibinfo {year}
  {2010})\BibitemShut {NoStop}%
\bibitem [{\citenamefont {Polchinski}(1984)}]{Polchinski:1983gv}%
  \BibitemOpen
  \bibfield  {author} {\bibinfo {author} {\bibfnamefont {J.}~\bibnamefont
  {Polchinski}},\ }\bibfield  {title} {\bibinfo {title} {{Renormalization and
  Effective Lagrangians}},\ }\href
  {https://doi.org/10.1016/0550-3213(84)90287-6} {\bibfield  {journal}
  {\bibinfo  {journal} {Nucl. Phys. B}\ }\textbf {\bibinfo {volume} {231}},\
  \bibinfo {pages} {269} (\bibinfo {year} {1984})}\BibitemShut {NoStop}%
\bibitem [{\citenamefont {Litim}(2005)}]{Litim:2005us}%
  \BibitemOpen
  \bibfield  {author} {\bibinfo {author} {\bibfnamefont {D.~F.}\ \bibnamefont
  {Litim}},\ }\bibfield  {title} {\bibinfo {title} {{Universality and the
  renormalisation group}},\ }\href
  {https://doi.org/10.1088/1126-6708/2005/07/005} {\bibfield  {journal}
  {\bibinfo  {journal} {JHEP}\ }\textbf {\bibinfo {volume} {07}},\ \bibinfo
  {pages} {005}},\ \Eprint {https://arxiv.org/abs/hep-th/0503096}
  {arXiv:hep-th/0503096} \BibitemShut {NoStop}%
\bibitem [{\citenamefont {Yabunaka}\ and\ \citenamefont
  {Delamotte}(2018)}]{Yabunaka:2018mju}%
  \BibitemOpen
  \bibfield  {author} {\bibinfo {author} {\bibfnamefont {S.}~\bibnamefont
  {Yabunaka}}\ and\ \bibinfo {author} {\bibfnamefont {B.}~\bibnamefont
  {Delamotte}},\ }\bibfield  {title} {\bibinfo {title} {{Why Might the Standard
  Large $N$ Analysis Fail in the O($N$) Model: The Role of Cusps in the Fixed
  Point Potentials}},\ }\href {https://doi.org/10.1103/PhysRevLett.121.231601}
  {\bibfield  {journal} {\bibinfo  {journal} {Phys. Rev. Lett.}\ }\textbf
  {\bibinfo {volume} {121}},\ \bibinfo {pages} {231601} (\bibinfo {year}
  {2018})},\ \Eprint {https://arxiv.org/abs/1807.04681} {arXiv:1807.04681
  [cond-mat.stat-mech]} \BibitemShut {NoStop}%
\bibitem [{\citenamefont {Litim}\ and\ \citenamefont
  {Trott}(2018)}]{Litim:2018pxe}%
  \BibitemOpen
  \bibfield  {author} {\bibinfo {author} {\bibfnamefont {D.~F.}\ \bibnamefont
  {Litim}}\ and\ \bibinfo {author} {\bibfnamefont {M.~J.}\ \bibnamefont
  {Trott}},\ }\bibfield  {title} {\bibinfo {title} {{Asymptotic safety of
  scalar field theories}},\ }\href {https://doi.org/10.1103/PhysRevD.98.125006}
  {\bibfield  {journal} {\bibinfo  {journal} {Phys. Rev. D}\ }\textbf {\bibinfo
  {volume} {98}},\ \bibinfo {pages} {125006} (\bibinfo {year} {2018})},\
  \Eprint {https://arxiv.org/abs/1810.01678} {arXiv:1810.01678 [hep-th]}
  \BibitemShut {NoStop}%
\bibitem [{\citenamefont {Cotler}\ and\ \citenamefont
  {Rezchikov}(2022)}]{Cotler:2022fze}%
  \BibitemOpen
  \bibfield  {author} {\bibinfo {author} {\bibfnamefont {J.}~\bibnamefont
  {Cotler}}\ and\ \bibinfo {author} {\bibfnamefont {S.}~\bibnamefont
  {Rezchikov}},\ }\href@noop {} {\bibinfo {title} {{Renormalization Group Flow
  as Optimal Transport}}} (\bibinfo {year} {2022}),\ \Eprint
  {https://arxiv.org/abs/2202.11737} {arXiv:2202.11737 [hep-th]} \BibitemShut
  {NoStop}%
\bibitem [{\citenamefont {DeWitt}(1965)}]{DeWitt:1965jb}%
  \BibitemOpen
  \bibfield  {author} {\bibinfo {author} {\bibfnamefont {B.~S.}\ \bibnamefont
  {DeWitt}},\ }\href@noop {} {\emph {\bibinfo {title} {Dynamical theory of
  groups and fields}}}\ (\bibinfo  {publisher} {Gordon and Breach},\ \bibinfo
  {address} {New York, USA},\ \bibinfo {year} {1965})\BibitemShut {NoStop}%
\bibitem [{\citenamefont {Abbott}(1982)}]{Abbott:1981ke}%
  \BibitemOpen
  \bibfield  {author} {\bibinfo {author} {\bibfnamefont {L.~F.}\ \bibnamefont
  {Abbott}},\ }\bibfield  {title} {\bibinfo {title} {{Introduction to the
  Background Field Method}},\ }\href
  {https://www.actaphys.uj.edu.pl/R/13/1/33/pdf} {\bibfield  {journal}
  {\bibinfo  {journal} {Acta Phys. Polon. B}\ }\textbf {\bibinfo {volume}
  {13}},\ \bibinfo {pages} {33} (\bibinfo {year} {1982})}\BibitemShut {NoStop}%
\bibitem [{\citenamefont {Braun}\ \emph {et~al.}(2019)\citenamefont {Braun},
  \citenamefont {Leonhardt},\ and\ \citenamefont {Pawlowski}}]{Braun:2018svj}%
  \BibitemOpen
  \bibfield  {author} {\bibinfo {author} {\bibfnamefont {J.}~\bibnamefont
  {Braun}}, \bibinfo {author} {\bibfnamefont {M.}~\bibnamefont {Leonhardt}},\
  and\ \bibinfo {author} {\bibfnamefont {J.~M.}\ \bibnamefont {Pawlowski}},\
  }\bibfield  {title} {\bibinfo {title} {{Renormalization group consistency and
  low-energy effective theories}},\ }\href
  {https://doi.org/10.21468/SciPostPhys.6.5.056} {\bibfield  {journal}
  {\bibinfo  {journal} {SciPost Phys.}\ }\textbf {\bibinfo {volume} {6}},\
  \bibinfo {pages} {056} (\bibinfo {year} {2019})},\ \Eprint
  {https://arxiv.org/abs/1806.04432} {arXiv:1806.04432 [hep-ph]} \BibitemShut
  {NoStop}%
\bibitem [{\citenamefont {Braun}\ \emph {et~al.}(2004)\citenamefont {Braun},
  \citenamefont {Schwenzer},\ and\ \citenamefont {Pirner}}]{Braun:2003ii}%
  \BibitemOpen
  \bibfield  {author} {\bibinfo {author} {\bibfnamefont {J.}~\bibnamefont
  {Braun}}, \bibinfo {author} {\bibfnamefont {K.}~\bibnamefont {Schwenzer}},\
  and\ \bibinfo {author} {\bibfnamefont {H.-J.}\ \bibnamefont {Pirner}},\
  }\bibfield  {title} {\bibinfo {title} {{Linking the quark meson model with
  QCD at high temperature}},\ }\href
  {https://doi.org/10.1103/PhysRevD.70.085016} {\bibfield  {journal} {\bibinfo
  {journal} {Phys. Rev. D}\ }\textbf {\bibinfo {volume} {70}},\ \bibinfo
  {pages} {085016} (\bibinfo {year} {2004})},\ \Eprint
  {https://arxiv.org/abs/hep-ph/0312277} {arXiv:hep-ph/0312277} \BibitemShut
  {NoStop}%
\bibitem [{\citenamefont {Herbst}\ \emph {et~al.}(2014)\citenamefont {Herbst},
  \citenamefont {Mitter}, \citenamefont {Pawlowski}, \citenamefont {Schaefer},\
  and\ \citenamefont {Stiele}}]{Herbst:2013ufa}%
  \BibitemOpen
  \bibfield  {author} {\bibinfo {author} {\bibfnamefont {T.~K.}\ \bibnamefont
  {Herbst}}, \bibinfo {author} {\bibfnamefont {M.}~\bibnamefont {Mitter}},
  \bibinfo {author} {\bibfnamefont {J.~M.}\ \bibnamefont {Pawlowski}}, \bibinfo
  {author} {\bibfnamefont {B.-J.}\ \bibnamefont {Schaefer}},\ and\ \bibinfo
  {author} {\bibfnamefont {R.}~\bibnamefont {Stiele}},\ }\bibfield  {title}
  {\bibinfo {title} {{Thermodynamics of QCD at vanishing density}},\ }\href
  {https://doi.org/10.1016/j.physletb.2014.02.045} {\bibfield  {journal}
  {\bibinfo  {journal} {Phys. Lett. B}\ }\textbf {\bibinfo {volume} {731}},\
  \bibinfo {pages} {248} (\bibinfo {year} {2014})},\ \Eprint
  {https://arxiv.org/abs/1308.3621} {arXiv:1308.3621 [hep-ph]} \BibitemShut
  {NoStop}%
\bibitem [{\citenamefont {Springer}\ \emph {et~al.}(2017)\citenamefont
  {Springer}, \citenamefont {Braun}, \citenamefont {Rechenberger},\ and\
  \citenamefont {Rennecke}}]{Springer2017}%
  \BibitemOpen
  \bibfield  {author} {\bibinfo {author} {\bibfnamefont {P.}~\bibnamefont
  {Springer}}, \bibinfo {author} {\bibfnamefont {J.}~\bibnamefont {Braun}},
  \bibinfo {author} {\bibfnamefont {S.}~\bibnamefont {Rechenberger}},\ and\
  \bibinfo {author} {\bibfnamefont {F.}~\bibnamefont {Rennecke}},\ }\bibfield
  {title} {\bibinfo {title} {{QCD-inspired determination of NJL model
  parameters}},\ }\href {https://doi.org/10.1051/epjconf/201713703022}
  {\bibfield  {journal} {\bibinfo  {journal} {EPJ Web Conf.}\ }\textbf
  {\bibinfo {volume} {137}},\ \bibinfo {pages} {03022} (\bibinfo {year}
  {2017})},\ \Eprint {https://arxiv.org/abs/1611.06020} {arXiv:1611.06020
  [hep-ph]} \BibitemShut {NoStop}%
\bibitem [{\citenamefont {Haas}\ \emph {et~al.}(2013)\citenamefont {Haas},
  \citenamefont {Stiele}, \citenamefont {Braun}, \citenamefont {Pawlowski},\
  and\ \citenamefont {Schaffner-Bielich}}]{PhysRevD.87.076004}%
  \BibitemOpen
  \bibfield  {author} {\bibinfo {author} {\bibfnamefont {L.~M.}\ \bibnamefont
  {Haas}}, \bibinfo {author} {\bibfnamefont {R.}~\bibnamefont {Stiele}},
  \bibinfo {author} {\bibfnamefont {J.}~\bibnamefont {Braun}}, \bibinfo
  {author} {\bibfnamefont {J.~M.}\ \bibnamefont {Pawlowski}},\ and\ \bibinfo
  {author} {\bibfnamefont {J.}~\bibnamefont {Schaffner-Bielich}},\ }\bibfield
  {title} {\bibinfo {title} {{Improved Polyakov-loop potential for effective
  models from functional calculations}},\ }\href
  {https://doi.org/10.1103/PhysRevD.87.076004} {\bibfield  {journal} {\bibinfo
  {journal} {Phys. Rev. D}\ }\textbf {\bibinfo {volume} {87}},\ \bibinfo
  {pages} {076004} (\bibinfo {year} {2013})},\ \Eprint
  {https://arxiv.org/abs/1302.1993} {arXiv:1302.1993 [hep-ph]} \BibitemShut
  {NoStop}%
\bibitem [{\citenamefont {Fefferman}(2020)}]{navier-stokes}%
  \BibitemOpen
  \bibfield  {author} {\bibinfo {author} {\bibfnamefont {C.~L.}\ \bibnamefont
  {Fefferman}},\ }\href
  {http://www.claymath.org/sites/default/files/navierstokes.pdf} {\bibinfo
  {title} {{Existence and smoothness of the Navier-Stokes equation}}} (\bibinfo
  {year} {2020}),\ \bibinfo {note} {[Online; accessed 2021.01.09]}\BibitemShut
  {NoStop}%
\bibitem [{\citenamefont {Gies}(2012)}]{Gies:2006wv}%
  \BibitemOpen
  \bibfield  {author} {\bibinfo {author} {\bibfnamefont {H.}~\bibnamefont
  {Gies}},\ }\bibfield  {title} {\bibinfo {title} {{Introduction to the
  Functional RG and applications to gauge theories}},\ }\href
  {https://doi.org/10.1007/978-3-642-27320-9\_6} {\bibfield  {journal}
  {\bibinfo  {journal} {Lect. Notes Phys.}\ }\textbf {\bibinfo {volume}
  {852}},\ \bibinfo {pages} {287} (\bibinfo {year} {2012})},\ \Eprint
  {https://arxiv.org/abs/hep-ph/0611146} {arXiv:hep-ph/0611146} \BibitemShut
  {NoStop}%
\bibitem [{\citenamefont {Pawlowski}\ \emph {et~al.}()\citenamefont
  {Pawlowski}, \citenamefont {Bonnet}, \citenamefont {Rechenberger},
  \citenamefont {Reichert},\ and\ \citenamefont {Wink}}]{PawlowskiScript}%
  \BibitemOpen
  \bibfield  {author} {\bibinfo {author} {\bibfnamefont {J.~M.}\ \bibnamefont
  {Pawlowski}}, \bibinfo {author} {\bibfnamefont {J.~A.}\ \bibnamefont
  {Bonnet}}, \bibinfo {author} {\bibfnamefont {S.}~\bibnamefont
  {Rechenberger}}, \bibinfo {author} {\bibfnamefont {M.}~\bibnamefont
  {Reichert}},\ and\ \bibinfo {author} {\bibfnamefont {N.}~\bibnamefont
  {Wink}},\ }\bibfield  {title} {\bibinfo {title} {{The functional
  renormalization group - applications to gauge theories and gravity}},\
  }\bibinfo {note} {unpublished lecture notes, in preparation}\BibitemShut
  {NoStop}%
\bibitem [{\citenamefont {Delamotte}(2012)}]{Delamotte:2007pf}%
  \BibitemOpen
  \bibfield  {author} {\bibinfo {author} {\bibfnamefont {B.}~\bibnamefont
  {Delamotte}},\ }\bibfield  {title} {\bibinfo {title} {{An introduction to the
  nonperturbative Renormalization Group}},\ }\href
  {https://doi.org/10.1007/978-3-642-27320-9_2} {\bibfield  {journal} {\bibinfo
   {journal} {Lect. Notes Phys.}\ }\textbf {\bibinfo {volume} {852}},\ \bibinfo
  {pages} {49} (\bibinfo {year} {2012})},\ \Eprint
  {https://arxiv.org/abs/cond-mat/0702365} {arXiv:cond-mat/0702365}
  \BibitemShut {NoStop}%
\bibitem [{\citenamefont {Blaizot}\ \emph {et~al.}(2021)\citenamefont
  {Blaizot}, \citenamefont {Pawlowski},\ and\ \citenamefont
  {Reinosa}}]{Blaizot:2021ikl}%
  \BibitemOpen
  \bibfield  {author} {\bibinfo {author} {\bibfnamefont {J.-P.}\ \bibnamefont
  {Blaizot}}, \bibinfo {author} {\bibfnamefont {J.~M.}\ \bibnamefont
  {Pawlowski}},\ and\ \bibinfo {author} {\bibfnamefont {U.}~\bibnamefont
  {Reinosa}},\ }\bibfield  {title} {\bibinfo {title} {{Functional
  renormalization group and 2PI effective action formalism}},\ }\href
  {https://doi.org/10.1016/j.aop.2021.168549} {\bibfield  {journal} {\bibinfo
  {journal} {Annals Phys.}\ }\textbf {\bibinfo {volume} {431}},\ \bibinfo
  {pages} {168549} (\bibinfo {year} {2021})},\ \Eprint
  {https://arxiv.org/abs/2102.13628} {arXiv:2102.13628 [hep-th]} \BibitemShut
  {NoStop}%
\bibitem [{\citenamefont {Rennecke}(2015)}]{Rennecke:2015lur}%
  \BibitemOpen
  \bibfield  {author} {\bibinfo {author} {\bibfnamefont {F.}~\bibnamefont
  {Rennecke}},\ }\emph {\bibinfo {title} {{The chiral phase transition of
  QCD}}},\ \href {https://doi.org/10.11588/heidok.00019205} {\bibinfo {type}
  {Phd thesis}},\ \bibinfo  {school} {University of Heidelberg} (\bibinfo
  {year} {2015})\BibitemShut {NoStop}%
\bibitem [{\citenamefont {Wilson}(1971{\natexlab{a}})}]{Wilson:1971bg}%
  \BibitemOpen
  \bibfield  {author} {\bibinfo {author} {\bibfnamefont {K.~G.}\ \bibnamefont
  {Wilson}},\ }\bibfield  {title} {\bibinfo {title} {{Renormalization group and
  critical phenomena. 1. Renormalization group and the Kadanoff scaling
  picture}},\ }\href {https://doi.org/10.1103/PhysRevB.4.3174} {\bibfield
  {journal} {\bibinfo  {journal} {Phys. Rev. B}\ }\textbf {\bibinfo {volume}
  {4}},\ \bibinfo {pages} {3174} (\bibinfo {year}
  {1971}{\natexlab{a}})}\BibitemShut {NoStop}%
\bibitem [{\citenamefont {Wilson}(1971{\natexlab{b}})}]{Wilson:1971dh}%
  \BibitemOpen
  \bibfield  {author} {\bibinfo {author} {\bibfnamefont {K.~G.}\ \bibnamefont
  {Wilson}},\ }\bibfield  {title} {\bibinfo {title} {{Renormalization group and
  critical phenomena. 2. Phase space cell analysis of critical behavior}},\
  }\href {https://doi.org/10.1103/PhysRevB.4.3184} {\bibfield  {journal}
  {\bibinfo  {journal} {Phys. Rev. B}\ }\textbf {\bibinfo {volume} {4}},\
  \bibinfo {pages} {3184} (\bibinfo {year} {1971}{\natexlab{b}})}\BibitemShut
  {NoStop}%
\bibitem [{\citenamefont {Wilson}(1979)}]{Wilson:1979qg}%
  \BibitemOpen
  \bibfield  {author} {\bibinfo {author} {\bibfnamefont {K.~G.}\ \bibnamefont
  {Wilson}},\ }\bibfield  {title} {\bibinfo {title} {{Problems in physics with
  many scales of length}},\ }\href
  {https://doi.org/10.1038/scientificamerican0879-158} {\bibfield  {journal}
  {\bibinfo  {journal} {Sci. Am.}\ }\textbf {\bibinfo {volume} {241}},\
  \bibinfo {pages} {158} (\bibinfo {year} {1979})}\BibitemShut {NoStop}%
\bibitem [{\citenamefont {Fu}\ \emph {et~al.}(2020)\citenamefont {Fu},
  \citenamefont {Pawlowski},\ and\ \citenamefont {Rennecke}}]{Fu:2019hdw}%
  \BibitemOpen
  \bibfield  {author} {\bibinfo {author} {\bibfnamefont {W.-j.}\ \bibnamefont
  {Fu}}, \bibinfo {author} {\bibfnamefont {J.~M.}\ \bibnamefont {Pawlowski}},\
  and\ \bibinfo {author} {\bibfnamefont {F.}~\bibnamefont {Rennecke}},\
  }\bibfield  {title} {\bibinfo {title} {{QCD phase structure at finite
  temperature and density}},\ }\href
  {https://doi.org/10.1103/PhysRevD.101.054032} {\bibfield  {journal} {\bibinfo
   {journal} {Phys. Rev. D}\ }\textbf {\bibinfo {volume} {101}},\ \bibinfo
  {pages} {054032} (\bibinfo {year} {2020})},\ \Eprint
  {https://arxiv.org/abs/1909.02991} {arXiv:1909.02991 [hep-ph]} \BibitemShut
  {NoStop}%
\bibitem [{\citenamefont {Greiner}\ and\ \citenamefont
  {Reinhardt}(1996)}]{Greiner:1996zu}%
  \BibitemOpen
  \bibfield  {author} {\bibinfo {author} {\bibfnamefont {W.}~\bibnamefont
  {Greiner}}\ and\ \bibinfo {author} {\bibfnamefont {J.}~\bibnamefont
  {Reinhardt}},\ }\href {https://doi.org/10.1007/978-3-642-61485-9} {\emph
  {\bibinfo {title} {{Field quantization}}}}\ (\bibinfo  {publisher} {Springer,
  Berlin Heidelberg},\ \bibinfo {year} {1996})\BibitemShut {NoStop}%
\bibitem [{\citenamefont {Goldstone}\ \emph {et~al.}(1962)\citenamefont
  {Goldstone}, \citenamefont {Salam},\ and\ \citenamefont
  {Weinberg}}]{Goldstone:1962es}%
  \BibitemOpen
  \bibfield  {author} {\bibinfo {author} {\bibfnamefont {J.}~\bibnamefont
  {Goldstone}}, \bibinfo {author} {\bibfnamefont {A.}~\bibnamefont {Salam}},\
  and\ \bibinfo {author} {\bibfnamefont {S.}~\bibnamefont {Weinberg}},\
  }\bibfield  {title} {\bibinfo {title} {{Broken symmetries}},\ }\href
  {https://doi.org/10.1103/PhysRev.127.965} {\bibfield  {journal} {\bibinfo
  {journal} {Phys. Rev.}\ }\textbf {\bibinfo {volume} {127}},\ \bibinfo {pages}
  {965} (\bibinfo {year} {1962})}\BibitemShut {NoStop}%
\bibitem [{\citenamefont {Wetterich}(2001)}]{Wetterich:2001kra}%
  \BibitemOpen
  \bibfield  {author} {\bibinfo {author} {\bibfnamefont {C.}~\bibnamefont
  {Wetterich}},\ }\bibfield  {title} {\bibinfo {title} {{Effective average
  action in statistical physics and quantum field theory}},\ }\href
  {https://doi.org/10.1142/S0217751X01004591} {\bibfield  {journal} {\bibinfo
  {journal} {Int. J. Mod. Phys. A}\ }\textbf {\bibinfo {volume} {16}},\
  \bibinfo {pages} {1951} (\bibinfo {year} {2001})},\ \Eprint
  {https://arxiv.org/abs/hep-ph/0101178} {arXiv:hep-ph/0101178} \BibitemShut
  {NoStop}%
\bibitem [{\citenamefont {Ising}(1925)}]{Ising:1925em}%
  \BibitemOpen
  \bibfield  {author} {\bibinfo {author} {\bibfnamefont {E.}~\bibnamefont
  {Ising}},\ }\bibfield  {title} {\bibinfo {title} {{Beitrag zur Theorie des
  Ferromagnetismus}},\ }\href {https://doi.org/10.1007/BF02980577} {\bibfield
  {journal} {\bibinfo  {journal} {Z. Phys.}\ }\textbf {\bibinfo {volume}
  {31}},\ \bibinfo {pages} {253} (\bibinfo {year} {1925})}\BibitemShut
  {NoStop}%
\bibitem [{\citenamefont {Canet}\ \emph
  {et~al.}(2003{\natexlab{a}})\citenamefont {Canet}, \citenamefont {Delamotte},
  \citenamefont {Mouhanna},\ and\ \citenamefont {Vidal}}]{Canet:2003qd}%
  \BibitemOpen
  \bibfield  {author} {\bibinfo {author} {\bibfnamefont {L.}~\bibnamefont
  {Canet}}, \bibinfo {author} {\bibfnamefont {B.}~\bibnamefont {Delamotte}},
  \bibinfo {author} {\bibfnamefont {D.}~\bibnamefont {Mouhanna}},\ and\
  \bibinfo {author} {\bibfnamefont {J.}~\bibnamefont {Vidal}},\ }\bibfield
  {title} {\bibinfo {title} {{Nonperturbative Renormalization Group approach to
  the Ising model: A derivative expansion at order $\partial^4$}},\ }\href
  {https://doi.org/10.1103/PhysRevB.68.064421} {\bibfield  {journal} {\bibinfo
  {journal} {Phys. Rev. B}\ }\textbf {\bibinfo {volume} {68}},\ \bibinfo
  {pages} {064421} (\bibinfo {year} {2003}{\natexlab{a}})},\ \Eprint
  {https://arxiv.org/abs/hep-th/0302227} {arXiv:hep-th/0302227} \BibitemShut
  {NoStop}%
\bibitem [{\citenamefont {Nambu}(1960)}]{Nambu:1960tm}%
  \BibitemOpen
  \bibfield  {author} {\bibinfo {author} {\bibfnamefont {Y.}~\bibnamefont
  {Nambu}},\ }\bibfield  {title} {\bibinfo {title} {{Quasiparticles and gauge
  invariance in the theory of superconductivity}},\ }\href
  {https://doi.org/10.1103/PhysRev.117.648} {\bibfield  {journal} {\bibinfo
  {journal} {Phys. Rev.}\ }\textbf {\bibinfo {volume} {117}},\ \bibinfo {pages}
  {648} (\bibinfo {year} {1960})}\BibitemShut {NoStop}%
\bibitem [{\citenamefont {Goldstone}(1961)}]{Goldstone:1961eq}%
  \BibitemOpen
  \bibfield  {author} {\bibinfo {author} {\bibfnamefont {J.}~\bibnamefont
  {Goldstone}},\ }\bibfield  {title} {\bibinfo {title} {{Field theories with
  superconductor solutions}},\ }\href {https://doi.org/10.1007/BF02812722}
  {\bibfield  {journal} {\bibinfo  {journal} {Nuovo Cim.}\ }\textbf {\bibinfo
  {volume} {19}},\ \bibinfo {pages} {154} (\bibinfo {year} {1961})}\BibitemShut
  {NoStop}%
\bibitem [{\citenamefont {Aoki}\ \emph {et~al.}(2018)\citenamefont {Aoki},
  \citenamefont {Kumamoto},\ and\ \citenamefont {Yamada}}]{Aoki:2017rjl}%
  \BibitemOpen
  \bibfield  {author} {\bibinfo {author} {\bibfnamefont {K.-I.}\ \bibnamefont
  {Aoki}}, \bibinfo {author} {\bibfnamefont {S.-I.}\ \bibnamefont {Kumamoto}},\
  and\ \bibinfo {author} {\bibfnamefont {M.}~\bibnamefont {Yamada}},\
  }\bibfield  {title} {\bibinfo {title} {{Phase structure of NJL model with
  weak renormalization group}},\ }\href
  {https://doi.org/10.1016/j.nuclphysb.2018.04.005} {\bibfield  {journal}
  {\bibinfo  {journal} {Nucl. Phys. B}\ }\textbf {\bibinfo {volume} {931}},\
  \bibinfo {pages} {105} (\bibinfo {year} {2018})},\ \Eprint
  {https://arxiv.org/abs/1705.03273} {arXiv:1705.03273 [hep-th]} \BibitemShut
  {NoStop}%
\bibitem [{\citenamefont {Braun}\ \emph {et~al.}(2017)\citenamefont {Braun},
  \citenamefont {Leonhardt},\ and\ \citenamefont {Pospiech}}]{Braun:2017srn}%
  \BibitemOpen
  \bibfield  {author} {\bibinfo {author} {\bibfnamefont {J.}~\bibnamefont
  {Braun}}, \bibinfo {author} {\bibfnamefont {M.}~\bibnamefont {Leonhardt}},\
  and\ \bibinfo {author} {\bibfnamefont {M.}~\bibnamefont {Pospiech}},\
  }\bibfield  {title} {\bibinfo {title} {{Fierz-complete NJL model study: Fixed
  points and phase structure at finite temperature and density}},\ }\href
  {https://doi.org/10.1103/PhysRevD.96.076003} {\bibfield  {journal} {\bibinfo
  {journal} {Phys. Rev. D}\ }\textbf {\bibinfo {volume} {96}},\ \bibinfo
  {pages} {076003} (\bibinfo {year} {2017})},\ \Eprint
  {https://arxiv.org/abs/1705.00074} {arXiv:1705.00074 [hep-ph]} \BibitemShut
  {NoStop}%
\bibitem [{\citenamefont {Braun}\ \emph {et~al.}(2018)\citenamefont {Braun},
  \citenamefont {Leonhardt},\ and\ \citenamefont {Pospiech}}]{Braun:2018bik}%
  \BibitemOpen
  \bibfield  {author} {\bibinfo {author} {\bibfnamefont {J.}~\bibnamefont
  {Braun}}, \bibinfo {author} {\bibfnamefont {M.}~\bibnamefont {Leonhardt}},\
  and\ \bibinfo {author} {\bibfnamefont {M.}~\bibnamefont {Pospiech}},\
  }\bibfield  {title} {\bibinfo {title} {{Fierz-complete NJL model study. II.
  Toward the fixed-point and phase structure of hot and dense two-flavor
  QCD}},\ }\href {https://doi.org/10.1103/PhysRevD.97.076010} {\bibfield
  {journal} {\bibinfo  {journal} {Phys. Rev. D}\ }\textbf {\bibinfo {volume}
  {97}},\ \bibinfo {pages} {076010} (\bibinfo {year} {2018})},\ \Eprint
  {https://arxiv.org/abs/1801.08338} {arXiv:1801.08338 [hep-ph]} \BibitemShut
  {NoStop}%
\bibitem [{\citenamefont {Braun}\ \emph {et~al.}(2020)\citenamefont {Braun},
  \citenamefont {Leonhardt},\ and\ \citenamefont {Pospiech}}]{Braun:2019aow}%
  \BibitemOpen
  \bibfield  {author} {\bibinfo {author} {\bibfnamefont {J.}~\bibnamefont
  {Braun}}, \bibinfo {author} {\bibfnamefont {M.}~\bibnamefont {Leonhardt}},\
  and\ \bibinfo {author} {\bibfnamefont {M.}~\bibnamefont {Pospiech}},\
  }\bibfield  {title} {\bibinfo {title} {{Fierz-complete NJL model study III:
  Emergence from quark-gluon dynamics}},\ }\href
  {https://doi.org/10.1103/PhysRevD.101.036004} {\bibfield  {journal} {\bibinfo
   {journal} {Phys. Rev. D}\ }\textbf {\bibinfo {volume} {101}},\ \bibinfo
  {pages} {036004} (\bibinfo {year} {2020})},\ \Eprint
  {https://arxiv.org/abs/1909.06298} {arXiv:1909.06298 [hep-ph]} \BibitemShut
  {NoStop}%
\bibitem [{\citenamefont {Canet}\ \emph
  {et~al.}(2003{\natexlab{b}})\citenamefont {Canet}, \citenamefont {Delamotte},
  \citenamefont {Mouhanna},\ and\ \citenamefont {Vidal}}]{Canet:2002gs}%
  \BibitemOpen
  \bibfield  {author} {\bibinfo {author} {\bibfnamefont {L.}~\bibnamefont
  {Canet}}, \bibinfo {author} {\bibfnamefont {B.}~\bibnamefont {Delamotte}},
  \bibinfo {author} {\bibfnamefont {D.}~\bibnamefont {Mouhanna}},\ and\
  \bibinfo {author} {\bibfnamefont {J.}~\bibnamefont {Vidal}},\ }\bibfield
  {title} {\bibinfo {title} {{Optimization of the derivative expansion in the
  nonperturbative Renormalization Group}},\ }\href
  {https://doi.org/10.1103/PhysRevD.67.065004} {\bibfield  {journal} {\bibinfo
  {journal} {Phys. Rev. D}\ }\textbf {\bibinfo {volume} {67}},\ \bibinfo
  {pages} {065004} (\bibinfo {year} {2003}{\natexlab{b}})},\ \Eprint
  {https://arxiv.org/abs/hep-th/0211055} {arXiv:hep-th/0211055} \BibitemShut
  {NoStop}%
\bibitem [{\citenamefont {Balog}\ \emph {et~al.}(2019)\citenamefont {Balog},
  \citenamefont {Chat\'e}, \citenamefont {Delamotte}, \citenamefont
  {Marohnic},\ and\ \citenamefont {Wschebor}}]{Balog:2019rrg}%
  \BibitemOpen
  \bibfield  {author} {\bibinfo {author} {\bibfnamefont {I.}~\bibnamefont
  {Balog}}, \bibinfo {author} {\bibfnamefont {H.}~\bibnamefont {Chat\'e}},
  \bibinfo {author} {\bibfnamefont {B.}~\bibnamefont {Delamotte}}, \bibinfo
  {author} {\bibfnamefont {M.}~\bibnamefont {Marohnic}},\ and\ \bibinfo
  {author} {\bibfnamefont {N.}~\bibnamefont {Wschebor}},\ }\bibfield  {title}
  {\bibinfo {title} {{Convergence of nonperturbative approximations to the
  Renormalization Group}},\ }\href
  {https://doi.org/10.1103/PhysRevLett.123.240604} {\bibfield  {journal}
  {\bibinfo  {journal} {Phys. Rev. Lett.}\ }\textbf {\bibinfo {volume} {123}},\
  \bibinfo {pages} {240604} (\bibinfo {year} {2019})},\ \Eprint
  {https://arxiv.org/abs/1907.01829} {arXiv:1907.01829 [cond-mat.stat-mech]}
  \BibitemShut {NoStop}%
\bibitem [{\citenamefont {Papenbrock}\ and\ \citenamefont
  {Wetterich}(1995)}]{Papenbrock:1994kf}%
  \BibitemOpen
  \bibfield  {author} {\bibinfo {author} {\bibfnamefont {T.}~\bibnamefont
  {Papenbrock}}\ and\ \bibinfo {author} {\bibfnamefont {C.}~\bibnamefont
  {Wetterich}},\ }\bibfield  {title} {\bibinfo {title} {{Two-loop results from
  improved one loop computations}},\ }\href
  {https://doi.org/10.1007/BF01556140} {\bibfield  {journal} {\bibinfo
  {journal} {Z. Phys. C}\ }\textbf {\bibinfo {volume} {65}},\ \bibinfo {pages}
  {519} (\bibinfo {year} {1995})},\ \Eprint
  {https://arxiv.org/abs/hep-th/9403164} {arXiv:hep-th/9403164} \BibitemShut
  {NoStop}%
\bibitem [{\citenamefont {Tetradis}\ and\ \citenamefont
  {Wetterich}(1994)}]{Tetradis:1993ts}%
  \BibitemOpen
  \bibfield  {author} {\bibinfo {author} {\bibfnamefont {N.}~\bibnamefont
  {Tetradis}}\ and\ \bibinfo {author} {\bibfnamefont {C.}~\bibnamefont
  {Wetterich}},\ }\bibfield  {title} {\bibinfo {title} {{Critical exponents
  from effective average action}},\ }\href
  {https://doi.org/10.1016/0550-3213(94)90446-4} {\bibfield  {journal}
  {\bibinfo  {journal} {Nucl. Phys. B}\ }\textbf {\bibinfo {volume} {422}},\
  \bibinfo {pages} {541} (\bibinfo {year} {1994})},\ \Eprint
  {https://arxiv.org/abs/hep-ph/9308214} {arXiv:hep-ph/9308214} \BibitemShut
  {NoStop}%
\bibitem [{\citenamefont {Bonanno}\ and\ \citenamefont
  {Zappala}(2001)}]{Bonanno:2000yp}%
  \BibitemOpen
  \bibfield  {author} {\bibinfo {author} {\bibfnamefont {A.}~\bibnamefont
  {Bonanno}}\ and\ \bibinfo {author} {\bibfnamefont {D.}~\bibnamefont
  {Zappala}},\ }\bibfield  {title} {\bibinfo {title} {{Towards an accurate
  determination of the critical exponents with the renormalization group flow
  equations}},\ }\href {https://doi.org/10.1016/S0370-2693(01)00273-8}
  {\bibfield  {journal} {\bibinfo  {journal} {Phys. Lett. B}\ }\textbf
  {\bibinfo {volume} {504}},\ \bibinfo {pages} {181} (\bibinfo {year}
  {2001})},\ \Eprint {https://arxiv.org/abs/hep-th/0010095}
  {arXiv:hep-th/0010095} \BibitemShut {NoStop}%
\bibitem [{\citenamefont {Bervillier}\ \emph {et~al.}(2007)\citenamefont
  {Bervillier}, \citenamefont {Juttner},\ and\ \citenamefont
  {Litim}}]{Bervillier:2007rc}%
  \BibitemOpen
  \bibfield  {author} {\bibinfo {author} {\bibfnamefont {C.}~\bibnamefont
  {Bervillier}}, \bibinfo {author} {\bibfnamefont {A.}~\bibnamefont
  {Juttner}},\ and\ \bibinfo {author} {\bibfnamefont {D.~F.}\ \bibnamefont
  {Litim}},\ }\bibfield  {title} {\bibinfo {title} {{High-accuracy scaling
  exponents in the local potential approximation}},\ }\href
  {https://doi.org/10.1016/j.nuclphysb.2007.03.036} {\bibfield  {journal}
  {\bibinfo  {journal} {Nucl. Phys. B}\ }\textbf {\bibinfo {volume} {783}},\
  \bibinfo {pages} {213} (\bibinfo {year} {2007})},\ \Eprint
  {https://arxiv.org/abs/hep-th/0701172} {arXiv:hep-th/0701172} \BibitemShut
  {NoStop}%
\bibitem [{\citenamefont {Litim}\ and\ \citenamefont
  {Zappala}(2011)}]{Litim:2010tt}%
  \BibitemOpen
  \bibfield  {author} {\bibinfo {author} {\bibfnamefont {D.~F.}\ \bibnamefont
  {Litim}}\ and\ \bibinfo {author} {\bibfnamefont {D.}~\bibnamefont
  {Zappala}},\ }\bibfield  {title} {\bibinfo {title} {{Ising exponents from the
  Functional Renormalisation Group}},\ }\href
  {https://doi.org/10.1103/PhysRevD.83.085009} {\bibfield  {journal} {\bibinfo
  {journal} {Phys. Rev. D}\ }\textbf {\bibinfo {volume} {83}},\ \bibinfo
  {pages} {085009} (\bibinfo {year} {2011})},\ \Eprint
  {https://arxiv.org/abs/1009.1948} {arXiv:1009.1948 [hep-th]} \BibitemShut
  {NoStop}%
\bibitem [{\citenamefont {Caillol}(2012)}]{Caillol:2012zz}%
  \BibitemOpen
  \bibfield  {author} {\bibinfo {author} {\bibfnamefont {J.-M.}\ \bibnamefont
  {Caillol}},\ }\bibfield  {title} {\bibinfo {title} {{The non-perturbative
  renormalization group in the ordered phase}},\ }\href
  {https://doi.org/10.1016/j.nuclphysb.2011.10.026} {\bibfield  {journal}
  {\bibinfo  {journal} {Nucl. Phys. B}\ }\textbf {\bibinfo {volume} {855}},\
  \bibinfo {pages} {854} (\bibinfo {year} {2012})},\ \Eprint
  {https://arxiv.org/abs/1109.4024} {arXiv:1109.4024 [cond-mat.stat-mech]}
  \BibitemShut {NoStop}%
\bibitem [{\citenamefont {Pangon}\ \emph {et~al.}(2011)\citenamefont {Pangon},
  \citenamefont {Nagy}, \citenamefont {Polonyi},\ and\ \citenamefont
  {Sailer}}]{Pangon:2009pj}%
  \BibitemOpen
  \bibfield  {author} {\bibinfo {author} {\bibfnamefont {V.}~\bibnamefont
  {Pangon}}, \bibinfo {author} {\bibfnamefont {S.}~\bibnamefont {Nagy}},
  \bibinfo {author} {\bibfnamefont {J.}~\bibnamefont {Polonyi}},\ and\ \bibinfo
  {author} {\bibfnamefont {K.~G.}\ \bibnamefont {Sailer}},\ }\bibfield  {title}
  {\bibinfo {title} {{Onset of symmetry breaking by the Functional RG
  method}},\ }\href {https://doi.org/10.1142/S0217751X11052839} {\bibfield
  {journal} {\bibinfo  {journal} {Int. J. Mod. Phys. A}\ }\textbf {\bibinfo
  {volume} {26}},\ \bibinfo {pages} {1327} (\bibinfo {year} {2011})},\ \Eprint
  {https://arxiv.org/abs/0907.0144} {arXiv:0907.0144 [hep-th]} \BibitemShut
  {NoStop}%
\bibitem [{\citenamefont {Pangon}(2012)}]{Pangon:2010uf}%
  \BibitemOpen
  \bibfield  {author} {\bibinfo {author} {\bibfnamefont {V.}~\bibnamefont
  {Pangon}},\ }\bibfield  {title} {\bibinfo {title} {{Structure of the broken
  phase of the sine-Gordon model using Functional Renormalization}},\ }\href
  {https://doi.org/10.1142/S0217751X12500145} {\bibfield  {journal} {\bibinfo
  {journal} {Int. J. Mod. Phys. A}\ }\textbf {\bibinfo {volume} {27}},\
  \bibinfo {pages} {1250014} (\bibinfo {year} {2012})},\ \Eprint
  {https://arxiv.org/abs/1008.0281} {arXiv:1008.0281 [hep-th]} \BibitemShut
  {NoStop}%
\bibitem [{\citenamefont {Wetterich}(1993{\natexlab{b}})}]{Wetterich:1991be}%
  \BibitemOpen
  \bibfield  {author} {\bibinfo {author} {\bibfnamefont {C.}~\bibnamefont
  {Wetterich}},\ }\bibfield  {title} {\bibinfo {title} {{The average action for
  scalar fields near phase transitions}},\ }\href
  {https://doi.org/10.1007/BF01474340} {\bibfield  {journal} {\bibinfo
  {journal} {Z. Phys. C}\ }\textbf {\bibinfo {volume} {57}},\ \bibinfo {pages}
  {451} (\bibinfo {year} {1993}{\natexlab{b}})}\BibitemShut {NoStop}%
\bibitem [{\citenamefont {Pawlowski}\ \emph {et~al.}(2018)\citenamefont
  {Pawlowski}, \citenamefont {Strodthoff},\ and\ \citenamefont
  {Wink}}]{Pawlowski:2017gxj}%
  \BibitemOpen
  \bibfield  {author} {\bibinfo {author} {\bibfnamefont {J.~M.}\ \bibnamefont
  {Pawlowski}}, \bibinfo {author} {\bibfnamefont {N.}~\bibnamefont
  {Strodthoff}},\ and\ \bibinfo {author} {\bibfnamefont {N.}~\bibnamefont
  {Wink}},\ }\bibfield  {title} {\bibinfo {title} {{Finite temperature spectral
  functions in the $O(N)$-model}},\ }\href
  {https://doi.org/10.1103/PhysRevD.98.074008} {\bibfield  {journal} {\bibinfo
  {journal} {Phys. Rev. D}\ }\textbf {\bibinfo {volume} {98}},\ \bibinfo
  {pages} {074008} (\bibinfo {year} {2018})},\ \Eprint
  {https://arxiv.org/abs/1711.07444} {arXiv:1711.07444 [hep-th]} \BibitemShut
  {NoStop}%
\bibitem [{\citenamefont {Tripolt}\ \emph {et~al.}(2014)\citenamefont
  {Tripolt}, \citenamefont {Strodthoff}, \citenamefont {von Smekal},\ and\
  \citenamefont {Wambach}}]{Tripolt:2013jra}%
  \BibitemOpen
  \bibfield  {author} {\bibinfo {author} {\bibfnamefont {R.-A.}\ \bibnamefont
  {Tripolt}}, \bibinfo {author} {\bibfnamefont {N.}~\bibnamefont {Strodthoff}},
  \bibinfo {author} {\bibfnamefont {L.}~\bibnamefont {von Smekal}},\ and\
  \bibinfo {author} {\bibfnamefont {J.}~\bibnamefont {Wambach}},\ }\bibfield
  {title} {\bibinfo {title} {{Spectral functions for the Quark-Meson Model
  phase diagram from the Functional Renormalization Group}},\ }\href
  {https://doi.org/10.1103/PhysRevD.89.034010} {\bibfield  {journal} {\bibinfo
  {journal} {Phys. Rev. D}\ }\textbf {\bibinfo {volume} {89}},\ \bibinfo
  {pages} {034010} (\bibinfo {year} {2014})},\ \Eprint
  {https://arxiv.org/abs/1311.0630} {arXiv:1311.0630 [hep-ph]} \BibitemShut
  {NoStop}%
\bibitem [{\citenamefont {Adams}\ \emph {et~al.}(1995)\citenamefont {Adams},
  \citenamefont {Berges}, \citenamefont {Bornholdt}, \citenamefont {Freire},
  \citenamefont {Tetradis},\ and\ \citenamefont {Wetterich}}]{Adams:1995cv}%
  \BibitemOpen
  \bibfield  {author} {\bibinfo {author} {\bibfnamefont {J.~A.}\ \bibnamefont
  {Adams}}, \bibinfo {author} {\bibfnamefont {J.}~\bibnamefont {Berges}},
  \bibinfo {author} {\bibfnamefont {S.}~\bibnamefont {Bornholdt}}, \bibinfo
  {author} {\bibfnamefont {F.}~\bibnamefont {Freire}}, \bibinfo {author}
  {\bibfnamefont {N.}~\bibnamefont {Tetradis}},\ and\ \bibinfo {author}
  {\bibfnamefont {C.}~\bibnamefont {Wetterich}},\ }\bibfield  {title} {\bibinfo
  {title} {{Solving nonperturbative flow equations}},\ }\href
  {https://doi.org/10.1142/S0217732395002520} {\bibfield  {journal} {\bibinfo
  {journal} {Mod. Phys. Lett. A}\ }\textbf {\bibinfo {volume} {10}},\ \bibinfo
  {pages} {2367} (\bibinfo {year} {1995})},\ \Eprint
  {https://arxiv.org/abs/hep-th/9507093} {arXiv:hep-th/9507093} \BibitemShut
  {NoStop}%
\bibitem [{\citenamefont {Schaefer}\ and\ \citenamefont
  {Pirner}(1999)}]{Schaefer:1999em}%
  \BibitemOpen
  \bibfield  {author} {\bibinfo {author} {\bibfnamefont {B.-J.}\ \bibnamefont
  {Schaefer}}\ and\ \bibinfo {author} {\bibfnamefont {H.-J.}\ \bibnamefont
  {Pirner}},\ }\bibfield  {title} {\bibinfo {title} {{Renormalization group
  flow and equation of state of quarks and mesons}},\ }\href
  {https://doi.org/10.1016/S0375-9474(99)00409-1} {\bibfield  {journal}
  {\bibinfo  {journal} {Nucl. Phys. A}\ }\textbf {\bibinfo {volume} {660}},\
  \bibinfo {pages} {439} (\bibinfo {year} {1999})},\ \Eprint
  {https://arxiv.org/abs/nucl-th/9903003} {arXiv:nucl-th/9903003} \BibitemShut
  {NoStop}%
\bibitem [{\citenamefont {Borchardt}\ and\ \citenamefont
  {Knorr}(2016)}]{Borchardt:2016pif}%
  \BibitemOpen
  \bibfield  {author} {\bibinfo {author} {\bibfnamefont {J.}~\bibnamefont
  {Borchardt}}\ and\ \bibinfo {author} {\bibfnamefont {B.}~\bibnamefont
  {Knorr}},\ }\bibfield  {title} {\bibinfo {title} {{Solving functional flow
  equations with pseudo-spectral methods}},\ }\href
  {https://doi.org/10.1103/PhysRevD.94.025027} {\bibfield  {journal} {\bibinfo
  {journal} {Phys. Rev. D}\ }\textbf {\bibinfo {volume} {94}},\ \bibinfo
  {pages} {025027} (\bibinfo {year} {2016})},\ \Eprint
  {https://arxiv.org/abs/1603.06726} {arXiv:1603.06726 [hep-th]} \BibitemShut
  {NoStop}%
\bibitem [{\citenamefont {{Wolfram Research{,}
  Inc.}}(2020)}]{Mathematica:12.1}%
  \BibitemOpen
  \bibfield  {author} {\bibinfo {author} {\bibnamefont {{Wolfram Research{,}
  Inc.}}},\ }\href {https://www.wolfram.com/mathematica} {\bibinfo {title}
  {{Mathematica, {V}ersion 12.1}}} (\bibinfo {year} {2020}),\ \bibinfo {note}
  {{Champaign, IL}}\BibitemShut {NoStop}%
\bibitem [{\citenamefont {Wink}(2019)}]{WinkHirschegg}%
  \BibitemOpen
  \bibfield  {author} {\bibinfo {author} {\bibfnamefont {N.}~\bibnamefont
  {Wink}},\ }\href {https://indico.gsi.de/event/8375/contributions/36534/}
  {\bibinfo {title} {{Resolving phase transitions with Discontinuous Galerkin
  methods}}},\ \bibinfo {howpublished} {{Talk at the EMMI Workshop Functional
  Methods in Strongly Correlated Systems}} (\bibinfo {year} {2019}),\ \bibinfo
  {note} {[Online; accessed 2020.11.24]}\BibitemShut {NoStop}%
\bibitem [{\citenamefont {Steil}\ \emph {et~al.}(2020)\citenamefont {Steil},
  \citenamefont {Koenigstein}, \citenamefont {Braun}, \citenamefont {Buballa},
  \citenamefont {Grossi}, \citenamefont {Rischke},\ and\ \citenamefont
  {Wink}}]{Koenigstein:2020Talk}%
  \BibitemOpen
  \bibfield  {author} {\bibinfo {author} {\bibfnamefont {M.~J.}\ \bibnamefont
  {Steil}}, \bibinfo {author} {\bibfnamefont {A.}~\bibnamefont {Koenigstein}},
  \bibinfo {author} {\bibfnamefont {J.}~\bibnamefont {Braun}}, \bibinfo
  {author} {\bibfnamefont {M.}~\bibnamefont {Buballa}}, \bibinfo {author}
  {\bibfnamefont {E.}~\bibnamefont {Grossi}}, \bibinfo {author} {\bibfnamefont
  {D.~H.}\ \bibnamefont {Rischke}},\ and\ \bibinfo {author} {\bibfnamefont
  {N.}~\bibnamefont {Wink}},\ }\href
  {https://www2.yukawa.kyoto-u.ac.jp/~erg2020/ERG2020_slides/steil_koenigstein.pdf}
  {\bibinfo {title} {{Numerical fluid dynamics for FRG-flow equations:
  Zero-dimensional QFTs as numerical test cases}}},\ \bibinfo {howpublished}
  {{Slides for the 10th International Conference on Exact Renormalization Group
  2020 (ERG2020)}} (\bibinfo {year} {2020}),\ \bibinfo {note} {[Online;
  accessed 2021.02.01]}\BibitemShut {NoStop}%
\bibitem [{\citenamefont {Ihssen}(2020)}]{Ihssen2020}%
  \BibitemOpen
  \bibfield  {author} {\bibinfo {author} {\bibfnamefont {F.~J.}\ \bibnamefont
  {Ihssen}},\ }\emph {\bibinfo {title} {{Low-energy effective models of QCD at
  finite temperatures and densities}}},\ \href@noop {} {\bibinfo {type}
  {master's thesis}},\ \bibinfo  {school} {University of Heidelberg} (\bibinfo
  {year} {2020})\BibitemShut {NoStop}%
\bibitem [{\citenamefont {Wink}(2020)}]{Wink:2020tnu}%
  \BibitemOpen
  \bibfield  {author} {\bibinfo {author} {\bibfnamefont {N.}~\bibnamefont
  {Wink}},\ }\emph {\bibinfo {title} {{Towards the spectral properties and
  phase structure of QCD}}},\ \href {https://doi.org/10.11588/heidok.00028503}
  {\bibinfo {type} {Phd thesis}},\ \bibinfo  {school} {University of
  Heidelberg} (\bibinfo {year} {2020})\BibitemShut {NoStop}%
\bibitem [{\citenamefont {Ehrenfest}(1933)}]{Ehrenfest1933}%
  \BibitemOpen
  \bibfield  {author} {\bibinfo {author} {\bibfnamefont {P.}~\bibnamefont
  {Ehrenfest}},\ }\bibfield  {title} {\bibinfo {title} {{Phasenumwandlungen im
  ueblichen und erweiterten Sinn, classifiziert nach den entsprechenden
  Singularitaeten des thermodynamischen Potentiales}},\ }\href
  {https://www.dwc.knaw.nl/DL/publications/PU00016385.pdf} {\bibfield
  {journal} {\bibinfo  {journal} {Proc. Royal Acad. Amsterdam}\ }\textbf
  {\bibinfo {volume} {36}},\ \bibinfo {pages} {153} (\bibinfo {year} {1933})},\
  \bibinfo {note} {supplement No 75b zu den Mitteilungen aus dem Kamerlingh
  Onnes-Institute, Leiden}\BibitemShut {NoStop}%
\bibitem [{\citenamefont {Coleman}(1985)}]{Coleman:1985rnk}%
  \BibitemOpen
  \bibfield  {author} {\bibinfo {author} {\bibfnamefont {S.~R.}\ \bibnamefont
  {Coleman}},\ }\href {https://doi.org/10.1017/CBO9780511565045} {\emph
  {\bibinfo {title} {{Aspects of Symmetry}: {Selected Erice Lectures}}}}\
  (\bibinfo  {publisher} {Cambridge University Press},\ \bibinfo {address}
  {Cambridge, U.K.},\ \bibinfo {year} {1985})\BibitemShut {NoStop}%
\bibitem [{\citenamefont {Wallace}\ and\ \citenamefont
  {Zia}(1974)}]{Wallace:1974dx}%
  \BibitemOpen
  \bibfield  {author} {\bibinfo {author} {\bibfnamefont {D.~J.}\ \bibnamefont
  {Wallace}}\ and\ \bibinfo {author} {\bibfnamefont {R.~K.~P.}\ \bibnamefont
  {Zia}},\ }\bibfield  {title} {\bibinfo {title} {{Gradient flow and the
  Renormalization Group}},\ }\href
  {https://doi.org/10.1016/0375-9601(74)90449-6} {\bibfield  {journal}
  {\bibinfo  {journal} {Phys. Lett. A}\ }\textbf {\bibinfo {volume} {48}},\
  \bibinfo {pages} {325} (\bibinfo {year} {1974})}\BibitemShut {NoStop}%
\bibitem [{\citenamefont {Wallace}\ and\ \citenamefont
  {Zia}(1975)}]{Wallace:1974dy}%
  \BibitemOpen
  \bibfield  {author} {\bibinfo {author} {\bibfnamefont {D.~J.}\ \bibnamefont
  {Wallace}}\ and\ \bibinfo {author} {\bibfnamefont {R.~K.~P.}\ \bibnamefont
  {Zia}},\ }\bibfield  {title} {\bibinfo {title} {{Gradient properties of the
  Renormalization Group Equations in multicomponent systems}},\ }\href
  {https://doi.org/10.1016/0003-4916(75)90267-5} {\bibfield  {journal}
  {\bibinfo  {journal} {Annals Phys.}\ }\textbf {\bibinfo {volume} {92}},\
  \bibinfo {pages} {142} (\bibinfo {year} {1975})}\BibitemShut {NoStop}%
\bibitem [{\citenamefont {Zamolodchikov}(1986)}]{Zamolodchikov:1986gt}%
  \BibitemOpen
  \bibfield  {author} {\bibinfo {author} {\bibfnamefont {A.~B.}\ \bibnamefont
  {Zamolodchikov}},\ }\bibfield  {title} {\bibinfo {title} {{Irreversibility of
  the flux of the Renormalization Group in a $2D$ field theory}},\ }\href
  {http://jetpletters.ru/ps/1413/article_21504.shtml} {\bibfield  {journal}
  {\bibinfo  {journal} {JETP Lett.}\ }\textbf {\bibinfo {volume} {43}},\
  \bibinfo {pages} {730} (\bibinfo {year} {1986})}\BibitemShut {NoStop}%
\bibitem [{\citenamefont {Zumbach}(1994{\natexlab{a}})}]{Zumbach:1994vg}%
  \BibitemOpen
  \bibfield  {author} {\bibinfo {author} {\bibfnamefont {G.}~\bibnamefont
  {Zumbach}},\ }\bibfield  {title} {\bibinfo {title} {{The Renormalization
  Group in the Local Potential Approximation and its applications to the $O(N)$
  model}},\ }\href {https://doi.org/10.1016/0550-3213(94)90011-6} {\bibfield
  {journal} {\bibinfo  {journal} {Nucl. Phys. B}\ }\textbf {\bibinfo {volume}
  {413}},\ \bibinfo {pages} {754} (\bibinfo {year}
  {1994}{\natexlab{a}})}\BibitemShut {NoStop}%
\bibitem [{\citenamefont {Zumbach}(1994{\natexlab{b}})}]{Zumbach:1994kc}%
  \BibitemOpen
  \bibfield  {author} {\bibinfo {author} {\bibfnamefont {G.}~\bibnamefont
  {Zumbach}},\ }\bibfield  {title} {\bibinfo {title} {{The Local Potential
  Approximation of the Renormalization Group and its applications}},\ }\href
  {https://doi.org/10.1016/0375-9601(94)90746-3} {\bibfield  {journal}
  {\bibinfo  {journal} {Phys. Lett. A}\ }\textbf {\bibinfo {volume} {190}},\
  \bibinfo {pages} {225} (\bibinfo {year} {1994}{\natexlab{b}})}\BibitemShut
  {NoStop}%
\bibitem [{\citenamefont {Rezzolla}\ and\ \citenamefont
  {Zanotti}(2018)}]{RezzollaZanotti:2013}%
  \BibitemOpen
  \bibfield  {author} {\bibinfo {author} {\bibfnamefont {L.}~\bibnamefont
  {Rezzolla}}\ and\ \bibinfo {author} {\bibfnamefont {O.}~\bibnamefont
  {Zanotti}},\ }\href
  {https://doi.org/10.1093/acprof:oso/9780198528906.001.0001} {\emph {\bibinfo
  {title} {{Relativistic hydrodynamics}}}}\ (\bibinfo  {publisher} {Oxford
  University Press},\ \bibinfo {address} {Oxford, England, UK},\ \bibinfo
  {year} {2018})\BibitemShut {NoStop}%
\bibitem [{\citenamefont {Felder}(1987)}]{Felder:1987}%
  \BibitemOpen
  \bibfield  {author} {\bibinfo {author} {\bibfnamefont {G.}~\bibnamefont
  {Felder}},\ }\bibfield  {title} {\bibinfo {title} {Renormalization group in
  the local potential approximation},\ }\href
  {https://doi.org/10.1007/BF01239018} {\bibfield  {journal} {\bibinfo
  {journal} {Comm. Math. Phys.}\ }\textbf {\bibinfo {volume} {111}},\ \bibinfo
  {pages} {101} (\bibinfo {year} {1987})}\BibitemShut {NoStop}%
\bibitem [{\citenamefont {Hasenfratz}\ and\ \citenamefont
  {Hasenfratz}(1986)}]{Hasenfratz:1985dm}%
  \BibitemOpen
  \bibfield  {author} {\bibinfo {author} {\bibfnamefont {A.}~\bibnamefont
  {Hasenfratz}}\ and\ \bibinfo {author} {\bibfnamefont {P.}~\bibnamefont
  {Hasenfratz}},\ }\bibfield  {title} {\bibinfo {title} {{Renormalization Group
  study of scalar field theories}},\ }\href
  {https://doi.org/10.1016/0550-3213(86)90573-0} {\bibfield  {journal}
  {\bibinfo  {journal} {Nucl. Phys. B}\ }\textbf {\bibinfo {volume} {270}},\
  \bibinfo {pages} {687} (\bibinfo {year} {1986})}\BibitemShut {NoStop}%
\bibitem [{\citenamefont {Braun}\ \emph
  {et~al.}(2021{\natexlab{b}})\citenamefont {Braun}, \citenamefont {Buballa},
  \citenamefont {Grossi}, \citenamefont {Koenigstein}, \citenamefont {Rischke},
  \citenamefont {Steil},\ and\ \citenamefont {Wink}}]{Koenigstein:2021}%
  \BibitemOpen
  \bibfield  {author} {\bibinfo {author} {\bibfnamefont {J.}~\bibnamefont
  {Braun}}, \bibinfo {author} {\bibfnamefont {M.}~\bibnamefont {Buballa}},
  \bibinfo {author} {\bibfnamefont {E.}~\bibnamefont {Grossi}}, \bibinfo
  {author} {\bibfnamefont {A.}~\bibnamefont {Koenigstein}}, \bibinfo {author}
  {\bibfnamefont {D.~H.}\ \bibnamefont {Rischke}}, \bibinfo {author}
  {\bibfnamefont {M.~J.}\ \bibnamefont {Steil}},\ and\ \bibinfo {author}
  {\bibfnamefont {N.}~\bibnamefont {Wink}},\ }\href@noop {} {\bibinfo {title}
  {{Reanalysis of the phase diagram of the quark meson model in local potential
  approximation via the FRG}}},\ \bibinfo {howpublished} {in preparation}
  (\bibinfo {year} {2021}{\natexlab{b}})\BibitemShut {NoStop}%
\bibitem [{\citenamefont {Bateman}(1915)}]{Bateman1915}%
  \BibitemOpen
  \bibfield  {author} {\bibinfo {author} {\bibfnamefont {H.}~\bibnamefont
  {Bateman}},\ }\bibfield  {title} {\bibinfo {title} {{Some recent researches
  on the motion of fluids}},\ }\href
  {https://doi.org/10.1175/1520-0493(1915)43<163:SRROTM>2.0.CO;2} {\bibfield
  {journal} {\bibinfo  {journal} {Monthly Weather Review}\ }\textbf {\bibinfo
  {volume} {43}},\ \bibinfo {pages} {163} (\bibinfo {year} {1915})}\BibitemShut
  {NoStop}%
\bibitem [{\citenamefont {Burgers}(1948)}]{Burgers1948}%
  \BibitemOpen
  \bibfield  {author} {\bibinfo {author} {\bibfnamefont {J.~M.}\ \bibnamefont
  {Burgers}},\ }\bibfield  {title} {\bibinfo {title} {{A mathematical model
  illustrating the theory of turbulence}},\ }\href
  {https://doi.org/10.1016/S0065-2156(08)70100-5} {\bibfield  {journal}
  {\bibinfo  {journal} {Advances in Applied Mechanics}\ }\textbf {\bibinfo
  {volume} {1}},\ \bibinfo {pages} {171 } (\bibinfo {year} {1948})}\BibitemShut
  {NoStop}%
\bibitem [{\citenamefont {Ames}(1992)}]{Ames:1992}%
  \BibitemOpen
  \bibfield  {author} {\bibinfo {author} {\bibfnamefont {W.~F.}\ \bibnamefont
  {Ames}},\ }\href@noop {} {\emph {\bibinfo {title} {{Numerical Methods for
  Partial Differential Equations}}}},\ \bibinfo {edition} {3rd}\ ed.,\ Computer
  science and scientific computing\ (\bibinfo  {publisher} {Academic Press},\
  \bibinfo {address} {Boston {[u.a.]}},\ \bibinfo {year} {1992})\BibitemShut
  {NoStop}%
\bibitem [{\citenamefont {Hesthaven}\ and\ \citenamefont
  {Warburton}(2007)}]{Hesthaven2007}%
  \BibitemOpen
  \bibfield  {author} {\bibinfo {author} {\bibfnamefont {J.~S.}\ \bibnamefont
  {Hesthaven}}\ and\ \bibinfo {author} {\bibfnamefont {T.}~\bibnamefont
  {Warburton}},\ }\href {https://doi.org/10.5555/1557392} {\emph {\bibinfo
  {title} {{Nodal Discontinuous Galerkin Methods: Algorithms, Analysis, and
  Applications}}}},\ \bibinfo {edition} {1st}\ ed.\ (\bibinfo  {publisher}
  {Springer Publishing Company, Incorporated},\ \bibinfo {year}
  {2007})\BibitemShut {NoStop}%
\bibitem [{\citenamefont {Kadanoff}(1966)}]{Kadanoff:1966wm}%
  \BibitemOpen
  \bibfield  {author} {\bibinfo {author} {\bibfnamefont {L.~P.}\ \bibnamefont
  {Kadanoff}},\ }\bibfield  {title} {\bibinfo {title} {{Scaling laws for Ising
  models near $T_c$}},\ }\href
  {https://doi.org/10.1103/PhysicsPhysiqueFizika.2.263} {\bibfield  {journal}
  {\bibinfo  {journal} {Physics Physique Fizika}\ }\textbf {\bibinfo {volume}
  {2}},\ \bibinfo {pages} {263} (\bibinfo {year} {1966})}\BibitemShut {NoStop}%
\bibitem [{\citenamefont {Lebowitz}(2008)}]{Lebowitz:2008}%
  \BibitemOpen
  \bibfield  {author} {\bibinfo {author} {\bibfnamefont {J.~L.}\ \bibnamefont
  {Lebowitz}},\ }\bibfield  {title} {\bibinfo {title} {{Time's arrow and
  Boltzmann's entropy}},\ }\href {https://doi.org/10.4249/scholarpedia.3448}
  {\bibfield  {journal} {\bibinfo  {journal} {Scholarpedia}\ }\textbf {\bibinfo
  {volume} {3}},\ \bibinfo {pages} {3448} (\bibinfo {year} {2008})},\ \bibinfo
  {note} {revision \#137152, [Online; accessed 2021.02.01]}\BibitemShut
  {NoStop}%
\bibitem [{\citenamefont {Zumbach}(1993)}]{Zumbach:1993zz}%
  \BibitemOpen
  \bibfield  {author} {\bibinfo {author} {\bibfnamefont {G.}~\bibnamefont
  {Zumbach}},\ }\bibfield  {title} {\bibinfo {title} {{Almost second order
  phase transitions}},\ }\href {https://doi.org/10.1103/PhysRevLett.71.2421}
  {\bibfield  {journal} {\bibinfo  {journal} {Phys. Rev. Lett.}\ }\textbf
  {\bibinfo {volume} {71}},\ \bibinfo {pages} {2421} (\bibinfo {year}
  {1993})}\BibitemShut {NoStop}%
\bibitem [{\citenamefont {Politzer}(1973)}]{Politzer:1973fx}%
  \BibitemOpen
  \bibfield  {author} {\bibinfo {author} {\bibfnamefont {H.~D.}\ \bibnamefont
  {Politzer}},\ }\bibfield  {title} {\bibinfo {title} {{Reliable perturbative
  results for strong interactions?}},\ }\href
  {https://doi.org/10.1103/PhysRevLett.30.1346} {\bibfield  {journal} {\bibinfo
   {journal} {Phys. Rev. Lett.}\ }\textbf {\bibinfo {volume} {30}},\ \bibinfo
  {pages} {1346} (\bibinfo {year} {1973})}\BibitemShut {NoStop}%
\bibitem [{\citenamefont {Gross}\ and\ \citenamefont
  {Wilczek}(1973{\natexlab{a}})}]{Gross:1973id}%
  \BibitemOpen
  \bibfield  {author} {\bibinfo {author} {\bibfnamefont {D.~J.}\ \bibnamefont
  {Gross}}\ and\ \bibinfo {author} {\bibfnamefont {F.~A.}\ \bibnamefont
  {Wilczek}},\ }\bibfield  {title} {\bibinfo {title} {{Ultraviolet behavior of
  non-Abelian gauge theories}},\ }\href
  {https://doi.org/10.1103/PhysRevLett.30.1343} {\bibfield  {journal} {\bibinfo
   {journal} {Phys. Rev. Lett.}\ }\textbf {\bibinfo {volume} {30}},\ \bibinfo
  {pages} {1343} (\bibinfo {year} {1973}{\natexlab{a}})}\BibitemShut {NoStop}%
\bibitem [{\citenamefont {Gross}\ and\ \citenamefont
  {Wilczek}(1973{\natexlab{b}})}]{Gross:1973ju}%
  \BibitemOpen
  \bibfield  {author} {\bibinfo {author} {\bibfnamefont {D.~J.}\ \bibnamefont
  {Gross}}\ and\ \bibinfo {author} {\bibfnamefont {F.~A.}\ \bibnamefont
  {Wilczek}},\ }\bibfield  {title} {\bibinfo {title} {{Asymptotically free
  gauge theories I}},\ }\href {https://doi.org/10.1103/PhysRevD.8.3633}
  {\bibfield  {journal} {\bibinfo  {journal} {Phys. Rev. D}\ }\textbf {\bibinfo
  {volume} {8}},\ \bibinfo {pages} {3633} (\bibinfo {year}
  {1973}{\natexlab{b}})}\BibitemShut {NoStop}%
\bibitem [{\citenamefont {Gross}\ and\ \citenamefont
  {Wilczek}(1974)}]{Gross:1974cs}%
  \BibitemOpen
  \bibfield  {author} {\bibinfo {author} {\bibfnamefont {D.~J.}\ \bibnamefont
  {Gross}}\ and\ \bibinfo {author} {\bibfnamefont {F.~A.}\ \bibnamefont
  {Wilczek}},\ }\bibfield  {title} {\bibinfo {title} {{Asymptotically free
  gauge theories II}},\ }\href {https://doi.org/10.1103/PhysRevD.9.980}
  {\bibfield  {journal} {\bibinfo  {journal} {Phys. Rev. D}\ }\textbf {\bibinfo
  {volume} {9}},\ \bibinfo {pages} {980} (\bibinfo {year} {1974})}\BibitemShut
  {NoStop}%
\bibitem [{\citenamefont {Rankine}(1870)}]{Rankine:1870}%
  \BibitemOpen
  \bibfield  {author} {\bibinfo {author} {\bibfnamefont {W.~J.~M.}\
  \bibnamefont {Rankine}},\ }\bibfield  {title} {\bibinfo {title} {{On the
  thermodynamic theory of waves of finite longitudinal disturbance}},\ }\href
  {https://doi.org/10.1098/rstl.1870.0015} {\bibfield  {journal} {\bibinfo
  {journal} {Phil. Trans. R. Soc.}\ }\textbf {\bibinfo {volume} {160}},\
  \bibinfo {pages} {277} (\bibinfo {year} {1870})}\BibitemShut {NoStop}%
\bibitem [{\citenamefont {Hugoniot}(1998)}]{Hugoniot:1887}%
  \BibitemOpen
  \bibfield  {author} {\bibinfo {author} {\bibfnamefont {P.-H.}\ \bibnamefont
  {Hugoniot}},\ }\bibinfo {title} {On the propagation of motion in bodies and
  in perfect gases in particular -- $\mathrm{I}$},\ in\ \href
  {https://doi.org/10.1007/978-1-4612-2218-7_7} {\emph {\bibinfo {booktitle}
  {Classic papers in shock compression science}}},\ \bibinfo {editor} {edited
  by\ \bibinfo {editor} {\bibfnamefont {J.~N.}\ \bibnamefont {Johnson}}\ and\
  \bibinfo {editor} {\bibfnamefont {R.}~\bibnamefont {Ch{\'e}ret}}}\ (\bibinfo
  {publisher} {Springer},\ \bibinfo {address} {New York},\ \bibinfo {year}
  {1998})\ pp.\ \bibinfo {pages} {161--243},\ \bibinfo {note} {translation of:
  Hugoniot, Pierre-Henri, Journal de l`\'Ecole Polytechnique, Vol. \textbf{57},
  pp. 3 -- 97 (1887)}\BibitemShut {NoStop}%
\bibitem [{\citenamefont {Banks}\ and\ \citenamefont
  {Martinec}(1987)}]{Banks:1987qs}%
  \BibitemOpen
  \bibfield  {author} {\bibinfo {author} {\bibfnamefont {T.}~\bibnamefont
  {Banks}}\ and\ \bibinfo {author} {\bibfnamefont {E.~J.}\ \bibnamefont
  {Martinec}},\ }\bibfield  {title} {\bibinfo {title} {{The Renormalization
  Group and String Field Theory}},\ }\href
  {https://doi.org/10.1016/0550-3213(87)90605-5} {\bibfield  {journal}
  {\bibinfo  {journal} {Nucl. Phys. B}\ }\textbf {\bibinfo {volume} {294}},\
  \bibinfo {pages} {733} (\bibinfo {year} {1987})}\BibitemShut {NoStop}%
\bibitem [{\citenamefont {Cardy}(1988)}]{Cardy:1988cwa}%
  \BibitemOpen
  \bibfield  {author} {\bibinfo {author} {\bibfnamefont {J.~L.}\ \bibnamefont
  {Cardy}},\ }\bibfield  {title} {\bibinfo {title} {{Is there a $c$-theorem in
  four dimensions?}},\ }\href {https://doi.org/10.1016/0370-2693(88)90054-8}
  {\bibfield  {journal} {\bibinfo  {journal} {Phys. Lett. B}\ }\textbf
  {\bibinfo {volume} {215}},\ \bibinfo {pages} {749} (\bibinfo {year}
  {1988})}\BibitemShut {NoStop}%
\bibitem [{\citenamefont {Osborn}(1989)}]{Osborn:1989td}%
  \BibitemOpen
  \bibfield  {author} {\bibinfo {author} {\bibfnamefont {H.}~\bibnamefont
  {Osborn}},\ }\bibfield  {title} {\bibinfo {title} {{Derivation of a four
  dimensional $c$-theorem for renormaliseable quantum field theories}},\ }\href
  {https://doi.org/10.1016/0370-2693(89)90729-6} {\bibfield  {journal}
  {\bibinfo  {journal} {Phys. Lett. B}\ }\textbf {\bibinfo {volume} {222}},\
  \bibinfo {pages} {97} (\bibinfo {year} {1989})}\BibitemShut {NoStop}%
\bibitem [{\citenamefont {Jack}\ and\ \citenamefont
  {Osborn}(1990)}]{Jack:1990eb}%
  \BibitemOpen
  \bibfield  {author} {\bibinfo {author} {\bibfnamefont {I.}~\bibnamefont
  {Jack}}\ and\ \bibinfo {author} {\bibfnamefont {H.}~\bibnamefont {Osborn}},\
  }\bibfield  {title} {\bibinfo {title} {{Analogs for the $c$-theorem for
  four-dimensional renormalizable field theories}},\ }\href
  {https://doi.org/10.1016/0550-3213(90)90584-Z} {\bibfield  {journal}
  {\bibinfo  {journal} {Nucl. Phys. B}\ }\textbf {\bibinfo {volume} {343}},\
  \bibinfo {pages} {647} (\bibinfo {year} {1990})}\BibitemShut {NoStop}%
\bibitem [{\citenamefont {Komargodski}\ and\ \citenamefont
  {Schwimmer}(2011)}]{Komargodski:2011vj}%
  \BibitemOpen
  \bibfield  {author} {\bibinfo {author} {\bibfnamefont {Z.}~\bibnamefont
  {Komargodski}}\ and\ \bibinfo {author} {\bibfnamefont {A.}~\bibnamefont
  {Schwimmer}},\ }\bibfield  {title} {\bibinfo {title} {{On Renormalization
  Group flows in four dimensions}},\ }\href
  {https://doi.org/10.1007/JHEP12(2011)099} {\bibfield  {journal} {\bibinfo
  {journal} {JHEP}\ }\textbf {\bibinfo {volume} {12}},\ \bibinfo {pages}
  {099}},\ \Eprint {https://arxiv.org/abs/1107.3987} {arXiv:1107.3987 [hep-th]}
  \BibitemShut {NoStop}%
\bibitem [{\citenamefont {Curtright}\ \emph {et~al.}(2012)\citenamefont
  {Curtright}, \citenamefont {Jin},\ and\ \citenamefont
  {Zachos}}]{Curtright:2011qg}%
  \BibitemOpen
  \bibfield  {author} {\bibinfo {author} {\bibfnamefont {T.~L.}\ \bibnamefont
  {Curtright}}, \bibinfo {author} {\bibfnamefont {X.}~\bibnamefont {Jin}},\
  and\ \bibinfo {author} {\bibfnamefont {C.~K.}\ \bibnamefont {Zachos}},\
  }\bibfield  {title} {\bibinfo {title} {{RG flows, cycles, and $c$-theorem
  folklore}},\ }\href {https://doi.org/10.1103/PhysRevLett.108.131601}
  {\bibfield  {journal} {\bibinfo  {journal} {Phys. Rev. Lett.}\ }\textbf
  {\bibinfo {volume} {108}},\ \bibinfo {pages} {131601} (\bibinfo {year}
  {2012})},\ \Eprint {https://arxiv.org/abs/1111.2649} {arXiv:1111.2649
  [hep-th]} \BibitemShut {NoStop}%
\bibitem [{\citenamefont {Haagensen}\ \emph {et~al.}(1994)\citenamefont
  {Haagensen}, \citenamefont {Kubyshin}, \citenamefont {Latorre},\ and\
  \citenamefont {Moreno}}]{Haagensen:1993by}%
  \BibitemOpen
  \bibfield  {author} {\bibinfo {author} {\bibfnamefont {P.~E.}\ \bibnamefont
  {Haagensen}}, \bibinfo {author} {\bibfnamefont {Y.}~\bibnamefont {Kubyshin}},
  \bibinfo {author} {\bibfnamefont {J.~I.}\ \bibnamefont {Latorre}},\ and\
  \bibinfo {author} {\bibfnamefont {E.}~\bibnamefont {Moreno}},\ }\bibfield
  {title} {\bibinfo {title} {{Gradient flows from an approximation to the Exact
  Renormalization Group}},\ }\href
  {https://doi.org/10.1016/0370-2693(94)91228-9} {\bibfield  {journal}
  {\bibinfo  {journal} {Phys. Lett. B}\ }\textbf {\bibinfo {volume} {323}},\
  \bibinfo {pages} {330} (\bibinfo {year} {1994})},\ \Eprint
  {https://arxiv.org/abs/hep-th/9310032} {arXiv:hep-th/9310032} \BibitemShut
  {NoStop}%
\bibitem [{\citenamefont {Generowicz}\ \emph {et~al.}(1997)\citenamefont
  {Generowicz}, \citenamefont {Harvey-Fros},\ and\ \citenamefont
  {Morris}}]{Generowicz:1997he}%
  \BibitemOpen
  \bibfield  {author} {\bibinfo {author} {\bibfnamefont {J.}~\bibnamefont
  {Generowicz}}, \bibinfo {author} {\bibfnamefont {C.}~\bibnamefont
  {Harvey-Fros}},\ and\ \bibinfo {author} {\bibfnamefont {T.~R.}\ \bibnamefont
  {Morris}},\ }\bibfield  {title} {\bibinfo {title} {{$C$ function
  representation of the Local Potential Approximation}},\ }\href
  {https://doi.org/10.1016/S0370-2693(97)00729-6} {\bibfield  {journal}
  {\bibinfo  {journal} {Phys. Lett. B}\ }\textbf {\bibinfo {volume} {407}},\
  \bibinfo {pages} {27} (\bibinfo {year} {1997})},\ \Eprint
  {https://arxiv.org/abs/hep-th/9705088} {arXiv:hep-th/9705088} \BibitemShut
  {NoStop}%
\bibitem [{\citenamefont {Forte}\ and\ \citenamefont
  {Latorre}(1998)}]{Forte:1998dx}%
  \BibitemOpen
  \bibfield  {author} {\bibinfo {author} {\bibfnamefont {S.}~\bibnamefont
  {Forte}}\ and\ \bibinfo {author} {\bibfnamefont {J.~I.}\ \bibnamefont
  {Latorre}},\ }\bibfield  {title} {\bibinfo {title} {{A proof of the
  irreversibility of Renormalization Group flows in four-dimensions}},\ }\href
  {https://doi.org/10.1016/S0550-3213(98)00631-2} {\bibfield  {journal}
  {\bibinfo  {journal} {Nucl. Phys. B}\ }\textbf {\bibinfo {volume} {535}},\
  \bibinfo {pages} {709} (\bibinfo {year} {1998})},\ \Eprint
  {https://arxiv.org/abs/hep-th/9805015} {arXiv:hep-th/9805015} \BibitemShut
  {NoStop}%
\bibitem [{\citenamefont {Codello}\ \emph {et~al.}(2014)\citenamefont
  {Codello}, \citenamefont {D'Odorico},\ and\ \citenamefont
  {Pagani}}]{Codello:2013iqa}%
  \BibitemOpen
  \bibfield  {author} {\bibinfo {author} {\bibfnamefont {A.}~\bibnamefont
  {Codello}}, \bibinfo {author} {\bibfnamefont {G.}~\bibnamefont {D'Odorico}},\
  and\ \bibinfo {author} {\bibfnamefont {C.}~\bibnamefont {Pagani}},\
  }\bibfield  {title} {\bibinfo {title} {{A Functional RG equation for the
  $c$-function}},\ }\href {https://doi.org/10.1007/JHEP07(2014)040} {\bibfield
  {journal} {\bibinfo  {journal} {JHEP}\ }\textbf {\bibinfo {volume} {07}},\
  \bibinfo {pages} {040}},\ \Eprint {https://arxiv.org/abs/1312.7097}
  {arXiv:1312.7097 [hep-th]} \BibitemShut {NoStop}%
\bibitem [{\citenamefont {Codello}\ \emph {et~al.}(2015)\citenamefont
  {Codello}, \citenamefont {D'Odorico},\ and\ \citenamefont
  {Pagani}}]{Codello:2015ana}%
  \BibitemOpen
  \bibfield  {author} {\bibinfo {author} {\bibfnamefont {A.}~\bibnamefont
  {Codello}}, \bibinfo {author} {\bibfnamefont {G.}~\bibnamefont {D'Odorico}},\
  and\ \bibinfo {author} {\bibfnamefont {C.}~\bibnamefont {Pagani}},\
  }\bibfield  {title} {\bibinfo {title} {{Functional and local Renormalization
  Groups}},\ }\href {https://doi.org/10.1103/PhysRevD.91.125016} {\bibfield
  {journal} {\bibinfo  {journal} {Phys. Rev. D}\ }\textbf {\bibinfo {volume}
  {91}},\ \bibinfo {pages} {125016} (\bibinfo {year} {2015})},\ \Eprint
  {https://arxiv.org/abs/1502.02439} {arXiv:1502.02439 [hep-th]} \BibitemShut
  {NoStop}%
\bibitem [{\citenamefont {Becker}\ and\ \citenamefont
  {Reuter}(2015)}]{Becker:2014pea}%
  \BibitemOpen
  \bibfield  {author} {\bibinfo {author} {\bibfnamefont {D.}~\bibnamefont
  {Becker}}\ and\ \bibinfo {author} {\bibfnamefont {M.}~\bibnamefont
  {Reuter}},\ }\bibfield  {title} {\bibinfo {title} {{Towards a $C$-function in
  $4D$ quantum gravity}},\ }\href {https://doi.org/10.1007/JHEP03(2015)065}
  {\bibfield  {journal} {\bibinfo  {journal} {JHEP}\ }\textbf {\bibinfo
  {volume} {03}},\ \bibinfo {pages} {065}},\ \Eprint
  {https://arxiv.org/abs/1412.0468} {arXiv:1412.0468 [hep-th]} \BibitemShut
  {NoStop}%
\bibitem [{\citenamefont {Becker}(2016)}]{Becker:2016zcn}%
  \BibitemOpen
  \bibfield  {author} {\bibinfo {author} {\bibfnamefont {D.}~\bibnamefont
  {Becker}},\ }\emph {\bibinfo {title} {{Asymptotically safe quantum gravity:
  Bimetric actions, boundary terms, $\&$ a C-function}}},\ \href
  {https://doi.org/10.25358/openscience-2831} {\bibinfo {type} {Phd thesis}},\
  \bibinfo  {school} {Johannes Gutenberg University Mainz} (\bibinfo {year}
  {2016})\BibitemShut {NoStop}%
\bibitem [{\citenamefont {Borchardt}\ and\ \citenamefont
  {Knorr}(2015)}]{Borchardt:2015rxa}%
  \BibitemOpen
  \bibfield  {author} {\bibinfo {author} {\bibfnamefont {J.}~\bibnamefont
  {Borchardt}}\ and\ \bibinfo {author} {\bibfnamefont {B.}~\bibnamefont
  {Knorr}},\ }\bibfield  {title} {\bibinfo {title} {{Global solutions of
  functional fixed point equations via pseudospectral methods}},\ }\href
  {https://doi.org/10.1103/PhysRevD.91.105011} {\bibfield  {journal} {\bibinfo
  {journal} {Phys. Rev. D}\ }\textbf {\bibinfo {volume} {91}},\ \bibinfo
  {pages} {105011} (\bibinfo {year} {2015})},\ \bibinfo {note} {[Erratum:
  Phys.Rev.D 93, 089904 (2016)]},\ \Eprint {https://arxiv.org/abs/1502.07511}
  {arXiv:1502.07511 [hep-th]} \BibitemShut {NoStop}%
\bibitem [{\citenamefont {Strodthoff}\ \emph {et~al.}(2012)\citenamefont
  {Strodthoff}, \citenamefont {Schaefer},\ and\ \citenamefont {von
  Smekal}}]{Strodthoff:2011tz}%
  \BibitemOpen
  \bibfield  {author} {\bibinfo {author} {\bibfnamefont {N.}~\bibnamefont
  {Strodthoff}}, \bibinfo {author} {\bibfnamefont {B.-J.}\ \bibnamefont
  {Schaefer}},\ and\ \bibinfo {author} {\bibfnamefont {L.}~\bibnamefont {von
  Smekal}},\ }\bibfield  {title} {\bibinfo {title} {{Quark-meson-diquark model
  for two-color QCD}},\ }\href {https://doi.org/10.1103/PhysRevD.85.074007}
  {\bibfield  {journal} {\bibinfo  {journal} {Phys. Rev. D}\ }\textbf {\bibinfo
  {volume} {85}},\ \bibinfo {pages} {074007} (\bibinfo {year} {2012})},\
  \Eprint {https://arxiv.org/abs/1112.5401} {arXiv:1112.5401 [hep-ph]}
  \BibitemShut {NoStop}%
\bibitem [{\citenamefont {Mitter}\ and\ \citenamefont
  {Schaefer}(2014)}]{Mitter:2013fxa}%
  \BibitemOpen
  \bibfield  {author} {\bibinfo {author} {\bibfnamefont {M.}~\bibnamefont
  {Mitter}}\ and\ \bibinfo {author} {\bibfnamefont {B.-J.}\ \bibnamefont
  {Schaefer}},\ }\bibfield  {title} {\bibinfo {title} {{Fluctuations and the
  axial anomaly with three quark flavors}},\ }\href
  {https://doi.org/10.1103/PhysRevD.89.054027} {\bibfield  {journal} {\bibinfo
  {journal} {Phys. Rev. D}\ }\textbf {\bibinfo {volume} {89}},\ \bibinfo
  {pages} {054027} (\bibinfo {year} {2014})},\ \Eprint
  {https://arxiv.org/abs/1308.3176} {arXiv:1308.3176 [hep-ph]} \BibitemShut
  {NoStop}%
\bibitem [{\citenamefont {Rennecke}\ and\ \citenamefont
  {Schaefer}(2017)}]{Rennecke:2016tkm}%
  \BibitemOpen
  \bibfield  {author} {\bibinfo {author} {\bibfnamefont {F.}~\bibnamefont
  {Rennecke}}\ and\ \bibinfo {author} {\bibfnamefont {B.-J.}\ \bibnamefont
  {Schaefer}},\ }\bibfield  {title} {\bibinfo {title} {{Fluctuation-induced
  modifications of the phase structure in $( 2 + 1 )$-flavor QCD}},\ }\href
  {https://doi.org/10.1103/PhysRevD.96.016009} {\bibfield  {journal} {\bibinfo
  {journal} {Phys. Rev. D}\ }\textbf {\bibinfo {volume} {96}},\ \bibinfo
  {pages} {016009} (\bibinfo {year} {2017})},\ \Eprint
  {https://arxiv.org/abs/1610.08748} {arXiv:1610.08748 [hep-ph]} \BibitemShut
  {NoStop}%
\bibitem [{\citenamefont {Lakaschus}\ \emph {et~al.}(2021)\citenamefont
  {Lakaschus}, \citenamefont {Buballa},\ and\ \citenamefont
  {Rischke}}]{Lakaschus:2020caq}%
  \BibitemOpen
  \bibfield  {author} {\bibinfo {author} {\bibfnamefont {P.}~\bibnamefont
  {Lakaschus}}, \bibinfo {author} {\bibfnamefont {M.}~\bibnamefont {Buballa}},\
  and\ \bibinfo {author} {\bibfnamefont {D.~H.}\ \bibnamefont {Rischke}},\
  }\bibfield  {title} {\bibinfo {title} {{Competition of inhomogeneous chiral
  phases and two-flavor color superconductivity in the NJL model}},\ }\href
  {https://doi.org/10.1103/PhysRevD.103.034030} {\bibfield  {journal} {\bibinfo
   {journal} {Phys. Rev. D}\ }\textbf {\bibinfo {volume} {103}},\ \bibinfo
  {pages} {034030} (\bibinfo {year} {2021})},\ \Eprint
  {https://arxiv.org/abs/2012.07520} {arXiv:2012.07520 [hep-ph]} \BibitemShut
  {NoStop}%
\bibitem [{\citenamefont {Fukushima}\ \emph {et~al.}(2011)\citenamefont
  {Fukushima}, \citenamefont {Kamikado},\ and\ \citenamefont
  {Klein}}]{Fukushima:2010ji}%
  \BibitemOpen
  \bibfield  {author} {\bibinfo {author} {\bibfnamefont {K.}~\bibnamefont
  {Fukushima}}, \bibinfo {author} {\bibfnamefont {K.}~\bibnamefont
  {Kamikado}},\ and\ \bibinfo {author} {\bibfnamefont {B.}~\bibnamefont
  {Klein}},\ }\bibfield  {title} {\bibinfo {title} {{Second-order and
  Fluctuation-induced First-order Phase Transitions with Functional
  Renormalization Group Equations}},\ }\href
  {https://doi.org/10.1103/PhysRevD.83.116005} {\bibfield  {journal} {\bibinfo
  {journal} {Phys. Rev. D}\ }\textbf {\bibinfo {volume} {83}},\ \bibinfo
  {pages} {116005} (\bibinfo {year} {2011})},\ \Eprint
  {https://arxiv.org/abs/1010.6226} {arXiv:1010.6226 [hep-ph]} \BibitemShut
  {NoStop}%
\bibitem [{\citenamefont {Fej\H{o}s}\ and\ \citenamefont
  {Patk\'os}(2021)}]{Fejos:2020lli}%
  \BibitemOpen
  \bibfield  {author} {\bibinfo {author} {\bibfnamefont {G.}~\bibnamefont
  {Fej\H{o}s}}\ and\ \bibinfo {author} {\bibfnamefont {A.}~\bibnamefont
  {Patk\'os}},\ }\bibfield  {title} {\bibinfo {title} {{Field dependence of the
  Yukawa coupling in the three flavor quark-meson model}},\ }\href
  {https://doi.org/10.1103/PhysRevD.103.056015} {\bibfield  {journal} {\bibinfo
   {journal} {Phys. Rev. D}\ }\textbf {\bibinfo {volume} {103}},\ \bibinfo
  {pages} {056015} (\bibinfo {year} {2021})},\ \Eprint
  {https://arxiv.org/abs/2011.08387} {arXiv:2011.08387 [hep-ph]} \BibitemShut
  {NoStop}%
\bibitem [{\citenamefont {Lakaschus}(2021)}]{Lakaschus:2021ewd}%
  \BibitemOpen
  \bibfield  {author} {\bibinfo {author} {\bibfnamefont {P.}~\bibnamefont
  {Lakaschus}},\ }\emph {\bibinfo {title} {{Inhomogeneous chiral condensates in
  low-energy color-superconductivity models of QCD}}},\ \href@noop {} {\bibinfo
  {type} {Phd thesis}},\ \bibinfo  {school} {Goethe University Frankfurt}
  (\bibinfo {year} {2021})\BibitemShut {NoStop}%
\bibitem [{\citenamefont {Godunov}(1959)}]{Godunov}%
  \BibitemOpen
  \bibfield  {author} {\bibinfo {author} {\bibfnamefont {S.~K.}\ \bibnamefont
  {Godunov}},\ }\bibfield  {title} {\bibinfo {title} {{Eine Differenzenmethode
  für die Näherungsberechnung unstetiger Lösungen der hydrodynamischen
  Gleichungen}},\ }\href@noop {} {\bibfield  {journal} {\bibinfo  {journal}
  {{Mat. Sb., Nov. Ser.}}\ }\textbf {\bibinfo {volume} {47}},\ \bibinfo {pages}
  {271} (\bibinfo {year} {1959})}\BibitemShut {NoStop}%
\bibitem [{\citenamefont {Roe}(1997)}]{ROE1981357}%
  \BibitemOpen
  \bibfield  {author} {\bibinfo {author} {\bibfnamefont {P.~L.}\ \bibnamefont
  {Roe}},\ }\bibfield  {title} {\bibinfo {title} {{Approximate Riemann solvers,
  parameter vectors, and difference schemes}},\ }\href
  {https://doi.org/10.1006/jcph.1997.5705} {\bibfield  {journal} {\bibinfo
  {journal} {Journal of Computational Physics}\ }\textbf {\bibinfo {volume}
  {135}},\ \bibinfo {pages} {250} (\bibinfo {year} {1997})}\BibitemShut
  {NoStop}%
\bibitem [{\citenamefont {Harten}\ \emph {et~al.}(1983)\citenamefont {Harten},
  \citenamefont {Lax},\ and\ \citenamefont {{van Leer}}}]{HLLE1}%
  \BibitemOpen
  \bibfield  {author} {\bibinfo {author} {\bibfnamefont {A.}~\bibnamefont
  {Harten}}, \bibinfo {author} {\bibfnamefont {P.~D.}\ \bibnamefont {Lax}},\
  and\ \bibinfo {author} {\bibfnamefont {B.}~\bibnamefont {{van Leer}}},\
  }\bibfield  {title} {\bibinfo {title} {{On upstream differencing and
  Godunov-type schemes for hyperbolic conservation laws}},\ }\href
  {https://doi.org/10.1137/1025002} {\bibfield  {journal} {\bibinfo  {journal}
  {SIAM Review}\ }\textbf {\bibinfo {volume} {25}},\ \bibinfo {pages} {35}
  (\bibinfo {year} {1983})}\BibitemShut {NoStop}%
\bibitem [{\citenamefont {Einfeldt}(1988)}]{HLLE2}%
  \BibitemOpen
  \bibfield  {author} {\bibinfo {author} {\bibfnamefont {B.}~\bibnamefont
  {Einfeldt}},\ }\bibfield  {title} {\bibinfo {title} {{On Godunov-type Methods
  for Gas Dynamics}},\ }\href {https://doi.org/10.1137/0725021} {\bibfield
  {journal} {\bibinfo  {journal} {SIAM Journal on Numerical Analysis}\ }\textbf
  {\bibinfo {volume} {25}},\ \bibinfo {pages} {294} (\bibinfo {year}
  {1988})}\BibitemShut {NoStop}%
\bibitem [{\citenamefont {Kurganov}\ and\ \citenamefont
  {Tadmor}(2000)}]{KTO2-0}%
  \BibitemOpen
  \bibfield  {author} {\bibinfo {author} {\bibfnamefont {A.}~\bibnamefont
  {Kurganov}}\ and\ \bibinfo {author} {\bibfnamefont {E.}~\bibnamefont
  {Tadmor}},\ }\bibfield  {title} {\bibinfo {title} {{New High-Resolution
  Central Schemes for Nonlinear Conservation Laws and Convection–Diffusion
  Equations}},\ }\href {https://doi.org/10.1006/jcph.2000.6459} {\bibfield
  {journal} {\bibinfo  {journal} {Journal of Computational Physics}\ }\textbf
  {\bibinfo {volume} {160}},\ \bibinfo {pages} {241 } (\bibinfo {year}
  {2000})}\BibitemShut {NoStop}%
\bibitem [{\citenamefont {Nessyahu}\ and\ \citenamefont {Tadmor}(1990)}]{NT}%
  \BibitemOpen
  \bibfield  {author} {\bibinfo {author} {\bibfnamefont {H.}~\bibnamefont
  {Nessyahu}}\ and\ \bibinfo {author} {\bibfnamefont {E.}~\bibnamefont
  {Tadmor}},\ }\bibfield  {title} {\bibinfo {title} {{Non-oscillatory central
  differencing for hyperbolic conservation laws}},\ }\href
  {https://doi.org/10.1016/0021-9991(90)90260-8} {\bibfield  {journal}
  {\bibinfo  {journal} {Journal of Computational Physics}\ }\textbf {\bibinfo
  {volume} {87}},\ \bibinfo {pages} {408 } (\bibinfo {year}
  {1990})}\BibitemShut {NoStop}%
\bibitem [{\citenamefont {{van Leer}}(1979)}]{MUSCL}%
  \BibitemOpen
  \bibfield  {author} {\bibinfo {author} {\bibfnamefont {B.}~\bibnamefont {{van
  Leer}}},\ }\bibfield  {title} {\bibinfo {title} {Towards the ultimate
  conservative difference scheme. v. a second-order sequel to godunov's
  method},\ }\href {https://doi.org/10.1016/0021-9991(79)90145-1} {\bibfield
  {journal} {\bibinfo  {journal} {Journal of Computational Physics}\ }\textbf
  {\bibinfo {volume} {32}},\ \bibinfo {pages} {101 } (\bibinfo {year}
  {1979})}\BibitemShut {NoStop}%
\bibitem [{\citenamefont {Lax}(1954)}]{LxF1}%
  \BibitemOpen
  \bibfield  {author} {\bibinfo {author} {\bibfnamefont {P.~D.}\ \bibnamefont
  {Lax}},\ }\bibfield  {title} {\bibinfo {title} {Weak solutions of nonlinear
  hyperbolic equations and their numerical computation},\ }\href
  {https://doi.org/10.1002/cpa.3160070112} {\bibfield  {journal} {\bibinfo
  {journal} {Communications on Pure and Applied Mathematics}\ }\textbf
  {\bibinfo {volume} {7}},\ \bibinfo {pages} {159} (\bibinfo {year}
  {1954})}\BibitemShut {NoStop}%
\bibitem [{\citenamefont {Friedrichs}(1954)}]{LxF2}%
  \BibitemOpen
  \bibfield  {author} {\bibinfo {author} {\bibfnamefont {K.~O.}\ \bibnamefont
  {Friedrichs}},\ }\bibfield  {title} {\bibinfo {title} {Symmetric hyperbolic
  linear differential equations},\ }\href
  {https://doi.org/10.1002/cpa.3160070206} {\bibfield  {journal} {\bibinfo
  {journal} {Communications on Pure and Applied Mathematics}\ }\textbf
  {\bibinfo {volume} {7}},\ \bibinfo {pages} {345} (\bibinfo {year}
  {1954})}\BibitemShut {NoStop}%
\bibitem [{\citenamefont {Tang}\ and\ \citenamefont
  {Tang}(2003)}]{KTmovingMesh}%
  \BibitemOpen
  \bibfield  {author} {\bibinfo {author} {\bibfnamefont {H.}~\bibnamefont
  {Tang}}\ and\ \bibinfo {author} {\bibfnamefont {T.}~\bibnamefont {Tang}},\
  }\bibfield  {title} {\bibinfo {title} {{Adaptive Mesh Methods for One- and
  Two-Dimensional Hyperbolic Conservation Laws}},\ }\href
  {https://doi.org/10.1137/S003614290138437X} {\bibfield  {journal} {\bibinfo
  {journal} {SIAM Journal on Numerical Analysis}\ }\textbf {\bibinfo {volume}
  {41}},\ \bibinfo {pages} {487} (\bibinfo {year} {2003})}\BibitemShut
  {NoStop}%
\bibitem [{\citenamefont {Harten}(1983)}]{HARTEN1983357}%
  \BibitemOpen
  \bibfield  {author} {\bibinfo {author} {\bibfnamefont {A.}~\bibnamefont
  {Harten}},\ }\bibfield  {title} {\bibinfo {title} {{High resolution schemes
  for hyperbolic conservation laws}},\ }\href
  {https://doi.org/https://doi.org/10.1016/0021-9991(83)90136-5} {\bibfield
  {journal} {\bibinfo  {journal} {Journal of Computational Physics}\ }\textbf
  {\bibinfo {volume} {49}},\ \bibinfo {pages} {357 } (\bibinfo {year}
  {1983})}\BibitemShut {NoStop}%
\bibitem [{\citenamefont {{Wikipedia contributors}}(2020)}]{wikiFluxLimiter}%
  \BibitemOpen
  \bibfield  {author} {\bibinfo {author} {\bibnamefont {{Wikipedia
  contributors}}},\ }\href {https://en.wikipedia.org/wiki/Flux_limiter}
  {\bibinfo {title} {{Flux limiter}}} (\bibinfo {year} {2020}),\ \bibinfo
  {note} {[Online; accessed 2020.10.15]}\BibitemShut {NoStop}%
\bibitem [{\citenamefont {Roe}(1986)}]{MinModRoe}%
  \BibitemOpen
  \bibfield  {author} {\bibinfo {author} {\bibfnamefont {P.~L.}\ \bibnamefont
  {Roe}},\ }\bibfield  {title} {\bibinfo {title} {{Characteristic-Based Schemes
  for the Euler Equations}},\ }\href
  {https://doi.org/10.1146/annurev.fl.18.010186.002005} {\bibfield  {journal}
  {\bibinfo  {journal} {Annual Review of Fluid Mechanics}\ }\textbf {\bibinfo
  {volume} {18}},\ \bibinfo {pages} {337} (\bibinfo {year} {1986})}\BibitemShut
  {NoStop}%
\bibitem [{\citenamefont {Levy}\ \emph {et~al.}(1999)\citenamefont {Levy},
  \citenamefont {Puppo},\ and\ \citenamefont {Russo}}]{WENO-C}%
  \BibitemOpen
  \bibfield  {author} {\bibinfo {author} {\bibfnamefont {D.}~\bibnamefont
  {Levy}}, \bibinfo {author} {\bibfnamefont {G.}~\bibnamefont {Puppo}},\ and\
  \bibinfo {author} {\bibfnamefont {G.}~\bibnamefont {Russo}},\ }\bibfield
  {title} {\bibinfo {title} {{Central WENO Schemes for hyperbolic systems of
  conservation laws}},\ }\href {https://doi.org/10.1051/m2an:1999152}
  {\bibfield  {journal} {\bibinfo  {journal} {Mathematical Modelling and
  Numerical Analysis}\ }\textbf {\bibinfo {volume} {33}},\ \bibinfo {pages}
  {547} (\bibinfo {year} {1999})}\BibitemShut {NoStop}%
\bibitem [{\citenamefont {Levy}\ \emph {et~al.}(2000)\citenamefont {Levy},
  \citenamefont {Puppo},\ and\ \citenamefont {Russo}}]{WENO-C2}%
  \BibitemOpen
  \bibfield  {author} {\bibinfo {author} {\bibfnamefont {D.}~\bibnamefont
  {Levy}}, \bibinfo {author} {\bibfnamefont {G.}~\bibnamefont {Puppo}},\ and\
  \bibinfo {author} {\bibfnamefont {G.}~\bibnamefont {Russo}},\ }\bibfield
  {title} {\bibinfo {title} {{Compact Central WENO Schemes for Multidimensional
  Conservation Laws}},\ }\href {https://doi.org/10.1137/S1064827599359461}
  {\bibfield  {journal} {\bibinfo  {journal} {Siam Journal on Scientific
  Computing}\ }\textbf {\bibinfo {volume} {22}} (\bibinfo {year} {2000})},\
  \Eprint {https://arxiv.org/abs/math/9911089} {arXiv:math/9911089 [cs.NA]}
  \BibitemShut {NoStop}%
\bibitem [{\citenamefont {Kurganov}\ and\ \citenamefont {Levy}(2000)}]{KTO3-0}%
  \BibitemOpen
  \bibfield  {author} {\bibinfo {author} {\bibfnamefont {A.}~\bibnamefont
  {Kurganov}}\ and\ \bibinfo {author} {\bibfnamefont {D.}~\bibnamefont
  {Levy}},\ }\bibfield  {title} {\bibinfo {title} {{A Third-Order Semi-Discrete
  Central Scheme for Conservation Laws and Convection-Diffusion Equations}},\
  }\href {https://doi.org/10.1137/S1064827599360236} {\bibfield  {journal}
  {\bibinfo  {journal} {SIAM Journal on Scientific Computing}\ }\textbf
  {\bibinfo {volume} {22}} (\bibinfo {year} {2000})},\ \Eprint
  {https://arxiv.org/abs/math/0002133} {arXiv:math/0002133 [math.NA]}
  \BibitemShut {NoStop}%
\bibitem [{\citenamefont {Jiang}\ and\ \citenamefont {Shu}(1996)}]{WENO2}%
  \BibitemOpen
  \bibfield  {author} {\bibinfo {author} {\bibfnamefont {G.-S.}\ \bibnamefont
  {Jiang}}\ and\ \bibinfo {author} {\bibfnamefont {C.-W.}\ \bibnamefont
  {Shu}},\ }\bibfield  {title} {\bibinfo {title} {{Efficient Implementation of
  Weighted ENO Schemes}},\ }\href {https://doi.org/10.1006/jcph.1996.0130}
  {\bibfield  {journal} {\bibinfo  {journal} {Journal of Computational
  Physics}\ }\textbf {\bibinfo {volume} {126}},\ \bibinfo {pages} {202 }
  (\bibinfo {year} {1996})}\BibitemShut {NoStop}%
\bibitem [{\citenamefont {Shi}\ \emph {et~al.}(2002)\citenamefont {Shi},
  \citenamefont {Hu},\ and\ \citenamefont {Shu}}]{WENO5}%
  \BibitemOpen
  \bibfield  {author} {\bibinfo {author} {\bibfnamefont {J.}~\bibnamefont
  {Shi}}, \bibinfo {author} {\bibfnamefont {C.}~\bibnamefont {Hu}},\ and\
  \bibinfo {author} {\bibfnamefont {C.-W.}\ \bibnamefont {Shu}},\ }\bibfield
  {title} {\bibinfo {title} {{A Technique of Treating Negative Weights in WENO
  Schemes}},\ }\href {https://doi.org/10.1006/jcph.2001.6892} {\bibfield
  {journal} {\bibinfo  {journal} {J. Comput. Phys.}\ }\textbf {\bibinfo
  {volume} {175}},\ \bibinfo {pages} {108–127} (\bibinfo {year}
  {2002})}\BibitemShut {NoStop}%
\bibitem [{\citenamefont {Kurganov}\ \emph {et~al.}(2007)\citenamefont
  {Kurganov}, \citenamefont {Petrova},\ and\ \citenamefont {Popov}}]{KTO5-0}%
  \BibitemOpen
  \bibfield  {author} {\bibinfo {author} {\bibfnamefont {A.}~\bibnamefont
  {Kurganov}}, \bibinfo {author} {\bibfnamefont {G.}~\bibnamefont {Petrova}},\
  and\ \bibinfo {author} {\bibfnamefont {B.}~\bibnamefont {Popov}},\ }\bibfield
   {title} {\bibinfo {title} {{Adaptive Semidiscrete Central-Upwind Schemes for
  Nonconvex Hyperbolic Conservation Laws}},\ }\href
  {https://doi.org/10.1137/040614189} {\bibfield  {journal} {\bibinfo
  {journal} {SIAM J. Scientific Computing}\ }\textbf {\bibinfo {volume} {29}},\
  \bibinfo {pages} {2381} (\bibinfo {year} {2007})}\BibitemShut {NoStop}%
\bibitem [{\citenamefont {Suresh}\ and\ \citenamefont {Huynh}(1997)}]{MP5}%
  \BibitemOpen
  \bibfield  {author} {\bibinfo {author} {\bibfnamefont {A.}~\bibnamefont
  {Suresh}}\ and\ \bibinfo {author} {\bibfnamefont {H.~T.}\ \bibnamefont
  {Huynh}},\ }\bibfield  {title} {\bibinfo {title} {{Accurate
  Monotonicity-Preserving Schemes with Runge–Kutta Time Stepping}},\ }\href
  {https://doi.org/https://doi.org/10.1006/jcph.1997.5745} {\bibfield
  {journal} {\bibinfo  {journal} {J. Comput. Phys.}\ }\textbf {\bibinfo
  {volume} {136}},\ \bibinfo {pages} {83 } (\bibinfo {year}
  {1997})}\BibitemShut {NoStop}%
\bibitem [{\citenamefont {Jourdan}\ \emph {et~al.}(2016)\citenamefont
  {Jourdan}, \citenamefont {Stoltz}, \citenamefont {Legoll},\ and\
  \citenamefont {Monasse}}]{KT-MP5}%
  \BibitemOpen
  \bibfield  {author} {\bibinfo {author} {\bibfnamefont {T.}~\bibnamefont
  {Jourdan}}, \bibinfo {author} {\bibfnamefont {G.}~\bibnamefont {Stoltz}},
  \bibinfo {author} {\bibfnamefont {F.}~\bibnamefont {Legoll}},\ and\ \bibinfo
  {author} {\bibfnamefont {L.}~\bibnamefont {Monasse}},\ }\bibfield  {title}
  {\bibinfo {title} {{An accurate scheme to solve cluster dynamics equations
  using a Fokker–Planck approach}},\ }\href
  {https://doi.org/https://doi.org/10.1016/j.cpc.2016.06.001} {\bibfield
  {journal} {\bibinfo  {journal} {Comput. Phys. Commun.}\ }\textbf {\bibinfo
  {volume} {207}},\ \bibinfo {pages} {170 } (\bibinfo {year}
  {2016})}\BibitemShut {NoStop}%
\bibitem [{\citenamefont {Liu}\ \emph {et~al.}(1994)\citenamefont {Liu},
  \citenamefont {Osher},\ and\ \citenamefont {Chan}}]{WENO}%
  \BibitemOpen
  \bibfield  {author} {\bibinfo {author} {\bibfnamefont {X.-D.}\ \bibnamefont
  {Liu}}, \bibinfo {author} {\bibfnamefont {S.}~\bibnamefont {Osher}},\ and\
  \bibinfo {author} {\bibfnamefont {T.}~\bibnamefont {Chan}},\ }\bibfield
  {title} {\bibinfo {title} {{Weighted Essentially Non-oscillatory Schemes}},\
  }\href {https://doi.org/10.1006/jcph.1994.1187} {\bibfield  {journal}
  {\bibinfo  {journal} {Journal of Computational Physics}\ }\textbf {\bibinfo
  {volume} {115}},\ \bibinfo {pages} {200 } (\bibinfo {year}
  {1994})}\BibitemShut {NoStop}%
\bibitem [{\citenamefont {M\"osta}\ \emph {et~al.}(2014)\citenamefont
  {M\"osta}, \citenamefont {Mundim}, \citenamefont {Faber}, \citenamefont
  {Haas}, \citenamefont {Noble}, \citenamefont {Bode}, \citenamefont
  {L\"offler}, \citenamefont {Ott}, \citenamefont {Reisswig},\ and\
  \citenamefont {Schnetter}}]{GRHydro}%
  \BibitemOpen
  \bibfield  {author} {\bibinfo {author} {\bibfnamefont {P.}~\bibnamefont
  {M\"osta}}, \bibinfo {author} {\bibfnamefont {B.~C.}\ \bibnamefont {Mundim}},
  \bibinfo {author} {\bibfnamefont {J.~A.}\ \bibnamefont {Faber}}, \bibinfo
  {author} {\bibfnamefont {R.}~\bibnamefont {Haas}}, \bibinfo {author}
  {\bibfnamefont {S.~C.}\ \bibnamefont {Noble}}, \bibinfo {author}
  {\bibfnamefont {T.}~\bibnamefont {Bode}}, \bibinfo {author} {\bibfnamefont
  {F.}~\bibnamefont {L\"offler}}, \bibinfo {author} {\bibfnamefont {C.~D.}\
  \bibnamefont {Ott}}, \bibinfo {author} {\bibfnamefont {C.}~\bibnamefont
  {Reisswig}},\ and\ \bibinfo {author} {\bibfnamefont {E.}~\bibnamefont
  {Schnetter}},\ }\bibfield  {title} {\bibinfo {title} {{GRHydro: A new open
  source general-relativistic magnetohydrodynamics code for the Einstein
  Toolkit}},\ }\href {https://doi.org/10.1088/0264-9381/31/1/015005} {\bibfield
   {journal} {\bibinfo  {journal} {Class. Quant. Grav.}\ }\textbf {\bibinfo
  {volume} {31}},\ \bibinfo {pages} {015005} (\bibinfo {year} {2014})},\
  \Eprint {https://arxiv.org/abs/1304.5544} {arXiv:1304.5544 [gr-qc]}
  \BibitemShut {NoStop}%
\bibitem [{\citenamefont {Kurganov}\ \emph {et~al.}(2001)\citenamefont
  {Kurganov}, \citenamefont {Noelle},\ and\ \citenamefont {Petrova}}]{KTO2-1}%
  \BibitemOpen
  \bibfield  {author} {\bibinfo {author} {\bibfnamefont {A.}~\bibnamefont
  {Kurganov}}, \bibinfo {author} {\bibfnamefont {S.}~\bibnamefont {Noelle}},\
  and\ \bibinfo {author} {\bibfnamefont {G.}~\bibnamefont {Petrova}},\
  }\bibfield  {title} {\bibinfo {title} {Semidiscrete central-upwind schemes
  for hyperbolic conservation laws and hamilton--jacobi equations},\ }\href
  {https://doi.org/10.1137/S1064827500373413} {\bibfield  {journal} {\bibinfo
  {journal} {SIAM J. Sci. Comput.}\ }\textbf {\bibinfo {volume} {23}},\
  \bibinfo {pages} {707–740} (\bibinfo {year} {2001})}\BibitemShut {NoStop}%
\bibitem [{\citenamefont {Liu}\ and\ \citenamefont
  {Tadmor}(2000)}]{LiuTadmore2000}%
  \BibitemOpen
  \bibfield  {author} {\bibinfo {author} {\bibfnamefont {X.-d.}\ \bibnamefont
  {Liu}}\ and\ \bibinfo {author} {\bibfnamefont {E.}~\bibnamefont {Tadmor}},\
  }\bibfield  {title} {\bibinfo {title} {{Third Order Nonoscillatory Central
  Scheme For Hyperbolic Conservation Laws}},\ }\href
  {https://doi.org/10.1007/s002110050345} {\bibfield  {journal} {\bibinfo
  {journal} {Numerische Mathematik}\ }\textbf {\bibinfo {volume} {79}}
  (\bibinfo {year} {2000})}\BibitemShut {NoStop}%
\bibitem [{\citenamefont {Jiang}\ and\ \citenamefont
  {Tadmor}(1997)}]{Jiang97non-oscillatorycentral}%
  \BibitemOpen
  \bibfield  {author} {\bibinfo {author} {\bibfnamefont {G.-s.}\ \bibnamefont
  {Jiang}}\ and\ \bibinfo {author} {\bibfnamefont {E.}~\bibnamefont {Tadmor}},\
  }\bibfield  {title} {\bibinfo {title} {{Non-oscillatory Central Schemes for
  Multidimensional Hyperbolic Conservation Laws}},\ }\href
  {https://doi.org/10.1137/S106482759631041X} {\bibfield  {journal} {\bibinfo
  {journal} {SIAM J. Sci. Comput}\ }\textbf {\bibinfo {volume} {19}},\ \bibinfo
  {pages} {1892} (\bibinfo {year} {1997})}\BibitemShut {NoStop}%
\bibitem [{\citenamefont {Chertock}\ \emph {et~al.}(2005)\citenamefont
  {Chertock}, \citenamefont {Kurganov},\ and\ \citenamefont
  {Rosenau}}]{Chertock2005}%
  \BibitemOpen
  \bibfield  {author} {\bibinfo {author} {\bibfnamefont {A.}~\bibnamefont
  {Chertock}}, \bibinfo {author} {\bibfnamefont {A.}~\bibnamefont {Kurganov}},\
  and\ \bibinfo {author} {\bibfnamefont {P.}~\bibnamefont {Rosenau}},\
  }\bibfield  {title} {\bibinfo {title} {{On degenerate saturated-diffusion
  equations with convection}},\ }\href
  {https://doi.org/10.1088/0951-7715/18/2/009} {\bibfield  {journal} {\bibinfo
  {journal} {Nonlinearity}\ }\textbf {\bibinfo {volume} {18}},\ \bibinfo
  {pages} {609} (\bibinfo {year} {2005})}\BibitemShut {NoStop}%
\bibitem [{\citenamefont {Jackson}(1998)}]{Jackson:1998nia}%
  \BibitemOpen
  \bibfield  {author} {\bibinfo {author} {\bibfnamefont {J.~D.}\ \bibnamefont
  {Jackson}},\ }\href
  {https://www.wiley.com/en-us/Classical+Electrodynamics%2C+3rd+Edition-p-9780471309321}
  {\emph {\bibinfo {title} {{Classical electrodynamics}}}},\ \bibinfo {edition}
  {3rd}\ ed.\ (\bibinfo  {publisher} {Wiley},\ \bibinfo {year}
  {1998})\BibitemShut {NoStop}%
\bibitem [{\citenamefont {Buckley}\ and\ \citenamefont
  {Leverett}(1942)}]{BuckleyLeverett:1942}%
  \BibitemOpen
  \bibfield  {author} {\bibinfo {author} {\bibfnamefont {S.}~\bibnamefont
  {Buckley}}\ and\ \bibinfo {author} {\bibfnamefont {M.~C.}\ \bibnamefont
  {Leverett}},\ }\bibfield  {title} {\bibinfo {title} {Mechanism of fluid
  displacement in sands},\ }\href {https://doi.org/10.2118/942107-G} {\bibfield
   {journal} {\bibinfo  {journal} {Trans.}\ }\textbf {\bibinfo {volume}
  {146}},\ \bibinfo {pages} {107 } (\bibinfo {year} {1942})}\BibitemShut
  {NoStop}%
\bibitem [{\citenamefont {Rajaraman}(1987)}]{Rajaraman:1982is}%
  \BibitemOpen
  \bibfield  {author} {\bibinfo {author} {\bibfnamefont {R.}~\bibnamefont
  {Rajaraman}},\ }\href
  {https://www.elsevier.com/books/solitons-and-instantons/rajaraman/978-0-444-87047-6}
  {\emph {\bibinfo {title} {{Solitons and instantons. An introduction to
  solitons and instantons in quantum field theory}}}}\ (\bibinfo  {publisher}
  {North-Holland},\ \bibinfo {address} {Amsterdam},\ \bibinfo {year}
  {1987})\BibitemShut {NoStop}%
\bibitem [{\citenamefont {Shifman}(1994)}]{Shifman:1994ee}%
  \BibitemOpen
  \bibinfo {editor} {\bibfnamefont {M.~A.}\ \bibnamefont {Shifman}},\ ed.,\
  \href@noop {} {\emph {\bibinfo {title} {{Instantons in gauge theories}}}}\
  (\bibinfo  {publisher} {World Scientific Publishing Co. Pt. Ltd.},\ \bibinfo
  {year} {1994})\BibitemShut {NoStop}%
\bibitem [{\citenamefont {Weinberg}(1972)}]{Weinberg:1972kfs}%
  \BibitemOpen
  \bibfield  {author} {\bibinfo {author} {\bibfnamefont {S.}~\bibnamefont
  {Weinberg}},\ }\href
  {https://www.wiley.com/en-us/Gravitation+and+Cosmology%3A+Principles+and+Applications+of+the+General+Theory+of+Relativity-p-9780471925675}
  {\emph {\bibinfo {title} {{Gravitation and Cosmology}: {Principles and
  Applications of the General Theory of Relativity}}}}\ (\bibinfo  {publisher}
  {John Wiley and Sons},\ \bibinfo {address} {New York},\ \bibinfo {year}
  {1972})\BibitemShut {NoStop}%
\bibitem [{\citenamefont {Misner}\ \emph {et~al.}(1973)\citenamefont {Misner},
  \citenamefont {Thorne},\ and\ \citenamefont {Wheeler}}]{Misner:1974qy}%
  \BibitemOpen
  \bibfield  {author} {\bibinfo {author} {\bibfnamefont {C.~W.}\ \bibnamefont
  {Misner}}, \bibinfo {author} {\bibfnamefont {K.~S.}\ \bibnamefont {Thorne}},\
  and\ \bibinfo {author} {\bibfnamefont {J.~A.}\ \bibnamefont {Wheeler}},\
  }\href
  {https://press.princeton.edu/books/hardcover/9780691177793/gravitation}
  {\emph {\bibinfo {title} {{Gravitation}}}}\ (\bibinfo  {publisher} {W. H.
  Freeman},\ \bibinfo {address} {San Francisco},\ \bibinfo {year}
  {1973})\BibitemShut {NoStop}%
\bibitem [{\citenamefont {Ryder}(2009)}]{Ryder:2009zz}%
  \BibitemOpen
  \bibfield  {author} {\bibinfo {author} {\bibfnamefont {L.}~\bibnamefont
  {Ryder}},\ }\href {https://doi.org/10.1017/CBO9780511809033} {\emph {\bibinfo
  {title} {{Introduction to general relativity}}}}\ (\bibinfo  {publisher}
  {Cambridge University Press},\ \bibinfo {year} {2009})\BibitemShut {NoStop}%
\bibitem [{\citenamefont {Baumgarte}\ and\ \citenamefont
  {Shapiro}(2010)}]{Baumgarte2010Jun}%
  \BibitemOpen
  \bibfield  {author} {\bibinfo {author} {\bibfnamefont {T.~W.}\ \bibnamefont
  {Baumgarte}}\ and\ \bibinfo {author} {\bibfnamefont {S.~L.}\ \bibnamefont
  {Shapiro}},\ }\href@noop {} {\emph {\bibinfo {title} {{Numerical
  Relativity}}}}\ (\bibinfo  {publisher} {Cambridge University Press},\
  \bibinfo {address} {Cambridge, England, UK},\ \bibinfo {year}
  {2010})\BibitemShut {NoStop}%
\bibitem [{\citenamefont {Alcubierre}(2008)}]{Alcubierre2008}%
  \BibitemOpen
  \bibfield  {author} {\bibinfo {author} {\bibfnamefont {M.}~\bibnamefont
  {Alcubierre}},\ }\href
  {https://doi.org/10.1093/acprof:oso/9780199205677.001.0001} {\emph {\bibinfo
  {title} {{Introduction to 3+1 Numerical Relativity}}}}\ (\bibinfo
  {publisher} {Oxford University Press},\ \bibinfo {address} {Oxford, England,
  UK},\ \bibinfo {year} {2008})\BibitemShut {NoStop}%
\bibitem [{\citenamefont {Grandclement}\ and\ \citenamefont
  {Novak}(2009)}]{Grandclement:2007sb}%
  \BibitemOpen
  \bibfield  {author} {\bibinfo {author} {\bibfnamefont {P.}~\bibnamefont
  {Grandclement}}\ and\ \bibinfo {author} {\bibfnamefont {J.}~\bibnamefont
  {Novak}},\ }\bibfield  {title} {\bibinfo {title} {{Spectral methods for
  numerical relativity}},\ }\href {https://doi.org/10.12942/lrr-2009-1}
  {\bibfield  {journal} {\bibinfo  {journal} {Living Rev. Rel.}\ }\textbf
  {\bibinfo {volume} {12}},\ \bibinfo {pages} {1} (\bibinfo {year} {2009})},\
  \Eprint {https://arxiv.org/abs/0706.2286} {arXiv:0706.2286 [gr-qc]}
  \BibitemShut {NoStop}%
\bibitem [{\citenamefont {Gourgoulhon}(2007)}]{Gourgoulhon:2007ue}%
  \BibitemOpen
  \bibfield  {author} {\bibinfo {author} {\bibfnamefont {E.}~\bibnamefont
  {Gourgoulhon}},\ }\href@noop {} {\bibinfo {title} {{3+1 formalism and bases
  of numerical relativity}}} (\bibinfo {year} {2007}),\ \Eprint
  {https://arxiv.org/abs/gr-qc/0703035} {arXiv:gr-qc/0703035} \BibitemShut
  {NoStop}%
\bibitem [{\citenamefont {Chen}(2001)}]{Chen:2001}%
  \BibitemOpen
  \bibfield  {author} {\bibinfo {author} {\bibfnamefont {G.-Q.~G.}\
  \bibnamefont {Chen}},\ }\bibfield  {title} {\bibinfo {title} {{Shock
  capturing and related numerical methods in computational fluid dynamics}},\
  }\href {https://www.emis.de/journals/AMUC/_vol-70/_no_1/_chen/chen.html}
  {\bibfield  {journal} {\bibinfo  {journal} {Acta Math. Univ. Comenianae}\
  }\textbf {\bibinfo {volume} {1}},\ \bibinfo {pages} {51} (\bibinfo {year}
  {2001})}\BibitemShut {NoStop}%
\bibitem [{\citenamefont {Pel\'aez}\ and\ \citenamefont
  {Wschebor}(2016)}]{Pelaez:2015nsa}%
  \BibitemOpen
  \bibfield  {author} {\bibinfo {author} {\bibfnamefont {M.}~\bibnamefont
  {Pel\'aez}}\ and\ \bibinfo {author} {\bibfnamefont {N.}~\bibnamefont
  {Wschebor}},\ }\bibfield  {title} {\bibinfo {title} {{Ordered phase of the
  $O(N)$ model within the nonperturbative renormalization group}},\ }\href
  {https://doi.org/10.1103/PhysRevE.94.042136} {\bibfield  {journal} {\bibinfo
  {journal} {Phys. Rev. E}\ }\textbf {\bibinfo {volume} {94}},\ \bibinfo
  {pages} {042136} (\bibinfo {year} {2016})},\ \Eprint
  {https://arxiv.org/abs/1510.05709} {arXiv:1510.05709 [cond-mat.stat-mech]}
  \BibitemShut {NoStop}%
\bibitem [{\citenamefont {Schaefer}\ \emph {et~al.}(2002)\citenamefont
  {Schaefer}, \citenamefont {Bohr},\ and\ \citenamefont
  {Wambach}}]{Schaefer:2001cn}%
  \BibitemOpen
  \bibfield  {author} {\bibinfo {author} {\bibfnamefont {B.-J.}\ \bibnamefont
  {Schaefer}}, \bibinfo {author} {\bibfnamefont {O.}~\bibnamefont {Bohr}},\
  and\ \bibinfo {author} {\bibfnamefont {J.}~\bibnamefont {Wambach}},\
  }\bibfield  {title} {\bibinfo {title} {{Finite temperature gluon condensate
  with renormalization group flow equations}},\ }\href
  {https://doi.org/10.1103/PhysRevD.65.105008} {\bibfield  {journal} {\bibinfo
  {journal} {Phys. Rev. D}\ }\textbf {\bibinfo {volume} {65}},\ \bibinfo
  {pages} {105008} (\bibinfo {year} {2002})},\ \Eprint
  {https://arxiv.org/abs/hep-th/0112087} {arXiv:hep-th/0112087} \BibitemShut
  {NoStop}%
\bibitem [{\citenamefont {Litim}(2002)}]{Litim:2002cf}%
  \BibitemOpen
  \bibfield  {author} {\bibinfo {author} {\bibfnamefont {D.~F.}\ \bibnamefont
  {Litim}},\ }\bibfield  {title} {\bibinfo {title} {{Critical exponents from
  optimized renormalization group flows}},\ }\href
  {https://doi.org/10.1016/S0550-3213(02)00186-4} {\bibfield  {journal}
  {\bibinfo  {journal} {Nucl. Phys. B}\ }\textbf {\bibinfo {volume} {631}},\
  \bibinfo {pages} {128} (\bibinfo {year} {2002})},\ \Eprint
  {https://arxiv.org/abs/hep-th/0203006} {arXiv:hep-th/0203006} \BibitemShut
  {NoStop}%
\bibitem [{\citenamefont {Benitez}\ \emph {et~al.}(2009)\citenamefont
  {Benitez}, \citenamefont {Blaizot}, \citenamefont {Chate}, \citenamefont
  {Delamotte}, \citenamefont {Mendez-Galain},\ and\ \citenamefont
  {Wschebor}}]{Benitez:2009xg}%
  \BibitemOpen
  \bibfield  {author} {\bibinfo {author} {\bibfnamefont {F.}~\bibnamefont
  {Benitez}}, \bibinfo {author} {\bibfnamefont {J.-P.}\ \bibnamefont
  {Blaizot}}, \bibinfo {author} {\bibfnamefont {H.}~\bibnamefont {Chate}},
  \bibinfo {author} {\bibfnamefont {B.}~\bibnamefont {Delamotte}}, \bibinfo
  {author} {\bibfnamefont {R.}~\bibnamefont {Mendez-Galain}},\ and\ \bibinfo
  {author} {\bibfnamefont {N.}~\bibnamefont {Wschebor}},\ }\bibfield  {title}
  {\bibinfo {title} {{Solutions of renormalization group flow equations with
  full momentum dependence}},\ }\href
  {https://doi.org/10.1103/PhysRevE.80.030103} {\bibfield  {journal} {\bibinfo
  {journal} {Phys. Rev. E}\ }\textbf {\bibinfo {volume} {80}},\ \bibinfo
  {pages} {030103} (\bibinfo {year} {2009})},\ \Eprint
  {https://arxiv.org/abs/0901.0128} {arXiv:0901.0128 [cond-mat.stat-mech]}
  \BibitemShut {NoStop}%
\bibitem [{\citenamefont {Benitez}\ \emph {et~al.}(2012)\citenamefont
  {Benitez}, \citenamefont {Blaizot}, \citenamefont {Chate}, \citenamefont
  {Delamotte}, \citenamefont {Mendez-Galain},\ and\ \citenamefont
  {Wschebor}}]{Benitez:2011xx}%
  \BibitemOpen
  \bibfield  {author} {\bibinfo {author} {\bibfnamefont {F.}~\bibnamefont
  {Benitez}}, \bibinfo {author} {\bibfnamefont {J.-P.}\ \bibnamefont
  {Blaizot}}, \bibinfo {author} {\bibfnamefont {H.}~\bibnamefont {Chate}},
  \bibinfo {author} {\bibfnamefont {B.}~\bibnamefont {Delamotte}}, \bibinfo
  {author} {\bibfnamefont {R.}~\bibnamefont {Mendez-Galain}},\ and\ \bibinfo
  {author} {\bibfnamefont {N.}~\bibnamefont {Wschebor}},\ }\bibfield  {title}
  {\bibinfo {title} {{Non-perturbative renormalization group preserving
  full-momentum dependence: implementation and quantitative evaluation}},\
  }\href {https://doi.org/10.1103/PhysRevE.85.026707} {\bibfield  {journal}
  {\bibinfo  {journal} {Phys. Rev. E}\ }\textbf {\bibinfo {volume} {85}},\
  \bibinfo {pages} {026707} (\bibinfo {year} {2012})},\ \Eprint
  {https://arxiv.org/abs/1110.2665} {arXiv:1110.2665 [cond-mat.stat-mech]}
  \BibitemShut {NoStop}%
\bibitem [{\citenamefont {Mitter}\ \emph {et~al.}(2015)\citenamefont {Mitter},
  \citenamefont {Pawlowski},\ and\ \citenamefont
  {Strodthoff}}]{Mitter:2014wpa}%
  \BibitemOpen
  \bibfield  {author} {\bibinfo {author} {\bibfnamefont {M.}~\bibnamefont
  {Mitter}}, \bibinfo {author} {\bibfnamefont {J.~M.}\ \bibnamefont
  {Pawlowski}},\ and\ \bibinfo {author} {\bibfnamefont {N.}~\bibnamefont
  {Strodthoff}},\ }\bibfield  {title} {\bibinfo {title} {{Chiral symmetry
  breaking in continuum QCD}},\ }\href
  {https://doi.org/10.1103/PhysRevD.91.054035} {\bibfield  {journal} {\bibinfo
  {journal} {Phys. Rev. D}\ }\textbf {\bibinfo {volume} {91}},\ \bibinfo
  {pages} {054035} (\bibinfo {year} {2015})},\ \Eprint
  {https://arxiv.org/abs/1411.7978} {arXiv:1411.7978 [hep-ph]} \BibitemShut
  {NoStop}%
\bibitem [{\citenamefont {Cyrol}\ \emph
  {et~al.}(2018{\natexlab{b}})\citenamefont {Cyrol}, \citenamefont {Mitter},
  \citenamefont {Pawlowski},\ and\ \citenamefont {Strodthoff}}]{Cyrol:2017qkl}%
  \BibitemOpen
  \bibfield  {author} {\bibinfo {author} {\bibfnamefont {A.~K.}\ \bibnamefont
  {Cyrol}}, \bibinfo {author} {\bibfnamefont {M.}~\bibnamefont {Mitter}},
  \bibinfo {author} {\bibfnamefont {J.~M.}\ \bibnamefont {Pawlowski}},\ and\
  \bibinfo {author} {\bibfnamefont {N.}~\bibnamefont {Strodthoff}},\ }\bibfield
   {title} {\bibinfo {title} {{Nonperturbative finite-temperature Yang-Mills
  theory}},\ }\href {https://doi.org/10.1103/PhysRevD.97.054015} {\bibfield
  {journal} {\bibinfo  {journal} {Phys. Rev. D}\ }\textbf {\bibinfo {volume}
  {97}},\ \bibinfo {pages} {054015} (\bibinfo {year} {2018}{\natexlab{b}})},\
  \Eprint {https://arxiv.org/abs/1708.03482} {arXiv:1708.03482 [hep-ph]}
  \BibitemShut {NoStop}%
\bibitem [{\citenamefont {Wilbraham}(1848)}]{Wilbraham:1848}%
  \BibitemOpen
  \bibfield  {author} {\bibinfo {author} {\bibfnamefont {H.}~\bibnamefont
  {Wilbraham}},\ }\bibfield  {title} {\bibinfo {title} {{On a certain periodic
  function}},\ }\href
  {https://books.google.de/books?id=jrq4aaaamaaj&pg=pa198&redir_esc=y#v=onepage&q&f=false}
  {\bibfield  {journal} {\bibinfo  {journal} {Cambridge and Dublin Mathematical
  Journal}\ }\textbf {\bibinfo {volume} {3}},\ \bibinfo {pages} {198} (\bibinfo
  {year} {1848})}\BibitemShut {NoStop}%
\bibitem [{\citenamefont {Gibbs}(1898)}]{Gibbs:1898}%
  \BibitemOpen
  \bibfield  {author} {\bibinfo {author} {\bibfnamefont {J.~W.}\ \bibnamefont
  {Gibbs}},\ }\bibfield  {title} {\bibinfo {title} {{Fourier's series}},\
  }\href {https://doi.org/10.1038/059200b0} {\bibfield  {journal} {\bibinfo
  {journal} {Nature}\ }\textbf {\bibinfo {volume} {59}},\ \bibinfo {pages}
  {200} (\bibinfo {year} {1898})}\BibitemShut {NoStop}%
\bibitem [{\citenamefont {Gibbs}(1899)}]{Gibbs:1899}%
  \BibitemOpen
  \bibfield  {author} {\bibinfo {author} {\bibfnamefont {J.~W.}\ \bibnamefont
  {Gibbs}},\ }\bibfield  {title} {\bibinfo {title} {{Fourier's series}},\
  }\href {https://doi.org/10.1038/059606a0} {\bibfield  {journal} {\bibinfo
  {journal} {Nature}\ }\textbf {\bibinfo {volume} {59}},\ \bibinfo {pages}
  {606} (\bibinfo {year} {1899})}\BibitemShut {NoStop}%
\bibitem [{\citenamefont {Boyd}(2001)}]{boyd2001chebyshev}%
  \BibitemOpen
  \bibfield  {author} {\bibinfo {author} {\bibfnamefont {J.~P.}\ \bibnamefont
  {Boyd}},\ }\href {https://depts.washington.edu/ph506/Boyd.pdf} {\emph
  {\bibinfo {title} {{Chebyshev and Fourier spectral methods}}}},\ \bibinfo
  {edition} {2nd}\ ed.\ (\bibinfo  {publisher} {Dover Publications},\ \bibinfo
  {address} {Mineola, New York},\ \bibinfo {year} {2001})\ \bibinfo {note} {1st
  ed. in 1969 by Springer-Verlag, Berlin Heidelberg, in the series
  \textit{Lecture Notes in Engeneering, Vol. 49}}\BibitemShut {NoStop}%
\bibitem [{\citenamefont {Lang}(2020)}]{Lang:plotgrid}%
  \BibitemOpen
  \bibfield  {author} {\bibinfo {author} {\bibfnamefont {L.}~\bibnamefont
  {Lang}},\ }\href
  {https://resources.wolframcloud.com/FunctionRepository/resources/PlotGrid}
  {\bibinfo {title} {{Wolfram Mathematica function repository: ResourceFunction
  PlotGrid}}} (\bibinfo {year} {2020}),\ \bibinfo {note} {[Online; accessed
  2020.10.08]}\BibitemShut {NoStop}%
\bibitem [{\citenamefont {Popkov}(2020)}]{Popkov:polygonmarker}%
  \BibitemOpen
  \bibfield  {author} {\bibinfo {author} {\bibfnamefont {A.}~\bibnamefont
  {Popkov}},\ }\href
  {https://resources.wolframcloud.com/FunctionRepository/resources/PolygonMarker}
  {\bibinfo {title} {{Wolfram Mathematica function repository: ResourceFunction
  PolygonMarker}}} (\bibinfo {year} {2020}),\ \bibinfo {note} {[Online;
  accessed 2020.10.08]}\BibitemShut {NoStop}%
\bibitem [{\citenamefont {Horvát}(2020)}]{Horvat:matex}%
  \BibitemOpen
  \bibfield  {author} {\bibinfo {author} {\bibfnamefont {S.}~\bibnamefont
  {Horvát}},\ }\href
  {http://szhorvat.net/pelican/latex-typesetting-in-mathematica.html} {\bibinfo
  {title} {{MaTeX}}} (\bibinfo {year} {2020}),\ \bibinfo {note} {[Online;
  accessed 2020.10.08]}\BibitemShut {NoStop}%
\bibitem [{\citenamefont {Collins}\ and\ \citenamefont
  {Vermaseren}(2016)}]{Collins:2016aya}%
  \BibitemOpen
  \bibfield  {author} {\bibinfo {author} {\bibfnamefont {J.~C.}\ \bibnamefont
  {Collins}}\ and\ \bibinfo {author} {\bibfnamefont {J.~A.}\ \bibnamefont
  {Vermaseren}},\ }\href@noop {} {\bibinfo {title} {{Axodraw version 2}}}
  (\bibinfo {year} {2016}),\ \Eprint {https://arxiv.org/abs/1606.01177}
  {arXiv:1606.01177 [cs.OH]} \BibitemShut {NoStop}%
\bibitem [{\citenamefont {Fornberg}(1988)}]{Fornberg1988}%
  \BibitemOpen
  \bibfield  {author} {\bibinfo {author} {\bibfnamefont {B.}~\bibnamefont
  {Fornberg}},\ }\bibfield  {title} {\bibinfo {title} {{Generation of finite
  difference formulas on arbitrarily spaced grids}},\ }\href
  {https://doi.org/10.1090/S0025-5718-1988-0935077-0} {\bibfield  {journal}
  {\bibinfo  {journal} {Math. Comput.}\ }\textbf {\bibinfo {volume} {51}},\
  \bibinfo {pages} {699} (\bibinfo {year} {1988})}\BibitemShut {NoStop}%
\bibitem [{\citenamefont {Abramowitz}\ and\ \citenamefont
  {Stegun}(1964)}]{abramowitz+stegun}%
  \BibitemOpen
  \bibfield  {author} {\bibinfo {author} {\bibfnamefont {M.}~\bibnamefont
  {Abramowitz}}\ and\ \bibinfo {author} {\bibfnamefont {I.~A.}\ \bibnamefont
  {Stegun}},\ }\href@noop {} {\emph {\bibinfo {title} {{Handbook of
  Mathematical Functions with Formulas, Graphs, and Mathematical Tables}}}},\
  \bibinfo {edition} {ninth dover printing, tenth gpo printing}\ ed.\ (\bibinfo
   {publisher} {Dover Publications Inc.},\ \bibinfo {address} {Mineola, NY,
  USA},\ \bibinfo {year} {1964})\BibitemShut {NoStop}%
\bibitem [{\citenamefont {Greiner}\ \emph {et~al.}(1995)\citenamefont
  {Greiner}, \citenamefont {Neise},\ and\ \citenamefont
  {Stöcker}}]{GreinerNeiseStoecker}%
  \BibitemOpen
  \bibfield  {author} {\bibinfo {author} {\bibfnamefont {W.}~\bibnamefont
  {Greiner}}, \bibinfo {author} {\bibfnamefont {L.}~\bibnamefont {Neise}},\
  and\ \bibinfo {author} {\bibfnamefont {H.}~\bibnamefont {Stöcker}},\ }\href
  {https://doi.org/10.1007/978-1-4612-0827-3} {\emph {\bibinfo {title}
  {{Thermodynamics and Statistical Mechanics}}}}\ (\bibinfo  {publisher}
  {Springer, New York},\ \bibinfo {year} {1995})\BibitemShut {NoStop}%
\bibitem [{\citenamefont {Niehof}(2021)}]{Niehof:2021}%
  \BibitemOpen
  \bibfield  {author} {\bibinfo {author} {\bibfnamefont {C.}~\bibnamefont
  {Niehof}},\ }\href@noop {} {\bibinfo {title} {{private communication}}}
  (\bibinfo {year} {2021})\BibitemShut {NoStop}%
\bibitem [{\citenamefont {Landau}(1937)}]{Landau:1937obd}%
  \BibitemOpen
  \bibfield  {author} {\bibinfo {author} {\bibfnamefont {L.~D.}\ \bibnamefont
  {Landau}},\ }\bibfield  {title} {\bibinfo {title} {{On the theory of phase
  transitions}},\ }\href
  {http://archive.ujp.bitp.kiev.ua/files/journals/53/si/53SI08p.pdf} {\bibfield
   {journal} {\bibinfo  {journal} {Zh. Eksp. Teor. Fiz.}\ }\textbf {\bibinfo
  {volume} {7}},\ \bibinfo {pages} {19} (\bibinfo {year} {1937})}\BibitemShut
  {NoStop}%
\bibitem [{\citenamefont {Braaten}\ and\ \citenamefont
  {Nieto}(1995)}]{Braaten:1995cm}%
  \BibitemOpen
  \bibfield  {author} {\bibinfo {author} {\bibfnamefont {E.}~\bibnamefont
  {Braaten}}\ and\ \bibinfo {author} {\bibfnamefont {A.}~\bibnamefont
  {Nieto}},\ }\bibfield  {title} {\bibinfo {title} {{Effective field theory
  approach to high temperature thermodynamics}},\ }\href
  {https://doi.org/10.1103/PhysRevD.51.6990} {\bibfield  {journal} {\bibinfo
  {journal} {Phys. Rev. D}\ }\textbf {\bibinfo {volume} {51}},\ \bibinfo
  {pages} {6990} (\bibinfo {year} {1995})},\ \Eprint
  {https://arxiv.org/abs/hep-ph/9501375} {arXiv:hep-ph/9501375} \BibitemShut
  {NoStop}%
\bibitem [{\citenamefont {Fister}\ and\ \citenamefont
  {Pawlowski}(2015)}]{Fister:2015eca}%
  \BibitemOpen
  \bibfield  {author} {\bibinfo {author} {\bibfnamefont {L.}~\bibnamefont
  {Fister}}\ and\ \bibinfo {author} {\bibfnamefont {J.~M.}\ \bibnamefont
  {Pawlowski}},\ }\bibfield  {title} {\bibinfo {title} {{Functional
  renormalization group in a finite volume}},\ }\href
  {https://doi.org/10.1103/PhysRevD.92.076009} {\bibfield  {journal} {\bibinfo
  {journal} {Phys. Rev. D}\ }\textbf {\bibinfo {volume} {92}},\ \bibinfo
  {pages} {076009} (\bibinfo {year} {2015})},\ \Eprint
  {https://arxiv.org/abs/1504.05166} {arXiv:1504.05166 [hep-ph]} \BibitemShut
  {NoStop}%
\end{thebibliography}%

\end{document}